\documentclass[12pt,twoside,a4paper]{ereport}

\usepackage{epsfig}
\usepackage{graphicx}
\usepackage{subfigure}
\usepackage[round]{natbib}
\usepackage{calc}
\usepackage{amsfonts}
\usepackage{amssymb}
\usepackage{amsmath}
\usepackage{color}
\usepackage{multirow}
\usepackage[geometry]{ifsym}
\usepackage{MnSymbol}
\usepackage{booktabs}
\usepackage{fullpage}
\usepackage{setspace} 
\usepackage[intoc,noprefix]{nomencl}
\usepackage[dvipsnames]{xcolor}
\usepackage{moreverb}
\usepackage{pifont}
\usepackage{lpic}
\usepackage{rotating}
\usepackage{xfrac}
\usepackage{fancyhdr}
\usepackage{pstricks}
\usepackage{enumerate} 

\RequirePackage{ifthen}
\renewcommand{\nomgroup}[1]{%
 \ifthenelse{\equal{#1}{R}}{\item[\textbf{Roman symbols}]}{%
  \ifthenelse{\equal{#1}{G}}{\item[\textbf{Greek symbols}]}{%
   \ifthenelse{\equal{#1}{A}}{\item[\textbf{Acronyms}]}{}}}}
\newcommand{\nomunit}[1]{%
\renewcommand{\nomentryend}{\hspace*{\fill}#1}}
\newlength{\uspc}
\fancypagestyle{plain}{%
	\fancyhead{}%
}

\graphicspath{{frontmatter/}{chapters/Figures/}}

\newcommand{\clearemptydoublepage}
           {\newpage\thispagestyle{empty}\cleardoublepage}

\def\tm{\leavevmode\hbox{$\rm {}^{TM}$}}

\begin{document}
\makenomenclature

% Title page etc ...
\begin{titlepage}
\begin{center}
\vspace{2cm}

\Huge{\bf Energy transfer and dissipation in equilibrium and nonequilibrium turbulence
\vspace{1cm}

\Large{by}\\
\vspace{0.5cm}

\LARGE{Pedro Manuel \\ da Silva Cardoso Isidro Valente}\\
\vspace{3.5cm}

\begin{figure}[h] 
	%\centering
    \setlength{\unitlength}{1cm}
    \begin{picture}(13,3.)
    \put(5.5,0.0){
    \includegraphics[bb=13 280 577 832,clip,width=50mm]{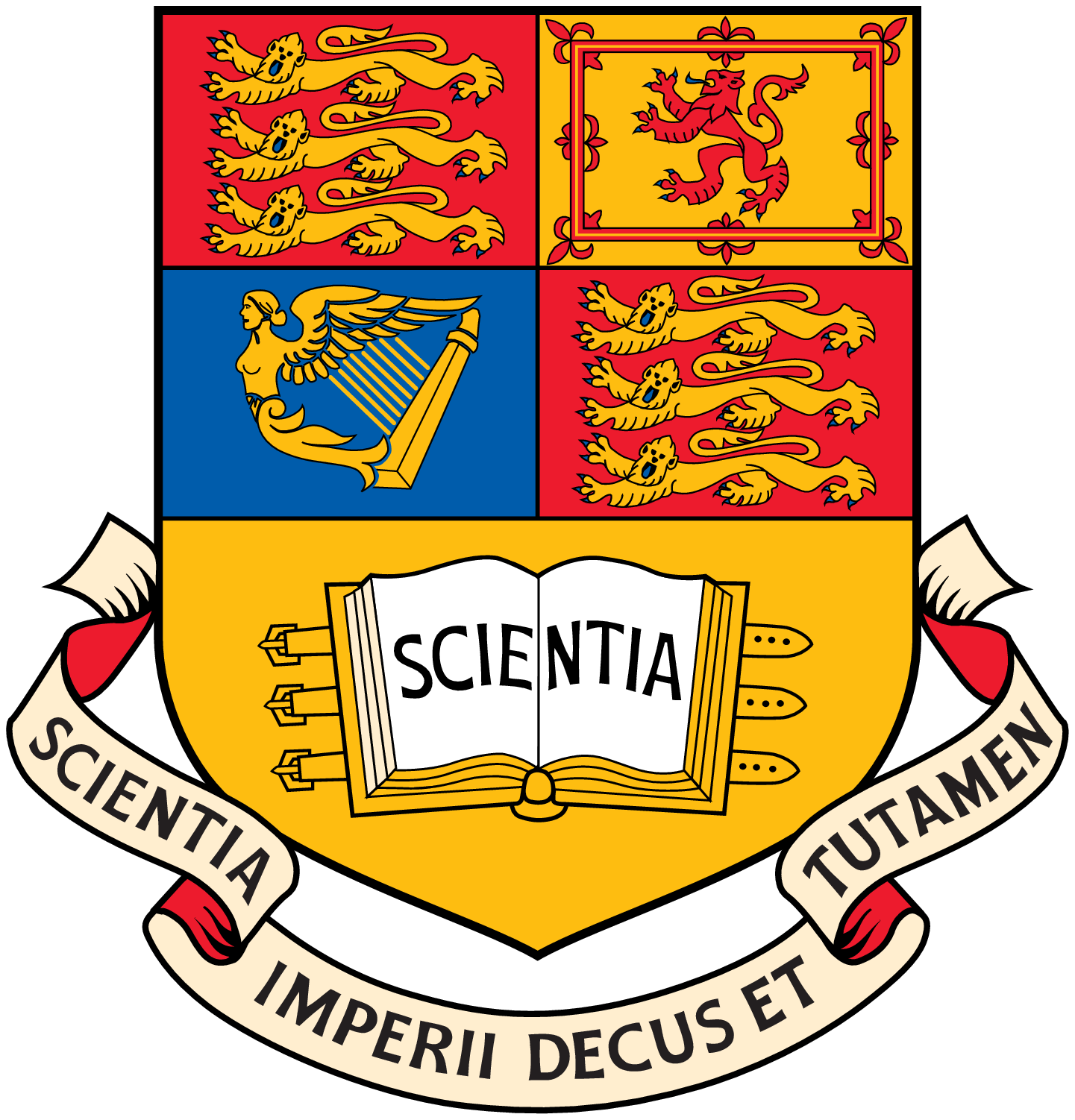}
                }
    \end{picture}
\end{figure}
\vspace{1.2cm}

\large{Department of Aeronautics\\
Imperial College of Science, Technology and Medicine\\
Prince Consort Road\\
London SW7 2BY\\}
\vspace{1.5cm}

\normalsize{This thesis is submitted for the degree of Doctor of Philosophy \\
of the Imperial College of Science, Technology and Medicine\\}
}
\vspace{0.2cm}
\large{2013}\\
\end{center}
\end{titlepage}

\addcontentsline{toc}{chapter}{Abstract}
\setcounter{page}{2}
\chapter*{{}}

\vspace{15mm}

I hereby declare that the work presented in this thesis is my own and  that contributions from other authors are appropriately acknowledged and referenced.
%and that contributions from other authors are duly acknowledged

\vspace{130mm}
\noindent \emph{The copyright of this thesis rests with the author and is made available under a Creative Commons Attribution Non-Commercial No Derivatives licence. Researchers are free to copy, distribute or transmit the thesis on the condition that they attribute it, that they do not use it for commercial purposes and that they do not alter, transform or build upon it. For any reuse or redistribution, researchers must make clear to others the licence terms of this work}
\chapter*{Abstract}

%This is it!!
The nonequilibrium dissipation behaviour discovered for decaying fractal square grid-generated turbulence is experimentally investigated using hot-wire anemometry in a wind tunnel.  
The previous results are consolidated and  benchmarked with turbulence generated by  regular square-mesh grids, designed to retain certain geometrical parameters of the fractal square grid.
This comparison shows that the nonequilibrium behaviour is manifested in both fractal square grid- and regular square-mesh grid-generated turbulence for a downstream region during the turbulence decay up to the first few multiples of the wake interaction distance.  For one of the regular grids it is shown that beyond this region there is a transition to the classical dissipation behaviour if the local turbulent Reynolds number is sufficiently high. 
A sharp conclusion can thus be drawn that this behaviour is more general than initially thought and therefore of much greater scientific and engineering significance. 

The nonequilibrium dissipation phenomena is further investigated by experimentally measuring the terms of an inhomogeneous von K\'arm\'an-Howarth-Monin equation. This equation is essentially a scale-by-scale energy transfer budget. From the data it is shown that the inhomogeneity of the turbulent flow does not tamper with the nonequilibrium phenomena and that the scaling of the nonlinear energy transfer, i.e. the transfer of energy to the small-scales, is out of balance with the dissipation. This imbalance leads to the growth of the small-scale advection to compensate for the increasing gap between the energy transferred and the energy dissipated.

For the highest Reynolds number data it is also shown that the nonequilibrium dissipation scaling appears to be consistent with the expectation that it is asymptotically independent of the viscosity (as the Reynolds number increases) and that the spectra exhibit a power-law range with the Kolmogorov-Obukhov exponent $-5/3$. These two observations are shown to be consistent. 
\chapter*{Preface}

Journal publications during the course of the work: \\

\begin{enumerate}[(i)]
\item P. C. Valente  \& J. C. Vassilicos, The decay of turbulence generated by a class of multi-scale grids, J. Fluid Mech., \textbf{687}, 300--340, 2011\\

\item P. C. Valente  \& J. C. Vassilicos, Comment on ``Dissipation and decay of fractal-generated turbulence'' [Phys. Fluids 19, 105108 (2007)], Phys. Fluids, \textbf{23}, 119101, 2011\\

\item P. C. Valente  \& J. C. Vassilicos, Dependence of decaying homogeneous isotropic turbulence on inflow conditions, Phys. Lett. A, \textbf{376}, 510--514, 2012\\

\item P. C. Valente  \& J. C. Vassilicos, Universal dissipation scaling for nonequilibrium turbulence, Phys. Rev. Lett. \textbf{108}, 214503, 2012 \\

\item A. R. Oxlade, P. C. Valente, B. Ganapathisubramani \& J. F. Morrison. Denoising of time-resolved PIV for accurate measurement of turbulence spectra and reduced error in derivatives. Exp. Fluids, \textbf{53}(5):1561Ð1575, 2012.\\

\item P. C. Valente  \& J. C. Vassilicos, The nonequilibrium region of grid-generated decaying turbulence. Part I: Flow characterisation (under preparation)\\

\item P. C. Valente  \& J. C. Vassilicos, The nonequilibrium region of grid-generated decaying turbulence. Part II: Scale-by-scale energy transfer budget (under preparation)\\
\end{enumerate}
\chapter*{Acknowledgements}

%I would like to take the Academy for this beautiful Oscar, and to the people at home that voted for me.
%I would like to say to all the people telling me I couldn't act, suck it!

First and foremost I am extremely grateful to Prof. Christos Vassilicos for being a truly outstanding supervisor. 
For sharing his valuable insight on turbulence physics. 
For always taking the time to discuss ideas and to comment on manuscripts. 
Overall, for his highly commendable attitude towards science. 
I am also very grateful to Prof. William K. George and Prof. Bharathram Ganapathisubramani for many scientific discussions and for providing valuable advice when I needed it. 

Very many thanks to my colleagues in the Aeronautics department. For numerous technical discussions. For the lunch, coffee and ``ciggy'' breaks when they were most needed. For making the time at Imperial College London so very worthwhile. A special thanks to Anthony Oxlade for the help with the performance testing of the anemometers and with the setup of the camera for the $2\times$XW apparatus and to Jovan Nedi\'{c} for the help with the refurbishment of the 3'x3' wind tunnel.

Furthermore, I would like to thank the Aeronautics workshop for their technical support. 
 To Andrew Wallace for his effort in re-building the entire test section of the 18''x18'' wind tunnel. To Alan Smith for the woodwork. To Ian Pardew and Mark Grant for machining the parts for the $2\times$XW apparatus in a record breaking time. A special thanks Roland Hutchins for coordinating the construction of the 18''x18'' test section and  the help in the manufacturing of the  $2\times$XW apparatus.

The help from Robert Jaryczewski (Dantec Dynamics UK) with the design and implementation of the anemometry performance tests and for his continued support is gratefully acknowledged. 

I would like to thank my wonderful family and friends for all their support and understanding. Many thanks to my good friend Pedro Viegas for the help  proofreading the thesis. 
I am particularly thankful to my dearest Luisa Pires with whom I shared frustrations and achievements on a daily basis and helped me make the most of each one.

Lastly, I am grateful for the financial support from ``Funda\c{c}\~{a}o para a Ci\^{e}ncia e a Tecnologia'' (grant SFRH/BD/61223/2009 -- cofinanced by POPH/FSE) without which this work would not have been possible.

%\include{frontmatter/dedication}
%\clearemptydoublepage
%\pagestyle{empty}
\nomenclature[AC]{CTA}{Constant-temperature anemometer}%
\nomenclature[ADAQ]{DAQ}{Data acquisition system}%
\nomenclature[APID]{PID}{Proportional-Integral-Derivative (feedback) controller}%
\nomenclature[AWS]{SW}{Single hot-wire}%
\nomenclature[AWX]{XW}{Cross-wire also denoted as X-probe}%
\nomenclature[AWZ]{$2\times$XW}{Arrangement of two parallel XW}%
\nomenclature[AFRN]{FRN}{Finite Reynolds number}%
\nomenclature[AFSG]{FSG}{Fractal square grid}%
\nomenclature[ARG]{RG}{Regular grid}%
\nomenclature[ARANS]{RANS}{Reynolds Averaged Navier Stokes}%
\nomenclature[ALES]{LES}{Large Eddy Simulation}%
\nomenclature[ADNS]{DNS}{Direct numerical simulation}%
\nomenclature[AARMS]{r.m.s.}{Root mean square}%

\nomenclature[Geps]{$\varepsilon$}{Turbulent kinetic energy dissipation rate per unit mass \nomunit{\makebox[\uspc][l]{$\mathrm{m}^2\,\mathrm{s}^{-3}$}}}%
\nomenclature[Glamb]{$\lambda$}{Taylor microscale defined as $\lambda \equiv \sqrt{5\nu \overline{q^2}/\varepsilon}$\nomunit{\makebox[\uspc][l]{$\mathrm{m}$}}}%
\nomenclature[Glamb2]{$\lambda^{\mathrm{iso}}$}{Taylor microscale defined as $\lambda^{\mathrm{iso}} \equiv \sqrt{15\nu \overline{u^2}/\varepsilon}$\nomunit{\makebox[\uspc][l]{$\mathrm{m}$}}}%
\nomenclature[Glamb3]{$\lambda_{\mathrm{mfp}}$}{Mean free path\nomunit{\makebox[\uspc][l]{$\mathrm{m}$}}}%
\nomenclature[Geta]{$\eta$}{Kolmogorov microscale\nomunit{\makebox[\uspc][l]{$\mathrm{m}$}}}%
\nomenclature[Gnu]{$\nu$}{Kinematic viscosity \nomunit{\makebox[\uspc][l]{$\mathrm{m}^2\,\mathrm{s}^{-1}$}}}%
\nomenclature[Gphi]{$\phi$}{Azimuthal angle  \nomunit{\makebox[\uspc][l]{rad}}}%
\nomenclature[Gtheta]{$\theta$}{Polar angle \nomunit{\makebox[\uspc][l]{rad}}}%
\nomenclature[Gsig]{$\sigma$}{Blockage ratio}%
\nomenclature[Gpi]{$\Pi$}{Scale-by-scale energy transfer term in \eqref{eq:KHMSimplified}\nomunit{\makebox[\uspc][l]{$\mathrm{m}^2\,\mathrm{s}^{-3}$}}}%
\nomenclature[Gpi2]{$\Pi\vert_{\mathrm{max}}$}{Maximum absolute value of the scale-by-scale energy transfer, $\Pi$ \nomunit{\makebox[\uspc][l]{$\mathrm{m}^2\,\mathrm{s}^{-3}$}}}%
\nomenclature[GDelta]{$\Delta x$, $\Delta y$, $\Delta z$}{Separation between the centres of the X-probes in $x$, $y$ \& $z$ \nomunit{\makebox[\uspc][l]{mm}}}%

\nomenclature[RCe]{$C_{\varepsilon}$}{Normalised energy dissipation rate, $C_{\varepsilon}\equiv \frac{\varepsilon \ell}{\left(\overline{u^2}\right)^{3/2}}$  or  $C_{\varepsilon} \equiv \frac{\varepsilon \ell}{\left(\overline{q^2}/3\right)^{3/2}}$}%
\nomenclature[RCe2]{$C_{\varepsilon}^{i(k)}$}{Normalised energy dissipation rate, $C_{\varepsilon}^{i(k)} \equiv \frac{\varepsilon L_{ii}^{(k)}}{\left(\overline{u^2}\right)^{3/2}}$ or $C_{\varepsilon}^{i(k)} \equiv \frac{\varepsilon L_{ii}^{(k)}}{\left(\overline{q^2}/3\right)^{3/2}}$}%
\nomenclature[RCe3]{$C_{\Pi}^{i(k)}$}{Normalised maximum energy transfer, $C_{\Pi}^{i(k)} \equiv -\frac{\Pi\vert_{\mathrm{max}} L_{ii}^{(k)}}{\left(\overline{q^2}/3\right)^{3/2}}$}%
\nomenclature[RL]{$\ell$}{Arbitrary length-scale\nomunit{\makebox[\uspc][l]{m}}}%
\nomenclature[Rl]{$l_{w}$}{Sensing length of the hot-wire probe \nomunit{\makebox[\uspc][l]{mm}}}%
\nomenclature[Rd]{$d_{w}$}{Diameter of the hot-wire filament \nomunit{\makebox[\uspc][l]{$\mu$m}}}%
\nomenclature[RT]{$T$}{Width of the test section of the wind tunnel \nomunit{\makebox[\uspc][l]{m}}}%
\nomenclature[RUi]{$U_{\infty}$}{Mean inlet velocity into the test section of the wind tunnel \nomunit{\makebox[\uspc][l]{$\mathrm{m\,s}^{-1}$}}}%
\nomenclature[RM]{$M$}{Mesh size \nomunit{\makebox[\uspc][l]{mm}}}%
\nomenclature[Rt0a]{$t_0$}{Largest lateral thickness of the grid mesh  \nomunit{\makebox[\uspc][l]{mm}}}%
\nomenclature[RZFa]{$f^{0\mathrm{dB}}_{\mathrm{cutoff}}$}{Cut-off frequency at the verge of attenuation \nomunit{\makebox[\uspc][l]{kHz}}}%
\nomenclature[RZFb]{$f^{-3\mathrm{dB}}_{\mathrm{cutoff}}$}{Cut-off frequency at the standard `$-3$dB' attenuation level \nomunit{\makebox[\uspc][l]{kHz}}}%
\nomenclature[RR2]{$Re_{\lambda}^{\mathrm{iso}}$}{Reynolds number based on $\lambda^{\mathrm{iso}}$, $Re_{\lambda}^{\mathrm{iso}}=\frac{u'\lambda^{\mathrm{iso}}}{\nu}$}%
\nomenclature[RR3]{$Re_{\lambda}$}{Reynolds number based on $\lambda$, $Re_{\lambda}=\sqrt{\frac{\overline{q^2}}{3}}\frac{\lambda}{\nu}$}%
\nomenclature[RR4]{${Re}_{L^{i(k)}}^{\mathrm{iso}}$}{Reynolds number based on $L_{ii}^{(k)}$, ${Re}_{L^{i(k)}}^{\mathrm{iso}}=\frac{u'L_{ii}^{(k)}}{\nu}$}%
\nomenclature[RR5]{$Re_{L^{i(k)}}$}{Reynolds number based on $L_{ii}^{(k)}$, ${Re}_{L^{i(k)}}=\sqrt{\frac{\overline{q^2}}{3}}\frac{L_{ii}^{(k)}}{\nu}$}%
\nomenclature[RR1]{$Re_{M}$}{Inflow/global Reynolds number, $Re_M= \frac{U_{\infty}M}{\nu}$}%
\nomenclature[RR0]{$Re$}{Reynolds number based on arbitrary length and velocity scales}%
\nomenclature[RK]{$K$}{Turbulent kinetic energy per unit mass \nomunit{\makebox[\uspc][l]{$\mathrm{m}^{2}\,\mathrm{s}^{-2}$}}}%
\nomenclature[RK2]{$\overline{q^2}$}{Twice the turbulent kinetic energy per unit mass, $2K=\overline{q^2}\equiv \overline{u_i u_i}$ \nomunit{\makebox[\uspc][l]{$\mathrm{m}^{2}\,\mathrm{s}^{-2}$}}}%
\nomenclature[RUia]{$U_i$}{Mean velocity in the $x_i$ direction, $U_1=U$, $U_2=V$ \& $U_3=W$ \nomunit{\makebox[\uspc][l]{$\mathrm{m}\,\mathrm{s}^{-1}$}}}%
\nomenclature[RUib]{$u_i$}{Fluctuating velocity in the $x_i$ direction, $u_1=u$, $u_2=v$ \& $u_3=w$ \nomunit{\makebox[\uspc][l]{$\mathrm{m}\,\mathrm{s}^{-1}$}}}%
\nomenclature[RUib]{$u'_i$}{R.m.s velocity in the $x_i$ direction, $u'_1=u'$, $u'_2=v'$ \& $u'_3=w'$ \nomunit{\makebox[\uspc][l]{$\mathrm{m}\,\mathrm{s}^{-1}$}}}%
\nomenclature[RUic]{$P,\,p,\,p'$}{Mean, fluctuating and r.m.s static pressure \nomunit{\makebox[\uspc][l]{Pa}}}%
\nomenclature[RL]{$L_{ii}^{(k)}$}{Integral-length scales based on the integral of $B_{ii}^{(k)}$ \nomunit{\makebox[\uspc][l]{m}}}%
\nomenclature[RB1]{$B_{ii}^{(k)}(\mathbf{X},r)$}{One-dimensional correlation function, $B_{ii}^{(k)}(\mathbf{X},r)\equiv B_{ii}(\mathbf{X},r_k)$ }%
\nomenclature[RB0]{$B_{ii}(\mathbf{X},\mathbf{r})$}{Three-dimensional correlation function }%
\nomenclature[RB2]{$B^*(r)$}{Spherical shell averaged correlation function }%
\nomenclature[RF]{$F_{ii}^{(k)}(k)$}{One-dimensional velocity spectra, $F_{ii}^{(k)}(k)\equiv F_{ii}(k_k)$ \nomunit{\makebox[\uspc][l]{$\mathrm{m}^3\,\mathrm{s}^{-2}\,\mathrm{rad}^{-1}$}}}%
\nomenclature[RF2]{$F_{ii}(\mathbf{k})$}{Three-dimensional velocity spectra \nomunit{\makebox[\uspc][l]{$\mathrm{m}^3\,\mathrm{s}^{-2}\,\mathrm{rad}^{-1}$}}}%
\nomenclature[RF3]{$E(k)$}{Spherical shell averaged three-dimensional velocity spectra \nomunit{\makebox[\uspc][l]{$\mathrm{m}^3\,\mathrm{s}^{-2}\,\mathrm{rad}^{-1}$}}}%
\nomenclature[Rk01]{$k$}{Longitudinal wavenumber, $k\equiv k_1$ and modulus, $k\equiv \vert\mathbf{k}\vert$ \nomunit{\makebox[\uspc][l]{rad m$^{-1}$ }}}%
\nomenclature[Rk00]{$\mathbf{k}$}{Wavenumber vector, $\mathbf{k}=(k_1,\,k_2,\,k_3)$ \nomunit{\makebox[\uspc][l]{rad m$^{-1}$ }}}%
\nomenclature[Rr01]{$r$}{Separation vector modulus, $r\equiv \vert\mathbf{r}\vert$ \nomunit{\makebox[\uspc][l]{m}}}%
\nomenclature[Rr00]{$\mathbf{r}$}{Separation vector, $\mathbf{r}=(r_1,\,r_2,\,r_3)$ \nomunit{\makebox[\uspc][l]{m}}}%
\nomenclature[RXX]{$\mathbf{X}$}{Position vector, $\mathbf{X}=(X_1,\,X_2,\,X_3)$ \nomunit{\makebox[\uspc][l]{m}}}%
\nomenclature[RX]{$x_i$}{Cartesian coordinate system, $x_1=x$, $x_2=y$ \& $x_3=z$ \nomunit{\makebox[\uspc][l]{m}}}%

%print nomenclature
% Command: makeindex thesis.nlo -s nomencl.ist -o thesis.nls
\printnomenclature[0.7in]
%\pagestyle{fancy}

% Table of contents etc ...
\addcontentsline{toc}{chapter}{Table of Contents}
\clearemptydoublepage
\tableofcontents
\addcontentsline{toc}{chapter}{List of Figures}
\clearemptydoublepage
\listoffigures
\addcontentsline{toc}{chapter}{List of Tables}
\listoftables

% The chapters
\clearemptydoublepage
\chapter{Introduction}
\label{chp:1}

%Turbulence is paramount in very many physical processes. Despite being generally hard to define, turbulence is easily identified as the  

``\emph{The word turbulence is used to describe diverse phenomena. Among these phenomena hydrodynamic
turbulence is one of the important, hard to describe, interesting and challenging problems.
Hydrodynamic turbulence occurs in a very wide variety of liquid and gas flows ranging from the mixing
of a cocktail to the behaviour of the atmosphere, from the blood flow in a vessel to the flow in tubes,
rivers, seas and the ocean, from thermal convection in a saucepan when soup is prepared to thermal
convection in stars, from air flows around pedestrians, automobiles and aircraft to liquid and gas flows
in technical devices.}'' \citep{Lvov91} 

%Why is turbulence research important? 
Turbulence is, indeed, paramount in very many physical processes and remains one of the great unsolved problems in (classical) physics. 
%But that does not, in itself, justify the immense  effort committed to turbulence research, ranging from applied mathematics and physics to almost every single engineering discipline. 
However, there are many turbulence models that are used on a daily basis for engineering design, geophysical and astrophysical studies and permit, to a certain extent, to circumvent the lack of fundamental knowledge. 
For example, it is currently possible to forecast weather with a reasonable accuracy (although this is partly due to the substantial increase of weather measuring stations); design functioning automobiles and aircraft and predict the lifecycle of stars, even though  turbulence phenomena plays a paramount role in each of these.
So, why is such an immense effort committed to turbulence research, ranging from applied mathematics and physics to almost every single engineering discipline?
 
In the author's viewpoint, the answer is threefold, (i) the need for improved predictability of turbulence phenomena, (ii) the natural pursuit of physical understanding and (iii) prospects of turbulence control. 
The first and third are, perhaps, the most financially driven and the second is an enabler to the other two. 
For example, in engineering applications, where turbulence is ubiquitous, improved predictability and/or control of the phenomena leads to better designs with higher efficiency and/or efficacy and consequently to a great reduction in the design cycle and operation costs. 
Improved meteorological and seismic forecasts (the latter pertaining to Earth's mantle convection) can diminish the impact of natural catastrophes.

Concerning `predictability', the prospects may be split into two categories. 
One strategy is the continued use of `simplified' numerical simulations where the turbulent flow field is, in some sense, averaged (such as RANS or LES) and the long term research goal is to obtain increasingly accurate and robust reduced-order turbulence models based on  the physical understanding of the phenomena. This has been, overwhelmingly, the most common approach in engineering and geophysics, although the current models are strongly empirical.  
On the other hand, one may perform direct numerical simulations (DNS) of the equations of motion (e.g. the Navier-Stokes for newtonian fluid turbulence which are generally considered to mimic the turbulence phenomena quite well).
However, even the most powerful supercomputers today can only simulate moderately low Reynolds numbers\footnote{The Reynolds number, $Re$, is a ratio between inertial and viscous forces. The $Re$ also characterises, in some sense, the number of degrees of freedom of the fluid flow.}. Nevertheless, the prospects for DNS are quite promising. 
Should Moore's law continue to hold, e.g. that the number of floating-point operations per second (FLOPS) doubles every one to two years \cite[see e.g.][]{Kurzweil2001}\footnote{Moore's law has been found to hold since the onset of computing up to this date \cite[]{Kurzweil2001}, however, the forecasts are quite controversial. Nevertheless, it is conceivable, through successive paradigm shifts in computing, from volumetric chips \cite[see e.g.][]{Ferry} to quantum computing \cite[see e.g.][]{Ekert}, that Moore's law may well persist for decades to come. },
and considering that (i) the current simulations are one to three orders of magnitude lower in Reynolds number than laboratory experiments \cite[see e.g.][]{Schlatter09,Sylvain2011}, (ii) the Reynolds number of laboratory experiments are one to three orders of magnitude lower than those found in industrial turbulent flows, (iii) the total simulated  time has to increase by at least an order of magnitude for statistical convergence and that (iv) the necessary floating-point operations increases with $(Re)^3$ \cite[]{Frisch:book}, then DNSs for most industrial applications may well be feasible within the next 50 -- 130 years in `supercomputers' and 10 -- 20 years thereafter in smaller industrial clusters.
These are two competing strategies which are also complementary.

Nonetheless, the pursuit for physical understanding will continue even if, in the long term, massive DNSs become common and cost-effective, since (i) geophysical (and astrophysical) simulations will continue to be inaccessible for many more decades, (ii) control strategies without fundamental knowledge are likely scarce and (iii) there are many synergies between turbulence research and other research fields. 
For example, cardiac fibrillation is a phenomenon of electrical turbulence in the heart leading to non-coordinated contractions of the muscle (see \citealp{NatureTurbHeart} for advances in low-energy control mechanisms alternative to the current defibrillators). Financial markets also share many remarkable properties with  turbulence phenomena \citep{NatureTurbFinance}, even though the extent of this analogy is debatable \citep{NatureTurbFinanceII}. Yet another pressing example of turbulence phenomena arises in the magnetic plasma confinement of nuclear fusion power-plants  \cite[see e.g.][]{RevModPhysFusion}. 

In the following, a brief introduction to the main concepts of turbulence theory, with emphasis on  the aspects pertaining to the research presented in this thesis, is given. 

%%%%%%%%%%%%%%%%%%%%%%%%%%%%%%%%%%%%%%%%%%%%%%%%%%%%%%%%%%%%%%%%%%%%%%%%%%%
\section{Turbulence theory: a brief introduction} \label{sec:theory}

\subsection{A conceptual picture of turbulence} \label{sec:introB}

\begin{figure}[ht!]
\centering
\fbox{\includegraphics[width=120mm]{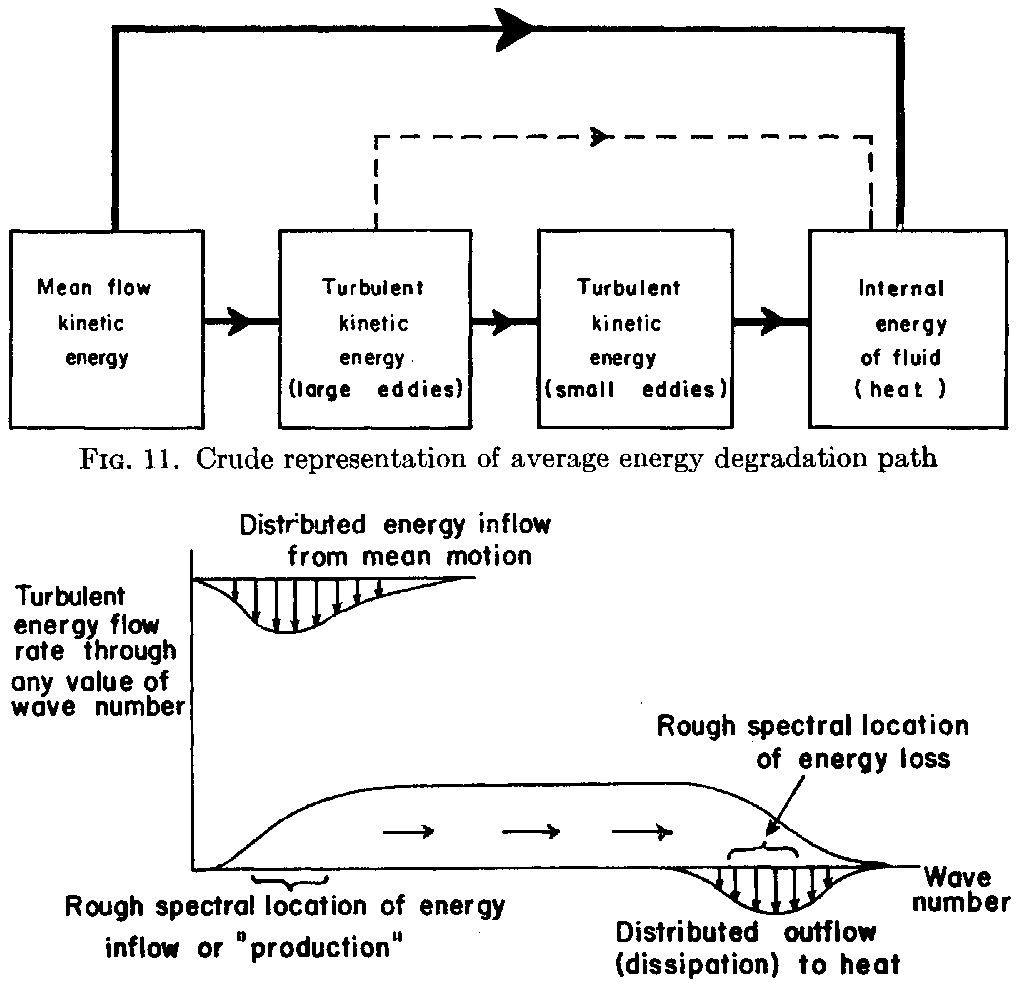}}
\caption[Conceptual picture of turbulence.]{Conceptual picture of turbulence, reprinted from \cite{CorrsinAmSci}.}
\label{fig:Corrsin}
\end{figure}

Firstly, it is instructive to consider the conceptual viewpoint of the phenomena since most theoretical approaches are, in essence, a formalisation of semi-empirical ideas of how turbulence behaves.  

The conceptual picture sketched by \cite{CorrsinAmSci} is presented in figure \ref{fig:Corrsin} and can be summarised as follows. 
The large-scale `eddies' of size $\ell$ and turbulent kinetic energy $K$ extract energy from the mean flow they are embedded in (via turbulent production from the mechanical work of the Reynolds stresses against the mean flow gradients). 
The turbulent kinetic energy then tends to be transferred, within a finite time, to smaller and smaller `eddies' of length $\ell^{(j)}$ and kinetic energy $K^{(j)}$ (the superscript $j$ denotes the multiple iterations), until these eddies are so small that $\sqrt{K^{(n)}}\ell^{(n)}\sim \nu$ (superscript $n$ denoting the smallest scales where the Reynolds number is of order unity) and they can very quickly lose their kinetic energy by linear viscous dissipation. 
This is the celebrated  nonlinear energy transfer mechanism of turbulence which is responsible for the highly increased dissipation rate of turbulent flows. 
In its absence, this kinetic energy would, otherwise, be dissipated very slowly via viscous diffusion at a rate proportional to $\nu/\ell^2$ \cite[]{TennekesLumley:book}. 
This conceptual picture, particularly that of a `cascade of energy', is often attributed to \cite{Richardson:book}. 
Note however that even though there is a net downscale energy transfer, this results from an imbalance between considerable transfers of energy upscale and downscale \cite[see][]{Piomelli1991,KCM03,Kaneda2009}. 
Therefore the arrows in figure \ref{fig:Corrsin} should really be bi-directional to highlight the fact that energy is continuously interchanged amongst the several `scales' of the flow.

Also note that, the single observation that there are many scales in turbulence does not, by itself, simplify the problem.
However, considering that the scale separation between energy containing and dissipative eddies is very large does simplify the theoretical approach, as will be seen in the next section.  
Treating the fluid as a continuum, which in turn enables the derivation of the Navier-Stokes using differential calculus, is another example of the usefulness of scale separation arguments.
A gaseous fluid is actually composed of individual molecules travelling at the r.m.s. speed $c_{\mathrm{th}}$ (for most fluids, $c_{\mathrm{th}}$ is of the order of the speed of sound, $c_{\mathrm{s}}$) and colliding with each other after travelling, on average, a certain length commonly denoted as the mean-free-path $\lambda_{\mathrm{mfp}}$. 
Yet, taking a sufficiently large ensemble of these molecules, they behave as a continuum with intrinsic transport properties such as the kinematic viscosity $\nu\sim c_{\mathrm{s}}\lambda_{\mathrm{mfp}}$, thermal conductivity and diffusivity. 
These are well known results from statistical mechanics.
Considering that the fluid is a continuum bears the implicit assumption that the smallest length-scale of the fluid flow is overwhelmingly larger than $\lambda_{\mathrm{mfp}}$ (this can be motivated to be a good approximation, particularly for incompressible flows, see e.g. \citealt{Frisch:book}, p. 110).

This conceptual picture of turbulence and the expectation that, with increasing scale separation, the process of energy extraction from the mean flow (and the general effect of inhomogeneities), the `energy cascade' and the energy dissipation become asymptotically self-governing (and related only by the net energy transferred) is key to construct physical and engineering models of the phenomena.
An example of one particularly successful phenomenological theory which is based on this conceptual picture is described below.

\subsection{Energy `cascade' and dissipation in turbulence} \label{sec:introB2}
In a seminal contribution to the theory of turbulence, \cite{K41a} employed a conceptual framework similar to Richardson's `cascade of energy' but considered ``the case of an arbitrary turbulent flow with very large Reynolds number''.
This lead to his hypothesis that owed to the very large scale separation between the  energy containing eddies and the small dissipative eddies, the latter are in statistical equilibrium and depend solely on the fluid viscosity, $\nu$ and on the kinetic energy that they receive from the large eddies (which ought to be equal to the dissipation, $\varepsilon$)\footnote{\cite{K41a} also put forward the hypothesis that the dissipative eddies are isotropic which renders their statistics universal functions of $\nu$ and $\varepsilon$.}.
From dimensional analysis one can determine the functional forms of the characteristic velocity-, time- and length-scales of the small dissipative eddies (the latter being commonly denoted as the Kolmogorov microscale, $\eta = (\nu^3/\varepsilon)^{1/4}$). 
Based on the same scale separation argument, \cite{K41a} also hypothesised that there must be an intermediate range of scales that are also not dependent on the large scale eddies, but nevertheless are sufficiently large not to be significantly affected by viscosity. 
Based, once more, on dimensional analysis this leads to the celebrated `inertial subrange' where the $2^{\mathrm{nd}}$-order structure functions follow a $r^{2/3}$ power-law and the velocity spectra follow a $k^{-5/3}$ power-law ($k$ is the wavenumber). 
Even with the subsequent refinements of the theory to, e.g., account for `intermittency' of the small scales \cite[]{K62}, the above ideas retain the gist of the mainstream theoretical approach to turbulence phenomena. 
 
Note that the theory by \cite{K41a,K41c} contains an implicit assumption (or `principle' and/or `law' depending on the viewpoint). 
Effectively, it is assumed that for increasingly large Reynolds number, $Re$ (based on some characteristic length and velocity scales of the turbulent flow as a whole), the dissipation at the small eddies becomes overwhelmingly larger than the direct viscous diffusion of the large eddies, i.e. \[\varepsilon \gg \nu K/\ell^2\] for $Re\gg 1$, where $K$ and $\ell$ are the kinetic energy  and a length scale (typically based on the integral of a velocity correlation function) characteristic of the turbulent velocity fluctuations.
In his analysis it is also implicit that, in the limit of $Re \gg 1$, the ratio between the size of the energy containing eddies and the size of the dissipative eddies is very large, i.e. $\ell/\ell^{(n)} \gg 1$\footnote{This can be motivated based on the second footnote of \cite{K41a} by noticing that an increasingly large $Re_{\ell} = \sqrt{K}\ell/\nu$ leads to a monotonic increase in scale separation since the small dissipative scales follow $\sqrt{K^{(n)}}\ell^{(n)}\sim \nu$ and thus $\ell/\ell^{(n)}\sim \sqrt{K^{(n)}/K}\,Re_{\ell}$ (one expects that $K^{(n)}<K$).}.
It should be noted how Kolmogorov negotiates the mathematical and conceptual difficulty of determining the behaviour of the dissipation in the limit  $Re \rightarrow \infty$, which ``is singular, so one cannot exclude the possibility that what is observed at whatever large but finite $Re$ can be very much different from what happens in the limit $Re \rightarrow \infty$ ($\nu \rightarrow 0$)'' \cite[p. 96]{Tsinober:book}.

A few years earlier, \cite{Taylor1935}, based on phenomenological considerations, had proposed that the rate of dissipation of  high Reynolds number turbulent flows should scale with $\sqrt{K}/\ell$ and thus $\varepsilon \sim K^{3/2}/\ell$, which
``is a remarkable formula, since it is completely independent of molecular viscosity'' \cite[p. 2]{Eyink}\footnote{Note that  \cite{Taylor1935,Taylor1935b} actually used the mesh size as the estimate of $\ell$. However, in his contributions, the integral scale is (erroneously) considered to be a definite fraction of the mesh size, i.e. constant throughout the decay.}. 
This formula takes into account the nonlinear dissipation mechanism of the turbulence but also assigns a specific time scale for the process, i.e. $\ell/\sqrt{K}$ (the same time scale for all high enough Reynolds numbers). 
Note that, \cite{K41b} arrived to a similar expression for the dissipation from the assumption that the large scale similarity of the  $2^{\mathrm{nd}}$-order structure functions (i.e. based on $K$ and $\ell$) can be extended to the inertial range.  
Together with the invariant proposed by \cite{Loitsyansky} it allowed Kolmogorov to make a quantitative estimation of the power-law exponent, $n$, of decaying homogeneous isotropic turbulence, i.e. $n=-10/7$.
In fact, the first two-equation model of turbulence was proposed by \cite{K42} based on his previous work \cite[]{K41a,K41b}.

The assumption $\varepsilon \sim K^{3/2}/\ell$ became standard in virtually all subsequent work, as evidenced, for example, in turbulence textbooks \cite[e.g.][]{Batchelor:book,TennekesLumley:book,Townsend:book,Frisch:book, Lesieur,MathieuScott,Pope,Cambon}.  \cite{TennekesLumley:book} introduce this scaling in their very first chapter with the words ``it is one of the cornerstone assumptions of turbulence theory''. 
\cite{Townsend:book} uses it explicitly in his treatment of free turbulent shear flows \cite[see page 197 in][]{Townsend:book} which includes wakes, jets, shear layers, etc. This scaling is also customarily used in theories of decaying homogeneous isotropic turbulence \cite[see][]{ Batchelor:book, Frisch:book, Rotta:book} and in analyses of wind tunnel realisations of such turbulence \cite[e.g.][]{batchelor1948decay, CC66} in the form
\begin{align}
%C_{\varepsilon}^{L_{11}^{(1)}} = \frac{\varepsilon\, L_{11}^{(1)}}{K^{3/2}} \sim \mathrm{constant}
C_{\varepsilon}^{1(1)} = \frac{\varepsilon\, L_{11}^{(1)}}{(\overline{u^2})^{3/2}} \sim \mathrm{constant}
\label{Eq:DissipationCoeff}
\end{align}
where $u'\equiv (\overline{u^2})^{1/2}$ is the r.m.s. of the longitudinal velocity fluctuation,  $L_{11}^{(1)}$ is the longitudinal integral scale and $C_{\varepsilon}^{1(1)}$ is a constant independent of time, space and Reynolds number when the Reynolds number is large enough ($u'$ and $L_{11}^{(1)}$ are the quantities measured with a single component sensor when the time varying signal is interpreted as spatially varying using Taylor's hypothesis, \citealp{Taylor1938}).
However, as \cite{Taylor1935} was careful to note, the constant $C_{\varepsilon}$($\equiv \varepsilon\, \ell/(\overline{u^2})^{3/2}$, where the superscript in $C_{\varepsilon}$ is dropped to indicate that an arbitrary characteristic length scale $\ell$ is used) does not need to be the same irrespective of the boundaries (initial conditions) where the turbulence is produced \cite[see][]{Burattini2005,MV2008,GV2009}.
 
In high Reynolds number self-preserving free turbulent shear flows, the cornerstone scaling $C_{\varepsilon}\sim \mathrm{constant}$ determines the entire dependence of $\varepsilon$ on the streamwise coordinate and ascertains its independence on Reynolds number \cite[see][]{Townsend:book}. 
It also specifies the relative size of the different length scales in the flow and how they vary with the Reynolds number. For example, the assumption that $C_{\varepsilon}^{1(1)}\sim \mathrm{constant}$ implies that the ratio between the integral-length scale and the Kolmogorov microscale (which is representative of the scale separation between energy containing and dissipative scales) varies as $L_{11}^{(1)}/\eta\sim {Re}_{L^{1(1)}}^{3/4}$ ( ${Re}_{L^{1(1)}} = u' L_{11}^{(1)} /\nu$).
This cornerstone scaling is also effectively used in turbulence models such as $K-\varepsilon$ \cite[see][]{Pope} and in LES \cite[see][]{Lesieur,Pope}.

As noted by \cite{Lumley92}, by 1992 there had not been too much detailed and comprehensive questioning of data to establish the validity of \eqref{Eq:DissipationCoeff} but he wrote: ``I hardly think the matter is really much in question''. 
He cited the data compilations of \cite{Sreeni84} which suggested that $C_{\varepsilon}^{1(1)}$ does become constant at $Re_{\lambda}^{\mathrm{iso}}$ ($= u'\lambda^{\mathrm{iso}}/\nu$) larger than about 50 for wind tunnel turbulence generated by various biplane square-mesh grids, but there seemed to be little else at the time ($\lambda^{\mathrm{iso}} = \sqrt{15 \nu u'^2/\varepsilon}$ is the `isotropic' Taylor microscale -- \citealp{Taylor1935}; in anisotropic turbulence and/or when the necessary data is available, it is preferred to use the general, i.e. anisotropic, definition of the Taylor microscale, $\lambda= \sqrt{5 \nu \overline{q^2}/\varepsilon}$ and its associated Reynolds number $Re_{\lambda} = \sqrt{\overline{q^2}/3}\, \lambda /\nu$, where $\overline{q^2}=\overline{u_i \, u_i} = 2K$; $\lambda^{\mathrm{iso}} = \lambda$ and $Re_{\lambda}^{\mathrm{iso}}=Re_{\lambda}$ in isotropic turbulence).
Since then, direct numerical simulations (DNSs) of high Reynolds number statistically stationary homogeneous isotropic turbulence have significantly strengthened support for the constancy of $C_{\varepsilon}$ at $Re_{\lambda}$ greater than about 150 (see compilation of data in \citealp{Burattini2005}, and \citealp{Sreeni98}). 
Other turbulent flows have also been tried in the past fifteen years or so such as various turbulent wakes and jets and
wind tunnel turbulence generated by active grids \cite[see][]{Sreeni95,Pearson,Burattini2005,MV2008} with some, perhaps less clear,
support of the constancy of $C_{\varepsilon}$ at large enough $Re_{\lambda}$ (perhaps larger than about 200 if the integral scale is defined appropriately, see \citealp{Burattini2005}) and also some clear indications that the high Reynolds number constant value of $C_{\varepsilon}$ is not universal, as indeed cautioned by \cite{Taylor1935}.

Nevertheless, the previous efforts towards the verification of dissipation scaling deserve some criticism. 
Arguably, the work of \cite{K41a,K41c} requires `only' a non-vanishing dissipation with increasingly large Reynolds numbers which can be verified by varying the `global' Reynolds number (i.e. the Reynolds number pertaining to the initial/boundary conditions where the turbulence is produced).
However, the expression \eqref{Eq:DissipationCoeff} effectively prescribes a time scale based on the r.m.s. velocity and integral-length scale. 
Clearly, the latter is a stronger statement than the former, in the sense that the latter implies the former, but the converse is not true.
To assess the validity of \eqref{Eq:DissipationCoeff} or $C_{\varepsilon}\sim \mathrm{constant}$ one must necessarily use a temporally/spatially evolving flow where the r.m.s. velocity and/or integral-length scale(s) vary as well as the local turbulent Reynolds number, e.g. $Re_{\lambda}$. 

Unfortunately, the acclaimed experimental and numerical evidence is mostly based on the variation of the global Reynolds number.
For example, the evidence presented by \cite{Sreeni84} is essentially  based on the data by \cite{Kistler} where the viscosity of the fluid is varied. 
\cite{Pearson} presents measurements of \eqref{Eq:DissipationCoeff} for several shear flows, but as far as the author can gather, the measurements are performed at a fixed spatial location and the (global) Reynolds number is varied by changing the mean  inflow velocity. 
The results from active grid experiments presented by \cite{GG2000} and \cite{Burattini2005} from the data of, respectively, \cite{MW1996} and \cite{Larssen} are also taken at a fixed spatial location and the Reynolds number is varied by changing the actuation (i.e. forcing) of the active elements of the grid. 
Finally, the acclaimed evidence from DNSs \cite[]{Sreeni98,Burattini2005} pertain to stationary homogeneous turbulence where the forcing relative to the viscosity is varied.
It will be shown in ch. \ref{chp:4} that measuring \eqref{Eq:DissipationCoeff} at a fixed location with varying global  Reynolds number, or conversely, setting the global Reynolds number and varying the spatial location can lead to significantly different results. \\

Note that the Taylor microscale ($\lambda$) is a mixed scale involving large- and small-scale quantities and its physical role has been the source of much debate. 
It is typically introduced in the literature as a convenient intermediate length scale which is larger than the Kolmogorov microscale ($\lambda/\eta \sim {Re}_{\lambda}^{1/2}$) and smaller than the integral-length scale ($L_{11}^{(1)}/\lambda^{\mathrm{iso}} \sim {Re}^{\mathrm{iso}}_{\lambda}$ if $C_{\varepsilon}^{1(1)}\sim \mathrm{constant}$) regardless of its physical relevance. 
However, it has been suggested in the literature that Taylor microscale is indeed characteristic of  some geometrical  properties of the turbulent flow field. 
\cite{Tennekes1968} suggested that $\lambda$ is the characteristic spacing of vortex sheets and consequently ``proportional to the average distance between zero crossings of a velocity fluctuation signal''. 
This proportionality has indeed been verified experimentally \cite[see][and references therein]{MV2008,Sreeni83} and received theoretical support \cite[see][]{MV2008,GV2009}. 
The thickness of turbulent/nonturbulent interfaces has also been found to be associated with $\lambda$ in laboratory and numerical experiments \cite[see][and references therein]{W2009,daSilva2010}.  

Some comments concerning the choice of length scale $\ell$ characterising the turbulence (and thus to be used in the normalisation $\varepsilon$) are also in order.
In experimental investigations, $\ell$ is typically taken to be the longitudinal integral-length scale $L_{11}^{(1)}$, leading to \eqref{Eq:DissipationCoeff}. In DNSs $\ell$ is an averaged integral-length scale derived from the spherical shell averaged three-dimensional energy spectra $E(k)$ \cite[]{MY75} and/or correlation function $R^{*}(r)$ as, 
\begin{equation}
L = \frac{\pi}{2}\frac{\int\limits_0^{\infty}\! k^{-1}E(k)\, dk}{\int\limits_0^{\infty}\! E(k)\, dk} = \frac{1}{R^{*}(0)}\int\limits_0^{\infty}\! R^*(r)\, dr,
\end{equation}
and thus the normalised energy dissipation may take the form,
\begin{equation}
C_{\varepsilon} = \frac{\varepsilon\, L}{\left(\frac{2}{3}K\right)^{3/2}}.
\label{eq:Ceps}
\end{equation}
For an homogeneous flow, $E(k)$ is the average over all possible directions of $F_{ii}(\mathbf{k})$ (summation over $i$ is implied) for a spherical shell of radius $k =|\mathbf{k}|$ and is the wavenumber space counterpart of the spherical shell averaged correlation function,   
\begin{equation}
E(k) \equiv \iint\limits_{k=|\mathbf{k}|}\!\frac{1}{2}F_{ii}(\mathbf{k})\,d\mathbf{k} = \frac{1}{\pi}\int\limits_0^{\infty}\!kr\,\sin(kr)\,R^*(r)\,dr; \hspace{7mm}R^*(r) = \iint\limits_{r=|\mathbf{r}|}\!R_{ii}(\mathbf{r})\,d\mathbf{r},
\end{equation}
where 
\begin{equation*}
R_{ii}(\mathbf{r}) =  \overline{u_i(\mathbf{X}-\mathbf{r}/2)\, u_i(\mathbf{X}+\mathbf{r}/2)} \hspace{3mm}\mathrm{and}\hspace{3mm} F_{ii}(\mathbf{k}) = \int e^{j\mathbf{k}.\mathbf{r}} R_{ii}(\mathbf{r}) d\, \mathbf{r}.
\end{equation*}
(Note that $\mathbf{k}$, $\mathbf{r}$ and $\mathbf{X}$ are, respectively, the wavenumber, separation and centroid position vectors and that for an homogeneous field the statistics are independent of the origin, $\mathbf{X}$; $j=\sqrt{-1}$.)

If the turbulence is also isotropic, $L_{11}^{(1)}$ is related to $L$ by,
\begin{equation*}
L_{11}^{(1)} \equiv \frac{1}{R_{11}^{(1)}(0)}\int\limits_0^{\infty}\! R_{11}^{(1)}(r)\, dr = \frac{3}{2} \frac{1}{R^{*}(0)}\int\limits_0^{\infty}\! R^*(r)\, dr = \frac{3}{2}L.
\end{equation*}
In the vast majority of the experimental investigations including shear flows \cite[e.g.][]{Sreeni95,Pearson}, data are acquired with one-component sensors which can only provide an estimate of $L_{11}^{(1)}$ and it is expected (or hoped) that the turbulence is not far from being isotropic so that the estimated integral-length scale is proportional to all other integral-length scales derived from the velocity field (including $L$) and therefore $L_{11}^{(1)}$ ought to be representative of the length of the large-scale eddies, even if only approximately. 
This topic is further discussed in ch. \ref{chp:3} where homogeneity and large-scale isotropy  are investigated.
Note also that there have been some attempts in defining other length-scales which, allegedly, better represent the size of the energy containing eddies and thus ought to be used instead of \eqref{Eq:DissipationCoeff} or \eqref{eq:Ceps} \cite[see][the discussion in \citealp{Burattini2005} and more recently \citealp{Mouri2012}]{MW1996,Pearson}. 
However, in the author's viewpoint, the support from data is still too meagre for these attempts to be considered more than speculative. \\

One of the main focuses of this thesis is to further investigate the behaviour of the normalised energy dissipation rate following the recent findings of \cite{SV2007} and \cite{MV2010} that reported a significant departure from the expected $C_{\varepsilon}^{1(1)}\sim \mathrm{constant}$ at the lee of a particular grid-design inspired by fractal objects. 
It is, therefore, appropriate to provide a brief historical context of grid-generated turbulence, which has been the most common experimental realisation of homogeneous isotropic turbulence.
Homogeneous isotropic turbulence is analytically simpler to tackle, but retains all the essential physics of turbulent energy transfer and dissipation mechanisms and thus is an excellent flow to test semi-empirical relations such as \eqref{Eq:DissipationCoeff} and  \eqref{eq:Ceps}.

\section[Experimental realisation of HIT]{Experimental realisation of homogeneous isotropic turbulence (HIT)} 
\label{sec:survey}
Ever since the seminal work by \cite{Taylor1935,Taylor1935b} on homogeneous isotropic turbulence, grid-generated turbulence became `canonical' as its experimental realisation / approximation.
Particularly, ``beyond the point where the `wind-shadow' has disappeared (and the turbulent motion) will depend only on the form and mesh size of the grid, and not on the cross-section of the bars or sheets from which it is constructed'' \cite[p. 440]{Taylor1935}.
In fact, \cite{Taylor1935,Taylor1935b}\footnote{According to \cite{Voyage}, L. Prandtl was actively involved in discussions  with G.I. Taylor about his theory and independently corroborated his findings with his experiments in collaboration with H. Reichardt.} used experimental data acquired by \cite{SimmonsSalter1934} at the National Physical Laboratory and by H.L. Dryden and co-workers at the National Bureau of Standards \cite[see data and references in][]{Taylor1935b} and provided the very first evidence substantiating his theory.

In fact, much of the research in the second half of the $20^{\mathrm{th}}$ century has been dedicated to the verification and articulation of the results from the seminal contributions of \cite{Taylor1935,Taylor1935b}, \cite{KH1938} and \cite{K41a,K41b,K41c}  which, in effect, originated a new research paradigm.
Turbulence generated by square-mesh grids in a wind tunnel has, undoubtedly, been one of experimental apparatus of choice in the experimental research.

\subsection{Regular (passive) grids}

\begin{table}[ht!]
\caption[Short survey of previous RG-generated turbulence experiments.]{Short survey of previous RG-generated turbulence experiments in a wind tunnel.
Further references are given in \cite{GC74,lavoie2007effects} and \cite{Ertunc2010}.}
\label{table:RGdata}
\centering
\rule{\linewidth}{.5pt}\vspace*{4mm}
\begin{tabular*}{\textwidth}{@{\extracolsep{\fill}}lcccccl}
Source  & Geometry & $\sigma$ & $M$  & $Re_M$ & $x/M$  & $Re_{\lambda}$ \\
        &          &  $(\%)$  & (mm)&$(\times 10^{-3})$&     &  \\
\midrule
\multirow{3}{*}{\cite{CorrsinMSC}}       & Round & 44 & 25.4 & 8.5   & 10--115 & -- \\
                         & Round & 44 & 25.4 & 17    & 10--115 & -- \\
\vspace{2mm}             & Round & 44 & 25.4 & 26    & 10--115 & -- \\
\multirow{4}{5cm}{\cite{BT1947,batchelor1948decay} \cite[see also][]{GN1957}} & Round & 34 & 12.5 & 17 & 20--130  & $\approx 20$ \\
                                   & Round  & 34 & 25.4 & 34   & 20--130 & $\approx 29$ \\
                                   & Round  & 34 & 25.4 & 68   & 20--60  & $\approx 41$ \\
\vspace{2mm}             & Round  & 34 & 50.8& 135  & 20--35   & $\approx 58$ \\
\multirow{4}{5cm}{\cite{Kistler}}    & Square & 34 & 171  & 137  & 23--60  & $\approx 200$ \\
                         & Square & 34 & 171  & 665  & 23--60  & $\approx 410$ \\
                         & Square & 34 & 171  & 1255 & 23--60  & $\approx 450$ \\
\vspace{2mm}             & Square & 34 & 171  & 2426 & 23--60  & $\approx 560$ \\
\vspace{2mm}\cite{U63}   & Square & 44 & 6.4  & 26.4 & 20--260 & -- \\
\multirow{4}{5cm}{\cite{CC66,CC71}}         & Square & 34 & 25.4 & 17   & 45--385 & 37--49 \\
                         & Square & 34 & 25.4 & 34   & 42--171 & 61--72 \\
                         & Square & 34 & 50.8 & 68   & 42--171 & -- \\
\vspace{2mm}             & Square & 34 & 101.6& 135  & 20--80  & -- \\
%\cite{SW1983}            & Round  & 34 & 25   & 5.1  & 40--130 & $26|_{\,x/M=100}$ \\
%\vspace{2mm}             & Round  & 34 & 25   & 9.5  & 40--130 & $37|_{\,x/M=100}$ \\
\multirow{4}{5cm}{\cite{ML1990}}   & Round  & 34 & 25.4 & 6    & 10--80  & $29|_{\,x/M=40}$ \\
                         & Round  & 34 & 25.4 & 10   & 10--80  & $36|_{\,x/M=40}$ \\
                         & Round  & 34 & 25.4 & 14   & 10--80  & $42|_{\,x/M=40}$ \\
\vspace{2mm}             & Round  & 34 & 50.8 & 12   &  5--70  & $44|_{\,x/M=40}$ \\
\multirow{2}{5cm}{\cite{JW1992}}   & Square & 34 & 25.4 & 24.4 & 40--160 & $74|_{\,x/M=62}$ \\
\vspace{2mm}             & Square & 34 & 101.6& 47.4 & 1--30   & -- \\
\multirow{3}{5cm}{\cite{lavoie2007effects}} & Square & 35 & 24.8 & 10.4 & 20--80  & $<70$ \\
                         & Round  & 35 & 24.8 & 10.4 & 20--80  & $<70$ \\
\vspace{2mm}             & Round  & 44 & 24.8 & 10.4 & 20--80  & $<70$ \\
\vspace{2mm}\cite{KD2010}& Square & 44 & 40.0 & 4    & 30--250 & 72--90 \\
\multirow{3}{5cm}{\cite{Ertunc2010}}  & Square & 36 & 10.0 & 5.3  & 4--110 & $34|_{\,x/M=16}$ \\
                         & Square & 36 & 10.0 & 8    & 4--110 & $50|_{\,x/M=16}$\\
                         & Square & 36 & 10.0 & 8    & 4--110 & $74|_{\,x/M=16}$
\end{tabular*}
\rule{\linewidth}{.5pt}\vspace*{1mm}
\end{table}

Most of the experiments on RG-generated turbulence follow \cite{Taylor1935} and take the mesh size, $M$, as the characteristic length of the grid for the normalisation of the downstream distance and for the definition of the `global' Reynolds number ($Re_M\equiv U_{\infty}M/\nu$).
The measurements, particularly those assessing the decay of the flow, are also typically restricted to $x/M\gtrapprox30$ \cite[]{CorrsinHandbook}, beyond which the `shadow' of the grid is thought to be negligible.

In table \ref{table:RGdata} a short survey of previous RG-generated turbulence experiments is presented. 
This survey is by no means complete, but as far as the author is aware, it is representative of the wind-tunnel experiments investigating turbulence generated by passive RGs.
In nearly 70 years of research, the design of the experiments is systematically very similar, following the (commonly accepted) guidelines to generate homogeneous (isotropic) turbulence \cite[]{CorrsinHandbook}. (Note that, even following these guidelines there is some controversy on the degree of homogeneity, see e.g. \cite{GN1957} and more recently \cite{Ertunc2010}.)
The grids are typically square-meshed, with either round or square bars and with a blockage ratio varying between $34\%<\sigma < 44\%$. 
The mesh size is typically an order of magnitude smaller than the tunnel's width leading to modest Reynolds numbers \cite[$Re_{\lambda}<100$, except the experiments by][on a pressurised wind tunnel]{Kistler} and the data for $x/M\lessapprox 30$ is typically discarded. 
Note that regular grid-generated turbulence closer to the grid ($x/M\lessapprox 30$) has nevertheless been used in previous experiments as a means to generate moderately high Reynolds number quasi-homogeneous free-stream turbulence to study its interaction with plane boundaries, bluff bodies, wings, etc. \cite[see e.g.][]{Graham,TH1977,Hunt1979,Hancock1980,HB1989}. However, these experiments do not cover as extensively the downstream evolution of the turbulent flow, namely the kinetic energy decay and scaling of the energy dissipation.

In fact, the process of generation of homogeneous turbulence whereby ``(...) the wakes of the individual bars (...) spread individually, and interact in some complicating way''\cite[]{GC74} seems to have received little attention prior to the advent of fractal grids. 
The notable exceptions are, (i) the experimental work by \cite{JW1992}, where it is presented for the first time (to the best of the author's knowledge) the longitudinal profiles, both halfway between bars and behind a bar, of the turbulence intensity near the grid ($x/M = \mathcal{O}(1)$, see table \ref{table:RGdata}), (ii) the low Reynolds number numerical simulations by \cite{Ertunc2010}\footnote{These authors also present experimental data, but those are restricted to $x/M>5$.}, where the same longitudinal profiles are presented together with estimates of the turbulent kinetic energy budget and (iii) the experimental data using Particle Image Velocimetry by \cite{cardesaetal12} where lateral profiles of single-point  statistics (up to $2^{\mathrm{nd}}$-order) are presented, together with estimates of the lateral correlation functions.

Nevertheless, as pointed out in \S \ref{sec:introB2}, the experimental data from RG-generated turbulence (far downstream, i.e. $x/M>30$) seem to support the validity of \eqref{Eq:DissipationCoeff}, at least within the experimental scatter \cite[]{Batchelor:book,Sreeni84}.
Recent experiments by \cite{KD2010} indicated a week dependence of $C_{\varepsilon}^{1(1)}$ with the downstream location. 
However, this small departure may be due to systematical errors in the measurements. (For example, \citealp{KD2010} integrate the correlation up to the first zero crossing, which may introduce a progressive bias to the integral scale estimates as the flow decays.)

%In the next section a short historical perspective of the work on fractal grids, which motivated the present work, is given.
\subsection{Active grids}
From the previous section, it became clear that the Reynolds number of the vast majority of the data acquired in the lee of RGs is quite small ($Re_{\lambda}<100$), posing difficulties in the comparison the results with the predictions from theories based on scale separation \cite[e.g.][]{K41a}. 

To overcome this limitation, experimentalists developed active grids, i.e. grids with moving elements and/or add mean momentum to the fluid \cite[and references therein]{GC74}.
However it seems that the early attempts were only modestly successful in increasing the Reynolds number \cite[p. 333]{MW1996}.
The first major advance in increasing the Reynolds number using an active grid was achieved by \cite{Makita91} using  randomly flapped plates superimposed on a bi-plane grid. 
 \cite{MW1996} and \cite{KCM03} used a similar active grid and were able to achieve turbulence with $Re_{\lambda}=\mathcal{O}(500)$, which is nearly an order of magnitude larger than that obtained with a typical RG on wind tunnels of similar size (see table \ref{table:RGdata}).
In fact,  $Re_{\lambda}=\mathcal{O}(10^3)$ seem to be possible \cite[]{Larssen}.
 
Note, however, that the homogeneity of active grid-generated turbulence has received very little attention.
This appears to be a significant gap in the literature since the low Reynolds numbers straddled by RGs are  a consequence of (i) the restriction of the measurements to $x/M>30$, where the turbulence intensity has decayed by nearly an order of magnitude \cite[see e.g.][]{JW1992}, and (ii) the use of small mesh sizes to be able to measure larger downstream locations relative to the mesh size. Both of these restrictions are aimed to achieve a high degree of homogeneity in the measured turbulence. Therefore, prior to future investigations on active grid-generated turbulence, the conclusion that these can reach a much higher Reynolds number than their passive counterparts must be considered as tentative.

Nevertheless, the data from these active grids also appear to support the validity of \eqref{Eq:DissipationCoeff}, at least for a fixed downstream location relative to the grid and varying $Re_{\lambda}$ via the active grid actuation.  For example, \cite{GG2000} presented data from \cite{MW1996} to shown that  $C_{\varepsilon} \approx 0.5$ for $140 < Re_{\lambda}^{\mathrm{iso}} < 500$, within the experimental scatter.
Similarly, \cite{Burattini2005} presented data from \cite{Larssen} that indicate that $C_{\varepsilon}^{1(1)} \approx 1.5$ for $400 < Re_{\lambda}^{\mathrm{iso}} < 1100$. 

\subsection{Fractal grids}
\cite{HV2007} published an exploratory study of wind tunnel grid-generated turbulence where they tried twenty-one different planar grids from three different families of passive fractal/multiscale grids: fractal cross grids, fractal I grids and fractal square grids (FSGs). 
Note that the designation of ``fractal'' is used in order to emphasise the fact that these grids have multiple scales which are self-similar (in order to reduce the number of parameters necessary to fully specify the design of a multiscale grid). 
Taking the fractal square grids as an example (see figures \ref{fig:grids}a,b), the fractal generating pattern is a square  which repeats itself four times in each iteration (in a total of four iterations, $N=4$). The side length of the square and bar width of the new iteration are fixed proportions of the side length and width of the larger square \cite[in the nomenclature of][these ratios are designated as $R_L$ and $R_t$, respectively]{HV2007}. 
To illustrate that this self-similar design leads to a fractal-like object consider that the bars composing the grid have zero thickness and an infinite number of iterations ($N\rightarrow \infty$).
Counting the number of boxes ($\mathcal{N}$) of size $\epsilon$ necessary to cover completely the perimeter of the grid and repeating the algorithm for various  $\epsilon$ one would notice that $\mathcal{N}(\epsilon) \sim \epsilon^{-D_f}$, where the exponent $D_f$ is commonly denoted as fractal dimension \cite[more precisely a capacity dimension, see][]{VH1991}. 
Choosing $\epsilon$ as $(R_L)^j$ (for the $j^\mathrm{th}$-iteration) it is easily confirmed that the total number of boxes to cover the perimeter is $4^j$ and therefore the fractal dimension characterising the perimeter is $D_f = -\log(4)/\log(R_L)$ \cite[]{HV2007}.  
Note that for $R_L=1/2$, $D_f$ takes its maximum value of 2 and if $N=\infty$ the perimeter fills the whole area, i.e. it is space-filling.
In practice the number of iterations is finite, due to e.g. manufacturing constraints, and the bars have a finite width which varies with the fractal iteration as mentioned above.
 
\cite{HV2007} also ascertained that the fractal dimension $D_f$ of these grids needs to take the maximal value $D_f = 2$ for least downstream turbulence inhomogeneity. 
They also identified some important grid-defining parameters (such as the thickness ratio $t_r$) and some of their effects on the flow, in particular on the Reynolds number $Re_{\lambda}^{\mathrm{iso}}$ which they showed can reach high values with some of these grids in small and conventional sized wind tunnels, comparable to values of $Re_{\lambda}^{\mathrm{iso}}$ achieved with active grids in similar wind tunnels and wind speeds. 
Their most interesting, and in fact intriguing, results were for their space-filling ($D_f = 2$) low-blockage (25\%) FSGs (see figure \ref{fig:grids}b). 
FSGs have therefore been the multiscale grids of choice in most subsequent works on multiscale/fractal-generated turbulence \cite[]{SV2007,nagata2008dns,nagata2008direct,SPSV2010,MV2010,suzuki,Sylvain2011}.
For the case of space-filling low-blockage FSGs, \cite{HV2007} found a protracted region between the grids and a distance $x_{\mathrm{peak}}$ downstream of the grid where the turbulence progressively builds up; and a decay region at $x>x_{\mathrm{peak}}$ where the turbulence continuously decays downstream. 
They reported a very fast turbulence decay which they fitted with an exponential and also reported very slow downstream growths of the longitudinal and lateral integral length-scales and of the Taylor microscale.

\cite{SV2007} concentrated their attention on the decay region of turbulence generated by space-filling low-blockage FSGs and confirmed the results of \cite{HV2007}.
In particular, they showed that $L_{11}^{(1)}/\lambda^{\mathrm{iso}}$ remains approximately constant whilst $Re_{\lambda}^{\mathrm{iso}}$ decays with downstream distance $x$ and they noted that this behaviour implies a fundamental break from \eqref{Eq:DissipationCoeff} where $C_{\varepsilon}^{1(1)}$ is constant. 
They also found that one-dimensional longitudinal energy spectra at different downstream centreline locations $x$ can be made to collapse with $u'$ and a single length-scale, as opposed to the two length-scales ($L_{11}^{(1)}$ and Kolmogorov microscale) required by Richardson-Kolmogorov phenomenology (see \S \ref{sec:introB2}). Finally, they also carried out
homogeneity assessments in terms of various profiles (mean flow, turbulence intensity, turbulence production rate) as well as some isotropy assessments.

\cite{MV2010} also worked on wind tunnel turbulence generated by space-filling low-blockage FSGs. 
They introduced the wake-interaction length-scale $x_*$ which is defined in terms of the largest length and thickness on the grid and they showed from their data that $x_{\mathrm{peak}} \approx 0.5 x_*$. 
They documented how very inhomogeneous and non-Gaussian the turbulent velocity statistics are in the production region near the grid and how homogeneous and Gaussian they appear by comparison beyond $0.5 x_*$. 
They confirmed the findings of \cite{HV2007} and \cite{SV2007} and added the observation that both $Re_{\lambda}^{\mathrm{iso}}$ and
$L_{11}^{(1)}/\lambda^{\mathrm{iso}}$ are increasing functions of the inlet velocity $U_{\infty}$. 
Thus, the value of $L_{11}^{(1)}/\lambda^{\mathrm{iso}}$ seems to be set by the inlet Reynolds number, in this case defined as $Re_{0}=U_{\infty}
x_{*}/\nu$ for example.

 \cite{MV2010} brought the two different single-scale turbulence decay behaviours of \cite{George1992} and
\cite{GeorgeWang2009} into a single framework which they used to analyse the turbulence decay in the downstream region beyond $x_{\mathrm{peak}} \approx 0.5x_*$. 
This allowed them to introduce and confirm against their data the notions that, in the decay region, the fast turbulence
decay observed by \cite{HV2007} and \cite{SV2007} may not be exponential but a fast decaying power-law and that $L_{11}^{(1)}$ and $\lambda^{\mathrm{iso}}$ are in fact increasing functions of $x$ which keep $L_{11}^{(1)}/\lambda^{\mathrm{iso}}$ approximately constant.

Note that, unlike the RGs where the choice of $M$ as the characteristic length of the grid was clear from the very beginning \cite[]{Taylor1935}, for the FSGs there  is a characteristic mesh size (i.e. the side-length of the squares) for each of the fractal iterations. 
\cite{HV2007} introduced an `effective mesh size', $M_{\mathrm{eff}}$, based on the fractal perimeter and the blockage ratio, $\sigma$. 
Their definition is appealing since it returns the conventional mesh size $M$ for a RG.
However, the turbulence decay only starts beyond $x/M_{\mathrm{eff}}>50$, which is in stark contrast with RG-generated turbulence where the decay typically starts at $x/M= \mathcal{O}(1)$ \cite[see e.g.][]{JW1992} and the `shadow' of the grid disappears before $x/M \approx 30$ \cite[]{CorrsinHandbook}.
On the other hand, the wake-interaction length-scale proposed by \cite{MV2010} has been shown to characterise the longitudinal extent of the production region.
However, this length-scale starkly weights the large scales of the grid (in contrast to $M_{\mathrm{eff}}$) and effectively neglects the role of the smaller fractal iterations (except, perhaps, on the proportionality constant between $x_*$ and $x_{\mathrm{peak}}$).
This uncertainty in the choice of the characteristic length of the grid has posed a significant challenge to design an experiment to faithfully compare FSG- and RG-generated turbulence and to understand role of the additional fractal iterations. \\

Lastly, it is worthwhile mentioning the work on the turbulent wakes generated by perforated plates \cite[see][and references therein and thereafter]{Castro1971}. 
In the lee of a perforated plate the shear layers generated at each of the perforations  interact with each other and with the shear layers generated by the plate boundaries. 
There is some similarity between the interaction of the shear layers originating from the array of perforations and those originating from the bars of a grid (one can actually have multiple sized perforations in order to simulate a multiscale object). 
Nevertheless, the perforated plates generate a wake-like flow (regardless of the blockage/solidity ratio) and the entrainment of the free-stream flow (either turbulent or non-turbulent depending on the upstream conditions) strongly influences the development/decay of the turbulent wake. 
The first fractal objects designed for wind-tunnel experiments were three-dimensional and, similarly to the perforated plates,  generated a wake-like turbulent flow with bleed air \cite[which passed through the porous fractal object, see][]{VassilicosQueirosConde,Staicu2003}.
However, there are several distinct physical phenomena simultaneously occurring in the turbulence generated by the 3D fractal grids. 
In the subsequent research it has been preferred to study separately the multiscale wake-interaction phenomena and the effect of the fractal (i.e. rough) edge on the generated turbulence. 
In addition, it has been preferred to use planar objects rather than 3D ones. 
The planar fractal grids that are used in this work are precisely dedicated to the study of turbulence generated by multiscale wake-interactions. 
The study of the effect of rough edges on the generation of turbulent wakes has been studied separately \cite[]{Jovan}. 
\cite{nedicetal12}  have also studied wakes with bleed air passing through a fractal grid  in the context of fractal spoilers for acoustic performance.

\section{Outline}\label{sec:outline}
The present thesis is structured as follows.
In ch. \ref{chp:2} the general details of the experimental apparatus used in the several experiments are given. 
These include assessments of the performance of the instrumentation and the details of the turbulence-generating grids to be investigated.   
In ch. \ref{chp:3} an overall assessment of the homogeneity and the large- and small-scale isotropy of the turbulent flows considered here is presented. 
This is followed in ch. \ref{chp:4} by the demonstration that the energy dissipation rate follows a nonclassical behaviour for a certain region of all decaying passive grid-generated turbulent flows (i.e. for both regular and multiscale/fractal grids), which is one of the main results of this thesis.
In ch. \ref{chp:5} it is shown how this nonclassical dissipation behaviour may be compatible with two length-scale theories of turbulence \cite[e.g. in the spirit of][]{K41a} and the observed $-5/3$ power-law spectrum for the highest Reynolds number data. 
Yet another of the main results of this thesis is presented in ch. \ref{chp:6} where measurements of the divergence of the triple structure function are presented (representing the net energy transfer to small scales) together with a scale-by-scale energy budget. 
The final discussions and conclusions are presented in the closing ch. \ref{chp:7}.

\clearemptydoublepage
\chapter{Experimental apparatus and setup}
\label{chp:2}

\section{Wind tunnels} \label{sec:WT}

\begin{figure}
\centering 
\includegraphics[width = 140mm]{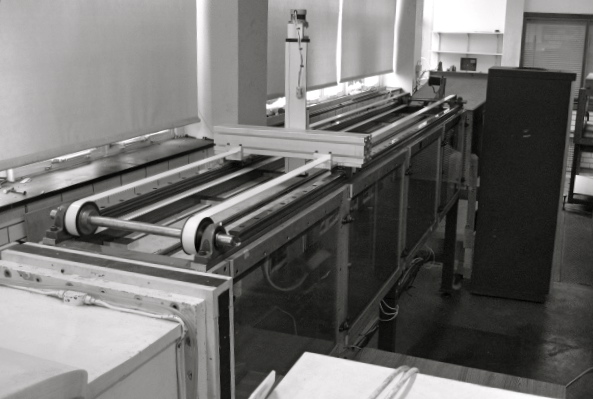}
\caption{Test section of the 18''x18'' wind tunnel.}
\label{fig:WT18}
\end{figure}

The experiments are performed in the 3'x3' closed circuit and the 18''x18'' blow-down wind tunnels at the Department of Aeronautics, Imperial College London. 

The 3'x3'  wind tunnel has a working section of 0.91 m x 0.91 m x 4.8 m, a contraction ratio of 9:1 and the free stream turbulence intensity (i.e. the ratio between the r.m.s and the mean longitudinal velocity) is about 0.05 \%. 
The test section of this wind tunnel was refurbished by the author and his colleague Jovan Nedi\'c to minimise air leakages from the confining walls of the test section.
The test section is equipped with a computer controlled traverse system with three degrees of freedom permitting longitudinal, vertical and horizontal motions. 
However, the span of the automated longitudinal traverse is $\approx 50$cm.
To cover the entire test section the traversing system can be manually traversed.

The 18''x18'' wind tunnel has a working section of 0.46 m x 0.46 m x 3.5\footnote{The data presented in \cite{VV2011}, and included in ch. \ref{chp:3}, was recorded on a modified configuration of this wind tunnel where an additional wooden section with $\approx\!1$m in length was added between the end of the contraction and the working test section to increase the total length of the tunnel to $\approx\!4.5$m.} m,  a contraction ratio of 8:1 and the free stream turbulence intensity is about 0.1 \% (a photograph of the test section is presented in figure \ref{fig:WT18}).
Between the end of the test section and the entrance of the diffuser (divergence angle of $\approx 5^{\circ}$ and expansion ratio of $\approx 1:2$), a grid is installed to maintain a slight overpressure throughout the test section (c.a. $8$Pa before the grid). 
The distance between the outlet of the diffuser and the back wall of the laboratory is about $2.5$m (see figure \ref{fig:WT18}).
The test section of this wind tunnel has been entirely re-built by the Aeronautics workshop, following the author's design, replacing the former wooden test section that was devoid of a probe traverse system.
The new test section is designed to have a continuous, $\approx 2$cm wide and $\approx 3.2$m long, slit on the ceiling to allow longitudinal and vertical traversing of the probes along the vertical mid-plane of the test section. 
The slit has a flexible rubber seal to prevent any air leakage.
The longitudinal traversing table is mostly made from off-the-shelf parts and is driven by a computer controlled stepper motor via a system of timing pulleys and belts (figure \ref{fig:WT18}). 
A vertical traversing table is mounted on the longitudinal one and is also driven by a computer controlled stepper motor. 
These are the two basic degrees of freedom of the installed traverse system which permit the positioning of the probe(s) on the vertical mid-plane. 
The minimum traversing step of the longitudinal and vertical traverses are $40\mu$m and $2.5\mu$m, respectively. 
Depending on the type of measurements, other traversing mechanisms are mounted on the basic traverse system (see \S \ref{sec:WTB}).

The inlet velocity $U_{\infty}$, on both wind tunnels, is imposed and stabilised with a PID feedback controller using the static pressure difference across the contraction and the temperature of the flow, both of which are measured with a Furness Controls micromanometer model FCO510.

The mean velocity profile entering both wind tunnel's test sections was verified to be a top-hat (to within $2\%$). 
%The blockage due to boundary layer growth on the test section's confining walls leads to an increase in the centreline mean velocity of up to $\approx\!4\%$ at the end of the test section. 

\subsection{Experiment-specific apparatus} \label{sec:WTB}
\begin{figure}[ht!]
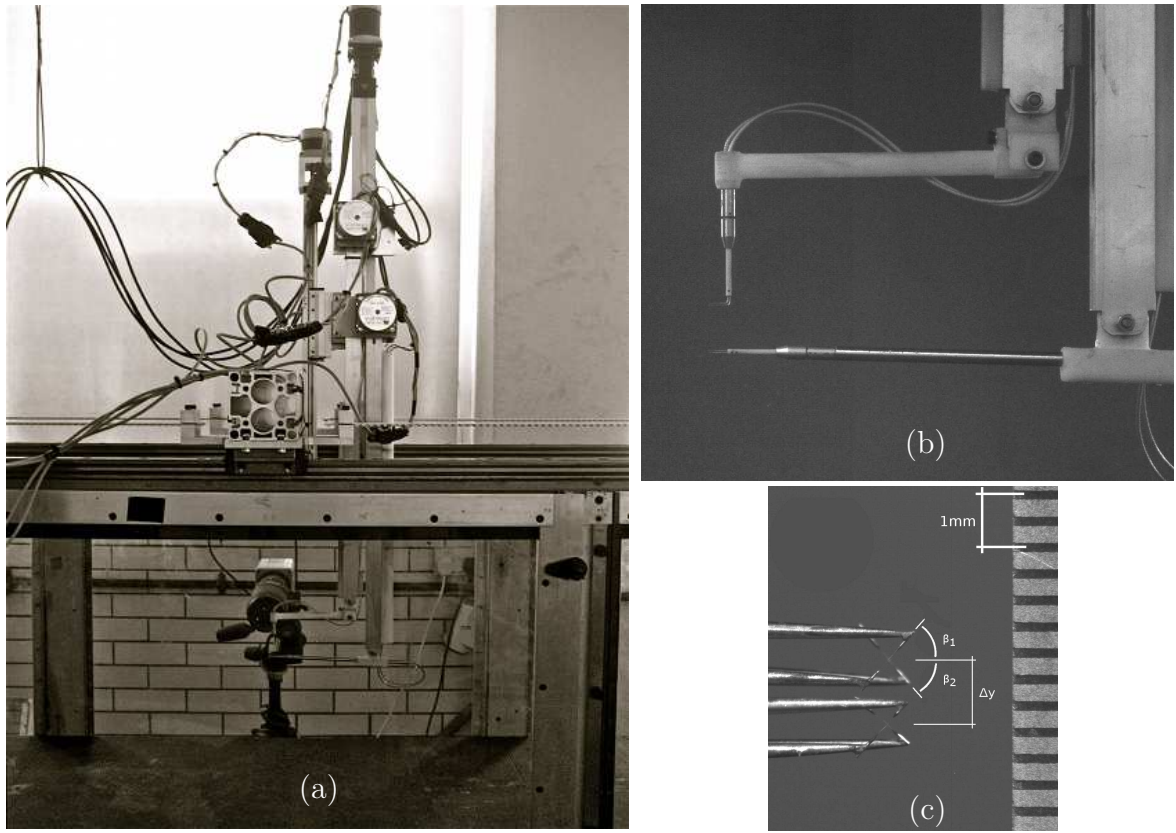

\centering
\begin{minipage}[c]{0.5\linewidth}
   \centering
   \begin{lpic}{2xXW(82mm)}
   \lbl{80,10;\textcolor{white}{(a)}}
   \end{lpic}
\end{minipage}%
\begin{minipage}[c]{0.5\linewidth}
   \centering 
   \begin{lpic}{TraverseImage(70mm)}
   \lbl{160,20;\textcolor{white}{(b)}}
   \end{lpic}
   \vspace*{0.5mm}
   \begin{lpic}{ProbesZoom(38mm)}
   \lbl{160,20;\textcolor{white}{(c)}}
   \end{lpic}
\end{minipage}
\caption[Traverse mechanism for the two-component/two-point measurements]{Traverse mechanism for the two-component/two-point measurements mounted on the 18''x18'' wind tunnel. (a) Overview of the setup, (b) close-up on the probe holders  and (c) detailed view of the probes. }
\label{fig:apparatus}
\end{figure}

For one-component measurements a streamlined structure holds the probe and is attached to the traverse systems of the 18''x18'' and 3'x3' wind tunnels.

For two-component measurements, a computer controlled pitching mechanism is attached to the basic traverse system for the calibration of the probe. 
For the 18''x18'' wind tunnel this mechanism is located outside of the test section and two thin metal struts  connect the pitching mechanism to the probe inside the test section (via the slit on the ceiling). 
For the 3'x3' wind tunnel this mechanism is located inside the test section and the ratio between its frontal area to the test section is $\approx 2\%$.

% Paragraph II - Probe configuration and apparatus
%A brief description of the apparatus is given here, but further details regarding the apparatus can be found in \cite{thesis}
For the two-component/two-point measurements the apparatus consists of two probes (each measuring the longitudinal and vertical velocity components, $u$ and $v$) mounted on a traverse mechanism controlling the vertical distance between the probes and their individual pitch angle (for \it in-situ \rm calibration), which, in turn, is mounted on the longitudinal/vertical traverse system table described in \S \ref{sec:WT} (figure \ref{fig:apparatus}a,b).
The vertical traverse mechanism has single degree of freedom, actuated by a computer controlled stepper motor, and can only displace the two probes symmetrically about their centroid (defined as the geometrical midpoint between the probes' centres; the minimum step is $5 \mu$m). 
Each of the two pitch angle traverses is also actuated by a stepper motor (with a 1:50 ratio gearbox) providing a minimum step angle of $0.018^{\circ}$.
All the stepper motors and gearboxes lie outside of the test section (figure \ref{fig:apparatus}a). 
In the case of the  pitch angle traverses the actuation is made via a timing belt running inside each of the two $1.2\times 2.5$cm rectangular section tubes (seen in figures \ref{fig:apparatus}a,b).

% Paragraph III - Probe separation measurement
The separation between the two X-probes is measured optically with the aid of an external camera (HiSense 4M camera fitted with a Sigma f/3.5 180mm macro lens and $1.4\times$ tele-converter; shown in figure \ref{fig:apparatus}a).
A calibration image is recorded for every set of measurements (i.e. one fixed centroid location and 23 different probe separations), such as the one shown in figure \ref{fig:apparatus}c (the original calibration image was cropped and annotated).
The typical field of view and pixel size are $14\times14$mm and $7\mu$m, respectively and the effective focal length is about $250$mm (note that the pixel size is $2-3$ times the wire diameter but it is sufficient to distinguish the wire from the background). 
The location of the centre of each probe is inferred from the images as the geometric interception between straight lines   connecting the extremities of the etched portion of the wires.
This differs slightly from the visual interception of the sensors (figure \ref{fig:apparatus}c) since the wires are slightly buckled due to the thermal load they are subjected during operation \cite[]{Perry82} as well as other residual stresses from the soldering/etching process.
The vertical separation between the two probes, $\Delta y$, is defined as the vertical distance between the two centres and $\Delta x$ is the downstream separation which should be zero. 
During the course of the experiments it was found that the overall precision of the prescribed vertical separation between the probes was typically $\pm 50\mu$m (i.e. over the three degrees of freedom) and the misalignment $\Delta x$ was typically smaller than $200\mu$m (in figure \ref{fig:apparatus}c $\Delta x=50\mu$m).
During the processing of the data the measured X-probes' location ($\Delta y$ is optically confirmed up to $\Delta y=10$mm) is taken into account (the misalignment $\Delta x$ is corrected with the aid of Taylor's hypothesis), but no noticeable difference was observed when no such corrections are applied. 
The minimum vertical separation between the probes is $\Delta y \approx 1.1$mm whereas the maximum possible separation is $\Delta y \approx 260$mm 

%%%%%%%%%%%%%%%%%%%%%%%%%%%%%%%%%%%%%%
\section{Turbulence generating grids} \label{sec:grids}
  
\begin{figure}[t!]
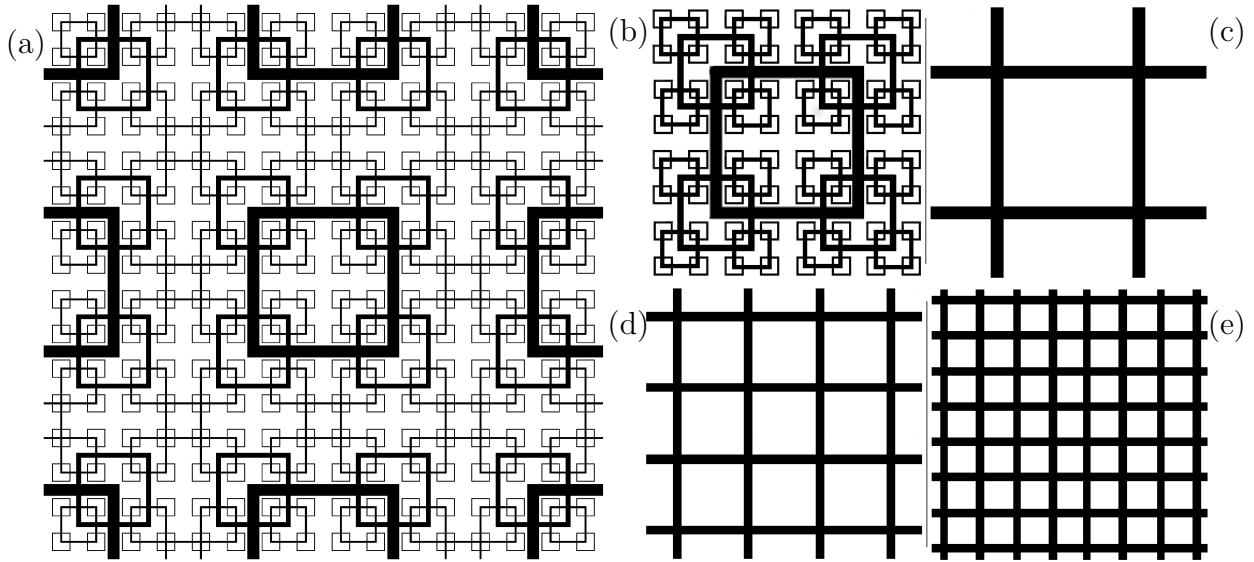

\begin{minipage}[c]{0.5\linewidth}
   \centering
   \begin{lpic}{FSG3x3(75mm)}
   \lbl{-4,160;(a)}
   \end{lpic}
\end{minipage}%
\begin{minipage}[c]{0.5\linewidth}
   \centering 
   \begin{lpic}{Grids18in(75mm)}
   \lbl{-5,175;(b)}
   \lbl{192,175;(c)}
   \lbl{-5,78;(d)}
   \lbl{192,78;(e)}
   \end{lpic}
\end{minipage}
\caption[Turbulence generating grids]{Turbulence generating grids, (a) FSG 3'x3' (b) FSG 18'' (c) RG230 (d) RG115 and (e) RG60. The figures are to scale.}
\label{fig:grids}
\end{figure}

\begin{table}
\caption[Geometric details of turbulence-generating grids]{Geometric details of turbulence-generating grids. $M$ and $t_0$ are distance between the bars (mesh size) and their thickness (for FSGs they refer to the largest bars), $d$ is the longitudinal (i.e. downstream) thickness of the bars and $\sigma$ is the blockage ratio. The value of $x_{\mathrm{peak}}$ for RG60 is taken from measurements of a very similar grid. The low-blockage space-filling FSGs, have four 'fractal iterations' and a thickness ratio, i.e the ratio between the thickest and thinnest bars, of $t_r=17$.}
\label{table:grids}
\rule{\linewidth}{.5pt}\vspace*{4mm}
\centering
\begin{tabular*}{0.9\textwidth}{@{\extracolsep{\fill}}lccccccc}
Grid &  & $M$ & $t_0$ & $d$ & $\sigma$ & $x_*$ & $x_{\mathrm{peak}}/x_*$ \\
        &  & (mm) & (mm) & (mm) & (\%) & (m)&  \\
\midrule
FSG3'x3'     & mono-planar & 228.7 & 19.2 & 5   & 25 & 2.72 & 0.43\\ 
FSG18''x18'' & mono-planar & 237.7 & 19.2 & 5   & 25 & 2.94 & 0.43\\ 
RG230 & mono-planar & 230    & 20    & 6   & 17 & 2.65 & 0.63\\
RG115 & mono-planar & 115    & 10    & 3.2   & 17 & 1.32& 0.63 \\
RG60   & bi-planar       & 60      & 10    & 10 & 32 & 0.36 & $\simeq 0.4$\\
\end{tabular*}
\rule{\linewidth}{.5pt}\vspace*{4mm}
\end{table}

% Grids
Data are recorded in the lee of five grids (figure \ref{fig:grids}) whose geometrical details are summarised in table \ref{table:grids}.

The low-blockage RGs (RG230 and RG115) have a square-mesh built from rectangular section bars with low aspect ratio ($t_0/d \approx 6.6$) and are mono-planar. 
The intermediate blockage RG60 is also square-meshed and is built from square sectioned bars in a bi-planar arrangement.

The low-blockage space-filling fractal square grids (SFGs) have 4 'fractal iterations' and a thickness ratio, i.e. the ratio between the widths of the thickest ($\equiv t_0$) and the thinnest bar,  of $t_{r}=17$ -- see figures \ref{fig:grids}a,b. 
The grids are space-filling because the fractal dimension of its   delimiting line takes the maximum value of $2$ over the range of  scales on the grid. In the limit of infinite number of fractal iterations the blockage ratio will tend to unity, without taking bar thickness into account. However with only four iterations and with finite bar thickness the grid's blockage ratio is roughly 25\%.
Further details of the fractal grids and their design can be found in \cite{MV2010} and \cite{HV2007} .

%%%%%%%%%%%%%%%%%%%%%%%%%%%%%%%%%%%%%%%%%%%%%%%%%%%%%%%%%%%%%%%%%%%%
\section{Thermal anemometry and data acquisition} \label{sec:CTA}

Thermal anemometry is a standard and widely used measurement technique which has been the subject of many analytical, experimental and numerical investigations.
Due to the technique's maturity, it is assumed that the reader has some familiarity with the subject and no review is given here, since it would either be too long or greatly lack in detail.
The only exception is the discussion on the electronic performance testing of these anemometry systems, see \S \ref{sec:SineWave}.
In any case, the reader may refer to any of the numerous monographs on thermal anemometry where detailed discussions of its operating principles, capabilities, limitations and caveats are given, such as \cite{CorrsinHandbook,Perry82,Bruun} and \cite{MeasurementBook}, ch. 3. 

All the data presented in this thesis are measured with an off-the-shelf thermal anemometry system, a DANTEC StreamLine CTA with four channels.
Each channel of the DANTEC StreamLine CTA system has an in-built signal conditioner, i.e. a buck-and-gain amplifier, a low-pass third order Butterworth filter and a high-pass filter (not used). 
Depending on the experiment, the anemometry system drives simultaneously one, two or four hot-wires, for one-, two-component single-point measurements and two-component/two-point measurements, respectively. 
All CTA channels are operated with a 1:20 bridge ratio.

\subsection{Hot-wire probes} \label{sec:Probes}

\begin{table}
\caption[Details on the hot-wires and cutoff frequencies]{Details on the hot-wires and cutoff frequencies. 
$l_w$ and $d_w$ is the sensing length and diameter of the hot-wires. 
$\Delta z$ is  the distance between the inclined wires in the z-direction (analogous to $\Delta x_{3}$ in the nomenclature of \cite{ZA96}, see their figure 1).
The cutoff frequencies are obtained for $U_{\infty}=10\mathrm{ms}^{-1}$ and $f^{-3\mathrm{dB}}_{\mathrm{cutoff}}$ is the cut-off frequency corresponding to $-3$dB signal attenuation and $f^{0\mathrm{dB}}_{\mathrm{cutoff}}$ is the highest frequency with negligible attenuation (see \S \ref{sec:SineWave}). }
\label{table:HW}
\rule{\linewidth}{.5pt}\vspace*{4mm}
\centering
\begin{tabular*}{0.9\textwidth}{@{\extracolsep{\fill}}lccccccc}
Probe & $l_{w}$ & $d_{w}$ & $\Delta z$  & Hot-wire  & Wire & $f^{-3\mathrm{dB}}_{\mathrm{cutoff}}$ & $f^{0\mathrm{dB}}_{\mathrm{cutoff}}$  \\
 & (mm) & ($\mu$m) &  (mm) & body & material & (kHz) & (kHz)   \\
\midrule
SW$5\mu$m & $\approx 1.0$ & 5.1 &  & 55P16 &  Pl-(10\%)Rh & $\approx 25$ & $\approx 12$ \\
SW$2.5\mu$m & $\approx 0.5$ & 2.5 &  & 55P01 & Pl-(10\%)Rh & $\approx 45$ & $\approx 20$  \\
SW$1\mu$m & $\approx 0.2$ & 1.0 &  & 55P11 & Pl-(10\%)Rh & $>50$ & $\approx 40$  \\
XW$5\mu$m & 1.25 & 5.0 & 1.0 &  55P51 & Tungsten & $\approx 30$ & $\approx 12$  \\
XW$2.5\mu$m (I) & $\approx 0.5$ & 2.5 & $\approx 0.5$ & 55P63 & Pl-(10\%)Rh & $\approx 45$ & $\approx 20$\\
XW$2.5\mu$m (II) & $\approx 0.5$ & 2.5 & $\approx 0.5$ & 55P61 & Pl-(10\%)Rh & $\approx 45$ & $\approx 20$\\
\end{tabular*}
\vspace*{1mm}\rule{\linewidth}{.5pt}
\end{table}

The details of the hot-wire probes used to acquire the present data are given in table \ref{table:HW}. 
Single hot-wires (SWs) measure the longitudinal velocity component, whereas two-component probes (XW or X-probe) have two inclined hot-wires and measure longitudinal and transverse velocity components.

All the sensors, except XW$5\mu$m, have in-house soldered and etched Platinum -- (10\%) Rhodium Wollaston wires on  standard probe bodies manufactured by Dantec Dynamics. XW$5\mu$m is an off-the-shelf sensor (55P51 from Dantec Dynamics).  

SWs, XW$5\mu$m and XW$2.5\mu$m (II) are used for the single-point experiments. 
XW$2.5\mu$m (I) and XW$2.5\mu$m (II) are used for the two-point/two-component experiments. These sensors are, respectively, the upper and lower X-probes shown in figures \ref{fig:apparatus}b,c.

For the two-point experiments, the two X-probes have a small incidence angle relative to the mean flow ($-1^{\circ}$ to $-3^{\circ}$ for upper probe and $1^{\circ}$ to $3^{\circ}$ for lower probe) to guarantee that probes' bodies remain free of contact for small separations.
The angle of the wires relative to the mean flow ($\beta_1$ \& $\beta_2$ in figure \ref{fig:apparatus}a) differ from the standard $\pm 45^{\circ}$ not only due to the incidence angle of the X-probes but also due to the manual soldering of the wires to the prongs. 
For the upper X-probe $\beta_1 = 48^{\circ}$ and $\beta_2=-50^{\circ}$ and for the lower probe $\beta_1 = 48^{\circ}$ and $\beta_2=-41^{\circ}$.

The probes are calibrated at the beginning and the end of each set of measurements using a fourth-order polynomial and a velocity-pitch map for the SW and XW measurements, respectively.
For the two-point/two-component measurements, at the start of the calibration procedure the X-probes are separated by $\Delta y = 55$mm, which was deemed sufficient to avoid aerodynamic interference between the probes whilst also being sufficiently far from the walls ($>130$mm) at all calibration incidence angles. 
The flow temperature variation from beginning of the first calibration to end of second calibration was typically less than $1^{\circ}$ thus avoiding the need for temperature corrections to the calibrations \cite[]{Perry82}.

\subsection{Probe resolution and mutual interference} \label{sec:resolution}

For the vast majority of the data acquired with a SW, the sensor's length is smaller than five Kolmogorov microscales, i.e.  $l_w/\eta < 5$. Consequently the estimated bias in the measured dissipation rate, caused by the lack of resolution,  is typically less than $10\%$ \cite[see e.g.][figure 19 and note that the errors related to the sampling rate are negligible in the present data]{B08}. 
Note that the resolution of the measurements increases slightly as the flow decays and the Kolmogorov microscale grows (since $\eta \sim \varepsilon^{-1/4}$ and $\varepsilon$ is decreasing). 
However, it has been estimated that the differences in the resolution of the sensor throughout the downstream extent of the measurements are small and therefore the bias in the dissipation estimates is approximately the same at the furthermost upstream and downstream measurement locations. 
Since this bias is relatively small and does not affect the functional form of the longitudinal profiles of the dissipation, it is preferred not to correct the dissipation estimates obtained from the SW data.

For the $2\times$XW data, however, there are many additional sources of error that need to be addressed. 
The common sources of error in the measurement of transverse velocity gradients using two parallel single- or X-probes are their finite resolution (individually and of the array), errors in the calibrations of the probes, electronic noise and mutual interference (\citealp{AntoniaBC84}, see also \citealp{Mestayer79,ZA96,ZA02}).  
Errors arising from differences in probe calibrations and electronic noise contamination on each of the probes were found to be negligible when the probe separation is larger than $3\eta$ \cite[p. 548 of][]{ZA95}, here $\Delta y > 4\eta$. 
Concerning thermal interference due to the proximity of the probes, \cite{AntoniaBC84}  measured quantities like $\overline{(\partial u/\partial x)^2}$ with both probes operating and observed that the quantities remained unchanged when one of the probes was switched off. The same tests were repeated here with the X-probes at the minimum separation and it is corroborated that there is no evidence of thermal interference.  

\begin{figure}
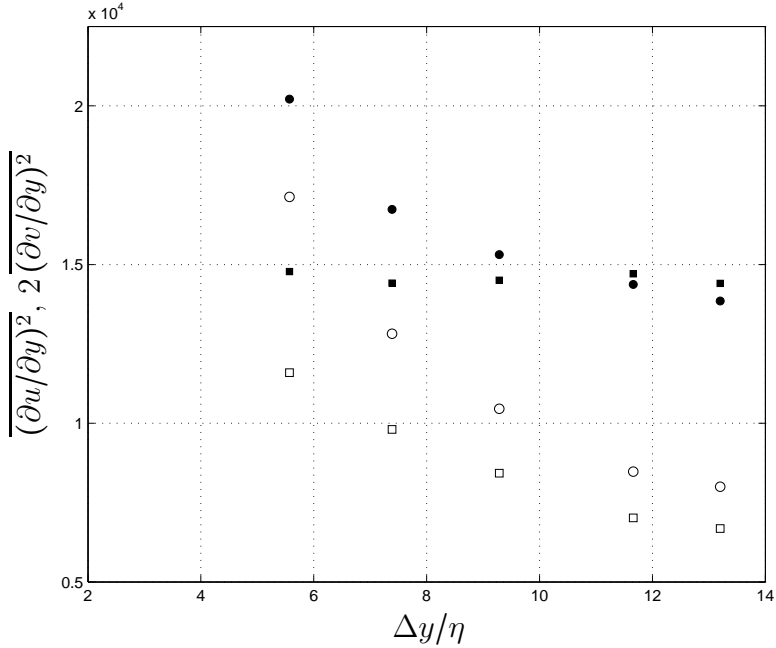

\centering
\begin{lpic}{CorrectionDissipation(100mm)}
\lbl[W]{2,78,90;$\overline{(\partial u /\partial y)^2}$, $2\,\overline{(\partial v /\partial y)^2}$}
\lbl[W]{92,2;$\Delta y /\eta$}
\end{lpic}
\caption[Dependence of mean square lateral velocity derivatives on X-probe separation]{Dependence of mean square lateral velocity derivatives (\protect\raisebox{-0.5ex}{\FilledSmallSquare} $\!|\!\!$ \protect\raisebox{-0.5ex}{\SmallSquare}) $\overline{(\partial u/\partial y)^2}$ and (\protect\raisebox{-0.5ex}{\FilledSmallCircle} $\!|\!\!$ \protect\raisebox{-0.5ex}{\SmallCircle}) $2\overline{(\partial v/\partial y)^2}$ on X-probe separation, $\Delta y/\eta$; (empty symbols) raw and (filled symbols) compensated measurements for the attenuation due to finite resolution.}
\label{fig:CorrDissipation}
\end{figure}

The aerodynamic interference due to the proximity of the X-probes depends on the configuration of the measurement apparatus and was investigated with the data from a precursory experiment (measuring RG115-generated turbulence). 
These experiments consisted of traversing, for each downstream location, the X-probes from $\Delta y=1.2$mm to $\Delta y=260$mm with the centroid positioned at (i) $(y,z)=(0,0)$mm, i.e. the centreline and (ii) $(y,z)=(-57.5,0)$mm, i.e. behind a bar. 
For each downstream location there is a region, $y=-130$mm to $73$mm which is measured twice.
In particular, the regions around $(y,z)=(0,0)$ and $(y,z)=(-57.5,0)$mm are measured when the probes are closely spaced, $\Delta y \approx 1.2$mm, and far apart, $\Delta y \approx 115$mm ($2\times 57.5$mm), thus allowing the assessment of aerodynamic interference on single point statistics.
The results show that for $\Delta y < 2$mm the error is never larger than 4\% in quantities like $U$, $u'$, $v'$, $\overline{(\partial u/\partial x)^2}$, $\overline{(\partial v/\partial x)^2}$. 
Higher order statistics such as the Skewness and Kurtosis of both velocity components are less influenced by the X-probes' proximity. 
However, the transverse component of the mean velocity, $V$, is severely influenced by the proximity of the X-probes and for $\Delta y<2$mm errors up to $\pm 0.5\mathrm{ms}^{-1}$ (inlet velocity, $U_{\infty}=10\mathrm{ms}^{-1}$) are observed and it may be responsible for the overestimation of the lateral mean square velocity derivative, $\overline{(\partial v/\partial y)^2}$ (see figure \ref{fig:CorrDissipation} and the discussion below).

The resolution of each individual X-probe is always better than $l_w\approx d_w = 3.5\eta$ leading to relatively small finite resolution correction factors to the mean square velocity streamwise derivatives  ($<15\%$).
The correction factors are defined as the ratio between finite resolution and actual values of the 
mean square velocity derivatives ($r_{1,1}\equiv \overline{(\partial u/\partial x)^2}^{\,m}/\overline{(\partial u/\partial x)^2}$ and $r_{2,1}\equiv \overline{(\partial v/\partial x)^2}^{\,m}/\overline{(\partial v/\partial x)^2}$, superscript $m$ indicating measured values, but are omitted throughout the thesis). 
In the present study, we use the DNS based correction factors obtained by \cite{B08}.
 
The finite separation between the two X-probes, in addition to the finite resolution of each probe and their aerodynamic interference, influences the estimation of the mean square velocity transverse derivatives, $\overline{(\partial u/\partial y)^2}$ and $\overline{(\partial v/\partial y)^2}$ \cite[]{ZA95,ZA96}. 
The correction factors, $r_{1,2}$($\equiv \overline{(\partial u/\partial y)^2}^{\,m}/\overline{(\partial u/\partial y)^2}$) and $r_{2,2}$ ($\equiv \overline{(\partial v/\partial y)^2}^{\,m}/\overline{(\partial v/\partial y)^2}$), are obtained from figure 3 in \cite{ZA96} to compensate the attenuation due to finite separation (the X-probe geometry for which the correction factors were obtained is not too dissimilar to the one used here).
The influence of the aerodynamic interference of the X-probes in the measurement of the velocity transverse derivatives is assessed by calculating the derivatives for $\Delta y = 1.2, 1.6, 2.0, 2.5$ and $3.0$mm, correcting for the different resolutions and comparing the results. 
If the aerodynamic interference is negligible and the finite separation is correctly compensated, the mean square velocity derivatives should be the same. 
In figure \ref{fig:CorrDissipation}, one example of such a comparison is made for measurements in the lee of the RG115 at  $x=2150$mm.
It is shown that the corrected $\overline{(\partial u/\partial y)^2}$ is indeed roughly independent of the X-probe separation, but $\overline{(\partial v/\partial y)^2}$ is not.
This may be due to the aerodynamic interference already observed in the spanwise mean velocity, $V$. 
Note that the value of $2\overline{(\partial v/\partial y)^2}$ seems to be tending towards the value of $\overline{(\partial u/\partial y)^2}$, which is the expected proportion for locally isotropic turbulence.
This topic is further discussed in \S \ref{sec:Eps}.

\subsection{Electronic testing} \label{sec:SineWave}

Hot-wire anemometry is often used to measure high frequency events in the turbulent field such as velocity derivative statistics. 
However, the ability to unbiasedly measure such statistics heavily relies upon keeping a sufficient high signal-to-noise ratio, at least up to $k\eta \approx 1$ (by keeping a sufficiently small spectral energy density noise floor) and assuring that the anemometer has an unbiased response up to that frequency. 
The topic of noise is dealt in the next section (\S \ref{sec:noise}).

Electronic testing is a practical way of assessing the frequency response of the anemometer to velocity perturbations. 
The maximum frequency response, or 'cutoff frequency' is commonly defined as the $-3$dB point, i.e. the frequency at which there has been an attenuation of 3dB between the square of the velocity fluctuation on the wire and the corresponding output square voltage of the anemometer.
This corresponds to the half-power attenuation ($10\log_{10}(1/2)\approx-3$dB), commonly used in the literature pertaining to electrical engineering and physics.
Two different, although in principle consistent, tests are usually used to determine the  cutoff frequency, the square-wave and the sine-wave tests \cite[]{F77}. 
These electronic tests consist of injecting a perturbation signal on one arm of the bridge and measure the system's response to it. 

Due to the simplicity of the square-wave test it became, overwhelmingly, the standard electronic test to estimate the 'cutoff frequency', both for off-the-shelf and in-house designed anemometry systems. 
However, from an experimentalist point of view, any attenuation beyond 0.1dB -- 0.2dB (2\% -- 5\%) in the measurement signal would be undesirable and attenuations approaching half-power would be unacceptable.
Unfortunately, the square-wave test only allows the estimation of the -3dB cutoff frequency\footnote{In principle, it also possible to extract a Bode plot with the square-wave test, but one would need a very fast digital acquisition system to be able to recover a good approximation to the perturbations spectrum, making it impractical. Furthermore, on the standard square-wave tests in-build in the anemometry systems one does not have access to the perturbation signal.}. 

\begin{figure}
\centering
% quick.m4
% To compile this macro into a tex file run the following in a shell
% m4 -I  circuit_macros pstricks.m4 libcct.m4 SineWaveCircuit.m4 > SineWaveCircuit.pic
% circuit_macros/dpic/dpic -p SineWaveCircuit.pic > SineWaveCircuit.tex 

\psset{unit=1in,cornersize=absolute,dimen=middle}%
\begin{pspicture}(-0.125,-0.02)(2.434722,0.826389)%
% dpic version 2012.07.14 option -p for PSTricks 0.93a or later
\psset{linewidth=0.8pt}%
\psset{linewidth=0.8pt}%
%\psset{noCurrentPoint}
\psset{arrowsize=1.1pt 4,arrowlength=1.64,arrowinset=0}%
\psline(0,0)(0,0.25)
\pscircle(0,0.375){0.125}
\psbezier(-0.0625,0.375)
(-0.061485,0.378179)(-0.059674,0.383807)(-0.058477,0.387506)
(-0.057279,0.391206)(-0.055288,0.39717)(-0.054052,0.400761)
(-0.052816,0.404351)(-0.050623,0.410214)(-0.049179,0.413789)
(-0.047735,0.417364)(-0.045391,0.422477)(-0.043971,0.425151)
(-0.04255,0.427825)(-0.040243,0.431358)(-0.038843,0.433002)
(-0.037443,0.434646)(-0.035162,0.436443)(-0.033774,0.436997)
(-0.032386,0.43755)(-0.030114,0.43755)(-0.028726,0.436997)
(-0.027338,0.436443)(-0.025057,0.434646)(-0.023657,0.433002)
(-0.022257,0.431358)(-0.01995,0.427825)(-0.018529,0.425151)
(-0.017109,0.422477)(-0.014765,0.417364)(-0.013321,0.413789)
(-0.011877,0.410214)(-0.009497,0.403807)(-0.008032,0.399552)
(-0.006568,0.395297)(-0.004161,0.388032)(-0.002685,0.383407)
(-0.001208,0.378783)(0.001208,0.371217)(0.002685,0.366593)
(0.004161,0.361968)(0.006568,0.354703)(0.008032,0.350448)
(0.009497,0.346193)(0.011877,0.339786)(0.013321,0.336211)
(0.014765,0.332636)(0.017109,0.327523)(0.018529,0.324849)
(0.01995,0.322175)(0.022257,0.318642)(0.023657,0.316998)
(0.025057,0.315354)(0.027338,0.313557)(0.028726,0.313003)
(0.030114,0.31245)(0.032386,0.31245)(0.033774,0.313003)
(0.035162,0.313557)(0.037443,0.315354)(0.038843,0.316998)
(0.040243,0.318642)(0.04255,0.322175)(0.043971,0.324849)
(0.045391,0.327523)(0.047735,0.332636)(0.049179,0.336211)
(0.050623,0.339786)(0.052816,0.345649)(0.054052,0.349239)
(0.055288,0.35283)(0.057279,0.358794)(0.058477,0.362494)
(0.059674,0.366193)(0.061485,0.371821)(0.0625,0.375)
\psline(0,0.5)(0,0.75)
\psline(0,0.75)(0.75,0.75)
\pscircle[fillstyle=solid,fillcolor=black](0.75,0.75){0.02}
\psline(0.75,0.75)(0.75,0.5)
(0.791667,0.479167)
(0.708333,0.4375)
(0.791667,0.395833)
(0.708333,0.354167)
(0.791667,0.3125)
(0.708333,0.270833)
(0.75,0.25)
(0.75,0)
\uput{2.5bp}[l](0.708333,0.375){$ 50\hspace*{0.5mm}\Omega$}
\pscircle[fillstyle=solid,fillcolor=black](0.75,0){0.02}
\psline(0.75,0.75)(1,0.75)
(1.020833,0.791667)
(1.0625,0.708333)
(1.104167,0.791667)
(1.145833,0.708333)
(1.1875,0.791667)
(1.229167,0.708333)
(1.25,0.75)
(1.5,0.75)
\uput{2.5bp}[u](1.125,0.791667){$ 2.5\hspace*{0.5mm}\mathrm{k}\Omega$}
\psline(1.5,0.75)(1.5,0.5)
(1.541667,0.479167)
(1.458333,0.4375)
(1.541667,0.395833)
(1.458333,0.354167)
(1.541667,0.3125)
(1.458333,0.270833)
(1.5,0.25)
(1.5,0)
\uput{2.5bp}[l](1.458333,0.375){$ \mathrm{HW}$}
\pscircle[fillstyle=solid,fillcolor=black](1.5,0){0.02}
\psline(1.5,0.75)(2.25,0.75)
\psline(2.25,0.75)(2.25,0.25)
\psline(2.25,0.25)(2.4,0.25)
\pscircle[fillstyle=solid,fillcolor=black](2.4,0.25){0.02}
\pscircle[fillstyle=solid,fillcolor=black](2.4,0){0.02}
\uput{2.5bp}[ur](2.4,0.203704){$ +$}
\uput{2.5bp}[r](2.4,0.125){$ \mathrm{CTA}$}
\uput{2.5bp}[dr](2.4,0.046296){$ -$}
\psline(2.4,0)(0,0)
\end{pspicture}%
\caption[Diagram of the circuit used to inject an electronic perturbation signal to the CTA]{Diagram of the circuit used to inject an electronic perturbation signal to the CTA. 
The output of the data acquisition system (NI-6229 USB) is used as signal generator. 
The amplitude of the generated signal is adjusted so that the perturbation signal into the CTA is less than 0.1V peak-to-peak. }
\label{fig:SineWaveCircuit}
\end{figure}
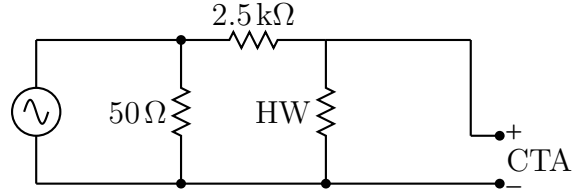

Conversely, the sine-wave test is a standard method to extract a complete Bode diagram of the system, but it is not readily available in the current state-of-the-art anemometry systems. 
Therefore, an external electronic perturbation method was implemented with the help of Dantec Dynamics (in particular Robert Jaryczewski) and the author's colleague, Anthony Oxlade.
The circuit diagram is shown in figure \ref{fig:SineWaveCircuit}. The main purpose of this electric circuit is to match the impedance between the CTA bridge and the signal generator. 
 
Note that, from the electronic circuit analysis of CTAs  \cite[][ch.3 and references therein]{F77,MeasurementBook} it is clear that the CTA bridges respond to the perturbation signal as well as its time derivative \cite[see eq. 12 in][]{F77}.
This contrasts with the response of the CTA bridge to resistance fluctuations which is insensitive to their rate of change.
Therefore, the Bode plot is expected to have a constant gain at lower frequencies (i.e. the bridge responds mostly to the signal rather than its derivative) and to be proportional to the square of the frequency ($\sim f^2$) at higher frequencies (i.e. the bridge responds mostly to the derivative of the signal). For even higher frequencies the response becomes attenuated due to the thermal capacity of the wire.
%(For the Bode plot, the input is the generated perturbation signal and the output is the CTA output signal.)

\begin{figure}[ht!]
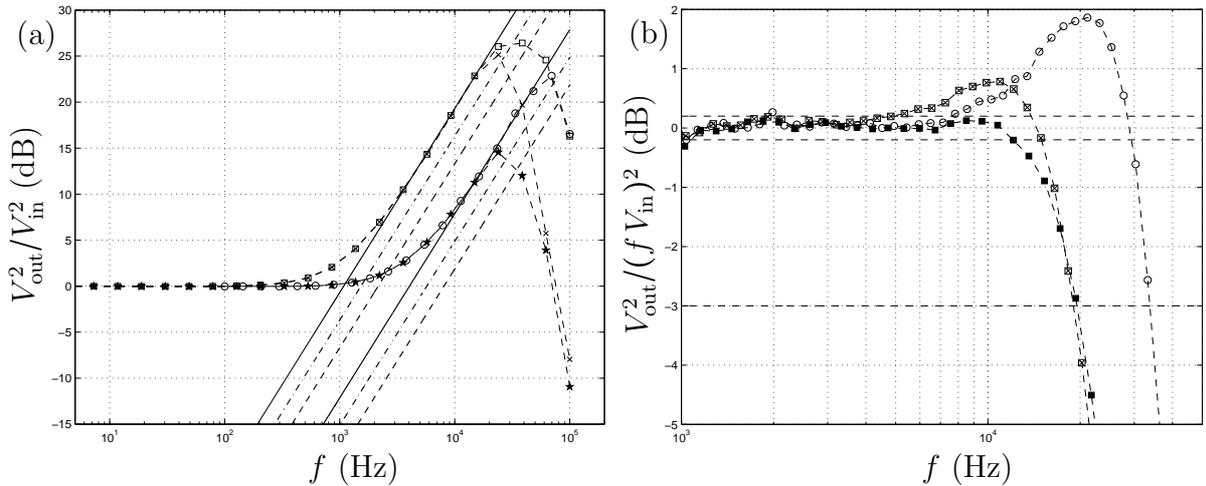

\centering
\begin{minipage}[c]{0.5\linewidth}
   \centering
   \begin{lpic}[b(1mm)]{BodePlotA(77mm)}
   \lbl{4,130;(a)}
   \lbl[W]{1,78,90;$V_{\mathrm{out}}^2/V_{\mathrm{in}}^2$ (dB)}
   \lbl[W]{100,0;$f$ (Hz)}
   \end{lpic}
\end{minipage}%
\begin{minipage}[c]{0.5\linewidth}
   \centering 
   \begin{lpic}[b(1mm)]{BodePlotB(75mm)}
   \lbl{4,130;(b)}
   \lbl[W]{1,78,90;$V_{\mathrm{out}}^2/(f\,V_{\mathrm{in}})^2$ (dB)}
   \lbl[W]{100,0;$f$ (Hz)}
   \end{lpic}
\end{minipage}
\caption[Performance assessment of the CTA using different hot-wires]{Performance assessment of the CTA using different hot-wires. (a) Sine-wave test using (\protect\raisebox{-0.5ex}{\SmallSquare} $\!|\!\!$ \protect\raisebox{-0.5ex}{\SmallCross}) SW$2.5\mu$m with a low-pass filter of 100kHz and 30kHz, respectively and  (\protect\raisebox{-0.5ex}{\SmallCircle} $\!|\!\!$ \protect\raisebox{0ex}{\ding{72}}) SW$1\mu$m with a low-pass filter of 100kHz and 30kHz, respectively. Solid lines are best fits to the $f^2$-region of the response and the dash-dotted and dashed lines are $-3$dB and $-6$dB offsets. 
(b) CTA response to white-noise perturbations, normalised by the best fit in the $f^2$-region, using XW$5\mu$m with different system setups, (\protect\raisebox{-0.5ex}{\FilledSmallSquare}) optimal setup, (\protect\raisebox{-0.5ex}{\rlap{\SmallSquare}\SmallCross}) incorrect inductance adjustment and (\protect\raisebox{-0.5ex}{\SmallCircle}) excessive bridge gain. The horizontal dashed lines reference gains of $\pm 0.2$dB and $-3$dB.}
\label{fig:SineWave}
\end{figure}

The typical Bode gain plot of the SW$2.5\mu$m and the SW$1\mu$m is shown in figure \ref{fig:SineWave} for two different analogue low-pass filters. 
Note that, the $\sim f^2$ region of the response depends on the feedback control system of the CTA, which is compensating for the response lag induced by the thermal capacitance of the wire.
Therefore, for thinner wires the $\sim f^2$ region is manifested at higher frequencies.
Note also that, during the experiments the low-pass filter is typically set at 30kHz, but to show the system's response with negligible influence from the signal conditioner the filtering frequency  is increased to 100kHz.

The $f^{0\mathrm{dB}}_{\mathrm{cutoff}}$ for the several probes are estimated from the sine-wave tests, whereas the $f^{-3\mathrm{dB}}_{\mathrm{cutoff}}$ are obtained from the in-built square-wave test (see table \ref{table:HW}). 
Note that the $f^{-3\mathrm{dB}}_{\mathrm{cutoff}}$ can also be recovered from the sine-wave tests from the abscissa of the interception between the $\sim f^2$ line, offset by $-3$dB, and the response curve of the system \cite[]{F77}.
However, unlike  \cite{F77}, it is found that the abscissa of the $-6$dB interception is closer to the frequency obtained from the in-built square-wave test.
From Dantec's measurement guide \cite[]{Dantec}, it seems that the $-3$dB refers to the ratio between the amplitude of the input and output signals rather than their square and thus their square-wave test results correspond to a $-6$dB attenuation ($\approx1/4^{\mathrm{th}}$) of the squared amplitude.
Furthermore, as it is shown in table \ref{table:HW}, it turns out that the highest frequency up to which there is a flat response is typically half the frequency of the reference $-3$dB cutoff (from square-wave tests),
\[ 
f^{0\mathrm{dB}}_{\mathrm{cutoff}}\approx \frac{f^{-3\mathrm{dB}}_{\mathrm{cutoff}}}{2} 
\]

Note also, the paramount effect of the bridge balance on the response of the system (figure \ref{fig:SineWave}b). 
When balancing the CTA bridge with a given hot-wire, the experimentalist is given the choice of the gain of the feedback amplifier and inductance compensation, amongst other bridge balancing parameters (which also depend on the design of the CTA bridges).
Using the sine-wave test (or equivalently a white-noise test, which leads to the same results) the experimentalist has the possibility to adjust the settings until an optimal response is obtained (filled squares in figure \ref{fig:SineWave}b). 
However, as it should be clear from figure \ref{fig:SineWave}b, all the non-optimal bridge balances lead to a higher  $f^{-3\mathrm{dB}}_{\mathrm{cutoff}}$ in expense of an overshoot and therefore the  $f^{0\mathrm{dB}}_{\mathrm{cutoff}}$ is actually reduced.
Unfortunately, with a square-wave test it is far from straightforward to know what the optimal response is (the shape of the square-wave response for three settings shown in figure \ref{fig:SineWave}b is remarkably similar) and therefore there is a tendency to choose the setting that gives the highest $f^{-3\mathrm{dB}}_{\mathrm{cutoff}}$ and, consequently, lower $f^{0\mathrm{dB}}_{\mathrm{cutoff}}$. 

\subsubsection{Comparison with other anemometry systems}

\begin{figure}[t!]
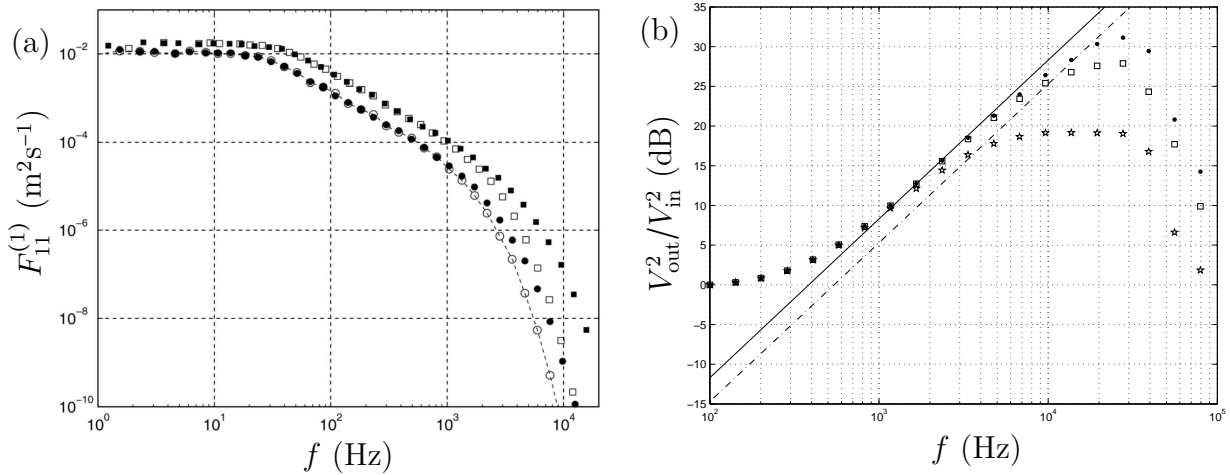

\centering
\begin{minipage}[c]{0.5\linewidth}
\centering
\begin{lpic}[l(-10mm)]{Fig1PoF(80mm)}
\lbl{30,380;(a)}
\lbl[W]{28,250,90;$F_{11}^{(1)}$ (m$^2$s$^{-1}$)}
\lbl[W]{320,6;$f$ (Hz)}
\end{lpic}
\end{minipage}%
\begin{minipage}[c]{0.5\linewidth}
\centering
\begin{lpic}[b(3mm)]{SineWaveTestAALabCh1(75mm)}
\lbl{0,130;(b)}
\lbl[W]{1,78,90;$V_{\mathrm{out}}^2/V_{\mathrm{in}}^2$ (dB)}
\lbl[W]{100,-1;$f$ (Hz)}
\end{lpic}
\end{minipage}
\caption[Performance comparison between two different anemometers]{(a) Comparison of the longitudinal energy spectra in the lee of FSG18''x18'' ($x=1850$mm) measured with two anemometry systems: 
(\protect\raisebox{-0.5ex}{\SmallCircle}) AN-1005, $U_{\infty}=10$ms$^{-1}$; 
(\protect\raisebox{-0.5ex}{\SmallSquare}) AN-1005, $U_{\infty}=15$ms$^{-1}$; 
(\protect\raisebox{-0.5ex}{\FilledSmallCircle}) Streamline, $U_{\infty}=10$ms$^{-1}$; 
(\protect\raisebox{-0.5ex}{\FilledSmallSquare}) Streamline, $U_{\infty}=15$ms$^{-1}$.
(b) Performance assessment of the AN-1005  for $U_{\infty}=10$ms$^{-1}$ using different signal amplifier gains ($G$),
(\protect\raisebox{-0.5ex}{\FilledSmallCircle}) $G=1$, 
(\protect\raisebox{-0.5ex}{\SmallSquare}) $G=5$ and 
(\protect\raisebox{0ex}{\ding{73}}) $G=12$. 
The solid line is a best fit to the $f^2$-region of the response and the dash-dotted line is a $-3$dB offset. } 
\label{fig:PoF}
\end{figure}

%\begin{figure}[t!]
%\centering
%\begin{lpic}{Fig2PoF(120mm)}
%\lbl[W]{5,60;$C_{\varepsilon}^{1(1)}$}
%\lbl[W]{78,5;${Re}_{\lambda}^{\mathrm{iso}}$}
%\end{lpic}
%\caption[Influence of anemometer performance on $C_{\varepsilon}^{1(1)}$ versus $Re_{\lambda}^{\mathrm{iso}}$]{ Normalised energy dissipation $C_{\varepsilon}^{1(1)}$ versus $Re_{\lambda}^{\mathrm{iso}}$  during the decay of turbulence generated by the FSG18''x18'' along the centreline for $1.5\mathrm{m} < x<3.2$m, measured with different anemometers: 
%(\protect\raisebox{-0.5ex}{\rlap{\Circle}\SmallCross}) 55M10, $U_{\infty}=5$ms$^{-1}$;
%(\protect\raisebox{-0.5ex}{\FilledSmallSquare}) 55M10, $U_{\infty}=15$ms$^{-1}$, 
%(\protect\raisebox{-0.5ex}{\SmallCircle}) Streamline, $U_{\infty}=10$ms$^{-1}$;
%(\protect\raisebox{-0.5ex}{\SmallSquare}) Streamline, $U_{\infty}=15$ms$^{-1}$;
%(\protect\raisebox{-0.5ex}{\FilledSmallTriangleRight}) AN-1005, $U_{\infty}=10$ms$^{-1}$; 
%(\protect\raisebox{-0.5ex}{\FilledSmallTriangleLeft}) AN-1005, $U_{\infty}=15$ms$^{-1}$. Solid and dashed lines follow $\sim Re_{\lambda}^{-1}$.} 
%\label{fig:PoF2}
%\end{figure}

Two other anemometry systems have been electronically tested and used to acquire data, namely, the A.A. Lab AN-1005 CTA system\footnote{Note that, the CTA channel used has the manufacturer's option 04 installed, i.e. high frequency response.} used by \cite{SV2007} and \cite{HV2007} and the DISA 55M10 CTA bridge with a DISA 55D26 signal conditioner used by \cite{MV2010}. 
These data are  used to investigate a discrepancy found between the experimental results presented by \cite{SV2007} and those presented by \cite{MV2010}. 
Specifically, \cite{SV2007} presented measurements along the centreline in the lee of three FSGs (one of the grids is FSG18''x18'' in the notation of this thesis) for $7\mathrm{ms}^{-1}<U_{\infty}<19\mathrm{ms}^{-1}$ and showed that for all the grids and inlet velocities their data followed $L_{11}^{(1)}/\lambda^{\mathrm{iso}}\approx 6.5\pm0.5$ with a range of  Reynolds numbers of $100<Re_{\lambda}^{\mathrm{iso}}<900$ (see their figure 9). 
Later,  \cite{MV2010} acquired data along the centreline in the lee of FSG18''x18'' using essentially the same experimental apparatus but with a different anemometry system and observed that $L_{11}^{(1)}/\lambda^{\mathrm{iso}} \approx 7.5,\,9.5,\,10.5$ for $U_{\infty} = 5.2,\,10,\,15\mathrm{ms}^{-1}$ (see their figure 15a). 
Even though the functional form of $L_{11}^{(1)}/\lambda^{\mathrm{iso}}$ throughout the decay is similar in the two experiments, i.e. approximately constant as $Re_{\lambda}^{\mathrm{iso}}$ varies, the numerical value of $L_{11}^{(1)}/\lambda^{\mathrm{iso}}$ is considerably different in the two experiments.
Furthermore,  $L_{11}^{(1)}/\lambda^{\mathrm{iso}}$ is an increasing function of $U_{\infty}$ in the data of \cite{MV2010} whereas in the data of \cite{SV2007} it is approximately independent of $U_{\infty}$. 
These discrepancies prompted the present assessment.

All the data are acquired in the 18''x18'' wind tunnel  in the lee of FSG18''x18'' and all the anemometers operate a SW$5\mu$m (see table \ref{table:HW}).
The experimental data obtained with the DISA 55M10 CTA closely match those obtained with the Streamline CTA system  (within the experimental error and excluding background noise).
However, comparing the energy spectra of the velocity signal acquired with the AN-1005 and the Streamline CTA, for the same downstream location and inlet velocity, it is observed that the spectra of the former data roll off faster by comparison with the latter (see figure \ref{fig:PoF}a). 
Using the sine-wave perturbation test described in \S \ref{sec:SineWave} it is found that the AN-1005 unit\footnote{At least the unit used by the author and by \citealp{HV2007,SV2007}. } exhibits problems in the in-built signal conditioning unit as well as in the in-built electronic perturbation test. 
In figure \ref{fig:PoF}b it is shown that the frequency response varies with the signal amplifier gain (which should not happen) and that for amplifier gains as low as 5--12 the response of system is severely attenuated for frequencies above 6kHz. 
On the other hand, the $f^{-3\mathrm{dB}}_{\mathrm{cutoff}}$ measured with the sine-wave test (without amplification of the output signal, i.e. $G=1$) is roughly 20kHz, which is much lower than the frequency response  inferred from the in-built electronic perturbation test ($f^{-3\mathrm{dB}}_{\mathrm{cutoff}}\approx  200$kHz). 
Lastly, it is noted that $f^{0\mathrm{dB}}_{\mathrm{cutoff}}$ is about 3-6kHz which is consistent with the steeper roll-off of the measured spectra.
%\cite{Hutchins} also reported problems concerning the frequency response of A.A. Lab Systems.

The immediate consequence of this high frequency attenuation (in fact a low-pass filter) is that the numerical values of the energy dissipation rate are underestimated and, conversely, the numerical values of the Taylor microscale are overestimated. 
As a result, the numerical values of $Re_{\lambda}^{\mathrm{iso}}$ are overestimated and the values of $C_{\varepsilon}^{1(1)}$ are underestimated, thus explaining the discrepancies between the data presented by \cite{SV2007} and \cite{MV2010}. %, as can be seen in figure \ref{fig:PoF2}. 
Nevertheless, other turbulent quantities do not meaningfully depend on high frequency statistics such as $U$, $u'$ and  $L_{11}^{(1)}$ and are not meaningfully affected by this high frequency attenuation.

\subsection{Noise and grounding of the instrumentation} \label{sec:noise}
There are three main noise contributions in the data acquired from hot-wire anemometry, (i) noise from the mains which power the instruments together with 'ground-loops', (ii) Johnson--Nyquist noise (or thermal noise) from the solid-state components of the anemometry system and (iii) quantisation noise from the discretisation/quantisation in the data acquisition (DAQ).

Starting with the latter, the quantisation error is related to the number of quantisation bits, $Q$ ($Q=16$ for the data acquisition card used). 
For a range $R$ of the quantisation sensor ($R=2$V for most SW$2.5\mu$m and SW$1\mu$m  measurements) the mean square noise is proportional to the the square of the resolution , i.e. $\overline{N^2}\propto(R/(2^Q-1))^2$ (for large signal resolution and assuming that the quantisation error is uniformly distributed $\overline{N^2}=1/12\,(R/(2^Q-1))^2$, \citealt{Widrow}). 
For the DAQ system used, the total mean square noise amounts to $\overline{N^2} \approx 1/2\,(R/(2^Q-1))^2$ (measured by short-circuiting the DAQ channel with a 50$\Omega$ resistor), which is larger than the estimation by \cite{Widrow}. Nevertheless, the total noise may take into account other DAQ noise sources other than quantisation. 

The quantisation error appears on the spectra of the measurements roughly as additive white-noise (since the quantisation error is well approximated to be delta correlated in time and approximately independent of the signal) with a level $\overline{N^2}/(f_{\mathrm{max}}-f_{\mathrm{min}})$ ($f_{\mathrm{max}}$ is half of the sampling frequency and $f_{\mathrm{min}}$ is the inverse of the total sampling time). 
Therefore, to decrease the quantisation noise floor one should, (i) use the minimum sensor range suitable for the measured analogue signal\footnote{Alternatively, one could maintain the sensor range and increase the gain of the signal conditioner amplifier. 
However, care must be taken with high gain amplifiers that may not have a constant gain over all the frequencies of interest.},
 (ii) sample as fast as possible and (iii) use acquisition cards with higher bit count. 

\begin{figure}
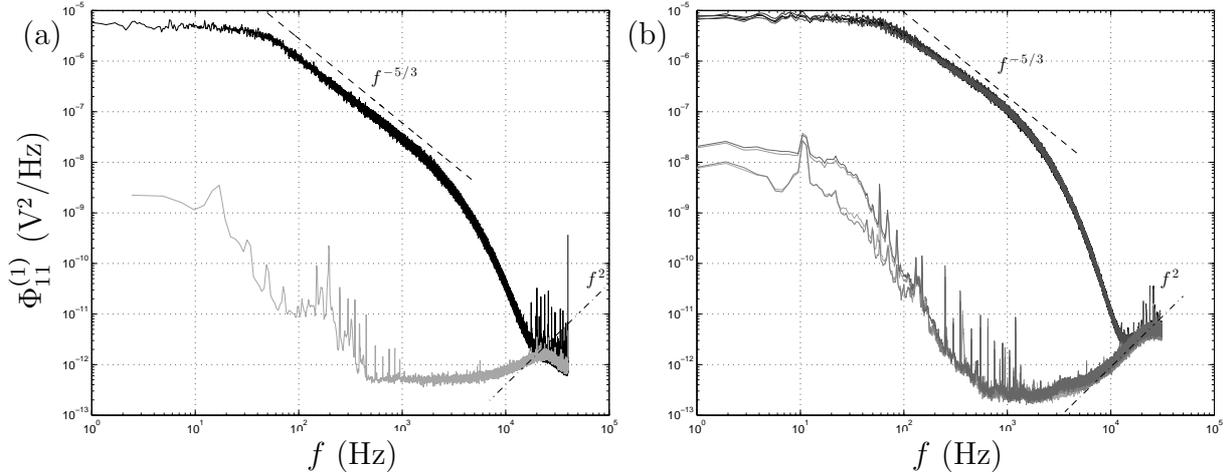

\centering
\begin{minipage}[c]{0.5\linewidth}
   \centering
   \begin{lpic}[b(1mm)]{NoiseSW(78mm)}
   \lbl{4,130;(a)}
   \lbl[W]{0,75,90;$\Phi_{11}^{(1)}$ (V$^2$/Hz)}
   \lbl[W]{100,1;$f$ (Hz)}
   \end{lpic}
\end{minipage}%
\begin{minipage}[c]{0.5\linewidth}
   \centering 
   \begin{lpic}[b(1mm)]{Noise2xXW(78mm)}
   \lbl{4,130;(b)}
   \lbl[W]{5,75,90;\hspace{8mm}${}^{}$\hspace{10mm}}
   \lbl[W]{100,1;$f$ (Hz)}
   \end{lpic}
\end{minipage}
\caption[Typical spectra of grid-generated versus free-stream turbulence]{(a) Spectra of the analogue signals acquired with  SW$2.5\mu$m measuring (grey) free-stream and (black) RG230-generated turbulence at the centreline, $x=1400$mm and $U_{\infty}=15$ms$^{-1}$ (DAQ range, $R=10$V and sampling rate of 80kHz).
(b)  Spectra of the four analogue signals acquired with  $2\times$XW$2.5\mu$m measuring (light grey) free-stream and (black and dark grey) RG115-generated turbulence at the centreline, $x=950$mm and $U_{\infty}=10$ms$^{-1}$ (DAQ range, $R=2$V and sampling rate of 62.5kHz). }
\label{fig:Noise}
\end{figure}

The mains noise and the ground-loop noise, on the other hand,  are manifested in the spectra as `spikes' (see figure \ref{fig:Noise}). 
The former is caused by an imperfect connection of the instruments (via the mains) to the `earth', i.e. without an infinite conductance, causing the ground to fluctuate (due to the return current of the many instruments and equipment of the building connected to the mains).  
The latter is due to slight electrical potential differences between the ground of the different instruments which cause a ground loop current (since the grounds are connected). 
In the present experiments, these two noise sources were minimised by (i) filtering the mains to which the CTA system is connected using a power conditioner (Oneac\textsuperscript{\textregistered} ConditionOne\textsuperscript{\textregistered} Series 1kVA), (ii) connecting the ground of the DAQ\footnote{The DAQ system is connected to the mains via an AC-DC transformer and it was observed that it was preferable to connect the DAQ directly to the mains rather than via the power conditioner.} directly to the ground of the CTA system and (iii) switching off all the stepper motors driving the traverse systems (which have a large return current to the ground) via a relay switch controlled by the DAQ.
Note that \cite{SV96} also reported a significant attenuation of the mains/ground loop noise using a power conditioner and an `uninterruptable power supply' (UPS). 
An UPS was also tested for the current setup (model: GE VH Series 1.5kVA UPS) but it was found that it did not further decreased the noise from the mains.  

Nevertheless, of all the noise sources, the Johnson--Nyquist noise \cite[]{Nyquist} turns out to be the most important at high frequencies. 
Even though the thermal noise of the solid state resistors can be shown to be `white' \cite[]{Nyquist}, the CTA bridge responds to these perturbations similarly than it does to the external sine-wave and/or white-noise perturbations (\S \ref{sec:SineWave}).
Therefore, at high frequencies this noise source follows $f^2$ (until the thermal capacitance of the wire leads to a significant attenuation of the response) and thus the spectral density of the noise is largest at frequencies where the turbulent signal has an exponentially decreasing spectral density (i.e. the dissipation region of the spectrum, see figure \ref{fig:Noise}). 
This appears to have been first noticed by  \cite{SV96}, but as pointed out by \cite{FF97} it was already implicit in the analysis of \cite{F77}.

There are many parameters influencing the onset of the $f^2$ region, e.g. the overheat ratio, the balance of the bridge and the mean convection velocity at the wire. 
For example, the latter effect is evident from the comparison between figure \ref{fig:Noise}a and figure \ref{fig:Noise}b. 
The onset of the $f^2$ region occurs at a slightly lower frequency for the latter (which leads to higher spectral densities) due to the fact that the mean convection velocity is smaller, not only because $U_{\infty}$ is two-thirds smaller but also because the wire is inclined by $45^{\circ}$ relative to the mean flow.
However, the largest effect can be achieved by changing the thermal capacity of the wire ($\propto d_w^2 l_w$, for a given wire material) and the best way to minimise the spectral density of this noise source, at high frequencies, is to use thinner and shorter wires (see figure \ref{fig:SineWave}a and discussion in \S \ref{sec:SineWave}). 

\section{Data acquisition and statistical convergence} \label{sec:DAQ}

The pressure and temperature measurements are digitally transferred to the computer using a parallel port. The analogue signal from the anemometers is sampled using a 16-Bit National Instruments NI-6229(USB) card.
The sampling frequency is chosen to be, at least higher than twice the analogue low-pass filtering frequency.
Typically the low-pass filter is set to 30kHz (and in some instances to 100kHz) and the sampling frequency is set to be at least 80kHz (up to 250kHz) for SW and XW experiments and  to 62.5kHz for $2\times$XW experiments (which is the maximum sampling frequency for four simultaneously acquired signals with the present hardware). 
The data acquisition and signal processing are performed with the commercial software MATLAB\tm.

The turbulent signals are typically acquired for 9min corresponding to $N=\mathcal{O}(10^5)$  independent samples (taking twice the integral-time scale as the characteristic time lag between independent samples). 
This is sufficient to converge the single-point statistics of interest in this work. 
Taking, for example, the RG60 data around the centreline ($x=1250$mm, $U_{\infty}=10$ms$^{-1}$, $N= 100\,800$, one of the X-probes from a $2\times$XW experiment) one can estimate the 95\% confidence intervals of the measurements  ($\pm 1.96 \sqrt{\mathrm{var}(m^{j}_r)}$; $m^{j}_r$ is the estimator of the r$^{\mathrm{th}}$ central moment of the $j^{\mathrm{th}}$ velocity component, see e.g. \citealt{BG96}). The 95\% confidence intervals estimated are, (i) $\pm0.02\%$ of $U_{\infty}$ for the mean velocity components, $U$ and $V$, (ii) $\pm 0.9\%$ of $\overline{u^2}$ and $\overline{v^2}$ for these second order moments and (iii) $\pm 1.4\%$ of $(\overline{u^2})^{3/2}$ for the third-order moments, $\overline{v^3}$ and $\overline{v^3}$.
Similarly, one can compute the confidence intervals for mixed moments such as $\overline{uv}$ and $\overline{vu^2}$ \cite[]{BG96}. The estimated $95\%$ confidence intervals of  $\overline{uv}$ and $\overline{vq^2}\approx \overline{v\,(u^2+2v^2)}$\footnote{Approximating $\overline{vw^2}$ as $\overline{vv^2}$.} are $0.6\%$ of $\sqrt{\overline{u^2\,}\overline{v^2}}$ and $0.7\%$ of  $\sqrt{\overline{v^2}}\,\overline{q^2}$, respectively.

\begin{figure}
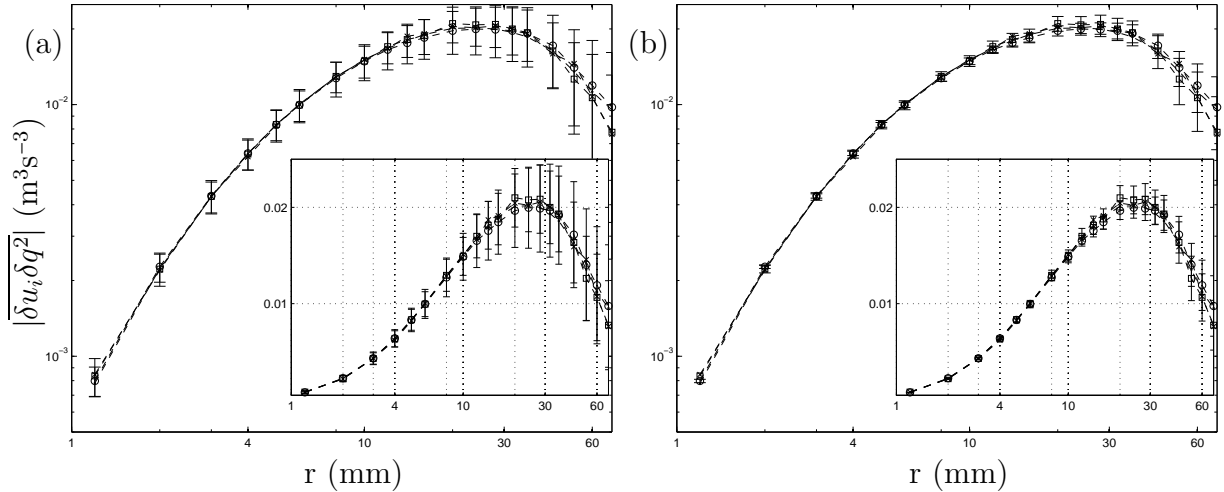

\centering
\begin{minipage}[c]{0.5\linewidth}
   \centering
   \begin{lpic}{ErrorBarsI(80mm)}
   \lbl{9,125;(a)}
   \lbl[W]{4,75,90;$|\overline{\delta u_i \delta q^2}|$ (m$^3$s$^{-3}$)}
   \lbl[W]{100,0;r (mm)}
   \end{lpic}
\end{minipage}%
\begin{minipage}[c]{0.5\linewidth}
   \centering 
   \begin{lpic}{ErrorBarsII(80mm)}
   \lbl{9,125;(b)}
   \lbl[W]{5,75,90;\hspace{8mm}${}^{}$\hspace{10mm}}
   \lbl[W]{100,0;r (mm)}
   \end{lpic}
\end{minipage}
\caption[Confidence intervals of $|\overline{\delta u_i \delta q^2}|$]{$|\overline{\delta u_i \delta q^2}|$ versus (\protect\raisebox{-0.5ex}{\SmallSquare} $\!|\!\!$ \protect\raisebox{-0.5ex}{\SmallCross}) longitudinal separations $(r_x,\,r_y)=(r,\,0)$ and (\protect\raisebox{-0.5ex}{\SmallCircle} $\!|\!\!$ \protect\raisebox{0ex}{\tiny{\FilledSmallCircle}}) transverse separations $(r_x,\,r_y)=(0,\,r)$; $95\%$ confidence intervals estimated with (a) global integral time-scale and (b) tailored integral time-scale characteristic of $\overline{\delta u_i \delta q^2}(r_x,\,r_y)$. Note that the same experiment is repeated twice and that (\protect\raisebox{-0.5ex}{\SmallSquare} $\!|\!\!$ \protect\raisebox{-0.5ex}{\SmallCircle}) represent exp. I and  (\protect\raisebox{-0.5ex}{\SmallCross} $\!|\!\!$ \protect\raisebox{0ex}{\tiny{\FilledSmallCircle}}) exp. II.}
\label{fig:ErrorBars}
\end{figure}

% Paragraph II - Convergence
For two-point statistics, such as those presented in ch. \ref{chp:6}, a similar analysis can be undertaken to show that the above acquisition time is sufficient.
One of the main statistics of interest for the two-point/two-component measurement is the triple structure function $\overline{\delta u_i \delta q^2}$ (where $\delta u_i\equiv u_i(\mathbf{X}+\mathbf{r}/2)- u_i(\mathbf{X}-\mathbf{r}/2)$), since it is related to the scale-by-scale energy transfer. 
It is also one of the most demanding in terms of statistical convergence, since it is a higher-order odd moment. 
The sampling variance ($\mathrm{var}(\overline{\delta u_i \delta q^2})$) can be estimated as,
\begin{equation}
\mathrm{var}(\delta u_i \delta q^2) = \frac{1}{N}\left(\overline{(\delta u_i \delta q^2)^2} - \overline{\delta u_i \delta q^2}^2\right).
\label{eq:vardudq2}
\end{equation}
%where $N$ is the number of independent samples. 
Note that $\overline{\delta u_i \delta q^2}$ is non-central statistical moment, hence equation 4 of \cite{BG96}, which is derived for central moments and was used above to assess the confidence intervals of the single-point statistics, has additional uncertainty terms which are not applicable, see \cite{KS58} for further details.

% Paragraph III - Repeatibility vs convergence
Measurements of  $\overline{\delta u_i \delta q^2}$, together with the estimated $95\%$ confidence intervals are shown in figure \ref{fig:ErrorBars}a (the same dataset for which the convergence of the single-point statistics was assessed is used here).
Notice, however, that the confidence intervals are undesirably large, which prompted an assessment of the repeatability (a somewhat more stringent test) of the measurements, by simply repeating the same measurement twice on two different days.
Remakably, the data from two different experiments overlay each other (figure \ref{fig:ErrorBars}a), and thus the confidence intervals are clearly overestimated.
This thus indicates that number of independent samples, $N$, based on the integral-time scale, are underestimated. 
Indeed, splitting the data into integral-time scale sized blocks and extracting a single sample of $\delta u_i \delta q^2(r_x,\,r_y)$ from each block leads to estimates of $\overline{\delta u_i \delta q^2}(r_x,\,r_y)$ with significantly more scatter, indicating that uncorrelated samples were lost. 
Instead of using the standard integral time-scale, one can define alternative de-correlation time-scales by taking the autocorrelation of $\delta u_i \delta q^2(r_x,\,r_y)$ at two times with varying lags and then integrating the resulting correlation functions. 
This methodology provides a tailored integral time-scale representative of the de-correlation length associated with  $\overline{\delta u_i \delta q^2}$ at each $(r_x,\,r_y)$, similar to what was done by \cite{YM2010}. 
Assuming that twice this tailored integral time-scale is the characteristic lag between independent samples of  $\delta u_i \delta q^2(r_x,\,r_y)$ one gets new estimates of $N(r_x,\,r_y)$, and consequently new confidence intervals, which are shown in figure \ref{fig:ErrorBars}b. 
(Note that, experiments performed with Particle Imagine Velocimetry obtain reasonable estimates of $\overline{\delta u_i \delta q^2}$, see e.g. \citealt{Moisy2011,Danaila2012}, even though the number of independent samples is $\mathcal{O}(10^3)$.)

The error bars of the spherically averaged divergence of $\overline{\delta u_i \delta q^2}$ (figures \ref{fig:KHMwInhomo} and \ref{fig:PeakValues}) include the $95\%$ confidence intervals plus the error due to the uncertainty of the vertical separation between the X-probe $\approx\!\pm 50\mu$m (see \S \ref{sec:WTB}). The two uncertainties are stacked with a standard propagation of error formula applied to the central differences scheme.

\section{Signal processing} \label{sec:processing}

The time-varying turbulent signal is converted into spatially-varying by means of Taylor's hypothesis\footnote{In \cite{VV2011} the local Taylor's hypothesis algorithm proposed by \cite{KMGC98} was employed. However, its use is discontinued since the author noticed that the proposed algorithm uses a resampling technique which acts itself as a low-pass filter. The latter may be the reason why \cite{KMGC98} found a better agreement with the correction to Taylor's hypothesis proposed by \cite{Lumley65}. Nevertheless, for low intensity flows such as those investigated here, both methodologies lead to similar results.} \cite[]{Taylor1938}.  Before Taylor's hypothesis is used the signal is digitally filtered at a wavenumber corresponding to $k\eta \ge 1.0$
(where $\eta\equiv (\nu^3/\varepsilon)^{1/4}$ is the Kolmogorov inner length-scale  and $k$ the wavenumber; the actual filtering frequency/wavenumber varies between $k\eta= 1.0$ and $k\eta =1.5$ depending on signal to noise ratio at these wavenumbers) using a $4^{th}$-order Butterworth filter to eliminate higher wavenumber noise.
 
The longitudinal and transverse spectra, $F^{(1)}_{11}(k)$ and $F^{(1)}_{22}(k)$, are calculated using an FFT based periodogram algorithm using a \emph{Hann} window with 50\% overlap and window length equivalent to at least $180$ integral-length scales. 

The mean square velocity derivatives in the streamwise direction are estimated from $F^{(1)}_{11}$ and $F^{(1)}_{22}$ as
\begin{equation*}
\overline{(\partial u/\partial x)^2} = \int_{k_{min}}^{k_{max}}k^2\,F^{(1)}_{11}(k)\, dk; \,\,\,
\overline{(\partial v/\partial x)^2} = \int_{k_{min}}^{k_{max}}k^2\,F^{(1)}_{22}(k)\, dk ,
\end{equation*}
where $k_{min}$ and $k_{max}$ are determined by the window length and the sampling frequency respectively. To negotiate the problem of low signal to noise ratios at high wavenumbers we follow \cite{antoniadissipation} and fit an exponential curve to the high wavenumber end of the spectra (it was checked that this does not change the mean square velocity derivatives by more than a few percent). 
The mean square velocity derivatives in the spanwise direction, $\overline{(\partial u/\partial y)^2}$ and $\overline{(\partial v/\partial y)^2}$, are estimated with finite differences.

%The Taylor microscale is calculated from twice the kinetic energy, $\overline{q^2}$, and the dissipation, $\varepsilon$($= \varepsilon^{\mathrm{iso,3}}$), as $\lambda = (5 \nu\, \overline{q^2}/\varepsilon)^{1/2}$ and the Kolmogorov microscale as $\eta = (\nu^3/\varepsilon)^{1/4}$.

%
%The longitudinal integral length-scales $L_{ii}^{(1)}$ are estimated as 
%\begin{equation*}
%L_{ii}^{(1)} = \frac{1}{B_{ii}^{(1)}(0)}\int_0^{r_L} B_{ii}^{(1)}(r) \, dr ,
%\end{equation*}
%where $B_{ii}^{(1)}(r) \equiv \overline{u_i()u_i(x+r)}/\overline{u(x)^2}$ is the
%auto-correlation function of the streamwise velocity fluctuations for
%streamwise separations $r$ and $r_L$ is maximum integration range
%taken to be about $10$ times the integral length scale. It was checked
%that (i) changing the integration limit $r_L$ by a factor between
%$2/3$ and $2$ has little effect on the numerical value of the integral
%scale and (ii) the choice of $r_L$, if large enough, does not
%influence the way that $L_u$ varies with downstream distance. The
%transverse integral scale is estimated in a similar way. 
\clearemptydoublepage
\chapter{Grid-generated turbulent flow}
\label{chp:3}

From the introduction (ch. \ref{chp:1}), the necessity to carry forward the research on FSG-generated turbulence and investigate the origin of the nonclassical dissipation behaviour should be clear. 
The first issue to be addressed is a \emph{ceteris paribus} comparison between RG- and FSG-generated turbulence. This is paramount to understand the role of additional fractal iterations in the observed behaviour.
This is followed by a comprehensive assessment of homogeneity and isotropy on comparable regions of the turbulent flows generated by RGs and FSGs. 
%Spectra measurements are shown in ch. \ref{chp:5}.

\section{Profiles of one-point statistics} \label{sec:topology}

%%%%%%%%%%%%%%%%%%%%%%%%%%%%%% Sub-Section %%%%%%%%%%%%%%%%%%%%%%%%%%
\subsection{Production and decay regions} \label{sec:wake}
\begin{figure}
\begin{minipage}[c]{0.5\linewidth}
   \centering
   \begin{lpic}{Thesis_TuAllGrids(78mm)}
   \lbl{6,130;(a)}
   \lbl[W]{100,0;$x/x_{\mathrm{peak}}$}   
   \lbl[W]{4,80,90;$u'_c/U_c$ (\%)}   
   \end{lpic}
\end{minipage}%
\begin{minipage}[c]{0.5\linewidth}
   \centering 
   \begin{lpic}{Thesis_IntegralScalesAllGrids(78mm)}
   \lbl{6,130;(b)}
   \lbl[W]{100,0;$x/x_{\mathrm{peak}}$}
   \lbl[W]{4,75,90;$L_{11}^{(1)}/M$}   
   \end{lpic}
\end{minipage}
\caption[Longitudinal profiles of $u'_c/U_{c}$ and $L_{11}^{(1)}$ along the centreline]{Longitudinal profiles along the centreline of (a) turbulence intensity and (b) integral length scale normalised by the mesh size for several turbulence generating grids. 
(\protect\raisebox{-0.5ex}{\FilledSmallSquare} $\!|\!\!$ \protect\raisebox{-0.5ex}{\SmallSquare}) RG230 at $U_{\infty}=10,\,20$ms$^{-1}$, respectively,
(\protect\raisebox{-0.5ex}{\SmallTriangleLeft} $\!|\!\!$ \protect\raisebox{-0.5ex}{\FilledSmallTriangleUp}) RG60 at $U_{\infty}=10,\,20$ms$^{-1}$, respectively,
(\ding{72} $|\!\!$ \ding{73}) FSG18''x18'' and FSG3'x3', respectively, at $U_{\infty}=15$ms$^{-1}$ and 
(\protect\raisebox{-0.5ex}{\rlap{\Circle}\FilledDiamondshape}) RG115 at $U_{\infty}=20$ms$^{-1}$.
Note that both abscissae and ordinates of the plots are in logarithmic coordinates.}
\label{fig:UmLu}
\end{figure}
% Production versus decay regions
The downstream evolution of the turbulent flow generated by the RGs, as well as by the FSGs, can be separated into two distinct regions, the production and the decay regions. 
The production region lies in the immediate vicinity downstream of the grid where the individual wakes generated by individual bars develop and interact.
This region extends as far downstream as where the wakes of the biggest bars interact, i.e. as far as that distance downstream where the width of these largest wakes is comparable to the largest mesh size $M$ \cite[][$M$ is the distance between parallel bars within a mesh]{MV2010}. 
There is only one mesh size in the case of RGs (\emph{cf.} figures \ref{fig:grids}c--e)  but for FSGs, which are made of many different square bar arrangements, i.e. meshes, of different sizes, $M$ refers to the largest mesh size (see figures \ref{fig:grids}a,b; see also table \ref{table:grids}).

In the case of the present RGs and of the particular type of space-filling low-blockage FSGs with high enough thickness ratio, $t_r$, (such as the ones studied here and previously by \citealp{SV2007} and \citealp{MV2010}; for the definition of $t_r$ see \S \ref{sec:grids}) the turbulent kinetic energy in the production region increases monotonically with downstream distance $x$ along the centreline until it reaches a maximum at $x=x_{\mathrm{peak}}$ (see figures \ref{fig:UmLu}a and \ref{fig:UmvsX}).
Further downstream, i.e. where $x>x_{\mathrm{peak}}$, the turbulence decays monotonically.
Along any other line parallel to the centreline the point downstream beyond which the turbulence decays monotonically occurs before $x_{\mathrm{peak}}$ as shown by \cite{JW1992} and \cite{Ertunc2010} for RGs and by \cite{MV2010} and \cite{Sylvain2011} for FSGs.
Note that this definition of production and decay separated by a plane perpendicular to the centreline located at  $x_{\mathrm{peak}}$ could be made more precise by defining the surface where the advection vanishes, $U_{k}\partial \overline{q^2}/\partial x_k = 0$ (i.e. where the turbulent kinetic energy, $K=\overline{q^2}/2$ is maximum).
Nevertheless, this distinction is only relevant to studies of the region upstream of $x_{\mathrm{peak}}$, whereas in the present thesis only the region downstream from $x_{\mathrm{peak}}$ is investigated. 

The downstream location of   $x_{\mathrm{peak}}$ for FSGs was shown to be proportional to a `wake-interaction length-scale', $x_* \equiv M^2/t_0$ \cite[]{MV2010} and the proportionality constant to be dependent on other grid details as well as upstream turbulence \cite[]{gomesfernandesetal12}.
The wake-interaction length-scale naturally extends to RGs as well.
Based on the data of \cite{JW1992}, $x_{\mathrm{peak}}\approx0.55 x_{*}$ for their RG with the largest $M$ (see table \ref{table:RGdata}) and for the present RG data, $x_{\mathrm{peak}}\approx 0.63 x_*$ for RG230 and RG115 and $x_{\mathrm{peak}}\simeq 0.4 x_*$ for RG60 (see table \ref{table:grids}).

Furthermore, \cite{MV2010} showed that the integral length scale is proportional to $M$, inline with what is typically found in RGs experiments \cite[see e.g.][]{CC71}. 
In figure  \ref{fig:UmLu}b the longitudinal profiles of  the longitudinal integral-length scale, $L_{11}^{(1)}$ normalised by $M$ are shown. 
The data corroborate that $L_{11}^{(1)}$ scales with $M$, but the grid geometry does play a role on the numerical value of their ratio and the growth rate of the former. 

The fact that $M$ collapses, to a first approximation, $L_{11}^{(1)}$ for a given grid geometry (RG or FSG), with different mesh sizes and blockage ratios (e.g. RG60 versus RG115 and RG230) and both near the grid and further downstream  (\emph{cf.} figure \ref{fig:UmLu}b), together with the fact that $M$ is the length-scale to be used in the definition of $x_*$(\footnote{Note that, $x_*= M(M/t_0)$ and therefore $M/t_0$ is a second parameter that should be taken into account in the normalisation of the downstream coordinate when comparing grids with different designs.
For a RG, $\sigma = t_0/M(2-t_0/M)$, and thus, for the vast majority of the RG experiments presented in the literature (see table \ref{table:RGdata}), the blockage ratios vary between $\approx30\%$ and $\approx 44\%$ and thus  $4<M/t_0<6$, therefore the normalisation with $M$ rather than with $x_{*}$ or $x_{\mathrm{peak}}$ yield unsuspectingly similar results.}), motivates the choice of $M$ as the length-scale of the grid for both RGs (as originally proposed by \citealp{Taylor1935}) and FSGs to be used in the definition of the `global' or `inlet' Reynolds number. 
The turbulent Reynolds number based on $L_{11}^{(1)}$, i.e. $Re_{L^{1(1)}} = u'L_{11}^{(1)}/\nu$,  is also proportional to the `inlet' or `global' Reynolds number based on $M$, i.e. $Re_M = U_{\infty}M/\nu$, for a given grid geometry at a given downstream location, since the r.m.s. velocity  scales with the mean inlet velocity as shown in figure \ref{fig:UmLu}a. 
However, unlike $Re_M$, $Re_{L^{1(1)}}$ varies with position.
%The thesis' focus is only on the decay region downstream of $x_{\mathrm{peak}}$ and up to a distance $x$ from the grid equal to the first few multiples of $x_{\mathrm{peak}}$.
%As it will be seen in the following chapter, in this region both RGs and FSGs exhibit a very different dissipation behaviour contrasting with the classical $\varepsilon \sim u'^3/L$, which is denoted as nonequilibrium dissipation behaviour (this terminology will be motivated in the subsequent chapters of the thesis). 
%For one regular grid, RG60, it will also be seen that the downstream extent of the measurements is sufficient to capture both the nonequilibrium and equilibrium regions and that the downstream extent of the former is about $5x_{\mathrm{peak}}$.

In the remainder of this chapter the homogeneity and isotropy of the RG- and FSG-generated turbulent flows are investigated.

%%%%%%%%%%%%%%%%%%%%%%%%%%%%%% Sub-Section %%%%%%%%%%%%%%%%%%%%%%%%%%
\subsection{Mean velocity deficit versus turbulent kinetic energy \mbox{decay}} \label{sec:profiles}
%%%%%% Figures %%%%%%%%%%%%%
\begin{figure}[ht!]
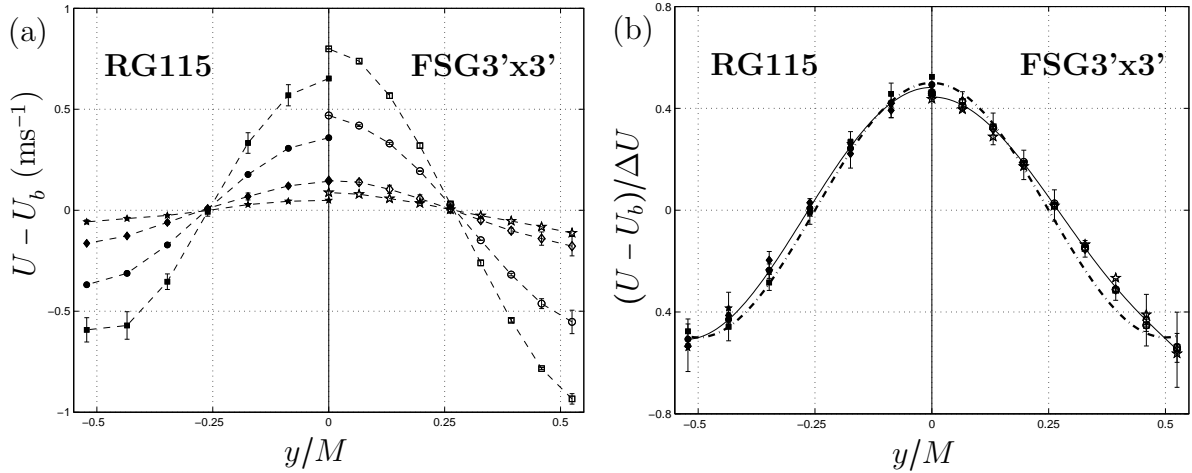

\begin{minipage}[c]{0.5\linewidth}
   \centering
   \begin{lpic}{UmMinusUbProfileRG115FSG3x3(78mm)}
   \lbl{10,130;(a)}
   \lbl{50,120;\bf{RG115}}
   \lbl{150,120;\bf{FSG3'x3'}}
   \lbl[W]{100,0;$y/M$}
   \lbl[W]{5,85;${}^{}_{}$\hspace{15mm}}         
   \lbl[W]{10,85,90;$U-U_b^{}$  (ms$^{-1}$)}         
   \end{lpic}
\end{minipage}%
\begin{minipage}[c]{0.5\linewidth}
   \centering 
   \begin{lpic}{UmMinusUbOverDeltaUProfileRG115FSG3x3(78mm)}
   \lbl{10,130;(b)}
   \lbl{50,120;\bf{RG115}}
   \lbl{150,120;\bf{FSG3'x3'}}
   \lbl[W]{100,0;$y/M$}
   \lbl[W]{9,75,90;$(U-U_b)/\Delta U^{}$}      
   \end{lpic}
\end{minipage}
\caption[Mean velocity transverse profiles]{ Mean velocity transverse profiles (normalised by the mean velocity deficit $\Delta U\equiv U(y=0)-U(y=\pm M/2)$ in (b)), for different downstream locations in the lee of RG115 (filled symbols) and FSG3'x3' (empty symbols). The bulk velocity, $U_b$ (see text for definition) is subtracted from the velocity profiles.
Downstream locations: (\protect\raisebox{-0.5ex}{\FilledSmallSquare} $\!|\!\!$ \protect\raisebox{-0.5ex}{\SmallSquare}) $x/x_{\mathrm{peak}}=1.4,\,1.5$; (\protect\raisebox{-0.5ex}{\FilledSmallCircle} $\!|\!\!$ \protect\raisebox{-0.5ex}{\SmallCircle}) $x/x_{\mathrm{peak}}=1.8,\,2.0$; (\protect\raisebox{-0.5ex}{\FilledDiamondshape} $\!|\!\!$ \protect\raisebox{-0.5ex}{\Diamondshape}) $x/x_{\mathrm{peak}}=2.8,\,3.0$; (\ding{72} $|\!\!$ \ding{73}) $x/x_{\mathrm{peak}}=3.7,\,3.5$.  All data are recorded at $U_{\infty}=15$ms$^{-1}$.
The dash-dotted line in figure (b) represents a cosine law, $\cos(\theta)/2$ with $\theta = [-\pi\,\,\,\,\pi]$ corresponding to $y =[-\mathrm{M}/2\,\,\,\,\mathrm{M}/2]$ and the solid lines represent a $6^{\mathrm{th}}$-order polynomial fit.
Error bars represent the departures from symmetry between the upper ($y/M>0$) and lower ($y/M<0$) half of the transverse measurements whereas the symbols represent the mean between the two.}
\label{fig:Um}
\end{figure}
\begin{figure}[ht!]
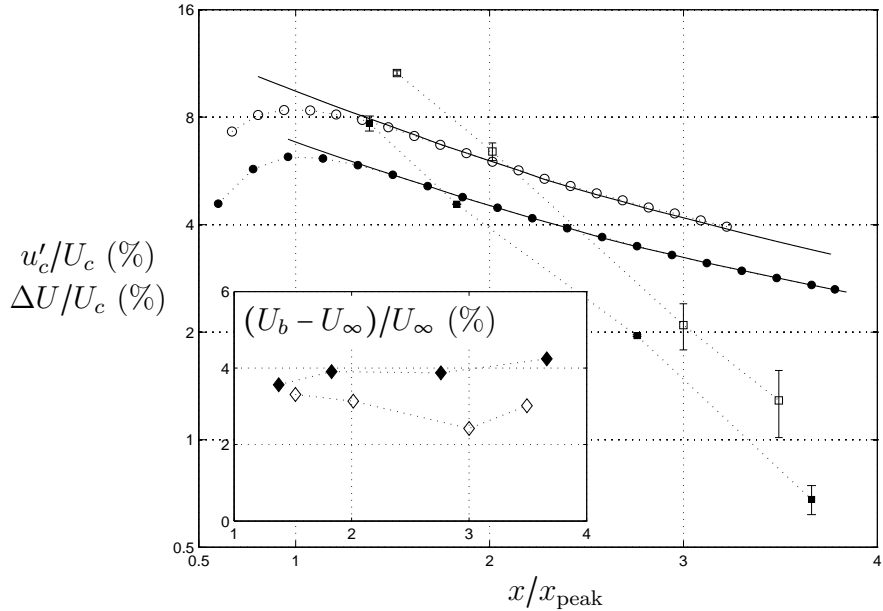

\centering
   \begin{lpic}{Thesis_DeltaUvsTuLongitudinalEvo(110mm)}
   \lbl[W]{118,2;$x/x_{\mathrm{peak}}$}
%   \lbl[W]{10,74;$\frac{u'_c}{U_c}$, $\frac{\Delta U}{U_c}$ (\%)}      
   \lbl[W]{10,74;${}_{}^{}$\hspace{20mm}}      
   \lbl[W]{10,80;$u'_c/U_c$ (\%)}      
   \lbl[W]{10,70;$\Delta U/U_c$ (\%)}      
   \lbl[W]{75,65;$(U_b-U_{\infty})/U_{\infty}$ (\%)}      
   \end{lpic}
\caption[Downstream evolution of $u'_c/U_c$ and  $\Delta U/U_c$]{Downstream evolution of 
(\protect\raisebox{-0.5ex}{\FilledSmallCircle} $\!|\!\!$ \protect\raisebox{-0.5ex}{\SmallCircle}) turbulence intensity at centreline, $u'_c/U_c$, 
(\protect\raisebox{-0.5ex}{\FilledSmallSquare} $\!|\!\!$ \protect\raisebox{-0.5ex}{\SmallSquare}) mean velocity deficit normalised by the centreline velocity, $\Delta U/U_c$ and 
(\protect\raisebox{-0.5ex}{\FilledDiamondshape} $\!|\!\!$ \protect\raisebox{-0.5ex}{\Diamondshape}) difference between bulk and free-stream velocities normalised by the free-stream velocity, $(U_b-U_{\infty})/U_{\infty}$ in the lee of the RG115 (filled symbols) and the FSG3'x3' (empty symbols). }
\label{fig:UmvsX}
\end{figure}
%%%%%% Figures %%%%%%%%%%%%%
% Velocity profiles
The mean velocity profiles as the flow decays resulting from the RG115 and the FSG3'x3' grids are compared first (see figure \ref{fig:Um}a). 
Note that the bulk velocity, $U_b\equiv 1/M \int_{-M/2}^{M/2}$ \!\! U dy, is subtracted from the mean velocity  profiles to compensate for the slight decrease in the effective area of the test section due to the blockage caused by the developing boundary-layers on the side-walls (see inset in figure \ref{fig:UmvsX}; note that $U_b>U_{\infty}$). 
(For wind-tunnels with mechanisms to compensate boundary-layer growth, e.g. a divergence test section, $U_b=U_{\infty}$.)
Normalising the profiles with the velocity deficit, $\Delta U$ (defined in the caption of figure \ref{fig:Um}), it can be seen that the mean profiles retain approximately the same shape as the velocity deficit decreases (figures \ref{fig:Um}b and \ref{fig:UmvsX}). 
In the RG115-generated turbulence case this profile is not very dissimilar from a cosine law.
However it does seem to have some slight deviation from a cosine law in the FSG3'x3'-generated turbulence case which must be attributable to the change in upstream conditions, i.e. grid geometry. 

% Wake-like flow
The decay of $\Delta U$ is much faster than decay of $u'$, see figure \ref{fig:UmvsX}. 
This starkly differs from a wake-like flow where scaling arguments suggest that $u' \sim \Delta U$ (\citealt{TennekesLumley:book}). 
% Turbulence decay
For reference, power-law fits to the decay of $\overline{u^2}$ are included in figure \ref{fig:UmvsX}. 
These are discussed in \S \ref{Sec:Decay}.

%%%%%%%%%%%%%%%%%%%%%%%%%%%%%% Sub-Section %%%%%%%%%%%%%%%%%%%%%%%%%%

\subsection{Profiles of 2$^{\mathrm{nd}}$-order one-point turbulence statistics}\label{sec:prof2nd}
%%%%%% Figures %%%%%%%%%%%%%
\begin{figure}[th!]
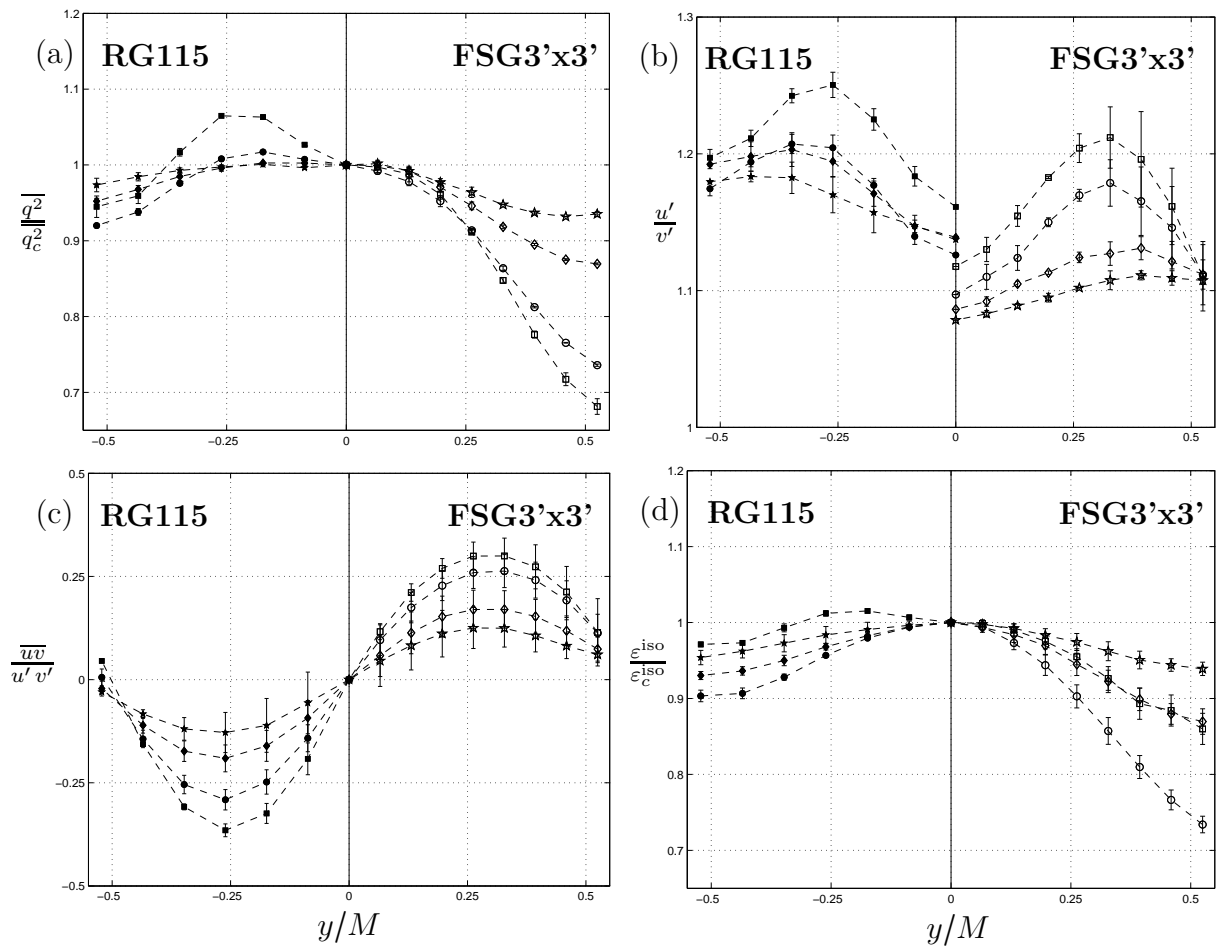

\centering
\begin{minipage}[c]{0.5\linewidth}
   \centering
   \begin{lpic}[b(-2mm)]{q2ProfileRG115FSG3x3(78mm)}
   \lbl{10,125;(a)}
   \lbl{40,125;\bf{RG115}}
   \lbl{150,125;\bf{FSG3'x3'}}
   \lbl[W]{5,75;$\frac{\overline{q^2}}{\overline{q_c^2}}$}        
   \lbl[W]{100,3;${}^{}$\hspace{5mm}} 
   \end{lpic}
\end{minipage}%
\begin{minipage}[c]{0.5\linewidth}
   \centering 
   \begin{lpic}[b(-2mm)]{uovervProfile(78mm)}
   \lbl{10,125;(b)}
   \lbl{40,125;\bf{RG115}}
   \lbl{150,125;\bf{FSG3'x3'}}
   \lbl[W]{11,75;\hspace{2mm}$\frac{u'}{v'}$}        
   \lbl[W]{100,3;${}^{}$\hspace{5mm}}    
   \end{lpic}
\end{minipage}
\begin{minipage}[c]{0.5\linewidth}
   \centering
   \begin{lpic}{uvProfile(78mm)}
   \lbl{10,125;(c)}
   \lbl{40,125;\bf{RG115}}
   \lbl{150,125;\bf{FSG3'x3'}}
   \lbl[W]{100,0;$y/M$}
   \lbl[W]{5,80;$\frac{\overline{uv}}{u'\,v'}$}           
   \end{lpic}
\end{minipage}%
\begin{minipage}[c]{0.5\linewidth}
   \centering 
   \begin{lpic}{EpsProfile(78mm)}
   \lbl{10,125;(d)}
   \lbl{40,125;\bf{RG115}}
   \lbl{150,125;\bf{FSG3'x3'}}
   \lbl[W]{100,0;$y/M$}
   \lbl[W]{5,70;${}^{}$}           
   \lbl[W]{8,80;$\frac{\varepsilon^{\mathrm{iso}}}{\varepsilon_c^{\mathrm{iso}}}$}           
   \end{lpic}
\end{minipage}
\caption[Reynolds stress transverse profiles  in the lee of RG115 and FSG3'x3']{Reynolds stress transverse profiles for different downstream locations in the lee of RG115 and FSG3'x3' ($\overline{q_c^2}$ and $\varepsilon_c^{\mathrm{iso}}$ are the centreline values of $\overline{q^2}$ and $\varepsilon^{\mathrm{iso}}$, respectively).  Symbols and error bars are described in the caption of figure \ref{fig:Um}. }
\label{fig:ReyStress}
\end{figure}
\begin{figure}
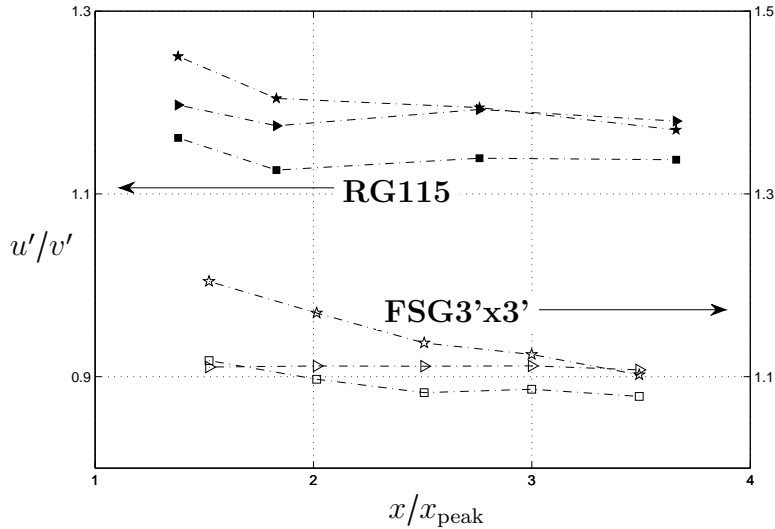

\centering
   \begin{lpic}{uovLongProfileRGFG(100mm)}
   \lbl[W]{105,2;$x/x_{\mathrm{peak}}$}
   \lbl[W]{5,75;\hspace{2mm}${}^{}$}        
   \lbl[W]{5,65;\hspace{2mm}${}^{}$}        
   \lbl[W]{5,68;\hspace{2mm}$u'/v'$}        
   \lbl[W]{95,82;\bf{RG115}}
   \lbl[W]{110,52;\bf{FSG3'x3'}}
   \end{lpic}
\caption[Downstream evolution of $u'/v'$ in the lee of RG115 and FSG3'x3']{Downstream evolution of the ratio between the longitudinal and spanwise rms velocities at  
(\protect\raisebox{-0.5ex}{\FilledSmallSquare} $\!|\!\!$ \protect\raisebox{-0.5ex}{\SmallSquare}) $y=0$, 
(\ding{72} $|\!\!$ \ding{73}) $y=-M/4$ and 
(\protect\raisebox{-0.5ex}{\FilledSmallTriangleRight} $\!|\!\!$ \protect\raisebox{-0.5ex}{\SmallTriangleRight}) $y=-M/2$ 
in the lee of the RG115 (filled symbols) and the FSG3'x3' (empty symbols).}
\label{fig:uov}
\end{figure}
%%%%%% Figures %%%%%%%%%%%%%

Previous experimental investigations on the turbulence generated by space-filling FSGs, e.g. \cite{MV2010}, reported that the flow field close to the grid is highly inhomogeneous. 
It was also observed that during the process of turbulent kinetic energy build up the turbulent flow is simultaneously homogenised by turbulent diffusion, and by the time it reaches a peak in turbulence intensity the flow has smoothed out most inhomogeneities.
\cite{SV2007} measured the turbulent kinetic energy production in various planes perpendicular to the mean flow along the centreline and observed that the turbulent production decreases rapidly just after the peak, i.e. where $ 1.0 <x/x_{\mathrm{peak}}< 1.7$ and that the turbulent energy production typically represents less than $30\%$ of the dissipation and never exceeds $20\%$ beyond this region.

\cite{MV2010} compared the characteristic time scales of the mean velocity gradients $(\partial U/\partial x)^{-1}$ and $(\partial U/\partial y)^{-1}$ (where $U$ is the streamwise mean velocity and $y$ is a coordinate along the horizontal normal to the streamwise direction) with the time scale associated with the energy-containing eddies and reached the conclusion that beyond the peak the mean gradient time scale is typically one to two orders of magnitude larger. 

Here, further measurements of  transverse profiles for FSG-generated turbulence are presented and compared with measurements for RG-generated turbulence.
The transverse profiles of the turbulent kinetic energy $\overline{q^2}/2 = \overline{u^2}/2 +  \overline{v^2}/2 + \overline{w^2}/2$ are estimated here as $\overline{u^2}/2 + \overline{v^2}$. 
This estimate relies on the assumption that $\overline{v^2} \approx \overline{w^2}$ in the decay region $x>x_{\mathrm{peak}}$ which is supported as a rough approximation by the direct numerical simulations of \cite{Sylvain2011} and the laboratory experiments of \cite{Nagata2012}.

% Reynolds stress profiles
The first observation is that the shape of the turbulent kinetic energy profile changes and becomes progressively more uniform as the turbulent flow decays (figure \ref{fig:ReyStress}a).
(The profiles are normalised by the centreline value to enhance variation of the kinetic energy with spanwise location.)
This is indeed what one expects of grid-generated turbulence which has a tendency to become asymptotically homogeneous with downstream distance. 
Another striking difference is that the profiles of the RG115-generated turbulence are more uniform than their counterparts of the FSG3'x3'-generated turbulence at similar downstream locations relative to $x_{\mathrm{peak}}$. 
This may be attributed to the additional turbulence generated by the wakes originating from the smaller squares of the FSG near the centreline (compare figure \ref{fig:grids}a with \ref{fig:grids}d near the centre of the grid), which increase  the kinetic energy in this region. 
A similar effect is manifested in the profiles of the isotropic dissipation estimate $\varepsilon^{\mathrm{iso}}=15\nu\overline{(\partial u/\partial x)^2}$ which is plotted in figure \ref{fig:ReyStress}d (normalised by the values of $\varepsilon^{\mathrm{iso}}$ on the centreline, $y=z=0$).
(Note, however, that the transverse profiles of the actual kinetic energy dissipation $\varepsilon$ divided by $\varepsilon^{\mathrm{iso}}$ may be different in the different turbulent flows considered here.) 
Even though the present FSG returns less homogeneity than the present RG, it does nevertheless seem to generate a slight improvement in the $u'/v'$ ratio (figure \ref{fig:ReyStress}b) which is one of the indicators of (an)isotropy (recall that $v'$ is the r.m.s. of the turbulent fluctuating velocity in the $y$ direction).
The downstream evolution of this ratio is presented in figure \ref{fig:uov} where it can also be seen that $u'/v'$ do not vary significantly during decay. This persistence of inter-component anisotropy during the turbulence decay is inline with the findings compiled in \S 3.9 of \cite{Townsend:book}.

% curvature of q^2 vs transport
Note that the curvature of the kinetic energy transverse profiles is associated with the lateral triple correlation transport, $\partial /\partial y \, \overline{vq^2}$.
This can be seen from an eddy diffusivity estimate $\overline{vq^2} = -\mathrm{D_T}\, \partial /\partial y \, \overline{q^2}$ which leads to $\partial /\partial y \, \overline{vq^2} = -\mathrm{D_T}\, \partial^2 /\partial y^2 \, \overline{q^2}$ if the eddy diffusivity $\mathrm{D_T}$ is independent of $y$.
Profiles of the triple correlation transport terms are presented in \S \ref{sec:homo} showing opposite net transport near the centreline at a distance of about $1.5x_{\mathrm{peak}}$ for the two grids (RG115 and FSG3'x3').
The opposite curvatures appearing near the centreline in the $\overline{q^2}$ transverse profiles at that distance from each grid directly relate to the opposite signs of the lateral triple correlation transports generated by the two grids at these locations.

% Shear stress.
Figure \ref{fig:ReyStress}c shows that differences in the Reynolds shear stress (normalised by the local $u'$ and $v'$) between the turbulence generated by the two grids are very tenuous. 
The numerical values of the normalised shear stress are similar and the shape of the profiles differs only slightly in the location of the peak and the numerical values at $y=M/2$.

%%%%%%%%%%%%%%%%%%%%%%%%%%%%%%%%%%%%%%%%%%%%%%%%%%%
\subsection{Wind tunnel confinement effects} \label{sec:confinement}
%%%%%% Figures %%%%%%%%%%%%%
\begin{figure}
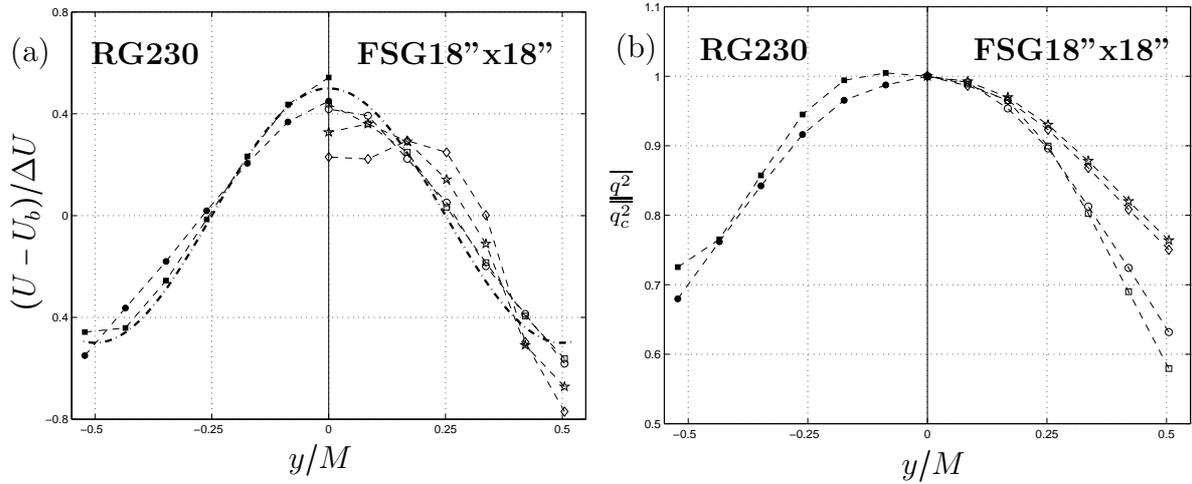

\centering
\begin{minipage}[c]{0.5\linewidth}
   \centering
   \begin{lpic}{UmMinusUbOverDeltaUProfileRG230FSG18(78mm)}
   \lbl{10,125;(a)}
   \lbl{45,125;\bf{RG230}}
   \lbl{140,125;\bf{FSG18''x18''}}
   \lbl[W]{100,0;$y/M$}
   \lbl[W]{9,75,90;$(U-U_b)/\Delta U^{}$}         
   \end{lpic}
\end{minipage}%
\begin{minipage}[c]{0.5\linewidth}
   \centering 
   \begin{lpic}{q2ProfileRG230FSG18(78mm)}
   \lbl{10,125;(b)}
   \lbl{45,125;\bf{RG230}}
   \lbl{140,125;\bf{FSG18''x18''}}
   \lbl[W]{7,80;$\frac{\overline{q^2}}{\overline{q_c^2}}$}        
   \lbl[W]{100,3;${}^{}$\hspace{5mm}} 
   \lbl[W]{100,0;$y/M$}    
   \end{lpic}
\end{minipage}%
\caption[Reynolds stress transverse profiles in the lee of RG230 and FSG18''x18'']{Reynolds stress transverse profiles for different downstream locations in the lee of the RG230 (filled symbols) and the FSG18''x18'' (empty symbols).Downstream locations: (\protect\raisebox{-0.5ex}{\FilledSmallSquare} $\!|\!\!$ \protect\raisebox{-0.5ex}{\SmallSquare}) $x/x_{\mathrm{peak}}=1.3,\,1.4$; (\protect\raisebox{-0.5ex}{\FilledSmallCircle} $\!|\!\!$ \protect\raisebox{-0.5ex}{\SmallCircle}) $x/x_{\mathrm{peak}}=1.8,\,1.8$; (\protect\raisebox{-0.5ex}{\Diamondshape}) $x/x_{\mathrm{peak}}=2.8$; (\ding{73}) $x/x_{\mathrm{peak}}=3.2$.  
The dash-dotted line on figure (a) represents a cosine law.}
\label{fig:ReyStressB}
\end{figure}
%%%%%% Figures %%%%%%%%%%%%%
%
%% background/motivation
%In VV11 (\S 3.2.3) some concerns were raised regarding the effect of the wind tunnel bounding walls on the measured turbulent flow.
%They compared the ratio of their integral length-scale and the tunnel width/height (estimated to range between 8-10) with DNS and other grid-turbulence experiments and it was argued that the effect of confinement could not, by itself, justify the outstanding properties of the measured flow, but, nonetheless, there could be an effect on the measured statistics, for example on the constancy of the ratio of $L_{11}^{(1)}/\lambda$ as the flow decayed. 

% what we are presenting
The effect of wind tunnel confinement is investigated here by comparing geometrically similar grids with different ratios between test section width/height ($T$) and the mesh size ($M$).
This is accomplished by (i) comparing mean profiles from FSG18''x18'' and the FSG3'x3' grid arrangements for which $M$  is approximately the same but the FSG3'x3' is a periodic extension of the FSG18''x18'' in a wind tunnel of double the size;
 and (ii) by comparing mean profiles from the RG230 and RG115 arrangements which are geometrically similar in the same wind tunnel where the mesh sizes differ by a factor 2.
These two comparisons provide an assessment of (i) the effect of generating large integral-length scales relative to the tunnel's cross section (which could influence, e.g., the downstream evolution of the $L_{11}^{(1)}/\lambda$ ratio, see \S \ref{sec:NoneqRG}) and (ii)  the difference between wakes interacting with each other whilst simultaneously interacting with the wall (as is the case for RG230- and FSG18''-generated turbulence) versus wakes interacting with each other in a quasi-periodic arrangement (as is the case of the centre regions of the RG115 and FSG3'x3'-generated turbulent flows). 
 
Note that \cite{HG1978} showed that the solid walls have a blocking action on the large-scale free-stream turbulence eddies adjacent to the wall up to a distance of the order of the integral-length scale (even in the absence of mean shear for the case of moving walls with a tangential velocity equal to that of the mean flow). 
Along the centreline the large-scale turbulence eddies generated by RG230 and FSG18Óx18Ó are 4 -- 5 integral-length scales away from the wall and it is reasonable to expect that they are not  directly influenced by it. 
However, along $y = \pm M/2$ the distance to the wall is only 2 -- 3 integral-length scales and it is conceivable that the turbulence eddies around this region are indeed influenced by the wall blocking mechanism  \cite[]{HG1978}.  
On the other hand, for the case of RG115 and FSG3Õx3Õ (where M/T is smaller by a factor of two by comparison to RG230 and FSG18Óx18Ó) the generated turbulence within the central mesh ($|y| \leq M/2$) is at least 6 -- 8 integral-length scales away from the wall and therefore no direct influence of the walls is expected.

% deformation of the velocity profiles
Comparing the normalised mean velocity profiles of RG230 and FSG18''x18'' in figure \ref{fig:ReyStressB}a with those of  RG115 and FSG 3'x3' in figure \ref{fig:Um}b, it is clear that the profiles corresponding to the grids with double the value of $M/T$ have lost the similarity with downstream position which characterises the profiles resulting from lower $M/T$ grids.
This effect  is more pronounced further downstream, in particular at the furthermost stations where the greatest departures from self-similar profile shape occurs (see figure \ref{fig:ReyStressB}a).
The two furthermost stations in the FSG18''x18'' case are at $x = 3650$mm and $x=4250$mm whereas the last measurement station in the RG230 case is $x=3050$mm.
It is therefore no surprise that the greatest deviations from self-similar mean flow profile are evidenced in the FSG18''x18'' case as the blockage induced by the boundary layers developing on the confining walls is greater at $x = 3650$mm and $x=4250$mm than at $x=3050$mm.

% deformation of the kinetic energy profiles
Turning to the effect on the kinetic energy profiles (figure \ref{fig:ReyStressB}b versus figure \ref{fig:ReyStress}a) it can be seen that there is a substantial decrease in the uniformity of the profiles across the transverse locations for the grids with higher $M/T$ (RG230 and FSG18''x18''). 
It also appears that this effect is felt throughout the decay and is as pronounced closer to $x_{\mathrm{peak}}$ as further downstream.
For example, the overshoot of kinetic energy off the centreline observed for the first measurement station ($x/x_{\mathrm{peak}}=1.4$) in the lee of RG115 (\emph{cf.} figure \ref{fig:ReyStress}a) is almost non-existent for the RG230 data at an identical downstream location relative to $x_{\mathrm{peak}}$ ($x/x_{\mathrm{peak}}=1.3$, \emph{cf.} figure \ref{fig:ReyStressB}b).
It is likely that this difference is a consequence of the wakes in the RG230 and FSG18''x18'' interacting with the wall instead of interacting with other wakes from a periodic bar, as in the case of RG115 and FSG3'x3'. 
These changes in the shape of the profiles lead to changes in the turbulent transport as is shown in \S \ref{sec:homo}.

%%%%%%%%%%%%%%%%%%%%%%%%%%%%%%%%%%%%%%%%%%%%%%%%%%%
\subsection{Turbulent transport and production} \label{sec:homo}

%%%%%% Figures %%%%%%%%%%%%%
\begin{figure}
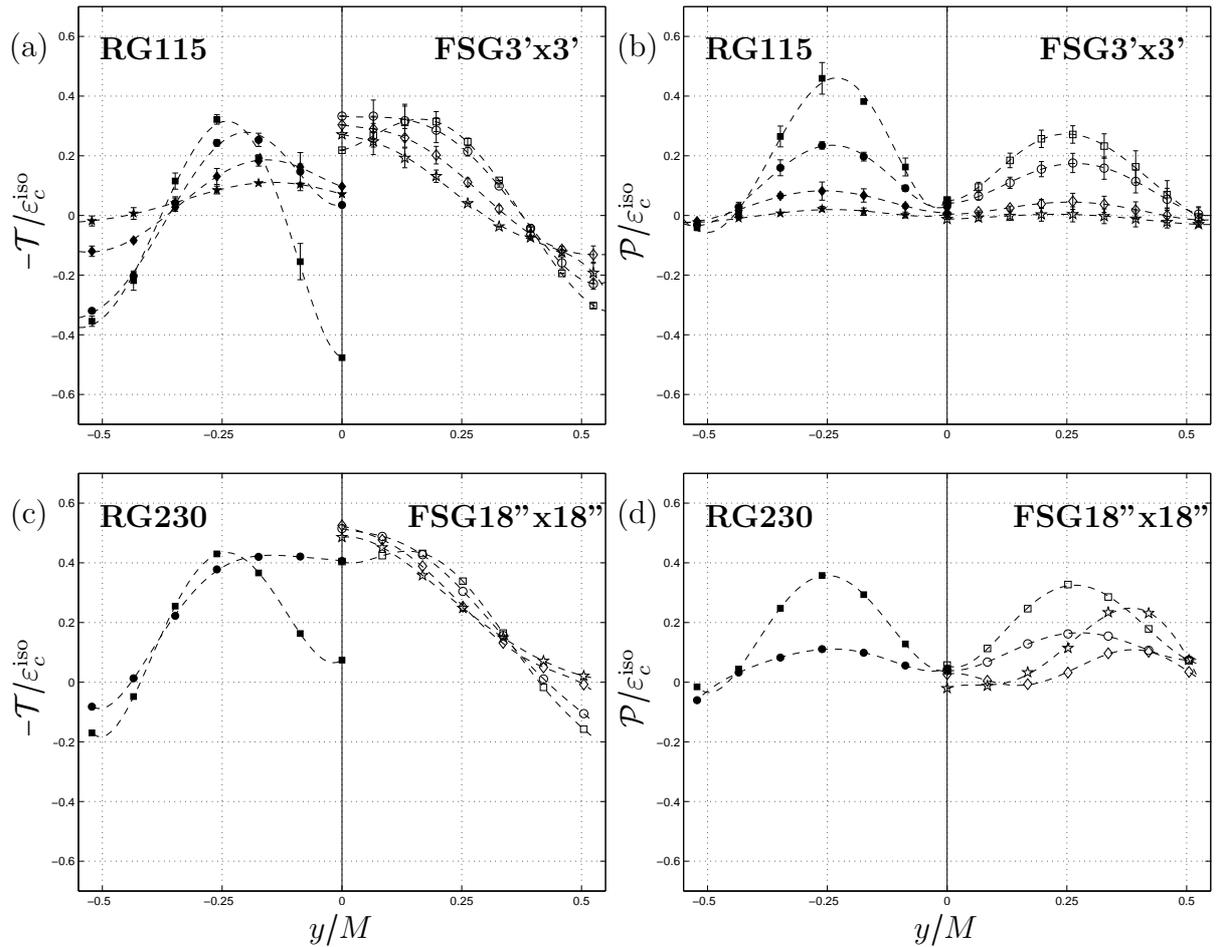

\centering
\begin{minipage}[c]{0.5\linewidth}
   \centering
   \begin{lpic}{TranspTransvProfile(80mm)}
   \lbl{8,125;(a)}
   \lbl{45,125;\bf{RG115}}
   \lbl{150,125;\bf{FSG3'x3'}}
   \lbl[W]{90,4;${}^{}$\hspace{15mm}}
   \lbl[W]{8,75,90;$-\mathcal{T}/\varepsilon_c^{\mathrm{iso}}$}   
   \end{lpic}
\end{minipage}%
\begin{minipage}[c]{0.5\linewidth}
   \centering 
   \begin{lpic}{ProdTransvProfile(80mm)}
   \lbl{9,125;(b)}
   \lbl{45,125;\bf{RG115}}
   \lbl{150,125;\bf{FSG3'x3'}}
   \lbl[W]{90,4;${}^{}$\hspace{15mm}}
   \lbl[W]{8,75,90;$\mathcal{P}/\varepsilon_c^{\mathrm{iso}}$}
   \end{lpic}
\end{minipage}
\begin{minipage}[c]{0.5\linewidth}
   \centering
   \begin{lpic}{TranspTransvProfileRG230FSG18(80mm)}
   \lbl{8,125;(c)}
   \lbl{45,125;\bf{RG230}}
   \lbl{150,125;\bf{FSG18''x18''}}
   \lbl[W]{102,1;$y/M$}   
   \lbl[W]{8,75,90;$-\mathcal{T}/\varepsilon_c^{\mathrm{iso}}$}   
   \end{lpic}
\end{minipage}%
\begin{minipage}[c]{0.5\linewidth}
   \centering 
   \begin{lpic}{ProdTransvProfileRG230FSG18(80mm)}
   \lbl{9,125;(d)}
   \lbl{45,125;\bf{RG230}}
   \lbl{150,125;\bf{FSG18''x18''}}
   \lbl[W]{102,1;$y/M$}   
   \lbl[W]{8,75,90;$\mathcal{P}/\varepsilon_c^{\mathrm{iso}}$}
   \end{lpic}
\end{minipage}%
\caption[Transverse profiles of turbulent  transport and  production in the lee of RG115, RG230, FSG18''x18'' and FSG3'x3']{Transverse profiles of turbulent (a, c) transport and (b, d) production for different downstream locations in the lee of RG115, RG230, FSG18''x18'' and FSG3'x3'. Symbols and error bars for the top plots (a, b) are described in the caption of figure \ref{fig:Um} and  symbols for the bottom plots (c, d) in figure \ref{fig:ReyStressB}.}
\label{fig:TKESpanProf}
\end{figure}

\begin{figure}[t!]
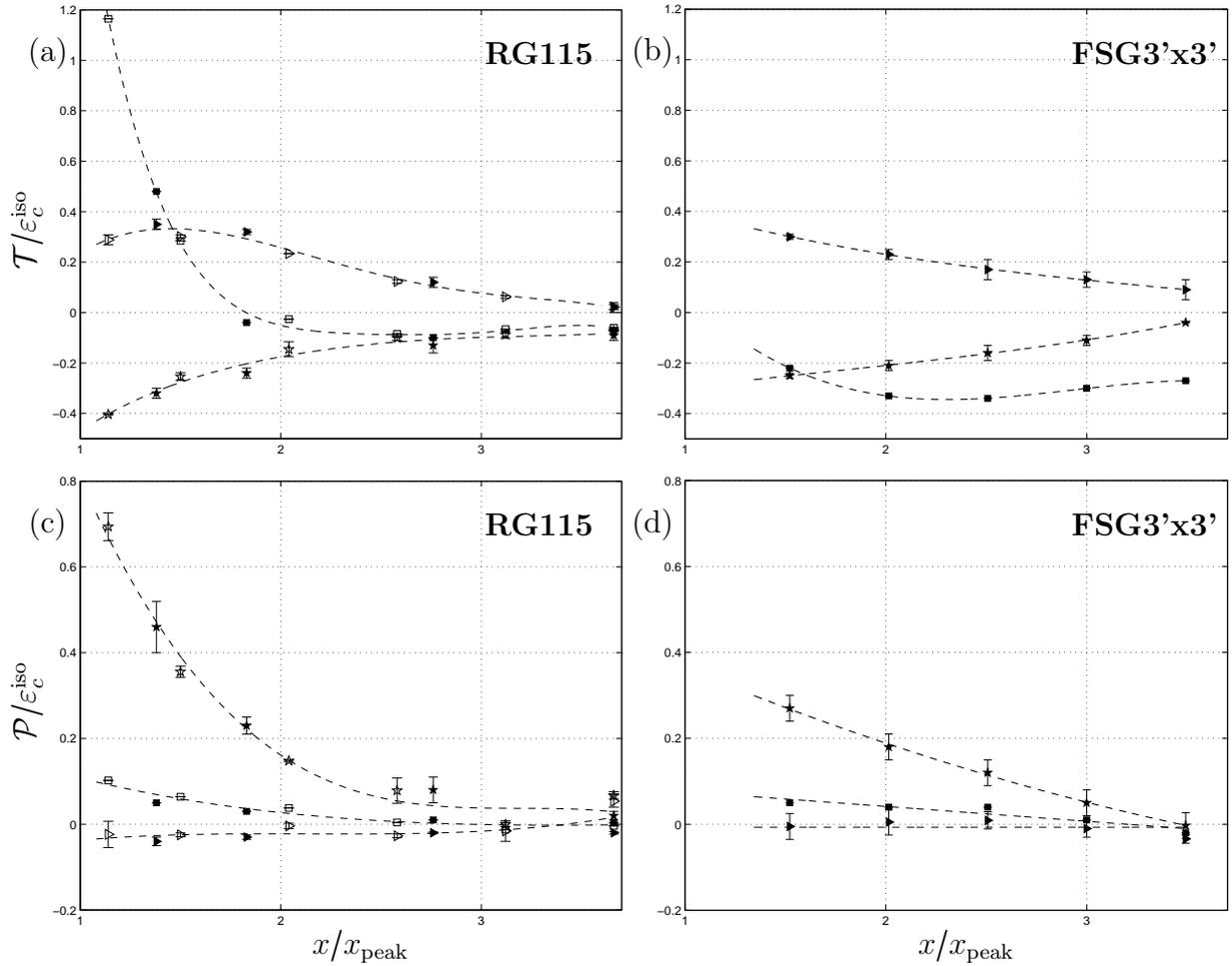

\centering
\begin{minipage}[c]{0.5\linewidth}
   \centering
   \begin{lpic}[b(-2mm)]{TranspLongProfileRG(80mm)}
   \lbl{8,125;(a)}
   \lbl{150,125;\bf{RG115}}
   \lbl[W]{90,4;${}^{}$\hspace{15mm}}
   \lbl[W]{2,75,90;$\mathcal{T}/\varepsilon_c^{\mathrm{iso}}$}
   \end{lpic}
\end{minipage}%
\begin{minipage}[c]{0.5\linewidth}
   \centering 
   \begin{lpic}[b(-2mm)]{TranspLongProfileFG(80mm)}
   \lbl{8,125;(b)}
   \lbl{150,125;\bf{FSG3'x3'}}
   \lbl[W]{90,4;${}^{}$\hspace{15mm}}
   \lbl[W]{5,75,90;${}^{}$\hspace{15mm}}
   \end{lpic}
\end{minipage}
\begin{minipage}[c]{0.5\linewidth}
   \centering
   \begin{lpic}{ProdLongProfileRG(80mm)}
   \lbl{8,125;(c)}
   \lbl{150,125;\bf{RG115}}
   \lbl[W]{100,2;$x/x_{\mathrm{peak}}$}   
   \lbl[W]{2,75,90;$\mathcal{P}/\varepsilon_c^{\mathrm{iso}}$}
   \end{lpic}
\end{minipage}%
\begin{minipage}[c]{0.5\linewidth}
   \centering 
   \begin{lpic}{ProdLongProfileFG(80mm)}
   \lbl{8,125;(d)}
   \lbl{150,125;\bf{FSG3'x3'}}
   \lbl[W]{100,2;$x/x_{\mathrm{peak}}$}   
   \lbl[W]{5,75,90;${}^{}$\hspace{15mm}}
   \end{lpic}
\end{minipage}%
\caption[Longitudinal profiles of turbulent transport and production in the lee of RG115 and FSG3'x3']{Longitudinal profiles of turbulent (a,b) transport and (c,d) production for three spanwise locations in the lee of RG115 and FSG3'x3', namely  
(\protect\raisebox{-0.5ex}{\FilledSmallSquare}) $y=0$,  
(\ding{72}) $y=M/4$ and 
(\protect\raisebox{-0.5ex}{\FilledSmallTriangleRight}) $y=M/2$. 
Additional data for RG115 from the $2 \times$XW experiments are added to (a,b) and represented with open symbols.
Error bars  are described in the caption of figure \ref{fig:Um}. Dashed lines are polynomial fits to the data.}
\label{fig:TKELongProf}
\end{figure}
\begin{figure}[t!]
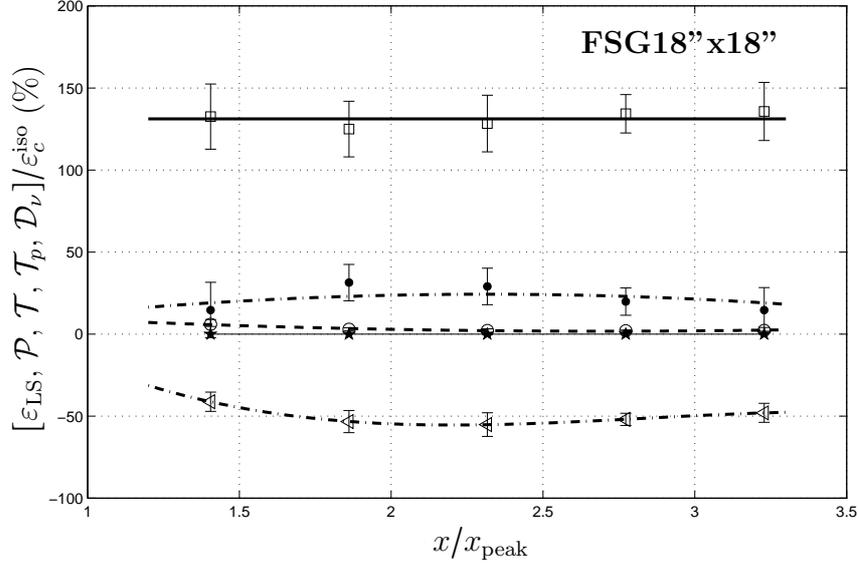

\centering
%trim option's parameter order: left bottom right top
\begin{lpic}{Thesis_HomogeneityPlotFSG18(140mm)}
\lbl{185,122;\bf{FSG18''x18''}}
\lbl[W]{140,5;$x/x_{\mathrm{peak}}$}
\lbl[W]{34,75,90;$[\varepsilon_{\mathrm{LS}},\,\mathcal{P},\,\mathcal{T},\,\mathcal{T}_p,\,\mathcal{D}_{\nu}]/\varepsilon_{c}^{\mathrm{iso}}$ (\%)}
\end{lpic}
\caption[Turbulent kinetic energy budget for FSG18''x18'' along the centreline]{ Turbulent kinetic energy budget \eqref{eq:TKE} normalised by the dissipation for the FSG18''x18'' at the centreline for $U_{\infty}=15ms^{-1}$, (\protect\raisebox{-0.5ex}{\SmallSquare})  $\varepsilon_{LS}$ - advection, 
(\protect\raisebox{-0.5ex}{\SmallCircle}) $\mathcal{P}$ - production, 
(\protect\raisebox{-0.5ex}{\SmallTriangleLeft}) $\mathcal{T}$ -  triple-correlation transport, 
(\protect\raisebox{-0.5ex}{\FilledSmallCircle}) $\mathcal{T}_p$ - pressure transport, 
(\protect\raisebox{0ex}{\ding{72}}) $\mathcal{D}_{\nu}$ - viscous diffusion.}
\label{Fig:CorrsinHomogeneityTKEBudget}
\end{figure}
%%%%%% Figures %%%%%%%%%%%%%
 
% intro
The focus is now directed to the estimates of the main terms of the single-point turbulent kinetic energy (T.K.E.) transport equation,
\begin{align}
\underbrace{\frac{U_{k}}{2}\frac{\partial\, \overline{q^{2}}}{\partial x_{k}}}_{\mathcal{A}} = 
\underbrace{-\overline{u_{i} u_{j}}\, \frac{\partial U_{i}}{\partial x_{j}}}_{\mathcal{P}} 
\underbrace{-\frac{\partial}{\partial x_{k}}\left( \frac{\overline{u_{k} q^{2}}}{2} + \frac{\overline{u_{k} p}}{\rho} \right)}_{\mathcal{T}} +
\underbrace{\frac{\nu}{2} \frac{\partial^{2} \overline{q^{2}}}{\partial x^2_{m}}}_{\mathcal{D}_{\nu}} -
\,\varepsilon.
\label{eq:TKE}
\end{align}
In the ideal approximation to homogeneous  wind tunnel turbulence the advection and the dissipation balance, $\mathcal{A}\approx \varepsilon$, and the other terms are comparatively negligible, $\mathcal{P},\, \mathcal{T},\, \mathcal{D}_{\nu} \ll \varepsilon$. 
The estimates of the turbulent transport and production can, therefore, be used to provide a quantitative assessment of the inhomogeneity of the flow, at least in terms of one-point, $2^{\mathrm{nd}}$-order, statistics. 
It is confirmed against all the data presented here that the Reynolds numbers are indeed sufficiently high ($Re_M=\mathcal{O}(10^5)$, $600<Re_{L^{1(1)}}<4000$) for the viscous diffusion $\mathcal{D}_{\nu}$ to be negligible relative to the turbulent dissipation, $\varepsilon$.
The turbulent kinetic energy dissipation, $\varepsilon$, is estimated using the isotropic surrogate evaluated at the centreline, $\varepsilon^{\mathrm{iso}}_c$.
This choice is motivated in \S \ref{sec:Eps}.

% Normalisation with dissipation, no estimate of pressure-velocity transport and axisymmetry
The turbulent transport and production terms are estimated in a cylindrical coordinate system ($r,\phi,x$), where the $x$-coordinate is the same as the $x$-coordinate of the Cartesian coordinate system.
This is done under the assumption of axisymmetry of the turbulent flow with respect to the centreline axis.
This assumption has been given substantial support for the decay region in the lee of FSGs by the wind tunnel measurements of \cite{Nagata2012}.
The turbulent production and triple velocity-correlation transport therefore take the form,
\begin{equation}
\mathcal{T} = - \frac{1}{r} \frac{\partial}{\partial r} \left(r\left(\frac{\overline{u_r q^2}}{2} + \frac{\overline{u_r p}}{\rho}\right) \right) - \frac{\partial}{\partial x} \left(\frac{\overline{u_1 q^2}}{2} + \frac{\overline{u_1 p}}{\rho}\right) 
\label{eq:T}
\end{equation}
and 
\begin{equation}
\mathcal{P} = - \overline{u_i u_1} \frac{\partial U_i}{\partial x} - \overline{u_i u_r} \frac{\partial U_i}{\partial r},
\label{eq:P}
\end{equation}
where $u_r$ and $u_1$ are the turbulent fluctuating velocity components aligned with $r$ and $x$, respectively.
The axisymmetry assumption allows the estimation of \eqref{eq:T} and \eqref{eq:P} by replacing $r$, $u_r$ and $\partial/\partial r$ with $y$, $v\equiv u_2$ and $\partial/\partial y$ in these equations.
The experimental apparatus does not allow the measurements of the pressure-velocity correlations and thus only triple velocity-correlation turbulent transport is estimated.
Hence, what is really calculated is:
\begin{equation}
\mathcal{T} = - \frac{1}{y} \frac{\partial}{\partial y} \left(y\left(\frac{\overline{v q^2}}{2}\right) \right) - \frac{\partial}{\partial x} \left(\frac{\overline{u q^2}}{2}\right), 
\label{eq:T2}
\end{equation}
where $\overline{q^2} = \overline{u^2}+2\overline{v^2}$ as in \S \ref{sec:prof2nd} and
\begin{equation}
\mathcal{P} = - \overline{u^2} \frac{\partial U}{\partial x} - \overline{u v} \frac{\partial U}{\partial y}.
\label{eq:P2}
\end{equation}
The conclusions in this subsection do not crucially depend on how good an approximation of the triple velocity-correlation transport and the turbulent production these two equation are. 
The quality of these approximations depends on how good the axisymmetry assumption is and on the impact of neglecting the pressure-velocity correlation. 
What is indeed addressed here are the differences between the flows generated by RGs and FSGs and the effects of the tunnel walls. 
For simplicity, $\mathcal{T}$ as defined by \eqref{eq:T2} is referred to as turbulent transport henceforth.

% transport
The spanwise profiles of turbulent transport normalised by $\varepsilon^{\mathrm{iso}}_c$ for the FSG3'x3'- and RG115-generated turbulence are shown in figure \ref{fig:TKESpanProf}a.
The profiles in the lee of the two grids are recorded at four comparable downstream locations relative to $x_{\mathrm{peak}}$.
The prominent differences for the two turbulent flows are striking. 
Whereas the transport near the centreline for the FSG3'x3' is always positive (i.e. net loss of T.K.E.) and amounts to roughly 30\% of the dissipation throughout the assessed region of decay, for the RG115 at  the measurement station closest to $x_{\mathrm{peak}}$ it amounts to $\approx -45\%$ (i.e. net gain of T.K.E.),  changes sign further downstream and at the farthest measurement station becomes  relatively small ($< 10\%$; see also figure  \ref{fig:TKELongProf}a).
These differences are very likely caused by the geometrical differences between the grids. 
Note, however, that $Re_M$ for the FSG3'x3' recordings is about twice those of the RG115.
Nevertheless, at these Reynolds numbers, the variation in $Re_M$ cannot by itself justify the observed differences.  

%production
Conversely, the differences in the spanwise profiles of turbulence production are more subtle (figure \ref{fig:TKESpanProf}b).
This observation is in-line with the more tenuous differences found in the spanwise profiles of $U$ and $\overline{uv}/u'v'$ (figures \ref{fig:Um}a,b and \ref{fig:ReyStress}c). 

% longitudinal profiles
The longitudinal profiles of the turbulence transport and production, at the centreline and at two parallel lines at $y=M/4,\,M/2$ can be found in figures \ref{fig:TKELongProf}a,b.
Complementary data from the $2\times$XW experiments (recorded at different downstream locations) are also included for the RG115 case.
Note that these data are recorded at a lower inlet velocity, $U_{\infty}=10\mathrm{ms}^{-1}$ and consequently at a lower $Re_M$, but seem to follow roughly the same longitudinal profiles.
However, the difference in $Re_M$ (the $Re_M$ of the $2\times$XW data is about $2/3$ of the $Re_M$ of the single XW data) is insufficient to draw definitive conclusions concerning the Reynolds number dependence of the distribution and magnitude of the turbulent transport and production.
For the centreline data acquired in the lee of FSG18''x18'', the longitudinal profiles of turbulent transport and production are supplemented with the estimated  large scale advection, $\mathcal{A}$, viscous diffusion, $\mathcal{D}_{\nu}$ and pressure transport, $\mathcal{T}_p$ (calculated from the balance of \eqref{eq:TKE}) and presented in figure \ref{Fig:CorrsinHomogeneityTKEBudget}.

% confinement
In \S \ref{sec:confinement} the effect of bounding wall confinement on the spanwise profiles of $U$ and $\overline{q^2}$ was demonstrated. 
%Two distinct effects of the bounding walls were identified, (i) decreasing ratio between the integral length-scale and the tunnel width as the integral length-scale grows and/or the boundary-layers on the confining walls thicken and decrease effective test section area, and (ii) wake interaction with the wall. 
%The effect of the former increases with increasing downstream location, whereas the effect of the latter is felt throughout the first few multiples of $x_{\mathrm{peak}}$, and possibly beyond that.
Figures \ref{fig:TKESpanProf}c,d can be compared to figures \ref{fig:TKESpanProf}a,b to assess the confinement effects in terms of turbulent transport and production. 
It is clear that the effect of confinement is more pronounced on the turbulent transport profiles. 
For the RGs this leads to a change in the direction of the transport at the centreline closer to the grid ($x/x_{\mathrm{peak}}\approx 1.4$) from $\mathcal{T}/\varepsilon^{\mathrm{iso}}_c \approx -45\%$ for RG115 to $\mathcal{T}/\varepsilon^{\mathrm{iso}}_c \approx 10\%$ for RG230 and an increase in the transport at $x/x_{\mathrm{peak}}\approx 1.8$ from $\mathcal{T}/\varepsilon^{\mathrm{iso}}_c \approx 5\%$ for RG115 to $\mathcal{T}/\varepsilon^{\mathrm{iso}}_c \approx 40\%$ for RG230.
For the FSGs the confinement leads to an improved `collapse' of the profiles with a value at the centreline of about $\mathcal{T}/\varepsilon^{\mathrm{iso}}_c \approx 45\%$ for the FSG18''x18'' (figure \ref{Fig:CorrsinHomogeneityTKEBudget}), c.a. $15\%$ higher than in FSG3'x3'.
The fact that this effect is felt throughout the decay leads to the hypothesis that it is caused by the influence of the wake/confining-wall interaction on the wake/wake interaction. 

On the other hand, the effect on the turbulent production is generally less pronounced, except far downstream for the FSG18''x18'' case where the profiles are severely distorted by comparison to the FSG3'x3' case, which can be attributed to the distortion in the mean velocity profiles (figure \ref{fig:ReyStressB}a). 
This is likely a consequence of the developing boundary layers on the confining walls as discussed in \S \ref{sec:confinement}. 
(Note that the last RG230 measurement is $x=3050$mm versus $x=4250$mm for FSG18''x18'', explaining why this effect is mostly seen for FSG18''x18''.) 

Turbulent transport and production are shown to become small ($<10\%$) beyond $x/x_{\mathrm{peak}}=3.5$ for the RG115-generated turbulence, regardless of the spanwise location (see figures \ref{fig:TKELongProf}a,c).
A similar observation can be made for the FSG3'x3'- and FSG18''x18''-generated turbulence (see figures \ref{fig:TKELongProf}b,d and \ref{Fig:CorrsinHomogeneityTKEBudget}), except for the turbulent transport around the centreline which has a substantially slower decay, perhaps only marginally faster than the dissipation and is persistent until the farthest downstream location measured.
Based on the present data it is shown that it is not due to confinement, although it is demonstrated that confinement does have a significant effect.

\subsection{Energy decay} \label{Sec:Decay}

The functional form of the turbulent kinetic energy decay is usually assumed to follow a power-law, which is mostly in agreement with the large database of laboratory and numerical experiments for both grid-generated turbulence and boundary-free turbulent flows
\begin{align}
\overline{u^{2}}\sim (x-x_{0})^{-n}.
\label{Eq:powerlaw}
\end{align}

Note, from the outset, that in the presence of significant variations of turbulence transport and production relative to the dissipation (as shown in the proceeding section) there is no \emph{a priori} reason for the decay to follow a power-law. 
Nevertheless, there are two data sets for which it may be argued that the power-law decay fits may bear some physical significance, namely (i) the far downstream data in the lee of RG60 where the data is closer to being homogeneous and (ii) the FSG18''x18'' data at the centreline, since the main inhomogeneity component, the lateral transport of kinetic energy, is approximately proportional to the dissipation in part due to the grid geometry (the FSG3'x3' data also exhibit this approximate proportionality), but also due to confinement effects (see \S \ref{sec:confinement} where it is shown that the approximate proportionality between $\mathcal{T}$ and $\varepsilon$ is better satisfied by  the FSG18''x18'' data than by the FSG3'x3' data).

\begin{figure}[t!]
\centering
\begin{minipage}[c]{0.5\linewidth}
   \centering
   \begin{lpic}{Lambda(85mm)}
   \lbl{8,100;(a)}
   \lbl[W]{85,3;$x$ (m)}
   \lbl[W]{11,65,90;$U\lambda^2$ (m$^3$s$^{-1}$)}
   \end{lpic}
\end{minipage}%
\begin{minipage}[c]{0.5\linewidth}
   \centering 
   \begin{lpic}{DecayMethodI(85mm)}
   \lbl{8,100;(b)}
   \lbl[W]{88,3;$x-x_0$ (m)}
   \lbl[W]{5,65,90;$\overline{u^2}/U_{\infty}^2$}
  \end{lpic}
\end{minipage}
\begin{minipage}[c]{0.5\linewidth}
   \centering
   \begin{lpic}{DecayMethodIII(85mm)}
   \lbl{8,100;(c)}
   \lbl[W]{88,3;$x-x_0$ (m)}
   \lbl[W]{5,65,90;$\overline{u^2}/U_{\infty}^2$}
   \end{lpic}
\end{minipage}%
\begin{minipage}[c]{0.5\linewidth}
   \centering 
   \begin{lpic}{DecayMethodIV(85mm)}
   \lbl{8,100;(d)}
   \lbl[W]{85,3;$x$ (m)}
   \lbl[W]{5,65,90;$\overline{u^2}/U_{\infty}^2$}
   \end{lpic}
\end{minipage}%
\caption[Decay of turbulence generated by RG60 and FSG18''x18'']{Decay of turbulence generated by RG60 and FSG18''x18''. (a) Linear growth of $U\lambda^2$ (b) power-law fit using method I, (c) power-law fit using method III (d) power-law fit using method IV. (\protect\raisebox{-0.5ex}{\FilledSmallSquare}) SFG18''x18'' at $U_{\infty}=10ms^{-1}$, (\protect\raisebox{-0.5ex}{\SmallCircle}) SFG18''x18'' at $U_{\infty}=15ms^{-1}$, (\protect\raisebox{-0.5ex}{\SmallTriangleLeft}) RG60 at $U_{\infty}=10ms^{-1}$, (\FilledSmallTriangleUp) RG60 at $U_{\infty}=15ms^{-1}$, (\protect\raisebox{-0.5ex}{\SmallTriangleRight}) RG60 at $U_{\infty}=20ms^{-1}$, (\protect\raisebox{0ex}{\ding{73}}) data from the Active-grid experiment by \cite{MW1996}.}
\label{Fig:Decay}
\end{figure}

Note also that for the FSG18''x18''  \cite{HV2007} found that the decay law appeared to be exponential and subsequently \cite{MV2010} proposed a convenient alternative functional form for the kinetic energy decay (and for the evolution of $\lambda$ when $U d \overline{u^{2}}/dx \propto \nu \overline{u^{2}}/\lambda^{2}$ is a good approximation) that is both consistent with the power-law decay and the exponential decay law proposed by \cite{GeorgeWang2009}:
\begin{equation}
\left\lbrace
\begin{aligned}
\lambda^2 &= \lambda_{0}^2  \left[1+\frac{4\nu a |c|}{l^2_{0} \,U_{\infty}}(x-x'_{0})\right]\\
\overline{u^{2}}&=\frac{2\, u'^2_{0}}{3}  \left[1+\frac{4\nu a |c|}{l^2_{0}\, U_{\infty}}(x-x'_{0})\right]^{(1-c)/2c}
\end{aligned}
\right.
\label{Eq:MVEquations}
\end{equation}
where $c<0$. In the limit of $c \rightarrow 0$ it asymptotes to an exponential decay with constant length-scales  throughout the decay,  but otherwise it is a power-law decay where $x_{0}$ is not the conventional virtual origin where the kinetic energy is singular. 
The two equations \eqref{Eq:powerlaw} \& \eqref{Eq:MVEquations} are equivalent with $n=(c-1)/2c$ and $x_{0} = x'_{0} - l^{2}_{0}\, U_{\infty}/(4\nu a c)$.

Determining the decay exponent directly from \eqref{Eq:powerlaw} is difficult, although feasible, since a nonlinear fit is generally needed to determine $n$ and $x'_{0}$ simultaneously. 
For a homogeneous (isotropic) turbulent decaying flow where advection balances dissipation it is possible to obtain a linear
equation for the Taylor microscale that can be used to determine the virtual origin, thus simplifying the task of determining the decay exponent. 
Using $\lambda^2=15\nu \overline{u^2}/\varepsilon$ in conjunction with the advection dissipation balance characteristic of
homogeneous isotropic turbulence ($3/2 \, U \partial \overline{u^{2}}/\partial x=-\varepsilon$) and assuming power-law
energy decay \eqref{Eq:powerlaw} one gets 
\begin{align}
\lambda^{2}=\frac{10 \, \nu}{n \, U}(x-x_{0}).
\label{Eq:lambda}
\end{align}
Note that for $\lambda^{2}$ to be linear the mean velocity has to be constant otherwise the linear relation holds for $U\lambda^2$. 

\begin{figure}[t!]
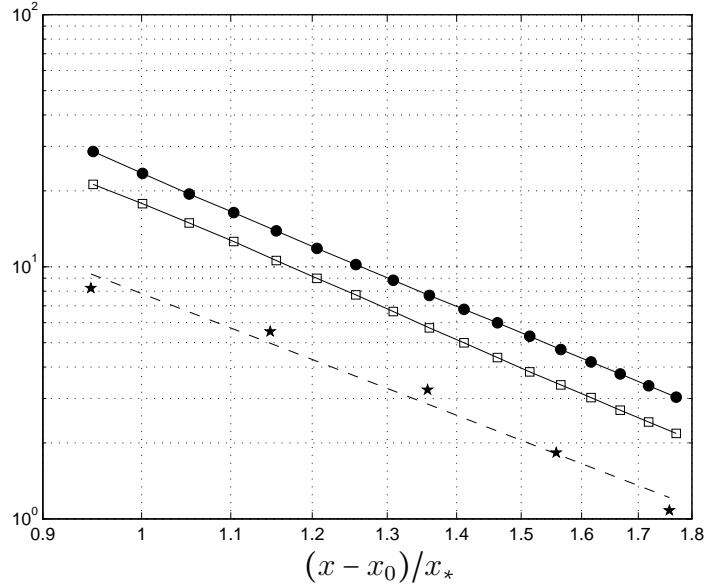
 
\centering
   \begin{lpic}{DecayVsDissipationVsDiffusion(110mm)}
   \lbl[W]{90,4;$(x-x_0)/x_*$}
   \end{lpic}
\caption[Downstream decay of $\frac{U}{2}\frac{\partial \overline{q^2}}{\partial x}$, $\varepsilon$ and $2\frac{\partial}{\partial y} \left( \frac{\overline{vq^2}}{2} \right)$ for FSG18''x18'' data]{Downstream decay of (\protect\raisebox{-0.5ex}{\FilledSmallCircle}) advection $\frac{U}{2}\frac{\partial \overline{q^2}}{\partial x}$, 
(\protect\raisebox{-0.5ex}{\SmallSquare}) dissipation $\varepsilon$ and
(\protect\raisebox{0ex}{\ding{72}}) transverse triple-correlation transport $2\frac{\partial}{\partial y} \left( \frac{\overline{vq^2}}{2} \right)$, for FSG18''x18''-generated turbulence at the centreline and for $U=15$ms$^{-1}$}
\label{DecayVsDissipationVsDiffusion} 
\end{figure}

As noted before, for the assessed decay region of the present FSGs at the centreline the transverse energy transport and dissipation remain approximately proportional to each other throughout the assessed decay region, particularly for the FSG18''x18''-generated turbulence due to the effect of the confining walls, whereas the production is negligible (see \S \ref{sec:confinement}).
Based on these results, \eqref{eq:TKE} reduces to 
\begin{equation}
\frac{U}{2}\frac{\partial\, \overline{q^{2}}}{\partial x} =  -\chi \, \varepsilon ,
\end{equation}
where $\chi = 1 + \mathcal{T}/\varepsilon \approx 1.4$ (figure \ref{Fig:CorrsinHomogeneityTKEBudget}c).
Therefore one can expect the decay exponent $n$ to be set by the dissipation rate $\varepsilon$ (irrespective of what sets the dissipation rate). 
In figure \ref{DecayVsDissipationVsDiffusion} the decay of the advection, the dissipation and the transverse triple-correlation transport, which are all measured independently, are plotted in logarithmic axes and they indeed seem to follow straight lines, i.e. power laws, with the same slope  thus supporting the present argument.

Therefore, for the decay region of the RG60- as well as the FSG18''x18''-generated turbulence at the centreline,  
\begin{equation}
U {d\over dx} \overline{u^{2}} \propto \nu \overline{u^{2}}/\lambda^{2}
\label{Eq:AdvectionTaylor}
\end{equation}
might be a good approximation, as is indeed supported by the  data which show that U$\lambda^2$ grows linearly with downstream location and even that $U\lambda^{2}$ versus $x$ collapses the data well for different inlet velocities $U_{\infty}$ (see figure \ref{Fig:Decay}a).

The decay exponents of \eqref{Eq:powerlaw} and \eqref{Eq:MVEquations} fitted to the  FSG18''x18''  data are estimated using four alternative methods and compared with the data from the RG60 for $x/M>18$: 
\begin{itemize}
\item Method I: linear fit to $U\lambda^2$ \eqref{Eq:lambda} to determine the  virtual origin followed by a linear fit to the logarithm of  \eqref{Eq:powerlaw} to determine the exponent $n$, as done by   \cite{HV2007}. \cite{antonia2003similarity} determined the virtual   origin in a similar fashion by plotting $\lambda^2/(x-x_{0})$ for  different $x_{0}$ and choosing the virtual origin yielding the  broadest plateau (which for their regular grid experiment was  $x_{0}\approx 0$). 

\item Method II: the linearised logarithm method proposed in  \cite{MV2010} to determine the unknowns in \eqref{Eq:MVEquations}.

\item Method III: direct application of a nonlinear least-squares  regression algorithm ('NLINFIT' routine in MATLAB\tm) to determine   the decay exponent and virtual origin simultaneously. This is   related to the method used by \cite{lavoie2007effects}, but further allowing the virtual origin to be determined by the algorithm. This method can be applied to \eqref{Eq:powerlaw} as well as  to \eqref{Eq:MVEquations}. Note that if applied to   \eqref{Eq:powerlaw} as it is done here, this fitting method does not necessarily yield a virtual origin compatible with \eqref{Eq:lambda}.

\item Method IV: assume the virtual origin coincides with the grid location and linearly fit the logarithm of \eqref{Eq:powerlaw}. This crude method typically yields biased estimates of the decay exponent, since there is no \emph{a priori} reason for the virtual origin to be zero. Nevertheless this is a robust method typically used to get first
order estimates of power law decay exponents in many flows  \cite[e.g. the active-grid data by][]{MW1996}. 
\end{itemize} 
 
\begin{table}[t!]
\caption{Decay exponents and virtual origin estimation using different methods}
\label{Table:DecayExponents}
\rule{\linewidth}{.5pt}\vspace*{4mm}
\centering
\begin{tabular*}{0.9\textwidth}{@{\extracolsep{\fill}}cccccccc}
Grid & U &  \multicolumn{2}{c}{Method I} & Method II & \multicolumn{2}{c}{Method III} & Method IV  \\
 & $(ms^{-1})$ & n & $x_{0}/x_*$ &  $(1+c)/2c$ &  n & $x_{0}/x_*$    \\
\midrule
RG60  & 10 & 1.32 & 0.18 & 4.34 & 1.25 & 0.53 & 1.36  \\
RG60  & 15 & 1.34 & 0.08 & 5.04  & 1.25 & 0.52 & 1.36  \\
RG60  & 20 & 1.32 & 0.06 & 5.47  & 1.21 & 0.63 & 1.33 \\
FSG18''x18''  & 10 & 2.57 & -0.31 & 7.10  & 2.51 & -0.28 & 1.93 \\
FSG18''x18''  & 15 & 2.53 & -0.28 & 8.01  & 2.41 & -0.22 & 1.95  \\
\end{tabular*}
\rule{\linewidth}{.5pt}\vspace*{4mm}
\end{table}

A main difference between these methods is the way of determining the virtual origin, which has an important influence on the decay exponent extracted. This inherent difficulty in accurately determining the decay exponent is widely recognised in the literature \cite[see e.g.][]{ML1990}.

\begin{figure}[t!]
\centering
%trim option's parameter order: left bottom right top
   \begin{lpic}{SGF_PowerLawVsExpo(110mm)}
   \lbl[W]{84,4;$x/x_*$}
   \lbl[W]{12,60;${}^{}$\hspace{5mm}}
   \lbl[W]{10,65;$\overline{u^2}$}
   \end{lpic}%\includegraphics[width=100mm]{SGF_PowerLawVsExpo}
\caption[Decay fits using methods II and III of turbulence generated by FSG18''x18'']{Turbulent kinetic energy decay of turbulence generated by the FSG18''x18'' fitted to \eqref{Eq:MVEquations} using method II (dashed-dot line) and method III (solid line). The data range used in method II is $0.6 < x/x_{*} < 1.1$, which corresponds to the streamwise region assessed by \cite{MV2010}. Streamwise data was taken at two fixed inlet velocities: (\protect\raisebox{-0.5ex}{\FilledSmallSquare}) $U_{\infty}=10ms^{-1}$, 
(\protect\raisebox{-0.5ex}{\SmallCircle}) $U_{\infty}=15ms^{-1}$. Notice that for $0.6 < x/x_{*} < 1.1$ the two methods appear to fit the data reasonably well, but further downstream the differences become evident.}
\label{Fig:Decay3}
\end{figure}

The decay data for the RG- and FSG-generated turbulence are well approximated by the curve fits obtained from
methods I \& III (see figures \ref{Fig:Decay}b \& \ref{Fig:Decay}c) and the numerical values of the exponents change only marginally (see table \ref{Table:DecayExponents}). 
On the other hand method IV also seems to fit the data reasonably well (see figures \ref{Fig:Decay}d) but the exponents retrieved for the FSG data are $n\approx 2$, slightly lower than the exponents predicted by the other methods, $n\approx 2.5$. 
The virtual origin which is forced to $x_0=0$ in method IV leads to a slight curvature in the $\log(\overline{u^2})$ versus $\log(x)$ data (almost imperceptible to the eye, compare the FSG data in figures \ref{Fig:Decay}c \& \ref{Fig:Decay}d) and a non-negligible bias in the estimated exponents. 
Nevertheless the difference in the power laws describing the measured RG- and FSG-generated turbulence is quite clear.
For completeness, the results from the experimental investigation by \cite{MW1996} on decaying active grid-generated turbulence are added in figure \ref{Fig:Decay}d. 
They applied a fitting method equivalent to method IV and reported a power-law fit yielding a decay exponent $n=1.21$. 
\cite{KCM03} employed the same method to their active grid-generated turbulence data and retrieved a similar result, $n=1.25$.

Note that there are residual longitudinal mean velocity gradients and therefore it is preferred to fit $\overline{u^2}$ data rather than $\overline{u^2}/U^2$ data. 
Nevertheless, it was checked that fitting $\overline{u^2}/U^2$ data does not meaningfully change the results nor the conclusions.

Concerning method II it can be seen (table \ref{Table:DecayExponents}) to be the most discrepant of the four methods yielding a much larger decay exponent. 
This method was proposed by \cite{MV2010} to fit the general decay law \eqref{Eq:MVEquations} and is based on the
linearisation of the logarithm appearing in the logarithmic form of \eqref{Eq:MVEquations}, i.e.
\begin{equation}
\begin{aligned}
\log(u'^2) = \log\left(\frac{2\, u'^{2}_{0}}{3}\right) + \left[ -\frac{1+c}{2c} \right] \log\left(1+\frac{4\nu a c}{l_{0}^2\, U_{\infty}}(x-x'_{0})\right).
\end{aligned}
\label{Eq:logMVEq}
\end{equation}
Linearisation of the second logarithm on the right hand side of \eqref{Eq:logMVEq} assumes  $4\nu a c/$ $(l_{0}^2 \, U_{\infty})(x-x'_{0})<<1$. 
This quantity, as has been confirmed in the present data, is indeed smaller than unity and for the farthest position, $4\nu a c/(l^2_{0} \, U_{\infty})(x-x_{0})\approx 0.3$, but the fact that this linearised method does not yield results comparable to methods I and III suggests that the linearisation of the logarithm may be an oversimplification. 
In figure \ref{Fig:Decay3} the kinetic energy decay data of turbulence generated by this FSG is shown along with the fitted curves obtained from methods II and III in a plot with a logarithmic ordinate and a linear abscissa. 
If in figure \ref{Fig:Decay3} the data taken at positions beyond $x/x_{*}\approx 1.05$ are excluded one can compare the present results with those presented in \cite{MV2010} where the data range was limited to $0.5<x/x_{*}<1.05$. 
Visually, for that range, the two different fitting methods appear to fit the data reasonably well and thus the linearisation of the logarithm in \eqref{Eq:logMVEq} seems to be justifiable. 
However, the two different fitting methods yield very different decay exponents because they also effectively yield different virtual origins: for example at $U_{\infty}=15ms^{-1}$ method III yields $(1+c)/(2c)\approx -2.4$ whereas method II yields $(1+c)/(2c)\approx -8.0$. 
If no data are excluded from figure \ref{Fig:Decay3}, it can clearly be seen that the two methods produce very different curves and very different decay exponents.

For the sake of completeness, the data acquired in the lee of the FSG3'x3' and the RG115 at the centreline are also fitted with a power-law using method III (plotted as solid lines in figure \ref{fig:UmvsX}). 
The decay exponents and virtual origins obtained are $n=3.0,\,2.4$ and $x_0/x_* = -0.6,\,-0.7$ for the FSG3'x3' and RG115 data, respectively. 
The differences in decay exponents and virtual origins between the FSG3'x3'  and FSG18''x18'' are very likely a consequence of the differences in the lateral transport profiles observed in   \S \ref{sec:homo}.

Nevertheless, the present data allow the inference  that the decay exponents for the present FSG- and RG-generated turbulence at the centreline in the region $1 < x/x_{\mathrm{peak}} < 4$ are consistently higher than those in
all boundary-free turbulent flows listed in table  4.1 (in p. 134) of \cite{TennekesLumley:book} and much higher (by a factor between 4/3 and 2) than those found in the assessed regions of decaying turbulence generated by active grids \cite[]{MW1996,KCM03} as well as RGs far downstream\footnote{It may be the case that the FSGs also return the usual power-laws far downstream. However, those measurements have not yet been possible since it would either require, (i) special facilities with, e.g., a much longer test section  or (ii) a FSG re-designed to have a considerably smaller $M$, so that $x_{*}$ and/or $x_{\mathrm{peak}}$ would be much smaller than the length of the test section. However, the latter would lead to a FSG where the smallest thickness, currently about 1mm, is too small to be manufactured by conventional fabrication methods.}, i.e  $x/x_{\mathrm{peak}} \geq 8$ (or $x/M \geq 20$, i.e. the region customarily assessed in the literature, see table \ref{table:RGdata}). \\

It might be interesting to note that in many boundary-free turbulent flows a conserved quantity such as $u'^2L^{M+1}=\mathrm{constant}$ exists. Look at table 4.1 of \cite{TennekesLumley:book} and note that $M=1,\,3,\,5,\,7$ for the four wakes, $M=-1$ for the mixing layer, $M=0,\,1$ for the jets and $M \ge 2$ for RG-grid turbulence (at least far downstream).
If the flow is also such that $Udu'^2/dx\propto -\varepsilon$ then $C_{\varepsilon}=\mathrm{constant}$ implies
 \begin{equation*}
 n=\frac{2(M+1)}{M+3} 
 \end{equation*}
 and $C_{\varepsilon}\sim Re_{\lambda}^{-1}$ implies 
 \begin{equation*}
 n=\frac{M+1}{2}
 \end{equation*}
 (which is larger than $n=2(M+1)/(M+3)$ provided that $M>1$). 
 Considering, for example, the range $ M \geq 2$, the exponent $n$ corresponding to $C_{\varepsilon}\sim Re_{\lambda}^{-1}$ is at least $5/4$ times larger than the exponent $n$ corresponding to $C_{\varepsilon} = \mathrm{constant}$, and is generally much larger. 
If $M=3$ or $M=4$ then $C_{\varepsilon} \sim Re_{\lambda}^{-1}$ implies $n=2$ or $n=2.5$, close to what is observed here for the RG- and FSG-generated turbulence in the  region $1 < x/x_{\mathrm{peak}} < 4$, whereas $C_{\varepsilon} \sim \mathrm{constant}$ implies $n=4/3$ or $n=10/7$.
 In the following chapter (ch. \ref{chp:4}) it is shown that $C_{\varepsilon}\sim Re_{\lambda}^{-1}$ is a good approximation for the dissipation behaviour of the RG- and FSG-generated turbulence in the  region $1 < x/x_{\mathrm{peak}} < 4$ at the centreline and conversely,  $C_{\varepsilon} \sim \mathrm{constant}$ is a good approximation for the downstream region $x/x_{\mathrm{peak}} > 5$, at least in the lee of RG60.
 
At this stage there is no proof that a conserved quantity such as $u'^{2}L^{M+1}=\mathrm{constant}$ exists for near field of RG- and/or FSG-generated turbulence, whereas for the far field it has been argued that there is \cite[see e.g.][]{KD2010,KD2011,V2011}\footnote{See also \cite{VV2011PLA} for further discussions.% The latter work, which was co-authored by the present researcher, is not further discussed in the context of this thesis
}. 
The previous paragraph is therefore only indicative and serves to illustrate how a $C_{\varepsilon}$ which is a decreasing
function of $Re_{\lambda}$ can cause the decay exponent to be significantly larger than a $C_{\varepsilon}$ which is constant during decay and can even return decay exponents comparable to the ones observed here. 
Of course the decaying turbulence studied in this work  is not homogeneous (nor isotropic), particularly because of the presence of transverse turbulent transport of turbulent kinetic energy and therefore of significant gradients of third-order one-point velocity correlations. 
As a consequence, a conserved quantity such as $u'^{2}L^{M+1}=\mathrm{constant}$, if it exists, cannot result from a two-point equation such as the von K\'arm\'an-Howarth equation \cite[]{KH1938} for homogeneous turbulence \cite[see][]{V2011}.

\section{Two-point large-scale anisotropy} \label{sec:Lu}
In the following, the large-scale anisotropy of one of the decaying turbulent flows is studied. 
Focus is given to RG115-generated turbulence for three reasons: (i) it is a better approximation to a periodic flow than  RG230-generated turbulence (\S\S \ref{sec:confinement}, \ref{sec:homo}); (ii) it has a larger $x_{\mathrm{peak}}$ value and therefore, as will be shown in ch. \ref{chp:4}, a longer region with a nonclassical dissipation behaviour than RG60-generated turbulence; (iii) the constancy of the integral-length scale to the Taylor microscale ratio, which is indicative of the nonclassical dissipation behaviour,  is improved by comparison to FSG-generated turbulence as is reported in ch. \ref{chp:4}.
Data for the farthest downstream location on the centreline of RG60 is also shown.
This data is from the region where the turbulent dissipation follows the classical behaviour (ch. \ref{chp:4}).
As $x/M\approx 51$, these data are as close to homogeneous and isotropic turbulence as any of the present datasets can be expected to get.

The downstream evolution of longitudinal and transverse correlations over both longitudinal and transverse separations is studied. 
Recall that these are defined as (no summation implied over the indices), 
\begin{equation}
B_{ii}^{(k)}(\mathbf{X},r) \equiv B_{ii}(\mathbf{X},r_k) = \frac{\overline{u_i(\mathbf{X}-r_k/2)\, u_i(\mathbf{X}+r_k/2)}}{\overline{u_i(\mathbf{X})\,u_i(\mathbf{X})}},
\end{equation}
where $r_k$ is the separation $r$ in the direction along the $x_k$-axis (with $x_1=x$, $x_2=y$, $x_3=z$) and $\mathbf{X}$ is the centroid position vector.

Data obtained with the $2\times$XW apparatus described in \S \ref{sec:WTB} are used to calculate $B_{11}^{(2)}$ and $B_{22}^{(2)}$.
In addition, $B_{11}^{(1)}$ and $B_{22}^{(1)}$ are calculated from the time-varying signal using Taylor's hypothesis.
These calculations are repeated for six downstream positions of $\mathbf{X}$ along the centreline which cuts through the centre of the central mesh (see figure \ref{fig:grids}d) and six downstream positions along the line ($y=-M/2$, $z=0$) which cuts through the lower bar of the central mesh.

Though of less importance for this thesis, it is nevertheless worth noting that these data can also be used to compute the scalar correlation function (with summation over $i$) $B_{ii}(\mathbf{X};r_1,\,r_2,\,0)$ in the $(r_1,\,r_2,\,0)$ plane if the assumption is made that $B_{ii}(\mathbf{X};r_1,\,r_2,\,0) =  B_{\parallel}(\mathbf{X};r_1,\,r_2,\,0) + 2B_{\perp\perp}(\mathbf{X};r_1,\,r_2,\,0)$.
By further assuming axisymmetry around the axis intercepting $\mathbf{X}$ and normal to the $(0,\,r_2,\,r_3)$ plane one can then map $B_{ii}(\mathbf{X};r_1,\,r_2,\,0)$ onto the spherical coordinates  ($R,\,\theta,\,\phi$) (where $\phi$ is the angle around the axis) and  extract an estimate of the spherical averaged correlation function,
\begin{equation}
B^*(\mathbf{X},r)\equiv \iint\limits_{r=|\mathbf{r}|}\!B_{ii}(\mathbf{X},\mathbf{r})\,d\mathbf{r}.
\end{equation}
The assessment of the two assumptions that are used to calculate $B^*$ lies beyond the scope of the present work as it concerns issues which are mostly peripheral to the main conclusions. 
Some support for these assumptions around the centreline can nevertheless be found in \cite{Sylvain2011} and \cite{Nagata2012}, though their validity around the ($y=-M/2$, $z=0$)  axis can be expected to be more doubtful. 

This section's main conclusions concern comparisons between the different longitudinal and transverse correlation functions and their associated integral-length scales $L_{ii}^{(k)}$  (no summation over $i$)
\begin{equation}
L_{ii}^{(k)}(\mathbf{X})=\frac{1}{B_{ii}^{(k)}(\mathbf{X},0)}\int\limits_0^{\infty}\!B_{ii}^{(k)}(\mathbf{X},r)\,dr.
\end{equation}
The integral-length scale
\begin{equation}
L(\mathbf{X}) \equiv \frac{1}{B^{*}(0)}\int\limits_0^{\infty}\! B^*(r)\, dr 
\end{equation}
is also calculated and compared with the other integral-length scales to check, for example, whether in spite of the two assumptions to estimate $B^*$, $3/2L=2L_{22}^{(1)}$ and/or $3/2L=L_{11}^{(1)}$ as is the case in incompressible isotropic turbulence \cite[]{MY75,Batchelor:book}.
The main checks, however, are to determine how far or close the turbulence is from incompressible isotropic relations $2L_{22}^{(1)}=L_{11}^{(1)}$, $2L_{11}^{(2)}=L_{22}^{(2)}$, $B_{11}^{(1)}=B_{22}^{(2)}$ and $B_{11}^{(2)}=B_{22}^{(1)}$ \cite[]{MY75,Batchelor:book}.
These checks do not rely on the two assumptions used in the estimation of $B^*$.

%%%%%% Figures %%%%%%%%%%%%%
\begin{figure}[t!]
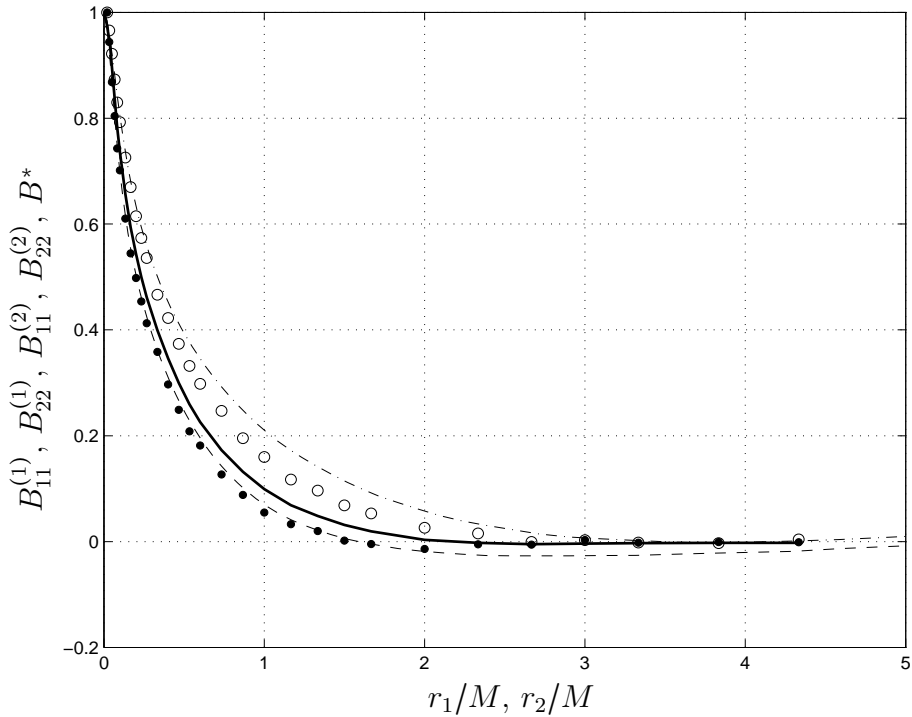

\centering
   \begin{lpic}{CompareCorrRG60_x=3050(120mm)}
   \lbl[W]{100,3;$r_1/M$, $r_2/M$}
   \lbl[W]{5,75,90;$B_{11}^{(1)}$, $B_{22}^{(1)}$, $B_{11}^{(2)}$, $B_{22}^{(2)}$, $B^{*}$}
   \end{lpic}
\caption[Comparison between the different correlation functions in the lee of RG60]{Comparison between the longitudinal and transverse correlation functions for longitudinal and transverse separations of turbulence generated by RG60. The centroid of the correlation functions is located at  the centreline $y=0$ and at $x=3050\mathrm{mm}$, corresponding to $x/x_{\mathrm{peak}}\approx 21$. (dash-dotted line)  $\mathrm{B}_{11}^{(1)}$, (dashed line)  $\mathrm{B}_{22}^{(1)}$, (\protect\raisebox{-0.5ex}{\FilledSmallCircle}) $\mathrm{B}_{11}^{(2)}$,  (\protect\raisebox{-0.5ex}{\SmallCircle})  $\mathrm{B}_{22}^{(2)}$ and (solid line) $B^*$. $L_{11}^{(1)}/M = 0.67$, $L_{22}^{(2)}/L_{11}^{(1)} = 0.74$, $2L_{22}^{(1)}/L_{11}^{(1)} = 0.82$, $\,2L_{11}^{(2)}/L_{11}^{(1)} = 0.91$ and $3/2\,L/L_{11}^{(1)} = 0.87$.}
\label{fig:B_RG60}
\end{figure}
\begin{figure}[ht!]
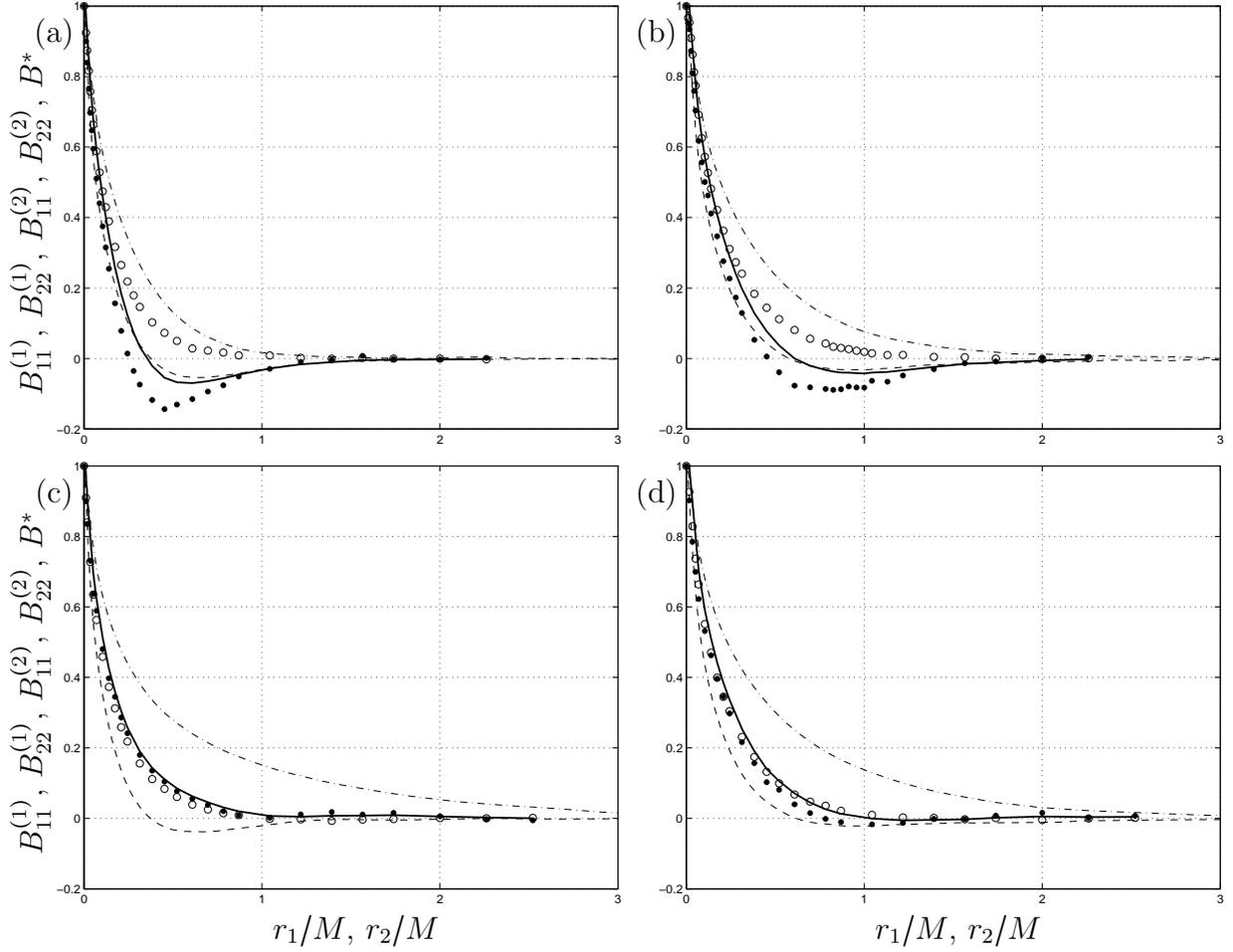

\centering
\begin{minipage}[c]{0.5\linewidth}
   \centering
   \begin{lpic}[b(-3mm)]{CompareCorrCentre_x=1250(80mm)}
   \lbl[W]{100,4;${}^{}$\hspace{100mm}}
   \lbl[W]{3,77,90;$B_{11}^{(1)}$, $B_{22}^{(1)}$, $B_{11}^{(2)}$, $B_{22}^{(2)}$, $B^{*}$\hspace{1mm}}
   \lbl{10,130;(a)}
   \end{lpic}
\end{minipage}%
\begin{minipage}[c]{0.5\linewidth}
   \centering 
   \begin{lpic}[b(-3mm)]{CompareCorrCentre_x=3050(80mm)}
   \lbl[W]{100,4;${}^{}$\hspace{100mm}}
   \lbl[W]{4.7,75,90;${}^{}$\hspace{60mm}}
   \lbl{10,130;(b)}
   \end{lpic}
\end{minipage}
\begin{minipage}[c]{0.5\linewidth}
   \centering
   \begin{lpic}{CompareCorrOff_x=1250(80mm)}
   \lbl[W]{100,1;$r_1/M$, $r_2/M$}
   \lbl[W]{3,77,90;$B_{11}^{(1)}$, $B_{22}^{(1)}$, $B_{11}^{(2)}$, $B_{22}^{(2)}$, $B^{*}$\hspace{1mm}}
   \lbl{10,130;(c)}
   \end{lpic}
\end{minipage}%
\begin{minipage}[c]{0.5\linewidth}
   \centering 
   \begin{lpic}{CompareCorrOff_x=3050(80mm)}
   \lbl[W]{4.7,75,90;${}^{}$\hspace{80mm}}
   \lbl[W]{100,1;$r_1/M$, $r_2/M$}
    \lbl{10,130;(d)}
 \end{lpic}
\end{minipage}%
\caption[Comparison between the different correlation functions in the lee of RG115]{Comparison between the longitudinal and transverse correlation functions for longitudinal and transverse separations of turbulence generated by RG115. The centroid of the correlation functions is located at, (a, b) the centreline $y=0$ and (c, d) behind a bar $y=-M/2$ for downstream locations, (a, c) $x=1250\mathrm{mm}$ and (b, d) $x=3050\mathrm{mm}$. Description of symbols/lines can be found in figure \ref{fig:B_RG60}.}
\label{fig:B_RG115}
\end{figure}
%\begin{figure}
%\centering
%\includegraphics[width=65mm]{LuProfile}
%\includegraphics[width=65mm]{LvProfile}
%\caption{Spanwise profiles of the integral length-scales based on the (a) longitudinal and (b) transverse correlations for longitudinal separations ($L_{11}^{(1)}$ and $L_{22}^{(1)}$, respectively) at different downstream locations at the lee of  RG115-generated turbulence. (\protect\raisebox{-0.5ex}{\SmallSquare}) $x=950$mm, (\protect\raisebox{-0.5ex}{\FilledSmallCircle}) $x=1250$mm, (\protect\raisebox{-0.5ex}{\Diamondshape}) $x=1700$mm, (\ding{72}) $x=2150$mm, (\protect\raisebox{-0.5ex}{\SmallCircle}) $x=2600$mm and (\protect\raisebox{-0.5ex}{\FilledSmallSquare}) $x=3050$mm.}
%\end{figure}
%%%%%% Figures %%%%%%%%%%%%%

Figure \ref{fig:B_RG60} shows the different correlation functions for RG60-generated turbulence at the farthest downstream location on the centreline where the turbulence is expected to be closest to homogeneous and isotropic.
For $r<2M$ the transverse correlations are roughly equal, $B_{22}^{(1)} \approx B_{11}^{(2)}$, whereas the longitudinal correlations, $B_{11}^{(1)}$ and $B_{22}^{(2)}$ are less so. 
For $r>2$M, $B_{22}^{(1)}$ seems to tend slowly to zero, contrasting with $B_{11}^{(2)}$.
These departures between $B_{11}^{(2)}$ and  $B_{22}^{(1)}$ for large $r$ may be related to the lack of validity of Taylor's hypothesis for long time differences, but may also be genuine departures from isotropy.

The ratio between the different integral-length scales is presented in the caption of figure  \ref{fig:B_RG60} (for isotropic turbulence these ratios are equal to one) and they indicate small, but non-negligible, departures from isotropy even at this relatively far downstream location ($x/M\approx 51$).

The RG115 data, on the other hand, show a larger departure from isotropy (figure \ref{fig:B_RG115} and table \ref{table:Lu}). 
The ratios $L_{22}^{(2)}/L_{11}^{(1)}$ and $2L_{22}^{(1)}/L_{11}^{(1)}$ do not show any tendency towards isotropy between $x/x_{\mathrm{peak}}\approx 1.1$ and $x/x_{\mathrm{peak}}\approx 3.7$ on the centreline as both remain about constant with values around $0.62$ and $0.54$ respectively.
The ratio $2L_{11}^{(2)}/L_{11}^{(1)}$ is even further away from the isotropic value 1 but grows quite steeply with streamwise distance $x$ on the centreline.
The very small values of this particular ratio reflect the prominent negative loop in  $B_{11}^{(2)}$ at the lower $x/x_{\mathrm{peak}}$ locations, a negative loop which progressively weakens as $x/x_{\mathrm{peak}}$ increases thereby yielding increasing the values of  $2L_{11}^{(2)}/L_{11}^{(1)}$.
This effect  also presumably explains the steep growth of $3/2L/L_{11}^{(1)}$ with increasing $x/x_{\mathrm{peak}}$ along the centreline because of the related negative loop in $B^*$ at the lower $x/x_{\mathrm{peak}}$ locations which also disappears with increasing $x/x_{\mathrm{peak}}$.
 
The decaying oscillation and related negative loop in the transverse correlation $B_{11}^{(2)}$ of longitudinal fluctuating turbulent velocities is likely a remnant of the periodicity of the grid leading to a peak in negative correlation mid way between bars, i.e. at $r=\pm M/2$ as indeed observed in figures \ref{fig:B_RG115}a,b.
The grid's periodicity can indeed leave a mark on the flow in the form of a transverse near-periodicity of its vortex shedding which disappears far downstream.
When this happens and the correlation function has a negative loop as a result, the integral-length scale obtained by integrating this correlation function loses its usual meaning as a spatial extent of correlation. 

Along the ($y=-M/2,\,z=0$) line crossing the lower bar of the central mesh of the grid the length-scale ratios are different.
Firstly, $L_{22}^{(2)}/L_{11}^{(1)}$ and $2L_{22}^{(1)}/L_{11}^{(1)}$ exhibit a significant increase with increasing $x$ which they do not exhibit on the centreline.
However, $L_{22}^{(2)}/L_{11}^{(1)}$ and $2L_{22}^{(1)}/L_{11}^{(1)}$ take significantly lower values than at the same streamwise positions along the centreline, indicating more anisotropy in the wake of the bar than along the centreline between bars. 
By a different measure, though, that of $2L_{11}^{(2)}/L_{11}^{(1)}$, the turbulence appears more isotropic in the wake of a bar than along the centreline because $2L_{11}^{(2)}/L_{11}^{(1)}$ is very much closer to 1 behind the bar.
It is clear that each of these ratios between the different length-scales and  $L_{11}^{(1)}$ show a different trend as the turbulence decays.
Nevertheless, it is also clear that $B_{11}^{(1)} > B_{22}^{(2)}$ for all separations $r$, and consequently $L_{11}^{(1)}>L_{22}^{(2)}$ at all streamwise positions accessed by the present measurements both behind the bar and along the centreline. 
This suggests that the large-scale eddies are elongated in the streamwise direction.
This had previously been observed for RG-generated turbulence by \cite{Graham} which estimated that $L_{33}^{(2)}/L_{33}^{(1)}\approx 3/4$ at $x/M\approx 11$, i.e. a $25\%$ elongation in the streamwise direction.
Also note that the PIV data of \cite{discettietal11}, taken along the centreline of a FSG similar to the FSG18''x18''  also suggest such an elongation of large-scale eddies in the streamwise direction (see their figure 14).
\\

\begin{table}[t!]
\caption[Integral-length scales of RG115-generated turbulence]{Several integral-length scales for different downstream location of RG115-generated turbulence. The different integral length-scales are normalised with $L_{11}^{(1)}$ with a pre-factor such that unity would correspond to isotropic incompressible turbulence.}
\label{table:Lu}
\rule{\linewidth}{.5pt}\vspace*{4mm}
\centering
\begin{tabular*}{0.9\textwidth}{@{\extracolsep{\fill}}cccccccc}
\vspace{1mm}$x/x_{\mathrm{peak}}$ &   & 1.1 & 1.5 & 2.0 & 2.6 & 3.1 & 3.7 \\
\vspace{.5mm}   \multirow{6}{*}{\begin{sideways}Centreline\end{sideways}} & $L_{11}^{(1)}/M$& 0.24 & 0.25 & 0.27 & 0.30 & 0.34 & 0.36 \\ 
\vspace{.5mm}                   & $L_{22}^{(2)}/L_{11}^{(1)}$     & 0.60 & 0.64 & 0.65 & 0.64 & 0.61 & 0.62 \\
\vspace{.5mm}                   & $2\,L_{22}^{(1)}/L_{11}^{(1)}$& 0.54 & 0.53 & 0.54 & 0.54 & 0.53 & 0.55 \\
\vspace{.5mm}                   & $2\,L_{11}^{(2)}/L_{11}^{(1)}$& 0.13 & 0.14 & 0.19 & 0.30 & 0.36 & 0.44 \\
\vspace{2mm}                    & $3/2\,L/L_{11}^{(1)}$              & 0.40 & 0.44 & 0.48 & 0.54 & 0.57 & 0.61 \\
\vspace{0.5mm}   \multirow{6}{*}{\begin{sideways}Behind bar\end{sideways}}& $L_{11}^{(1)}/M$ & 0.40 & 0.40 & 0.42 & 0.40 & 0.43 & 0.44 \\
\vspace{.5mm}                  & $L_{22}^{(2)}/L_{11}^{(1)}$    & 0.38 & 0.40 & 0.40 & 0.46 & 0.44 & 0.47 \\
\vspace{.5mm}                  & $2\,L_{22}^{(1)}/L_{11}^{(1)}$& 0.31 & 0.34 & 0.36 & 0.43 & 0.44 & 0.49 \\
\vspace{.5mm}                  & $2\,L_{11}^{(2)}/L_{11}^{(1)}$& 0.95 & 0.94 & 0.88 & 0.91 & 0.86 & 0.84 \\
\vspace{.5mm}                  & $3/2\,L/L_{11}^{(1)}$              & 0.75 & 0.75 & 0.74 & 0.79 & 0.74 & 0.75 \\
\end{tabular*}
\rule{\linewidth}{.5pt}\vspace*{4mm}
\end{table}

The presented data raises the issue of the appropriateness of $L_{11}^{(1)}$ and $L_{11}^{(1)}/\sqrt{\overline{q^2}}$ (or $L_{11}^{(1)}/u'$) as the characteristic length- and time-scales of the large eddies, which are typically used for the normalisation of $\varepsilon$.
For turbulent flows with varying anisotropy with downstream location, such as the present flow(s) and likely many free shear flows,  the growth rate of the various integral-length scales and their corresponding time scales will be different and therefore, care must be taken interpreting these surrogate normalisations of the energy dissipation rate. 
This issue is further discussed in the following chapter.

\section{Small-scale anisotropy} \label{sec:Eps}

The attention is now turned to the (an)isotropy of the small scales.
These are customarily assessed by comparing the ratios between the various mean square velocity derivatives with the isotropic benchmark \cite[see e.g.][]{George91}.
For example,
\begin{equation}
K_1 = 2\frac{\overline{\left(\partial u/\partial x\right)^2}}{\overline{\left(\partial v/\partial x\right)^2}}, \hspace{5mm} K_2 = 2\frac{\overline{\left(\partial v/\partial y\right)^2}}{\overline{\left(\partial u/\partial y\right)^2}}, \hspace{5mm}
K_3 = 2\frac{\overline{\left(\partial u/\partial x\right)^2}}{\overline{\left(\partial u/\partial y\right)^2}},
\label{eq:localiso}
\end{equation}
should all be unity for a locally isotropic flow \cite[see][where all the velocity derivative ratios are determined for an isotropic turbulent field]{Taylor1935}. 

There have been many experimental investigations of local isotropy in canonical turbulent shear flows \cite[see e.g.][and references therein]{BAS1987,George91,SV94}, but it seems that in this context grid-generated turbulence has not attracted much attention, perhaps because local isotropy is thought to be guaranteed.  
However, the experimental data by \cite{Tsinober92} for RG-generated turbulence (see table \ref{tab:RGTsinober}) suggest significant departures from local isotropy (to the best of the author's knowledge no other assessment of local isotropy for RGs can be found in the literature).  
The data have significant scatter making it difficult to discern trends as the flow decays, particularly for $K_2$, but it seems that $K_1$ and $K_3$ are about constant between $6<x/x_{\mathrm{peak}}<31$ with numerical values surrounding $1.4$ and $0.8$ respectively. 
Nevertheless, this isolated experiment has yet to receive corroboration and therefore one cannot exclude the possibility that these results are an artifice of measurement error and/or bias, which is plausible taking into account the complexity and lack of maturity of the multicomponent hot-wire sensor used.  

Recently, \cite{gomesfernandesetal12} presented estimates of $K_1$ and $K_3$ in the lee of three FSGs along the centreline up to a downstream region of about $4x_{\mathrm{peak}}$.
Their data for the FSG similar to the present FSG18''x18'' indicate that $K_1$ and $K_3$ are approximately constant beyond $x_{\mathrm{peak}}$ with  numerical values of about $1.2$ and $1.1$, respectively.

However, the validity of the approximations of local isotropy (or locally axisymmetry) to the decaying turbulence generated by square-mesh grids is a peripheral topic to the present work. 
For example, one may easily argue that the Reynolds numbers straddled in the present experiments, and generally in grid-generated turbulence experiments (see e.g. table \ref{table:RGdata}) is far too low for the approximation to local isotropy to be expected to hold. 
What is, in fact, the main concern here is to assess how the anisotropy of the small scales varies as the turbulent flow decays and/or $Re_{\lambda}$ changes. 
If the ratios $K_1$, $K_2$ and $K_3$ (and/or ratios than can be formed with the other components of the mean square velocity derivative tensor) vary significantly during the turbulence decay and/or with $Re_{\lambda}$, then the surrogate isotropic dissipation estimate, $\varepsilon^{\mathrm{iso}} = 15\nu \overline{(\partial u/\partial x)^2}$, obtained from 
one-component measurements (e.g. with a single hot-wire)  is not representative of the true turbulent kinetic energy dissipation, $\varepsilon$, as usually assumed.
This would bear severe consequences, not only for the present work, but also for turbulence research in general since the overwhelming majority of the dissipation estimates found in the literature are indeed estimates of the surrogate $\varepsilon^{\mathrm{iso}}$, typically obtained with a single hot-wire\footnote{Multicomponent hot-wires and particle image velocimetry (PIV) have many resolution and/or noise issues and inhibit accurate and reliable measurements of several components of the mean square velocity derivative tensor \cite[see also the discussion in][where an exact filter is proposed for the unavoidable noise contaminating PIV measurements]{Anthony2012}.}.
In any case, the data of \cite{Tsinober92} and \cite{gomesfernandesetal12} do not suggest that there are significant variations of the anisotropy ratios.  \\
   
In the present thesis, the $2\times$XW apparatus described in \S \ref{sec:WTB} is used to measure $\overline{\left(\partial u/\partial x\right)^2}$, $\overline{\left(\partial v/\partial x\right)^2}$, $\overline{\left(\partial u/\partial y\right)^2}$ and $\overline{\left(\partial v/\partial y\right)^2}$ for the RG60- and RG115-generated turbulent flows along the centreline, which in turn allow the estimation of the ratios \eqref{eq:localiso}.
These are presented in tables \ref{tab:RG60} and \ref{tab:RG115}, respectively. 
Note that, as will be shown in the following chapter, the downstream extent for the RG115 data corresponds to a region where the dissipation behaves in a nonclassical way whereas for the RG60 data it corresponds to the classically expected behaviour, thus allowing their direct comparison. 
Furthermore, the Reynolds numbers, $Re_{\lambda}$ of the turbulence generated by the two grids are comparable  at the  respective measurement locations (due to the large difference in blockage ratio between the grids). 
Also, the Kolmogorov microscales $\eta$ are comparable, which is beneficial since the same apparatus can be used for both experiments without penalising resolution. 

The first observation is that the ratios $K_1$  and $K_3$ are roughly constant during the turbulence decay for both the  RG60 and the RG115 data, in-line with the observations from the data of \cite{gomesfernandesetal12}. 
The numerical values of $K_1\approx 1.09\,\&\,1.04$ and $K_3\approx 0.8\,\&\,0.72$ for the RG60 and RG115 data respectively suggest that the RG115 data are closer to the isotropic benchmark in terms of the $K_1$ ratio but conversely, the RG60 data are closer to unity in terms of the $K_3$ ratio. 
The ratio $K_2$ increases away from unity, particularly for the RG60 data, which could be an indication of increasing anisotropy as the flow decays. However, $K_2$ is a ratio involving $\overline{(\partial v/\partial y)^2}$ whose measurement is strongly contaminated by aerodynamic interference (see \S \ref{sec:resolution}) and therefore the results are likely artificial.

Comparing with the data of \cite{gomesfernandesetal12}, it is clear that the present  numerical values of $K_1$ are always closer to the isotropic benchmark. 
Curiously, the ratio $K_3$ for the present data are $20\%$ to $30\%$ smaller than 1 whereas for the data of \cite{gomesfernandesetal12} they are about $10\%$ higher than unity. 
These differences may be attributable to the different inflow conditions, e.g. grid geometry, free-stream turbulence and $Re_M$, but may also be an artifice of measurement bias.  
In any case, the present data and those of \cite{gomesfernandesetal12} support the hypothesis that the small-scale anisotropy remains approximately constant, regardless of the behaviour of the energy dissipation as the turbulent flow decays (see ch. \ref{chp:4}).\\

The present data allow the calculation of several estimates of the turbulent dissipation.
In particular, the four mean square velocity derivative components are necessary and sufficient to estimate the dissipation in a locally axisymmetric turbulent flow \cite[]{George91}.
Even though the present data do not allow the test for local axisymmetry, one might nevertheless expect a locally axisymmetric dissipation estimate to be closer to the actual dissipation rate than the isotropic dissipation estimate.
%
 % RG60
\begin{table}[ht!]
\caption[Turbulence statistics for the RG60]{Turbulence statistics for the RG60. The dissipation estimate $\varepsilon^{iso,3}$ is used to compute $Re_{\lambda}$, $\lambda$ and $\eta$, whereas $\varepsilon^{iso}$ is used to compute $Re_{\lambda}^{\mathrm{iso}}$.}
\label{tab:RG60}
\rule{\linewidth}{.5pt}\vspace*{4mm}
\centering
\begin{tabular*}{0.9\textwidth}{@{\extracolsep{\fill}}rcccccc}
Location & $1250$ & $1700$ & $2150$ & $2600$ & $3050$ \\
$x/x_{\mathrm{peak}}$ & $8.5$ & $11.5$ & $15.6$ & $17.6$ & $20.7$ \\
$Re_{\lambda}^{\mathrm{iso}}$ & $106$ & $102$ & $98$ & $94$ & $91$ \\
$Re_{\lambda}$ & $91$ & $86$ & $82$ & $80$ & $77$ \\
$\overline{q^2}\,(\mathrm{m^2s^{-2}})$  & $0.43$ & $0.28$ & $0.20$ & $0.16$ & $0.13$ \\
$\lambda \, (\mathrm{mm})$ & $3.6$ & $4.2$ & $4.8$ & $5.3$ & $5.6$ \\
$\eta\, (\mathrm{mm})$ & $0.19$ & $0.23$ & $0.27$ & $0.30$ & $0.32$ \\
$\varepsilon^{iso}$   & 2.19 &  0.99 & 0.55 & 0.36 & 0.25 \\% Eps_iso 15*nu/dudx2
$\varepsilon^{iso,2}$& 2.05 &  0.92 & 0.51 & 0.33 & 0.24 \\% Eps_iso2 nu(3dudx2+6dvdx2)
$\varepsilon^{iso,3}$& $2.51\pm0.07$ &  $1.18\pm0.02$ & $0.66\pm0.02$ & $0.42\pm0.01$ & $0.30\pm0.01$ \\% Eps_iso3 nu(dudx2+2dvdx2+4*dudy2+2dvdy2)
$\varepsilon^{axi}$  & $2.75\pm 0.19$ &  $1.32\pm0.08$ & $0.77\pm0.07$ & $0.51\pm0.05$ & $0.36\pm0.04$  \\% Eps_axi
$d/\eta \approx l_w/\eta$ & 2.6  & 2.2 & 1.9 & 1.7 & 1.6 \\% d/eta~l/eta
$(\Delta y\approx 1.2\mathrm{mm})/\eta$& 6.4 &  5.4   &  4.8  & 4.0   & 3.8 \\% dy/eta
$(\Delta y\approx 2.0\mathrm{mm})/\eta$& 10.5 &  9.0   &  7.5  & 6.7   & 6.2 \\% dy/eta
%$\overline{\left(\partial v/\partial x \right)^2}/ 2\overline{\left(\partial u/\partial x\right)^2} $& 0.92 &  0.92 & 0.92 & 0.92 & 0.91 \\% dvdx2r/2dudx2r
%$\overline{\left(\partial u/\partial y\right)^2}/2\overline{\left(\partial v/\partial y\right)^2} $ & $0.86\pm 0.10$ &  $0.84\pm 0.09$ & $0.78\pm 0.11$ & $0.70\pm 0.11$ & $0.72\pm 0.12$ \\% dudy2r./2dvdy2r
%$\overline{\left(\partial u/\partial y\right)^2}/\overline{\left(\partial v/\partial x\right)^2}$ &  $1.31\pm 0.05$ &  $1.38\pm 0.03$ & $1.38\pm 0.01$ & $1.31\pm 0.04$ & $1.35\pm 0.01$ \\% dudy2r./dvdx2r
$K_1$& 1.09 & 1.09 & 1.08 & 1.09 & 1.09 \\
$K_2$& $1.18\pm 0.13$ &  $1.21\pm 0.13$ & $1.31\pm 0.18$ & $1.46\pm 0.22$ & $1.41\pm 0.23$ \\
$K_3$& $0.83\pm 0.03$ &  $0.79\pm 0.02$ & $0.79\pm 0.01$ & $0.83\pm 0.03$ & $0.81\pm 0.01$ \\
$u'/v'$ & $1.07$ & $1.07$ & $1.06$ & $1.06$ & $1.06$ \\
%$r_{1,1}$& 0.939 & 0.951 & 0.956 & 0.960 & 0.962 \\% r_dudx2
%$r_{2,1}$& 0.881 &  0.903 & 0.915 & 0.924 & 0.930 \\% r_dvdx2
%$r_{1,2}\,|\Delta y(2)$& 0.616 &  0.698 & 0.756 & 0.794 & 0.820 \\% r_dudy2
%$r_{2,2}\,|\Delta y(2)$& 0.713 &  0.780 & 0.825 & 0.855 & 0.874 \\% r_dvdy2
\end{tabular*} 
\rule{\linewidth}{.5pt}\vspace*{4mm}
\end{table} 
%
% RG115
\begin{table}[ht!]
\caption[Turbulence statistics for the RG115]{Turbulence statistics for the RG115. The dissipation estimate $\varepsilon^{iso,3}$ is used to compute $Re_{\lambda}$, $\lambda$ and $\eta$, whereas $\varepsilon^{iso}$ is used to compute $Re_{\lambda}^{\mathrm{iso}}$.}
\label{tab:RG115}
\rule{\linewidth}{.5pt}\vspace*{4mm}
\centering
\begin{tabular*}{0.9\textwidth}{@{\extracolsep{\fill}}rcccccc}
Location & $1250$ & $1700$ & $2150$ & $2600$ & $3050$ \\
$x/x_{\mathrm{peak}}$ & $1.5$ & $2.0$ & $2.6$ & $3.1$ & $3.7$ \\
$Re_{\lambda}^{\mathrm{iso}}$  & $156$ & $140$ & $133$ & $124$ & $116$ \\
$Re_{\lambda}$  & $114$ & $105$ & $98$ & $91$ & $88$ \\
$\overline{q^2}\,(\mathrm{m^2s^{-2}})$  & $0.78$ & $0.51$ & $0.36$ & $0.26$ & $0.20$ \\
$\lambda \, (\mathrm{mm})$ & $3.3$ & $3.8$ & $4.3$ & $4.7$ & $5.2$ \\
$\eta\, (\mathrm{mm})$ & $0.16$ & $0.19$ & $0.22$ & $0.25$ & $0.28$ \\
$\varepsilon^{iso}\,(\mathrm{m^{2}s^{-3}})$  &  4.23 & 2.06 & 1.17 & 0.70 & 0.45 \\% Eps_iso 15*nu/dudx2
$\varepsilon^{iso,2}\,(\mathrm{m^{2}s^{-3}})$&  4.08 & 2.02 & 1.13 & 0.67 & 0.43 \\% Eps_iso2 nu(3dudx2+6dvdx2)
$\varepsilon^{iso,3}\,(\mathrm{m^{2}s^{-3}})$&  $5.21\pm0.24$ & $2.59\pm0.12$ & $1.48\pm0.02$ & $0.89\pm0.02$ & $0.55\pm0.04$\\% Eps_iso3 nu(dudx2+2dvdx2+4*dudy2+2dvdy2)
$\varepsilon^{axi}\,(\mathrm{m^{2}s^{-3}})$  &  $5.44\pm0.58$ & $2.71\pm0.21$ & $1.59\pm0.07$ & $0.97\pm0.06$ & $0.62\pm0.09$  \\% Eps_axi
$d/\eta \approx l_w/\eta$ & 3.2 & 2.6 & 2.3 & 2.0 & 1.8 \\% d/eta~l/eta
$(\Delta y\approx 1.2\mathrm{mm})/\eta$& 7.7 & 6.3 & 5.5 & 5.7 & 4.2 \\% dy/eta
$(\Delta y\approx 2.0\mathrm{mm})/\eta$& 12.5& 10.8 & 9.3 & 8.1 & 7.2 \\% dy/eta
%$\overline{\left(\partial v/\partial x \right)^2}/ 2\overline{\left(\partial u/\partial x\right)^2} $& 0.96 & 0.97 & 0.96 & 0.95 & 0.96 \\% dvdx2r/2dudx2r
%$\overline{\left(\partial u/\partial y\right)^2}/2\overline{\left(\partial v/\partial y\right)^2} $ &  $0.99\pm0.05$ & $1.00\pm0.13$ & $0.94\pm0.07$ & $0.92\pm0.09$ & $0.85\pm0.15$\\% dudy2r./2dvdy2r
%$\overline{\left(\partial u/\partial y\right)^2}/\overline{\left(\partial v/\partial x\right)^2}$ &  $1.43\pm0.02$ & $1.43\pm0.09$ & $1.45\pm 0.02$ & $1.48\pm0.03$ & $1.36\pm0.05$\\% dudy2r./dvdx2r
$K_1$ & 1.05 & 1.03 & 1.04 & 1.05 & 1.04 \\
$K_2$ &  $1.01\pm0.05$ & $1.00\pm0.13$ & $1.06\pm0.08$ & $1.09\pm0.11$ & $1.20\pm0.21$\\
$K_3$ &  $0.73\pm0.01$ & $0.72\pm0.04$ & $0.72\pm0.01$ & $0.71\pm0.01$ & $0.77\pm0.02$\\
$u'/v'$ & $1.18$ & $1.15$ & $1.16$ & $1.16$ & $1.14$ \\
%$r_{1,1}$&  0.922 & 0.936 & 0.948 & 0.953 & 0.958\\% r_dudx2
%$r_{2,1}$&  0.850 & 0.877 & 0.897 & 0.909 & 0.920\\% r_dvdx2
%$r_{1,2}\,|\Delta y(2)$& 0.543 & 0.620 & 0.680 & 0.719 & 0.773 \\% r_dudy2
%$r_{2,2}\,|\Delta y(2)$& 0.650 & 0.716 & 0.766 & 0.797 & 0.839 \\% r_dvdy2
\end{tabular*} 
\rule{\linewidth}{.5pt}\vspace*{4mm}
\end{table} 
\begin{table}[ht!]
\caption[Data from \cite{Tsinober92} for turbulence generated by a RG]{Data from \cite{Tsinober92} for turbulence generated by a square-mesh RG with round bars, $M=60$ and $\sigma = 44\%$. The value of $x_{\mathrm{peak}}$ used here is the estimate made by \cite{gomesfernandesetal12} for these data, since it was not given in the original paper.}
\label{tab:RGTsinober}
\rule{\linewidth}{.5pt}\vspace*{4mm}
\centering
\begin{tabular*}{0.9\textwidth}{@{\extracolsep{\fill}}rccccccc}
Location & $480$ & $1020$ & $1800$ & $2280$ & $3840$ & $5400$ \\
$x/x_{\mathrm{peak}}$ & $2.8$ & $5.9$ & $10.3$ & $13.1$ & $22.1$ & $31.0$ \\
$Re_{\lambda}$ & $96$ & $88$ & $74$ & $67$ & $82$ & $63$ \\
$\overline{q^2}\,(\mathrm{m^2s^{-2}})$  & $0.64$ & $0.15$ & $0.07$ & $0.05$ & $0.03$ & $0.02$ \\
$K_1$ & 1.47 & 1.57 & 1.34 & 1.44 & 2.08 & 1.37 \\
$K_2$ & 0.95 & 0.75 & 0.69 & 0.80 & 0.44 & 0.64 \\
$K_3$ & 1.05 & 0.93 & 0.83 & 0.84 & 0.79 & 0.82 \\
$u'/v'$ & $1.20$ & $0.84$ & $1.22$ & $1.23$ & $1.52$ & $1.24$ \\
\end{tabular*} 
\rule{\linewidth}{.5pt}\vspace*{4mm}
\end{table} 
The data are used to calculate four estimates of the dissipation, namely 
\begin{equation}
\left\lbrace
\begin{aligned}
\varepsilon^{\mathrm{iso}} &\equiv 15\nu \overline{(\partial u/\partial x)^2};\\
 \varepsilon^{\mathrm{iso,2}} &\equiv \nu(3\overline{(\partial u/\partial x)^2}+6\overline{(\partial v/\partial x)^2});\\
\varepsilon^{\mathrm{iso,3}} &\equiv \nu(\overline{(\partial u/\partial x)^2}+2\overline{(\partial v/\partial x)^2}+4\overline{(\partial u/\partial y)^2}+2\overline{(\partial v/\partial y)^2});\\
\varepsilon^{\mathrm{axi}} &\equiv \nu(-\overline{(\partial u/\partial x)^2}+2\overline{(\partial v/\partial x)^2}+2\overline{(\partial u/\partial y)^2}+8\overline{(\partial v/\partial y)^2}).
\end{aligned}
\right.
\end{equation}
The first estimate, $\varepsilon^{\mathrm{iso}}$ is the widely used isotropic dissipation estimate where all the kinematic constraints of locally isotropy are implied. 
The second estimate ($\varepsilon^{\mathrm{iso,2}}$) is very similar to the first, but with one less isotropy relation, namely that $K_1 = 1$.
The last estimate ($\varepsilon^{\mathrm{axi}}$) is the locally axisymmetric estimate  \cite[]{George91}. 
However, this dissipation estimate heavily weights $\overline{(\partial v/\partial y)^2}$ whose measurement is, as noted before, strongly contaminated by aerodynamic interference (see \S \ref{sec:resolution}).
To overcome this limitation, a more reliable estimate ($\varepsilon^{\mathrm{iso,3}}$) is proposed by transferring part of the weight of $\overline{(\partial v/\partial y)^2}$ to $\overline{(\partial u/\partial y)^2}$ assuming that $\overline{(\partial u/\partial y)^2}\approx 2\overline{(\partial v/\partial y)^2}$, which is a kinematic constraint based on isotropy. 

The several dissipation estimates are presented in tables \ref{tab:RG60} and \ref{tab:RG115}.
Taking $\varepsilon^{\mathrm{iso,3}}$ as the benchmark, it is noticeable that throughout the turbulence decay, the isotropic dissipation estimate, $\varepsilon^{\mathrm{iso}}$, underestimates dissipation rate by $20\% \pm 2\%$ for the RG115 data and $15\%\pm 2\%$ for the RG60 data, whereas the estimate $\varepsilon^{\mathrm{iso,2}}$, underestimates dissipation rate by $30\% \pm 3\%$ for the RG115 data and $26\%\pm 3\%$ for the RG60 data
This motivates the choice, in this thesis, of $\varepsilon^{\mathrm{iso}}$ rather than $\varepsilon^{\mathrm{iso,2}}$ as the prime dissipation estimate whenever the additional data needed to estimate $\varepsilon^{\mathrm{iso,3}}$ are not available. 
Most importantly, the observation that $\varepsilon^{\mathrm{iso}}$, $\varepsilon^{\mathrm{iso,2}}$ and $\varepsilon^{\mathrm{iso,3}}$ (and $\varepsilon^{\mathrm{axi}}$ within the scatter) remain approximately proportional throughout the decay lead to the expectation that these are also approximately proportional to the actual dissipation rate.
Therefore, using either of the dissipation estimates to infer, for example, on the behaviour of the normalised energy dissipation rate (see the following chapter) leads to curves with the same functional form but offset from one another.\\

\cleardoublepage
\section{Summary}
In the first part of this chapter the longitudinal and lateral profiles of single-point statistics  are presented for two RGs and two FSGs for a downstream extent of $1<x/x_{\mathrm{peak}}<4$. 
It is shown that the turbulent transport and production relative to the dissipation are particularly pronounced around $x\approx x_{\mathrm{peak}}$ (for all spanwise locations and for all assessed RGs and FSGs) but significantly decrease through the first few multiples of $x_{\mathrm{peak}}$ and become small for the farthest downstream measurement station, $x\approx 4 x_{\mathrm{peak}}$. 
An outstanding exception is the turbulent transport along the centreline in the lee of the FSGs which remains  a non-negligible fraction of the dissipation throughout the assessed region. 
It is also demonstrated that the lateral profiles of single-point statistics, including the turbulent transport and production, vary with the grid geometry and are influenced by the confining wind tunnel walls when the mesh size is large compared to the width/height of the test section.

The decay of the mean square longitudinal velocity fluctuations along the centreline in the lee of RG115, FSG18''x18'' and FSG3'x3' (for the same downstream region $1<x/x_{\mathrm{peak}}<4$) are fitted with power-laws and it is observed that the decay exponents are always $n\gtrapprox 2$, which are much larger values than the usual power-law exponents found in the literature for RG- and active grid-generated turbulence, typically $1 \lessapprox n \lessapprox 1.5$. 
Nevertheless, the homogenous kinetic energy balance between advection and dissipation is not satisfied for this assessed region.
However, for the FSG18''x18'' data, and to some extent for the FSG3'x3' data as well, the advection is approximately proportional to the dissipation along the centreline. 
This is due to the fact that the contribution to the kinetic energy balance from turbulence production is negligible along the centreline (even though off-centreline it is not) and the contribution from turbulent transport is roughly a constant fraction of  the dissipation. Given the approximate proportionality between advection and dissipation it is argued that the power-law exponent of the decaying turbulence along the centreline can be compared with the values for homogeneous freely decaying turbulence.
As will be clear in the following chapter,  there are stark differences in the behaviour of the normalised energy dissipation for this region which are consistent with the higher decay exponents observed.

The RG115-generated turbulence  is further investigated in terms of  large- and small-scale (an)isotropy.
It is shown that the small-scales do not follow the isotropic relations between the measured components of the mean square velocity gradient tensor. 
Nevertheless, these ratios stay approximately constant during the assessed region of the decay, in-line with the findings of  \cite{gomesfernandesetal12} for a FSG similar to FSG18''x18'' and with the present RG60-generated turbulence data acquired for much higher $x/x_{\mathrm{peak}}$.
On the other hand, it is found that the ratios between the various integral-length scales exhibit significant departures from isotropy and that some of these ratios vary significantly throughout the assessed region. 
In the following section attention is devoted to the behaviour of the normalised energy dissipation as the flow decays and the influence of the large-scale anisotropy variations.

\clearemptydoublepage
\chapter{Energy dissipation scaling}
\label{chp:4}

% paragraph 1
In this chapter it is shown that the nonclassical energy dissipation behaviour observed during the decay of FSG-generated turbulence, i.e. that $C_{\varepsilon}^{1(1)} \sim f(Re_M)/Re_{\lambda}$ instead of the classically expected $C_{\varepsilon}^{1(1)} \sim \mathrm{constant}$ (see ch. \ref{chp:1}), is also observed in the lee of RG-generated turbulence. 
Therefore, this nonclassical energy dissipation behaviour is not exceptional to the very special class of inflow conditions defined by FSGs and is, in fact, more general rendering this nonclassical behaviour of general scientific and engineering significance.

\section[Nonclassical dissipation scaling in RG-generated turbulence]{Nonclassical dissipation behaviour in turbulence generated by RGs}   \label{sec:NoneqRG}

%figure 1
\begin{figure}[ht]
\centering
\includegraphics[trim = 45 0 12 17, clip=true,width=120mm]{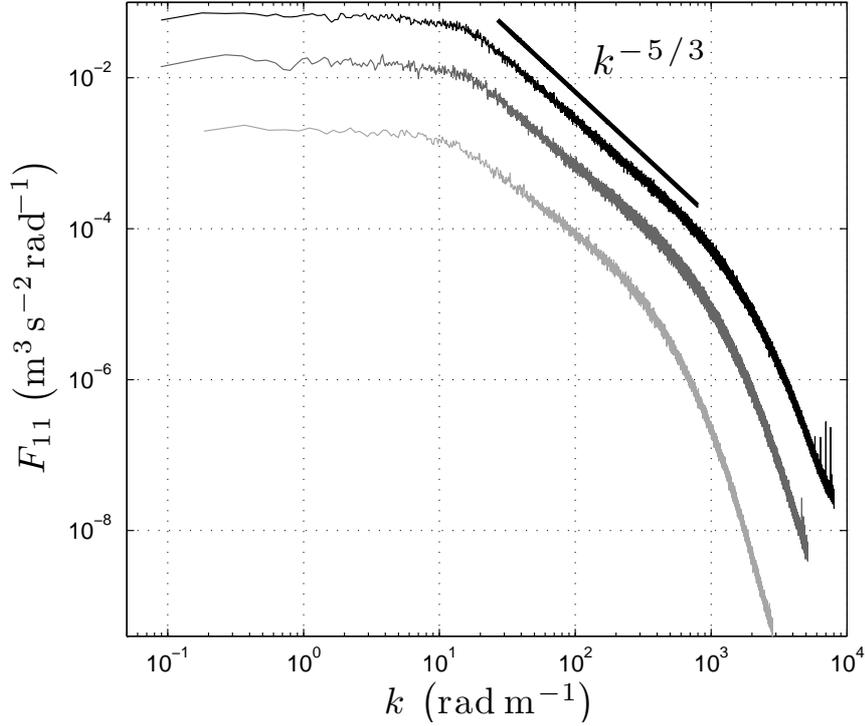}
\caption[Longitudinal energy spectra $F_{11}^{(1)}$ of RG230-generated turbulence]{ Longitudinal energy density spectra $F_{11}^{(1)}$ per wavenumber $k$ of turbulence generated by RG230 for (black) $U_{\infty}=20 \mathrm{m}\mathrm{s}^{-1}$, $x/x_{\mathrm{peak}}=1.01$, (dark grey) $U_{\infty}=10 \mathrm{m}\mathrm{s}^{-1}$, $x/x_{\mathrm{peak}}=1.01$ and (light grey) $U_{\infty}=5 \mathrm{m}\mathrm{s}^{-1}$, $x/x_{\mathrm{peak}}=1.90$.}
\label{fig0}
\end{figure}
% figure 2
\begin{figure}
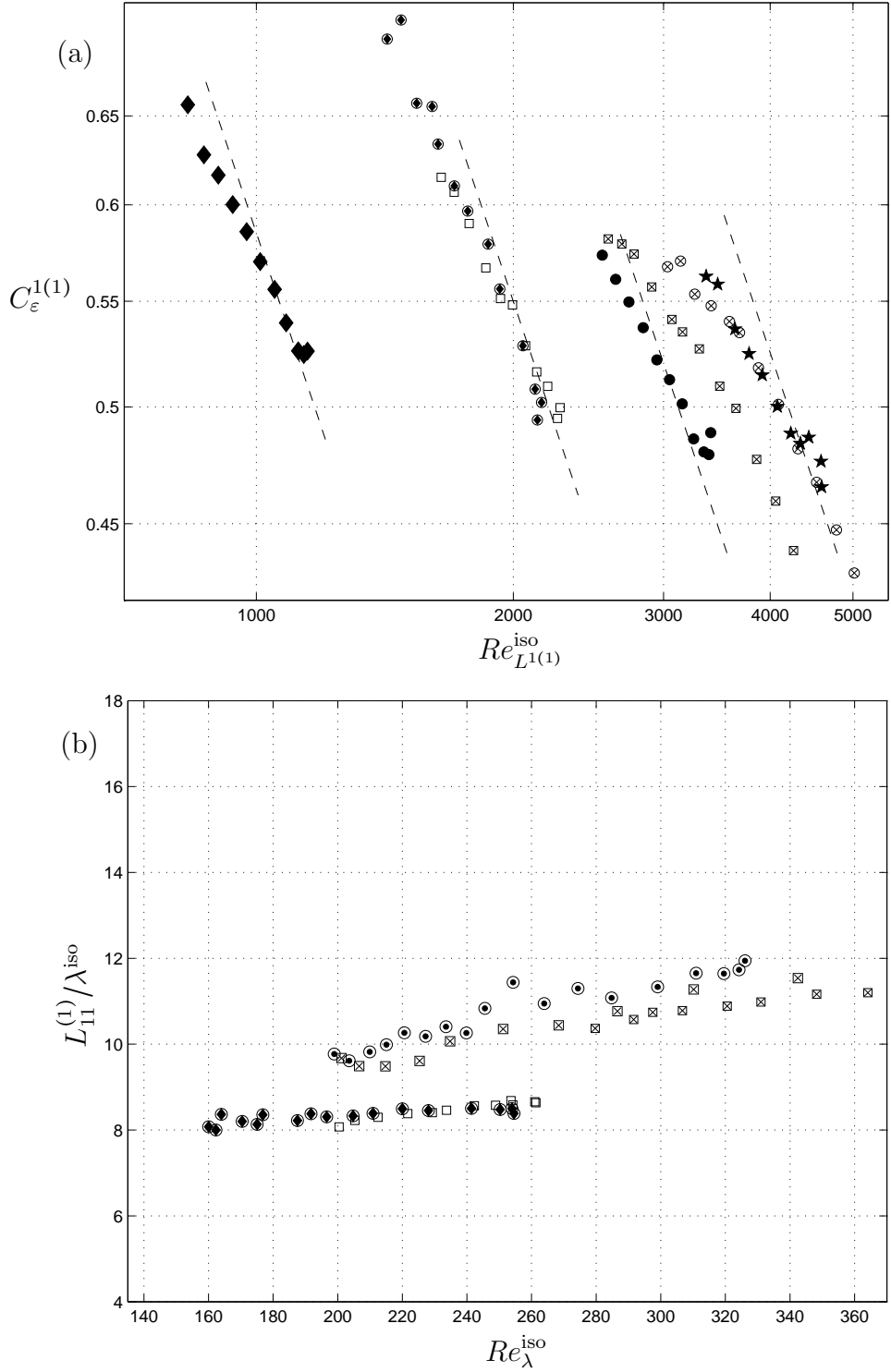

\centering
\begin{minipage}[c]{\linewidth}
   \centering 
   \begin{lpic}{ThesisCepsReLRG230RG115FSG18(120mm)}
   \lbl{5,125;(a)}
   \lbl[W]{-1,75;$C_{\varepsilon}^{1(1)}$}
   \lbl[W]{98,1;${Re}_{L^{1(1)}}^{\mathrm{iso}}$}
    \end{lpic}
    \vspace{4mm}
\end{minipage}
\begin{minipage}[c]{\linewidth}
   \centering 
   \begin{lpic}{ThesisCepsConfinement(120mm)}
   \lbl{5,129;(b)}
   \lbl[W]{4,78,90;$L_{11}^{(1)}/\lambda^{\mathrm{iso}}$}
   \lbl[W]{96,3;${Re}_{\lambda}^{\mathrm{iso}}$}
    \end{lpic}
\end{minipage}
\caption[$C_{\varepsilon}^{1(1)}$ versus ${Re}_{L^{1(1)}}^{\mathrm{iso}}$ in the lee of FSG18''x18'', RG230 and RG115]{(a) Normalised energy dissipation $C_{\varepsilon}^{1(1)}$ versus local Reynolds number ${Re}_{L^{1(1)}}^{\mathrm{iso}}$ of turbulence generated by FSG18''x18'', RG230 \& RG115  for different inflow Reynolds numbers $Re_M$. The dashed lines follow $\propto \!\! 1/{Re}_{L^{1(1)}}^{\mathrm{iso}}$ for different $Re_M$ and the $Re_{\lambda}^{\mathrm{iso}}$ values of these data range between 140 and 418. (b) Ratio between the longitudinal integral-length scale to the Taylor microscale versus the local Reynolds number, $Re_{\lambda}^{\mathrm{iso}}$, of turbulence generated by FSG18''x18'' versus FSG3'x3' and RG230 versus RG115 at similar $Re_M$ for each pair. The symbols are described in table \ref{Table:Results}.}
\label{fig1}
 \end{figure}
 \begin{table}[ht!]
\caption{ Overview of the SW measurements in the lee of FSG18''x18'', FSG3'x3', RG230, RG115 and RG60.}
\rule{0.6\linewidth}{.5pt}\vspace*{4mm}
\centering
\begin{tabular*}{0.5\textwidth}{@{\extracolsep{\fill}}lcccc}
Grid & Symbol & $U_{\infty}$ & $Re_{M}$& y \\
 &&$\left(\mathrm{ms}^{-1}\right)$& $\left(\times 10^{3}\right)$ \\
\hline
\multirow{2}{*}{FSG18''x18''} &\scriptsize{\rlap{\Square}\SmallCross}& $15.0$ & $237$ & \multirow{2}{*}{0}  \\
&\scriptsize{\rlap{\Circle}\SmallCross}& $17.5$ & $277$ & \\ \\

\multirow{2}{*}{FSG3'x3'} &\scriptsize{\rlap{\BigCircle}\FilledSmallCircle}& $15.0$ & $237$ & \multirow{1}{*}{0}  \\
&\scriptsize{\rlap{\BigSquare}\FilledSmallCircle}& $15.0$ & $277$ & $-M/2$\\ \\

\multirow{4}{*}{RG230} & \FilledDiamondshape & $5.0$ & $77$ & \multirow{4}{*}{0}  \\
&\SmallSquare& $10.0$ & $153$ &\\
&\FilledSmallCircle& $15.0$ & $230$ & \\
%&\Diamondshape& $17.5$ & $268$ & \\
&\Large{$\filledstar$}& $20.0$ & $307$ & \\  \\

\multirow{2}{*}{RG115} &\scriptsize{\rlap{\Circle}\FilledDiamondshape}& $20.0$ & $153$ & 0  \\ 
&\ding{74}& $20.0$ & $153$ & $M/2$  \\ \\

\multirow{3}{*}{RG60} & \SmallTriangleLeft& $10.0$ & $40$  & \multirow{3}{*}{0} \\
&\FilledSmallTriangleUp& $15.0$ & $60$ &\\
&\SmallTriangleRight & $20.0$ & $80$ & \\
\end{tabular*}
\label{Table:Results}
\rule{0.6\linewidth}{.5pt}\vspace*{4mm}
\end{table}

First, the dissipation scalings of the decaying turbulence originating from RG230 and FSG18''x18'' are compared. 
The range of Reynolds numbers throughout the measurement stations is $140\leq Re_{\lambda}^{\mathrm{iso}}\equiv (\overline{u^2})^{1/2}{\lambda}^{\mathrm{iso}}/\nu \leq418$. All these Reynolds numbers are large enough for a significant separation to exist between the large, energy containing, eddies and the smallest dissipative eddies. 
Indeed, the scale separation for the highest Reynolds number is $L_{11}^{(1)}/\eta \approx 460$ and for the lowest Reynolds number is $L_{11}^{(1)}/\eta \approx 140$. 
(Note that, data recorded between a grid and its corresponding $x_{\mathrm{peak}}$ are excluded as this study is confined to decaying turbulence.)
The measured one-dimensional longitudinal energy spectra $F_{11}^{(1)}$ exhibit clear power-laws over more than a decade with an exponent close to Kolmogorov's $-5/3$, at least for $Re_M \ge 2.3\! \times\! 10^{5}$ and $Re_{\lambda}^{\mathrm{iso}} \ge 250$ (see figure \ref{fig0} where only RG230 spectra are plotted for brevity and clarity; FSG18''x18'', RG115 and RG60 spectra can be found in ch. \ref{chp:5}). 

Both for RG230 and FSG18''x18'', the one-dimensional form of the cornerstone assumption of turbulence theory, $C_{\varepsilon}^{1(1)}\equiv \varepsilon L_{11}^{(1)}/(\overline{u^2})^{3/2} \approx \mathrm{constant}$, does not hold in this region where the turbulence decays (between about 1.3m from the grid and the end of the test section) at these Reynolds numbers (see figure \ref{fig1}a). %(Note that throughout this chapter the streamwise fluctuating velocity component is used in the definition of the normalised energy dissipation.)
Instead, for any fixed $Re_M$, $C_{\varepsilon}^{1(1)} \sim 1/{Re}^{\mathrm{iso}}_{L^{1(1)}}$ (as one moves along $x$; ${Re}^{\mathrm{iso}}_{L^{i(k)}} \equiv (\overline{u^2})^{1/2}{L_{ii}^{(k)}}/\nu$) is a good qualitative approximation (in figure  \ref{fig1} each set of symbols corresponds to one $Re_M$ and one grid, see table  \ref{Table:Results}; $Re_{L^{1(1)}}^{\mathrm{iso}}$ decreases as $x$ increases). 

Note that the ratio between the integral-length scale and the Taylor microscale is directly related to the normalised energy dissipation rate,
%From the general anisotropic definitions of the Taylor microscale $\lambda$($\equiv (5 \nu \overline{q^2}/\varepsilon)^{1/2}$) and $Re_{\lambda}$($\equiv (\overline{q^2}/3)^{1/2}\lambda/\nu$) it follows that (no summation over $i$ is implied),
\begin{equation}
%C_{\varepsilon}^{i(k)}\equiv\frac{\varepsilon L_{ii}^{(k)}}{(\overline{q^2}/3)^{\,3/2}} = \frac{15}{Re_{\lambda}}\frac{L_{ii}^{(k)}}{\lambda},
C_{\varepsilon}^{i(k)}\equiv\frac{\varepsilon L_{ii}^{(k)}}{(\overline{u^2})^{\,3/2}} = \frac{15}{Re_{\lambda}^{\mathrm{iso}}}\frac{L_{ii}^{(k)}}{\lambda^{\mathrm{iso}}}=\frac{15}{Re_{L^{i(k)}}^{\mathrm{iso}}}\left(\frac{L_{ii}^{(k)}}{\lambda^{\mathrm{iso}}}\right)^2,
\label{eq:Ceps}
\end{equation}
where $L_{ii}^{(k)}/(\overline{u^2})^{1/2}$ are the various time-scales corresponding to the different integral-length scales.
Therefore, the observation that $C_{\varepsilon}^{1(1)} \sim 1/{Re}^{\mathrm{iso}}_{L^{1(1)}}$ as the flow decays for a fixed $Re_M$ is  equivalent to $C_{\varepsilon}^{1(1)} \sim 1/{Re}^{\mathrm{iso}}_{\lambda}$ and $L_{11}^{(1)}/\lambda^{\mathrm{iso}} \approx \mathrm{constant}$ \cite[figure \ref{fig1}a,b; see also][]{MV2010}.
%\footnote{Instead of using the longitudinal fluctuating velocity component, $u$, on could have used twice the kinetic energy, $\overline{q^2}=\overline{u_j\,u_j}$ and generalise \eqref{eq:Ceps} to read
%\begin{equation*}
%C_{\varepsilon}^{i(k)}\equiv\frac{\varepsilon L_{ii}^{(k)}}{(\overline{q^2}/3)^{\,3/2}} = \frac{15}{Re_{\lambda}}\frac{L_{ii}^{(k)}}{\lambda}=\frac{15}{Re_{L^{i(k)}}}\left(\frac{L_{ii}^{(k)}}{\lambda}\right)^2,
%\label{eq:CepsB}
%\end{equation*}
%where  $\lambda\equiv (5 \nu \overline{q^2}/\varepsilon)^{1/2}$, $Re_{\lambda}\equiv (\overline{q^2}/3)^{1/2}\lambda/\nu$ and $Re_{L^{i(k)}}\equiv (\overline{q^2}/3)^{1/2}L_{ii}^{(k)}/\nu$. 
%}

At the furthest downstream locations which correspond to the lowest ${Re}_{L^{1(1)}}^{\mathrm{iso}}$ values for each $Re_M$ in figure \ref{fig1}, there is a slight departure from $C_{\varepsilon}^{1(1)} \sim 1/{Re}_{L^{1(1)}}^{\mathrm{iso}}$ and $L_{11}^{(1)}/\lambda^{\mathrm{iso}} \approx \mathrm{constant}$, particularly for the FSG data. 
For the FSG18''x18'' and FSG3'x3'  centreline data a mixed behaviour of the type, $L_{11}^{(1)}/\lambda^{\mathrm{iso}} \sim \mathrm{A} + \mathrm{B}\,{Re}_{\lambda}^{\mathrm{iso}}$, with $A$ and $B$ being two numerical constants, seems to be a better approximation (see figure \ref{fig1}b)\footnote{It  has been verified that this behaviour is not due to any misalignments of the probe relative to the centreline.}.
This issue is further discussed in \S \ref{sec:offcentre}. 
However, as can be seen from the comparison between the FSG3'x3' versus FSG18''x18'' data and RG115 versus RG230 data in figure \ref{fig1}b, confinement does not meaningfully change the slope of $L_{11}^{(1)}/\lambda^{\mathrm{iso}}$ versus ${Re}_{\lambda}^{\mathrm{iso}}$.
Therefore the confining walls, which are observed to have a strong influence on e.g. turbulence transport (see \S \ref{sec:confinement}), do not seem to meaningfully influence the behaviour of $C_{\varepsilon}^{1(1)}$.

Note also that unless otherwise stated, the isotropic energy dissipation estimate  $\varepsilon^{\mathrm{iso}}$ is used as a surrogate for the true kinetic energy dissipation. 
This is motivated by the finding in ch. \ref{chp:3} that the anisotropy ratios $K_1$ and $K_3$ do not meaningfully vary during the measured decay regions of RG115- and RG60-generated turbulence (corresponding respectively to regions of classical and nonclassical dissipation behaviour). % which leads to a systematic underestimate of $\varepsilon$, without meaningfully biasing its functional form during the turbulence decay. %\footnote{The ratio $K_2 = 2\overline{(\partial v/\partial y)^2}/\overline{(\partial u/\partial y)^2} $ appears to increase slightly during decay, particularly for the RG60 data. However,  as it was commented on \S \ref{sec:resolution} the estimates of $\overline{(\partial v/\partial y)^2}$ were the most prune to errors arising from aerodynamical interference between the probes.}. 
In particular, $\varepsilon^{\mathrm{iso}}$ systematically underestimates $\varepsilon$ (taking $\varepsilon^{\mathrm{iso,3}}$ as the reference estimate) by $\approx 24\%$ and $18\%$ for the RG115 and the RG60 data, respectively (\emph{cf.} tables \ref{tab:RG60} and \ref{tab:RG115}). 
The data presented by \cite{gomesfernandesetal12} concerning small-scale anisotropy in the lee of a FSG similar to the present FSG18''x18''  leads to a similar conclusion.
Such a systematical bias of the estimates just offsets the $L_{ii}^{(k)}/\lambda$ versus $Re_{\lambda}$ curves but does not meaningfully change the functional form.
  
 %%%%%%%
\section{Dependence on the global/inflow Reynolds number} 

When, instead of keeping $Re_M$ fixed and varying $x$, one keeps $x$ fixed and varies $Re_M$, one then finds a very different dependence of $C_{\varepsilon}^{1(1)}$ on Reynolds number; asymptotically independent of it for both RG230 and FSG18''x18'' as $Re_M$ increases. 
%figure 4
\begin{figure}[t!]
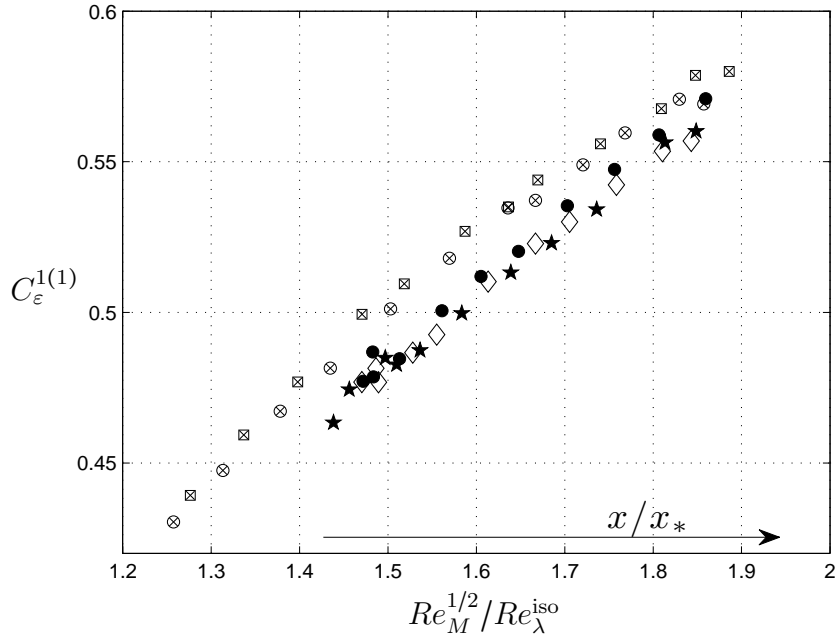

\centering
\begin{lpic}{Figure4(120mm)}
\lbl[W]{8,62;$C_{\varepsilon}^{1(1)}$}
\lbl[W]{84,5;\hspace{2mm}$Re_M^{1/2}/Re_{\lambda}^{\mathrm{iso}}$\hspace{2mm}}
\end{lpic}
\caption[ $C_{\varepsilon}^{1(1)}$ versus  $Re_M^{1/2}/Re_{\lambda}^{\mathrm{iso}}$ for RG230- and FSG18''x18''-generated turbulence]{Normalised energy dissipation $C_{\varepsilon}^{1(1)}$ versus the Reynolds number ratio $Re_M^{1/2}/Re_{\lambda}^{\mathrm{iso}}$ of turbulence generated by RG230 and FSG18''x18''  for different inflow Reynolds numbers $Re_M$. }
\label{fig2} 
\end{figure}
Keeping the usual expectation that $C_{\varepsilon}^{1(1)}$ is independent of $\nu$ at high enough $Re_M$ (which may be close to, but not exactly, true, see the discussion in \citealp{MV2008}), then these two different dependencies on Reynolds number can be reconciled by
\begin{equation}
C_{\varepsilon}^{1(1)} \propto \frac{Re_M}{{Re}^{\mathrm{iso}}_{L^{1(1)}}} \propto \frac{Re_M^{1/2}}{{Re}^{\mathrm{iso}}_{\lambda}}
\label{eq1}
\end{equation}
because $u'/U_{\infty}$ and $L_{11}^{(1)}/M$ are independent of $Re_M$ to leading order at high enough Reynolds numbers. 
Note that $C_{\varepsilon}^{1(1)}\! \propto\! Re_{M}/{Re}^{\mathrm{iso}}_{L^{1(1)}}$ is equivalent to $L_{11}^{(1)}/\lambda^{\mathrm{iso}} \sim Re_{M}^{1/2}$ and therefore to $C_{\varepsilon}^{1(1)} \! \propto\! Re_{M}^{1/2}/{Re}_{\lambda}^{\mathrm{iso}}$.
This equation is fairly well supported by the present data both for FSG18''x18'' and RG230 at $Re_{M} \ge 2.3\!\times\!10^{5}$  (figure \ref{fig2}) but with a grid-dependent constant of proportionality in (\ref{eq1}). 

\begin{figure}[t!]
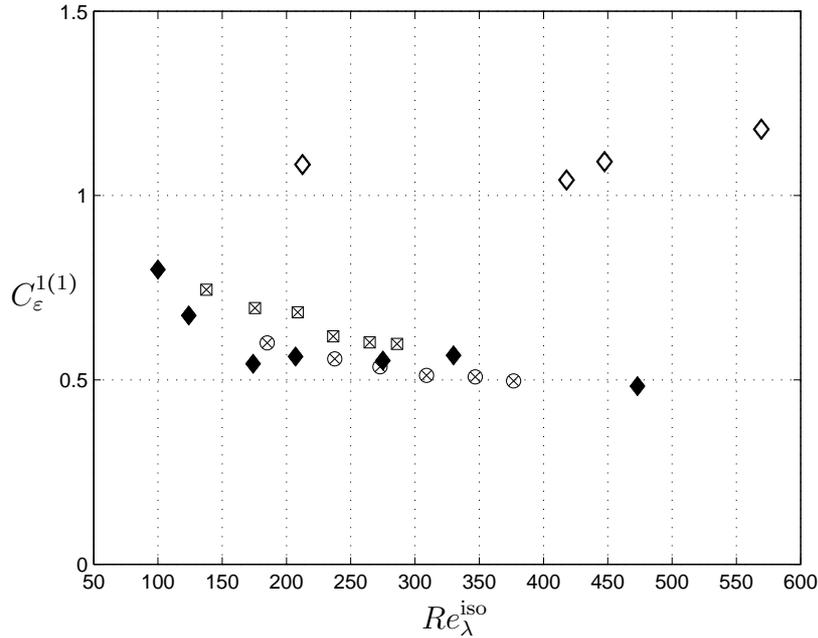
 
\centering
\begin{lpic}{Figure_ch3_3(120mm)}
\lbl[W]{13,61;$C_{\varepsilon}^{1(1)}$}
\lbl[W]{84,4;\hspace{6mm}${Re}_{\lambda}^{\mathrm{iso}}$\hspace{6mm}}
\end{lpic}
\caption[$C_{\varepsilon}^{1(1)}$ versus   ${Re}^{\mathrm{iso}}_{\lambda}$ with ${Re}^{\mathrm{iso}}_{\lambda}$ changing as a function of $Re_M$ for several grids]{Normalised energy dissipation rate $C_{\varepsilon}^{1(1)}$ versus   ${Re}^{\mathrm{iso}}_{\lambda}$ with ${Re}^{\mathrm{iso}}_{\lambda}$ changing as a function of $Re_M$ for a fixed streamwise downstream location for FSG18''x18''-, active grid- and RG-generated turbulence. 
For the FSG18''x18'' data the inlet Reynolds number is changed by varying the free-stream speed between $5\mathrm{ms}^{-1} < U_{\infty} < 17.5 \mathrm{ms}^{-1}$ at two streamwise downstream positions, 
(\protect\raisebox{-0.5ex}{\rlap{\Circle}\SmallCross}) $x/x_*=0.63$ and
(\protect\raisebox{-0.5ex}{\rlap{\Square}\SmallCross}) $x/x_*=1.04$.
(\protect\raisebox{-0.5ex}{\FilledDiamondshape}) Active grid data is taken from table 1 of \cite{GG2000} which is based on the experimental data by \cite{MW1996} (\citealp{GG2000} computed the longitudinal and the transverse integral scales from the spectra, but their latter estimate yielded less scatter, hence we assume isotropy and use twice the transverse integral scale). 
(\protect\raisebox{-0.5ex}{\Diamondshape}) RG data from the data compilation by \cite{Sreeni84}, figure 1 (only data by \cite{Kistler} is used since no other experiment with more than one data point had $Re_{\lambda}>100$).} 
\label{Fig:CepsRe0}
\end{figure}

This can be further verified by plotting $C_{\varepsilon}^{1(1)}$ for different $Re_M$ (by varying $U_{\infty}$) at two fixed streamwise downstream positions from the FSG18''x18''. 
The data presented in figure \ref{Fig:CepsRe0} suggest that $C_{\varepsilon}^{1(1)}$ is roughly constant beyond $Re_{\lambda} \approx 200$. 
This high $Re_M$ behaviour of $C_{\varepsilon}^{1(1)}$ is very comparable to that found with RGs and active-grids at similar Reynolds numbers (figure \ref{Fig:CepsRe0}) and more generally with other boundary-free turbulent flows such as various wakes \cite[see e.g.][]{Burattini2005, Pearson} and DNS of forced stationary homogeneous turbulence \cite[see data compilations by][]{Sreeni98,Burattini2005}. 
However, even though for very high Reynolds numbers $C_{\varepsilon}^{1(1)}$ is expected to become invariant (to a first approximation at least) to changes in viscosity, free-stream velocity and $M$, the fundamental difference with the presently assessed region of the RG230- and FSG18''x18''-generated turbulence is that the numerical value of  $C_{\varepsilon}^{1(1)}$ is different for each streamwise downstream location.

\section[Compatibility with prior evidence of $C_{\varepsilon}\sim \mathrm{constant}$]{Compatibility with prior evidence of $\mathbf{C_{\varepsilon}\sim constant}$} \label{sec:RG60eqnoneq}

Equation \eqref{eq1} may appear to clash with the experimental and numerical evidence supporting the classical expectation that $C_{\varepsilon}$ is approximately independent of both $x$ (and/or time) and $Re_M$ for stationary and decaying turbulence.
However, as is discussed in \S \ref{sec:introB2}, in most of the evidence aiming to verify and/or support $C_{\varepsilon}\sim \mathrm{constant}$,  $C_{\varepsilon}$ is not assessed along coordinates with spatially (and/or time) varying $Re_{\lambda}$, and in the few cases where this is done, $Re_{\lambda}$ is insufficiently high throughout the measured decay region for there to be a clear separation of scales (for all downstream locations) and for the support of $C_{\varepsilon}\sim \mathrm{constant}$ to be convincing. 
For example, the measurements of $C_{\varepsilon}^{1(1)}$ along the decay of RG-generated turbulence at a  fixed $Re_M$ by \cite{CC71}, presented in figure 1 of \cite{Sreeni84}, consist of data acquired at the lee of two RGs with different mesh sizes\footnote{Note that in figure 1 of \cite{Sreeni84} the same symbol is used for both datasets.}. The Reynolds number of these experiments ranges between $35<{Re}^{\mathrm{iso}}_{\lambda}<75$ and there is no clear trend of $C_{\varepsilon}^{1(1)}$ against ${Re}^{\mathrm{iso}}_{\lambda}$.
The same conclusion can be reached from the experimental and numerical data compilations presented by \cite{Burattini2005}. 
Their experimental estimates of $C_{\varepsilon}^{1(1)}$ along the decay of  RG-generated turbulence for a fixed $Re_M$ (i.e. the three datasets in their figure 1 where ${Re}^{\mathrm{iso}}_{\lambda}<100$) indicate that it decreases (and in one case increases) with ${Re}^{\mathrm{iso}}_{\lambda}$.
Similarly, the DNS data for decaying turbulence presented in the figure 3 of  \cite{Burattini2005} show a strong decrease with $Re_{\lambda}$ (although $Re_{\lambda}<150$).
Finally, the experimental data acquired in the lee of a RG and two multiscale grids presented by \cite{KD2010,KD2011} show $C_{\varepsilon}^{1(1)}$ slowly increasing with ${Re}^{\mathrm{iso}}_{\lambda}$ with $65<{Re}^{\mathrm{iso}}_{\lambda}<100$\footnote{See also the discussion in \cite{VV2011PLA} about the possible dependence of the power-law decay exponent and $C_{\varepsilon}^{1(1)}$ on initial conditions, based on the data by \cite{KD2011}.}.

%figure 5
\begin{figure}
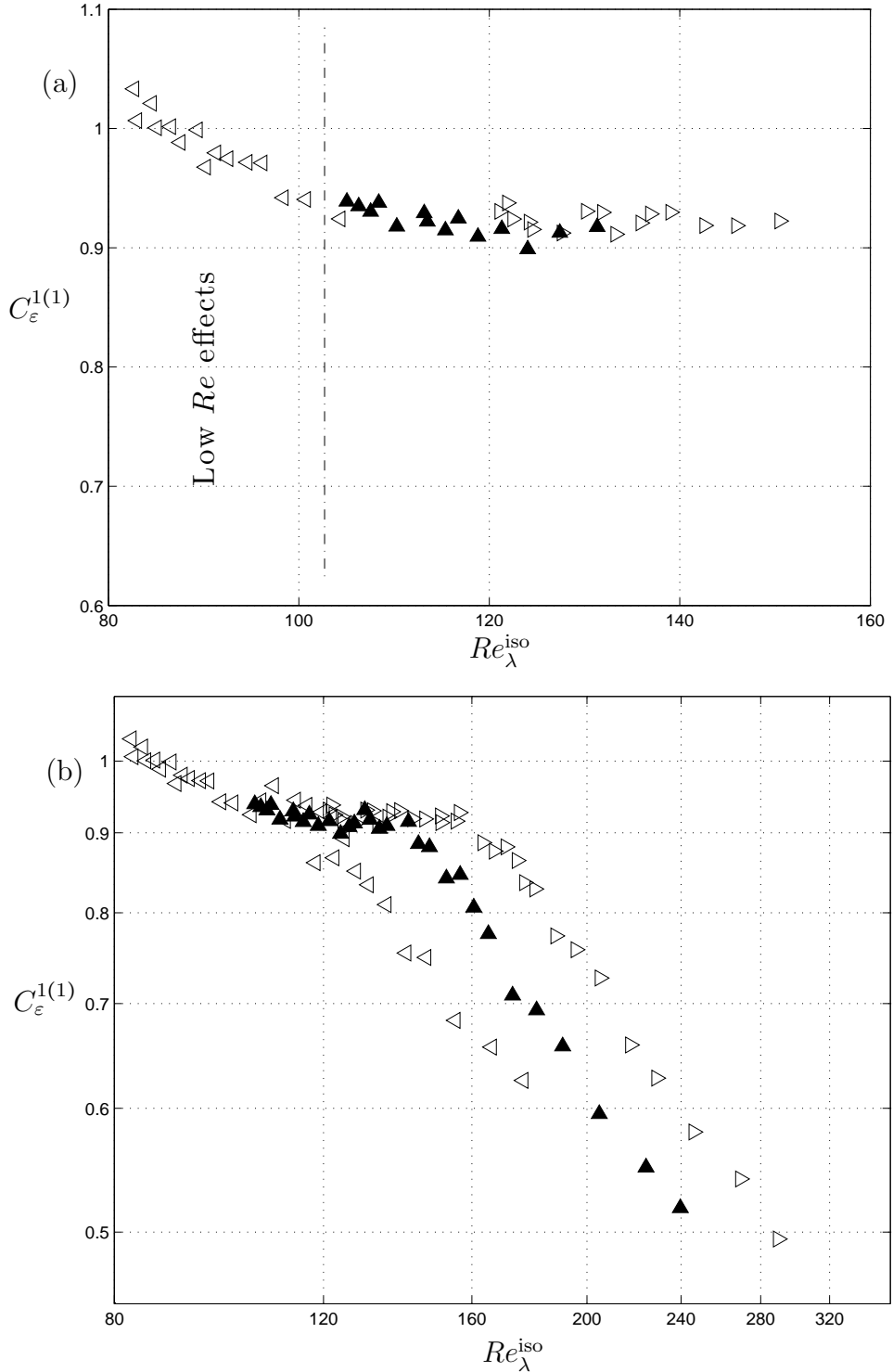

\centering
\begin{minipage}[c]{\linewidth}
   \centering
   \begin{lpic}{ThesisCepsRG60eq(120mm)}
   \lbl{3,120;(a)}
   \lbl[W]{0,74;$C_{\varepsilon}^{1(1)}$}
   \lbl[W]{95,2;$Re_{\lambda}^{\mathrm{iso}}$}
   \end{lpic}
   \vspace{4mm}
\end{minipage}
\begin{minipage}[c]{\linewidth}
   \centering 
   \begin{lpic}{ThesisCepsRG60eqnoneq(120mm)}
   \lbl{3,120;(b)}
   \lbl[W]{0,74;$C_{\varepsilon}^{1(1)}$}
   \lbl[W]{95,1;$Re_{\lambda}^{\mathrm{iso}}$}
   \end{lpic}
\end{minipage}
\caption[$C_{\varepsilon}^{1(1)}$ versus $Re_{\lambda}^{\mathrm{iso}}$ for RG60-generated turbulence]{ Normalised energy dissipation $C_{\varepsilon}^{1(1)}$ versus $Re_{\lambda}^{\mathrm{iso}}$ for RG60-generated turbulence. The downstream extent of the measurements is (a) $10x_{\mathrm{peak}}<x<29x_{\mathrm{peak}}$ and (b) $1.8x_{\mathrm{peak}}<x<21x_{\mathrm{peak}}$. The axes in (b) are logarithmically spaced.}
\label{fig:CepsRG60}
\end{figure}

This insufficient amount of evidence found in the current literature prompted the measurement of the turbulence in the lee of a typical high blockage ratio square-mesh RG (RG60; see \S \ref{sec:grids}). 
Indeed, for $x/M \geq 25$ and for sufficiently high $Re_M$ such that the local $Re_{\lambda}$ is greater than about 100, the measurements indicate that $C_{\varepsilon}^{1(1)}$ is approximately independent of both $x$ and $Re_M$ (see figure \ref{fig:CepsRG60}a), in apparent clash with equation \eqref{eq1}.
This is a distance greater than about $10x_{\mathrm{peak}}$ from the grid because $x_{\mathrm{peak}} \approx 0.15\mathrm{m}$ for RG60. 
However, \eqref{eq1} has so far been established for decaying turbulence originating from RG230 and FSGs up to downstream distances of less than about $4x_{\mathrm{peak}}$ ($x_{\mathrm{peak}}$ takes much greater values for these grids, see table \ref{table:grids}). 

\begin{figure}
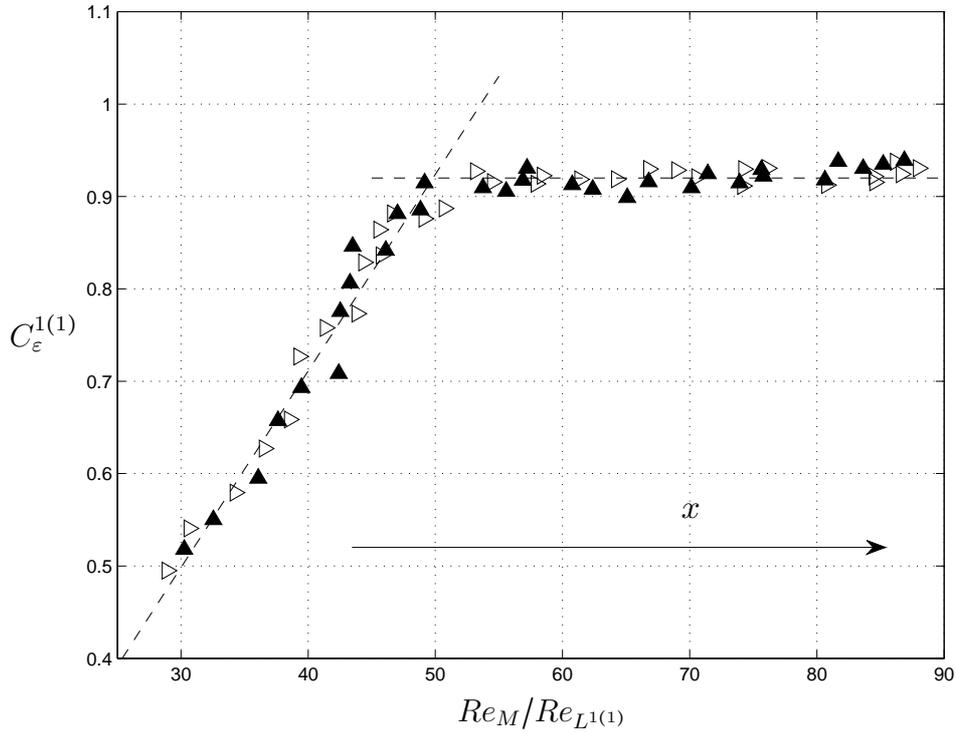

\centering
\begin{lpic}{ThesisCepsReMReLRG60all(120mm)}
\lbl[W]{0,75;$C_{\varepsilon}^{1(1)}$}
\lbl[W]{95,2;$Re_M/Re_{L^{1(1)}}$}
\end{lpic}
\caption[$C_{\varepsilon}^{1(1)}$ versus $Re_M/Re_{L^{1(1)}}$ for RG60-generated turbulence]{Normalised energy dissipation $C_{\varepsilon}^{1(1)}$ versus the Reynolds number ratio $Re_M/Re_{L^{1(1)}}$ for RG60-generated turbulence. }
\label{fig4} 
\end{figure}

It is therefore reasonable to investigate whether \eqref{eq1} and its equivalent relation $L_{11}^{(1)}/\lambda^{\mathrm{iso}} \sim Re_{M}^{1/2}$ hold at distances below a few multiples of $x_{\mathrm{peak}}$ from the RG60 grid. 
In figure \ref{fig:CepsRG60}b the data from figure \ref{fig:CepsRG60}a is complemented with measurements much closer to the grid (although always restricted to the decay region) and indeed it is found that $C_{\varepsilon}^{1(1)}\sim f(Re_M)/{Re}^{\mathrm{iso}}_{\lambda}$ in the region between $1.8 x_{\mathrm{peak}}$ and $5x_{\mathrm{peak}}$ (where ${Re}^{\mathrm{iso}}_{\lambda}$ takes the largest values).
Replotting the RG60 data for the two highest $Re_M$ so as to directly compare with \eqref{eq1}, one obtains figure \ref{fig4}. 
Equation \eqref{eq1} is a fairly good representation of the data up to $Re_{M}/Re_{L^{1(1)}} = 50$, i.e. in the turbulent decay region closest to the grid up to $x \approx 5x_{\mathrm{peak}}$. 
At streamwise distances larger than $5x_{\mathrm{peak}}$ where $Re_{M}/Re_{L^{1(1)}}$ is larger than 50, $C_{\varepsilon}^{1(1)}$ becomes approximately independent of both $x$ and $Re_M$ \footnote{It has been brought to the attention of the author \cite[see p. 824 in][]{UCL2010} that the difference in the dissipation behaviour for $x< 5x_{\mathrm{peak}}$ may be related to an additional sensitivity to initial conditions owing to the possibility that the Lagrangian time scale (e.g. $L_{11}^{(1)}/u'$) is small relative to the travel time ($\sim x/U$). This can be verified by computing the 'number of eddy turnovers' beyond $x_{\mathrm{peak}}$ as $\int_{x_{\mathrm{peak}}/U}^{x/U} u'/L_{11}^{(1)} d\tau$ using the data presented in figures  \ref{fig:UmLu}a,b. It turns out that the nonclassical dissipation behaviour takes about 4 eddy turnovers and that the classical dissipation region between $5x_{\mathrm{peak}}$ and $24x_{\mathrm{peak}}$ (for the RG60 case) takes another 3 -- 4 eddy turnovers. The small difference in eddy turnovers between the classical and nonclassical dissipation regions and the fact that the transition between the two behaviours is sharp (see figure \ref{fig4}) leads to the conjecture that the nonclassical behaviour is not directly linked to a memory effect.}

%For example, \cite{K41b} arrives to the celebrated turbulent kinetic energy power-law decay with the exponent $n=10/7$ by considering decaying turbulence such that $C_{\varepsilon}\sim \mathrm{constant}$ during decay and that the large-eddies satisfy the Loitsyansky invariant \cite[]{Loitsyansky}\footnote{The same analysis would lead to $n=6/5$ considering the Birkoff-Saffman invariant \cite[][see also \citealp{V2011}]{Saffman67}.}.

%%figure 5
%\begin{figure}
%\centering
%\begin{lpic}{Figure5(125mm)}
%\lbl[W]{10,62,90;$L_{11}^{(1)}/\lambda^{\mathrm{iso}}$}
%\lbl[W]{84,4;$Re_{\lambda}^{\mathrm{iso}}$}
%\end{lpic}
%%\includegraphics[trim = 15 6 20 17, clip=true,width=120mm]{Figure5}
%\caption{\label{fig3}  $L/\lambda$ versus the local Reynolds number $Re_{\lambda}$ of turbulence generated by the RG60 for different $Re_M$ and by RG115 \& RG230 for the same $Re_M$. The dashed line follows $C_{\varepsilon}/15\, Re_{\lambda}$ with $C_{\varepsilon}=0.92$.}
%\end{figure}

% paragraph 12
The present data and those of \cite{MV2010} conspire to form the conclusion that, irrespective of the turbulence generating grid (figure \ref{fig:grids}) and for high enough $Re_M$,
\begin{equation}
C_{\varepsilon}^{1(1)} \approx \left\lbrace
\begin{array}{l r}
 C_{1}(\star) \,\,Re_M/Re_{L^{1(1)}} & \quad \text{$x_{\mathrm{peak}} < x < x_{e}$} \\
 C_{2}(\star)& \quad \text{$x > x_{e}$} 
\end{array}
\right. ,
\label{eq2}
\end{equation}
along the centreline, where $x_{e} \approx 5x_{\mathrm{peak}}$ for RG60\footnote{One might expect $x_e$ to scale with $x_{\mathrm{peak}}$ for other grids as well}, $C_1$ and $C_2$ are dimensionless constants which only depend on inlet/boundary geometry (type of fractal/regular grid, $\sigma$, etc; see also \citealp{MV2008}). 
However, the present RG115, RG230, FSG18''x18'' and FSG3'x3' data and those of  \cite{MV2010} do not allow these expectations to be tested, nor do they allow the exploration of how $x_{e}/x_{\mathrm{peak}}$ may depend on inlet/boundary conditions.
RG230, FSG18''x18'' and FSG3'x3', in particular, act as magnifying lenses which make the region exhibiting the nonclassical dissipation behaviour to be longer than the entire tunnel test section's length. 
Equations \eqref{eq1} and more generally $C_{\varepsilon} = f(Re_{M})/Re_{L^{1(1)}}$ which also covers lower values of $Re_M$, are approximately true in the nonclassical dissipation behaviour region, irrespective of flow/turbulence profile details which differ from grid to grid. 
The FSGs are magnifying lenses with added capabilities for tailoring flow and turbulence profiles which go beyond variations in $\sigma$ as illustrated in ch. \ref{chp:3}.

%%%%%%%
\section{Off-centreline behaviour} \label{sec:offcentre}
\begin{figure}
\centering
\includegraphics[width=120mm]{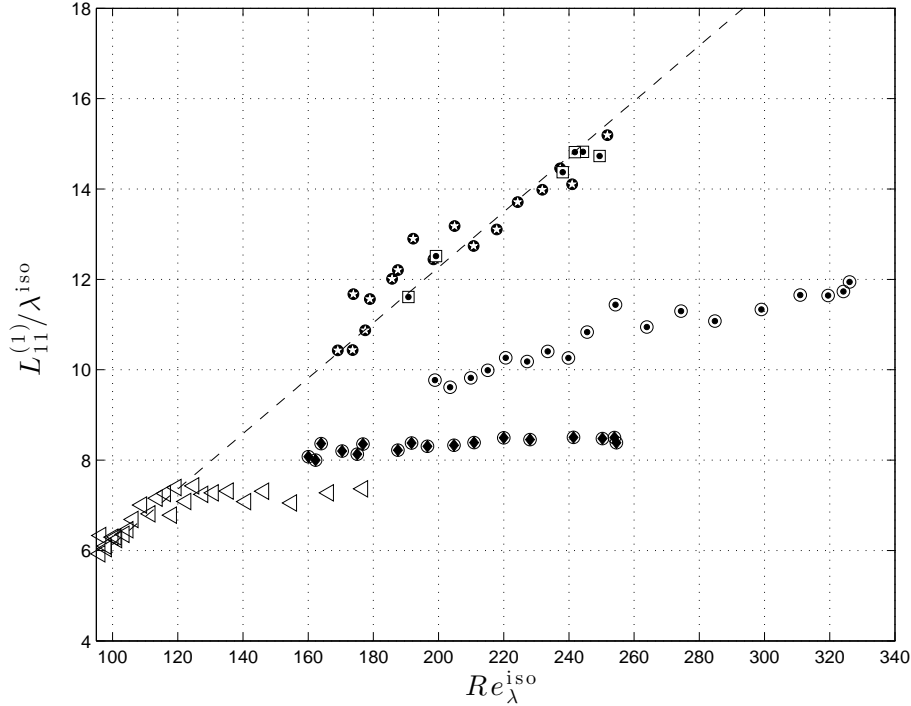}
\caption[$L_{11}^{(1)}/\lambda^{\mathrm{iso}}$ versus $Re_{\lambda}^{\mathrm{iso}}$ for centreline and off-centreline data]{Downstream evolution of $L_{11}^{(1)}/\lambda^{\mathrm{iso}}$ versus $Re_{\lambda}^{\mathrm{iso}}$ for centreline and off-centreline data. Symbols are described in table \ref{Table:Results}.
The dashed line follows $B/15\,{Re}^{\mathrm{iso}}_{\lambda}$, with $B=0.92$. }
\label{fig:LlambdaB}
\end{figure}

So far, only centreline data has been considered, i.e. data along the longitudinal lines intercepting the grid half-way between the bars ($y=z=0$). 
Data following the parallel lines intercepting the bars of the grid ($y=\pm M/2$, $z=0$) are presented in figure \ref{fig:LlambdaB}.  
Outstandingly, the RG115 and FSG3'x3' data behind the bar  $(y=\pm M/2, z=0)$ clearly follow  $L_{11}^{(1)}/\lambda \sim {Re}_{\lambda}^{\mathrm{iso}}$ which corresponds to a classical type of behaviour where $C_{\varepsilon}^{1(1)}$ is independent of ${Re}^{\mathrm{iso}}_{\lambda}$. 
For the longitudinal line in-between the centreline and the bars of the grid ($y=\pm M/4$, $z=0$) a mixed behaviour of the type $L_{11}^{(1)}/\lambda^{\mathrm{iso}} \sim \mathrm{A} + \mathrm{B}\,{Re}^{\mathrm{iso}}_{\lambda}$ is observed (see figure \ref{fig:Llambda}a where RG115 data are plotted; A, B are two numerical constants). 
Therefore the data suggests that the dissipation behaviour, as assessed by the ratio $L_{11}^{(1)}/\lambda$ or equivalently $C_{\varepsilon}^{1(1)}$, follows the classical expectation  $L_{11}^{(1)}/\lambda \sim {Re}_{\lambda}^{\mathrm{iso}}$ but only along the longitudinal lines intercepting the bars of the grid ($y=\pm M/2$, $z=0$).
For all the other parallel lines ($-M/2<y < M/2$, $z=0$) the energy dissipation exhibits a nonclassical behaviour, which for the particular case of the centreline is well approximated by $L_{11}^{(1)}/\lambda \sim \mathrm{constant}$.
This discussion is continued in the following section where additional data are presented.

As previously noted, the FSG3'x3' data along the centreline appears to exhibit a departure from $L_{11}^{(1)}/\lambda^{\mathrm{iso}} \approx \mathrm{constant}$. 
The mixed behaviour observed for the RG115 data along the intermediate line ($y=\pm M/4$, $z=0$), i.e. $L_{11}^{(1)}/\lambda^{\mathrm{iso}} \sim \mathrm{A} + \mathrm{B}\,{Re}^{\mathrm{iso}}_{\lambda}$, is actually a better approximation.
The cause for this mixed behaviour is unclear, but it is plausible that it may be a consequence of the turbulence generated by the wakes originating from the additional fractal iterations.
These wakes will naturally interact closer to the grid (see \S \ref{sec:wake}) and it is possible that the turbulence they generate `transition' earlier  to classical energy dissipation behaviour (i.e. at a downstream distance $x$ closer to the grid) in proportion to the smaller  wake-interaction length-scales. 
If the observation that the extent of the nonclassical dissipation behaviour region is $\approx 2x_{*}$ (\S \ref{sec:RG60eqnoneq}) can be extrapolated to the turbulence originating from the two smallest size meshes, then the fraction of the turbulent flow they generate `transition' at $x\approx 1.2 x_{\mathrm{peak}}$ and $x\approx 2.2 x_{\mathrm{peak}}$, thus leading to the mixed behaviour. 
This is, however, no more than a tentative conceptual explanation.

The centreline data for the RG60 is also included for reference.  
It is perhaps curious to note that the value of $C_{\varepsilon}^{1(1)}$ in the classical dissipation scaling region of RG60-generated turbulence is similar to that behind the bar of RG115 and of FSG3'x3' in the region ($1<x/x_{\mathrm{peak}}<4$).
This may be, perhaps, no more than a coincidence.

%%%%%%%
\section{Role of large-scale anisotropy}
 %%%%%%%%%%%% figures %%%%%%%%%%%%%%%%
\begin{figure}
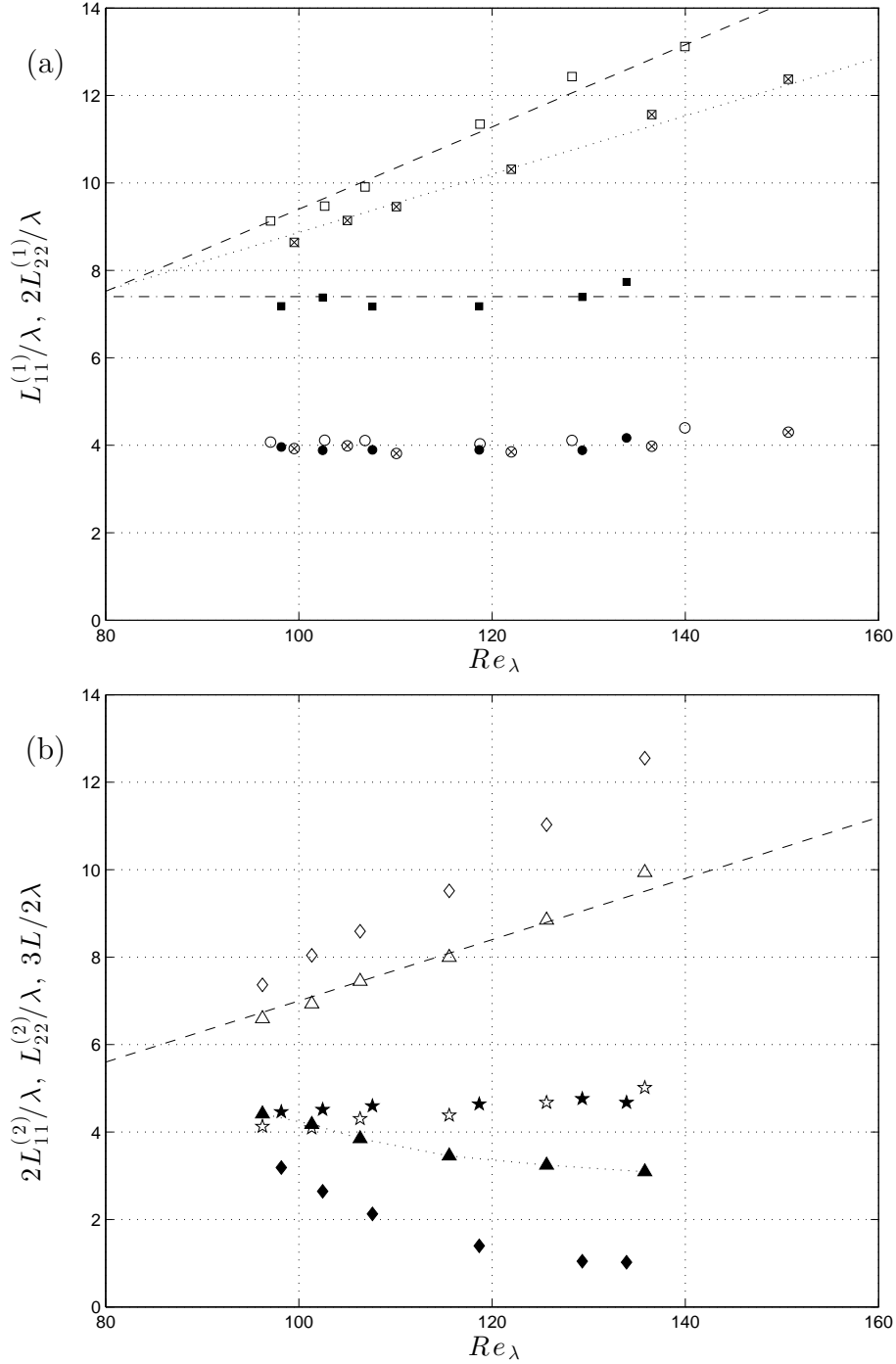

\centering
\begin{minipage}[c]{1\linewidth}
   \centering
   \begin{lpic}{CepsLongL(120mm)}
   \lbl{8,125;(a)}
   \end{lpic}
\end{minipage}
\begin{minipage}[c]{1\linewidth}
   \centering 
   \begin{lpic}{CepsTransL(120mm)}
   \lbl{8,125;(b)}
   \end{lpic}
\end{minipage}
\caption[Ratio between the different integral-length scales and $\lambda$ versus $Re_{\lambda}$ for RG115]{Downstream evolution of the ratio between the integral-length scales  based on the longitudinal and transverse correlations for (a) longitudinal and (b) transverse separations and the Taylor microscale versus $Re_{\lambda}$.  
Data are acquired in the lee of RG115 at $U_{\infty}=10$ms$^{-1}$. 
(\protect\raisebox{-0.5ex}{\FilledSmallSquare} $\!|\!\!$ \protect\raisebox{-0.5ex}{\rlap{\SmallSquare}\SmallCross} $\!|\!\!$ \protect\raisebox{-0.5ex}{\SmallSquare}) $L_{11}^{(1)}/\lambda$ for $y/M=0,\,-0.25,\,-0.5$, (\protect\raisebox{-0.5ex}{\FilledSmallCircle} $\!|\!\!$ \protect\raisebox{-0.4ex}{\rlap{\Circle}\SmallCross} $\!|\!\!$ \protect\raisebox{-0.5ex}{\SmallCircle}) $L_{22}^{(1)}/\lambda$ for $y/M=0,\,-0.25,\,-0.5$, (\protect\raisebox{-0.5ex}{\FilledDiamondshape} $\!|\!\!$ \protect\raisebox{-0.5ex}{\Diamondshape}) $L_{11}^{(2)}/\lambda$ for $y/M=0,\,-0.5$, (\ding{72} $\!|\!\!$ \ding{73}) $L_{22}^{(2)}/\lambda$ for $y/M=0,\,-0.5$ and (\protect\raisebox{-0.5ex}{\FilledSmallTriangleUp} $\!|\!\!$ \protect\raisebox{-0.5ex}{\SmallTriangleUp}) $2L/\lambda$ for $y/M=0,\,-0.5$.
The dashed line follows $B/15\,Re_{\lambda}$, with (a) $B=1.4$ and (b) $B=1.0$. 
The dotted line in (a) follows $A + B/15\,Re_{\lambda}$  with $A=2.2$ and $B=1.0$. }
\label{fig:Llambda}
\end{figure}
%%%%%%%%%%%% figures %%%%%%%%%%%% 

The several integral-length scales obtained from the data of the $2\times$XW experiments in the lee of RG115 and RG60 (see \S \ref{sec:Lu}) are now used to explore the effect of large-scale anisotropy on the behaviour of the normalised energy dissipation. % (recall that these experiments are performed at $U_{\infty}=10$ms$^{-1}$).
Since two velocity components are available for these data\footnote{The mean square of the third unmeasured velocity component, $\overline{w^2}$, is assumed to be approximately equal to the measured spanwise component $\overline{v^2}$.}, the anisotropic definitions of the Taylor microscale, $\lambda$($\equiv (5 \nu \overline{q^2}/\varepsilon)^{1/2}$) and the turbulent Reynolds numbers $Re_{\lambda}$($\equiv (\overline{q^2}/3)^{1/2}\lambda/\nu$) and $Re_{L^{i(k)}}$($\equiv (\overline{q^2}/3)^{1/2}L_{ii}^{(k)}/\nu$) are used.
Equation \eqref{eq:Ceps} can then be modified to read
\begin{equation}
C_{\varepsilon}^{i(k)}\equiv\frac{\varepsilon L_{ii}^{(k)}}{(\overline{q^2}/3)^{\,3/2}} = \frac{15}{Re_{\lambda}}\frac{L_{ii}^{(k)}}{\lambda}=\frac{15}{Re_{L^{i(k)}}}\left(\frac{L_{ii}^{(k)}}{\lambda}\right)^2.
\label{eq:CepsB}
\end{equation}

For isotropic turbulence there is no ambiguity in the choice of integral-length scale to be used in \eqref{eq:CepsB} since $3/2 L = L_{11}^{(1)} = L_{22}^{(2)} = 2L_{22}^{(1)} = 2L_{11}^{(2)}$ (and so forth) and therefore the customarily used $L_{11}^{(1)}$ can be replaced by any of the other integral-length scales without changing the qualitative behaviour of $C_{\varepsilon}^{i(k)}$.

However, when the large-scales are ``elongated''/anisotropic and characterised by different integral-length scales in different directions as found in \S \ref{sec:Lu}, then the dependence of $C_{\varepsilon}$ on Reynolds numbers may depend on the choice of length-scale in its definition.

Turning now to the data and starting with $L_{11}^{(1)}/\lambda$ and $L_{22}^{(1)}/\lambda$ versus $Re_{\lambda}$ along the centreline  (figure \ref{fig:Llambda}a) it is clear that both ratios are approximately constant throughout the assessed region of the decay ($1<x/x_{\mathrm{peak}}<4$).
This behaviour was already reported in \S \ref{sec:NoneqRG} by plotting $L_{11}^{(1)}/\lambda^{\mathrm{iso}}$ (\emph{cf.} figure \ref{fig1}) and is now found to extend to $L_{22}^{(1)}/\lambda$. 
A remarkable new finding which is reported here for the first time (in the near-field decay region of RG115) is that this behaviour occurs along three different streamwise lines with the same numerical constant for $L_{22}^{(1)}/\lambda$ (the
centreline $(y=0, z=0)$ and the lines $(y=-M/4, z=0)$ and $(y=-M/2, z=0)$; see figure \ref{fig:Llambda}a). 
This is in stark contrast with the off-centreline behaviour of $L_{11}^{(1)}/\lambda^{\mathrm{iso}}$ presented in the previous section where along the streamwise line $(y=M/2, z=0)$ in the lee of the bars, $L_{11}^{(1)}/\lambda \sim Re_{\lambda}$ corresponding to the classical dissipation behaviour. 
It is interesting that the classical behaviour for $L_{11}^{(1)}/\lambda$ and $C_{\varepsilon}^{1(1)}$ is associated with $L_{22}^{(1)}/\lambda \sim \mathrm{constant}$ and $C_{\varepsilon}^{2(1)} \sim 1/Re_{\lambda}$ in the near-field decay region of RG115 turbulence. 
Clearly the large eddies become less anisotropic as one probes them by moving downstream along the $(y=\pm M/2, z=0)$ line because $L_{11}^{(1)}/L_{22}^{(1)}$ decreases proportionally to $Re_{\lambda}$ as $Re_{\lambda}$ decreases. 
The author is not aware of any other relation such as $L_{11}^{(1)}/L_{22}^{(1)} \sim Re_{\lambda}$ in the literature to describe the large-scale anisotropy's dependence on $Re_{\lambda}$. 
It will be worth revisiting canonical free shear flows such as wakes and jets in future studies because, to the author's knowledge, only measurements of $C_{\varepsilon}^{1(1)}$ have been reported in such flows in support of $C_{\varepsilon}^{1(1)} \sim \mathrm{constant}$ for high enough Reynolds numbers \cite[e.g. see][]{Sreeni95,Pearson,Burattini2005}.
It will be interesting to know whether $C_{\varepsilon}^{2(1)} \sim 1/Re_{\lambda}$ and $L_{11}^{(1)}/L_{22}^{(1)} \sim Re_{\lambda}$ also hold in such flows or whether these relations are only valid in grid-generated turbulence. 
Note that  the downstream distances of our measurements relative to the bar thickness range within $95 < x/t_{0} <305$ which would be typically considered the far wake.

Considering now the integral-length scales based on the transverse separations (figure \ref{fig:Llambda}b), the results show that $L_{22}^{(2)}/\lambda \approx \mathrm{constant}$ for both the centreline and behind the bar. 
This observation behind the bar also leads to the observation that $L_{11}^{(1)}/L_{22}^{(2)}\sim Re_{\lambda}$. 
On the other hand, at the centreline $L_{11}^{(2)}/\lambda$ increases  as the flow, and the local Reynolds numbers, decay. (This would also imply that $C_{\varepsilon}^{1(2)}$ grows faster than $Re_{\lambda}^{-1}$ with decreasing $Re_{\lambda}.)$

Note that, using the definition of the Kolmogorov microscale, $\eta$($\equiv (\nu^3/\varepsilon)^{1/4}$), and of the Taylor microscale, it follows directly that $\ell/\eta \propto \ell/\lambda\, Re_{\lambda}^{1/2}$. 
Therefore, for $\ell/\lambda$ increasing faster than $Re_{\lambda}^{-1/2}$ as $Re_{\lambda}$ decreases, $\ell/\eta$ increases during decay.
It was checked that $L_{11}^{(2)}/\lambda$ and $L/\lambda$ increase faster than $Re_{\lambda}^{-1/2}$ which leads to the unusual situation where $L_{11}^{(2)}/\eta$ and $L/\eta$ increase during decay. 
The author is unable, at this point, to give a definitive explanation for this behaviour, but as discussed in \S \ref{sec:Lu} it may be related to periodic shedding from the bars which is contaminating the correlation functions, in particular $B_{11}^{(2)}$.

\section{Summary}
In this section it is shown that the nonclassical behaviour of the turbulent kinetic energy dissipation, previously found in the lee of FSGs during decay, i.e. $C_{\varepsilon}^{1(1)}  = f(Re_M)/Re_{\lambda}$, is also manifested in decaying RG-generated turbulence.
In the one case of RG60 where the grid geometry is such that $x_{\mathrm{peak}}$ is small relative to the streamwise extent of the tunnel ($x_{\mathrm{max}}/x_{\mathrm{peak}}\approx 21$) it is shown  that beyond about $x/x_{\mathrm{peak}}\approx 5$ the nonclassical dissipation behaviour transitions to the classically expected $C_{\varepsilon}^{1(1)}  \approx \mathrm{constant}$, i.e. both independent of $x$ and $Re_M$ whenever the Reynolds number is sufficiently high, but with a constant dependent on inflow/boundary conditions. 
%(Even though it was not possible to acquire data sufficiently far from the FSGs, due to the large value of $x_{\mathrm{peak}}$ relative to the tunnel's length, to observe the return to the classical behaviour it is, nevertheless, plausible that the classical dissipation behaviour is recovered far downstream.)
The present data also support the possibility that for large $Re_M$ the functional form of the nonclassical dissipation behaviour follows $C_{\varepsilon}^{1(1)} = C_1 Re_M/Re_{L^{1(1)}}$, with $C_1$ dependent on initial/boundary conditions and therefore becomes, to a first approximation at least, independent of viscosity. 
Lastly, it is shown that the normalised energy dissipation behaviour, using $L_{11}^{(1)}$ as the characteristic length-scale, is starkly different behind a bar than along the centreline. 
In fact, behind the bar, the data suggests that the dissipation scaling is compatible with the classical behaviour. 
However, outstandingly, if $L_{22}^{(1)}$ or $L_{22}^{(2)}$ are used instead, the data strongly suggests that along any longitudinal line (at least those which lie in the plane $z=0$) the normalised energy dissipation follows a specific nonclassical behaviour of the type $L_{22}^{(1)}/\lambda \sim L_{22}^{(2)}/\lambda \sim \mathrm{constant}$.

\clearemptydoublepage
\chapter{Single versus two length-scale dynamics}
\label{chp:5}

The experimental observation that $L_{11}^{(1)}/\lambda=\mathrm{constant}$ as the local Reynolds number ($Re_{\lambda}$ or $Re_{L^{1(1)}}$, see preceding chapter) decays, is consistent with predictions from single-length scale theories of turbulence decay. 
In the introduction, single-length scale theories were reviewed and compared with the mainstream two-length scales theories \cite[]{K41a}. 
In this chapter, experimental evidence supporting the latter is given. 

\section{Single-length scale theories}
In the seminal work of \cite{KH1938}, the ``equation for the propagation of correlation'' for homogeneous turbulence was derived. 
In its isotropic form it reads,

\begin{equation}
\frac{\partial (f \overline{u^2})}{\partial t} + 2(\overline{u^2})^{3/2} \left(\frac{\partial h}{\partial r} + \frac{4}{r}h\right) = 2\nu \left(\frac{\partial^2 f}{\partial r^2} + \frac{4}{r}\frac{\partial f}{\partial r}\right),
\label{eq:KH}
\end{equation}
where $f(r,t)\equiv \overline{u_{\parallel}(\mathbf{X,t})u_{\parallel}(\mathbf{X}+\mathbf{r},t)}/\overline{u^2}$ and $h(r,t)\equiv \overline{u_{\parallel}(\mathbf{X},t)^2u_{\parallel}(\mathbf{X}+\mathbf{r},t)}/(\overline{u^2})^{3/2}$ with $u_{\parallel}$ being the velocity component parallel to $\mathbf{r}$; $\mathbf{X}$ and $\mathbf{r}$ are, respectively, the position and separation vectors (due to homogeneity the statistics are independent on the former) and  $r=|\mathbf{r}|$.

In the same paper, the authors presented one set of solutions to \eqref{eq:KH} (which is obviously not a closed-form expression from the outset) pertaining to decaying turbulence at the lee of a grid for high Reynolds numbers. 
This was accomplished by introducing a self-preservation \emph{ansatz}. 
Namely, that $f(r,t)$ and $h(r,t)$ are functions of only one variable $\xi = r/\ell(x)$, where $\ell$ is a length-scale which is, in principle, dependent on the mesh size and the downstream location.
These solutions lead to $\ell \propto \lambda$ and to $Re_{\lambda}=\mathrm{constant}$ during decay\footnote{Note that, \cite{KH1938} also presented solutions for partial self-preservation, but these implicitly introduce an additional length-scale.} \cite[see also][]{SB1992}.  
However,  $Re_{\lambda}=\mathrm{constant}$ implies that the kinetic energy decays as $\overline{u^2}\propto x^{-1}$.
Nearly all experimental and numerical data indicate that $\overline{u^2}\propto x^{-n}$ with $n>1$, except the `isolated' experiments by \cite{batchelor1948decay} and later by \cite{Kistler}.

Later, \cite{Sedov:paper,Sedov:book} showed that it is possible to build complete self-preserving solutions with $\overline{u^2}\propto x^{-n}$ with $n>1$ \cite[see the review by][where it is shown that Sedov's results lead to unphysical results at high Reynolds numbers]{SB1992}.
Note that, in general, the self-preservation \emph{ansatz} leads to a proportionality between the Taylor microscale, $\lambda$ and the integral-length scale, $L$ and both can be taken as dynamically relevant length-scales\footnote{In fact, any of the length-scales derived from the three-dimensional energy spectra as $\lambda^{n}\sim (\int_0^{\infty}\!k^{2(n-1)}\,dk/\int_0^{\infty}\!k^{2n}\,dk)^{1/2}$, with $n$ being any positive integer number, are proportional to the Taylor microscale $\lambda\equiv \lambda^{(1)}$, and thus can be used as a dynamically relevant length-scale.}.

However, after the seminal contributions by \cite{K41a,K41b,K41c}, it appears that complete self-preservation \emph{ans\"{a}tze} became heterodox. 
Kolmogorov's two-length scale theory for high Reynolds number turbulence, in particular the introduction of dynamically relevant length and velocity scales for the dissipative range ($\eta\equiv (\nu^3/\varepsilon)^{1/4}$ and $v_k\equiv (\varepsilon\nu)^{1/4}$, respectively), received significant support from experiments in very many turbulent flows \cite[see e.g.][figure 9]{SV94}. 
(Note that, it is generally accepted that the dynamically relevant length and velocity scales for the energy containing range are the integral length-scale and the square-root of the kinetic energy.)
Therefore, Kolmogorov's theory is fundamentally incompatible with self-preserving turbulence decay, since $\lambda/\eta \propto Re_{\lambda}^{1/2}$ (unless, of course, $Re_{\lambda}=\mathrm{constant}$ during decay, which is the solution found by \citealt{KH1938}).

Nevertheless, the topic of freely decaying turbulence continued to be riddled with controversy and open questions (even up to the present day) and the possibility of self-preserving decay, perhaps due to it's elegant conceptual framework, was never discarded completely. 

In the early 90's, \cite{George1992} proposed an alternative approach to self-preservation which led to decay laws compatible with experimental evidence and also overcame some of the problems of Sedov's approach \cite[]{SB1992}.
In summary, the theory by \cite{George1992} starts with the single-length scale \emph{ansatz}, i.e. that the second- and third-order correlation functions depend on a single variable $\xi = r/\ell(x)$, but does not assume that the magnitudes of the $2^{\mathrm{nd}}$- and  $3^{\mathrm{rd}}$-order correlations scale as $\overline{u^2}$ and $(\overline{u^2})^{3/2}$, respectively\footnote{Note that, the theory was actually developed in spectral space for Lin's equation \cite[]{lin1947remarks}, but the concepts and results are equivalent for isotropic turbulence, see e.g. \citealt{antonia2003similarity}.}. 
From this \emph{ansatz} and from the constraints of the equation (e.g. the $2^{\mathrm{nd}}$-order correlation evaluated at $r=0$ is just $\overline{u^2}$) George deduces that, (i) $\overline{u^2}\sim x^{-n}$ and $Re_{\lambda}\sim x^{(1-n)/2}$ with $n>1$ ($n=1$ would correspond to infinite Reynolds number), (ii) the scaling of the $2^{\mathrm{nd}}$- and  $3^{\mathrm{rd}}$-order correlations are $\overline{u^2}$ and $(\overline{u^2})^{3/2}/Re_{\lambda}$, respectively (the latter is the main departure from the previous theories) and (iii) $L/\lambda=\mathrm{constant}$ during decay.

\section{Testing the self-preservation \emph{ansatz}} \label{sec:SelfSimEvidence}
From the outset, it is clear that the decaying turbulence presented here seems to support two of the predictions from the self-preserving decay theory of \cite{George1992}. Namely, (i)  the decay follows a power law with $n>1$ (\S \ref{Sec:Decay}) and (ii) $L_{11}^{(1)}/\lambda \approx \mathrm{constant}$ during decay (ch. \ref{chp:4}).
However, our decaying turbulence is not homogenous nor isotropic (ch. \ref{chp:3}), hence there is no \emph{a priori} reason for the theory to apply.
Nevertheless, it is conceivable that the theory could be generalised to take into account the additional (inhomogeneity) terms in the von-K\'{a}rm\'{a}n-Howarth (or Lin) equation. 
On the other hand, anisotropy can be dealt with by taking spherical averages of the correlation functions. 
In fact, this is implicitly done in Lin's equation, where the three-dimensional energy and energy transfer spectra are used. 

With these caveats in mind, one might expect that the dissipation range of turbulence is approximately isotropic (or at least that the anisotropy doesn't significantly vary during the decay, as suggested by our data, see \S \ref{sec:Eps}) and therefore the one-dimensional velocity spectrum for high frequencies/ large wavenumbers should exhibit a single-length scale behaviour if the turbulence itself is decaying in a self-preserving way. Similarly, the longitudinal structure functions and correlations for small separation would exhibit the same behaviour. 

The preliminary data by \cite{SV2007} and \cite{MV2010} suggested that the one-dimensional spectra, at different downstream locations during the turbulence decay, reasonably overlaid each other for all wavenumbers if the axes were normalised using $\lambda$ and $\overline{u^2}$  \cite[see][figures 21 and 22]{MV2010}.
However, the collapse was equally good at large wavenumbers using Kolmogorov's variables, $\eta$ and $v_k$ \cite[see][figure 27]{MV2010}, even though the $Re_{\lambda}$ decreased from $\approx 200$ to $\approx 150$  \cite[\emph{cf.}][figure 15b]{MV2010}. 
As stated above, both pairs of variables cannot, simultaneously, collapse the spectra at large wavenumbers, unless  $Re_{\lambda}$ stays constant during decay.
Therefore, it is clear that one of the overlaid spectra can only be apparent.

It is noted in passing, that testing an \emph{ansatz} from the visual collapse of experimental or numerical data is less than ideal and, in the author's opinion, it is the main reason why there is so much controversy surrounding this subject. 

\subsection{Spread estimate of (improperly) normalised spectra}
A simple method of estimating the necessary range of Reynolds numbers $Re_{\lambda}$ for the collapse of the normalised spectra in logarithmic coordinates to be meaningful is now presented where it is shown that the collapse (or spread) at two streamwise locations is only significant if the logarithm of the respective Reynolds numbers' ratio is large, typically $\log\left(Re_{\lambda_1}/Re_{\lambda_2}\right) > 1/4$. The starting point in this methodology is the assumption that a given scaling is correct (e.g. Kolmogorov or single-length scalings) which then allows the quantification of the spread for a given $Re_{\lambda}$ range of any other attempted normalisation.

%\begin{figure}
%\centering
%%trim option's parameter order: left bottom right top
%\includegraphics[trim=10 20 80 110, clip=true, width=120mm]{SketchSpectralSpread}
%\caption[Sketch of two spectra normalised with outer variables spreading at high wavenumbers]{Sketch of two spectra at two streamwise positions $x=\xi_1$ and $x=\xi_2$ normalised with outer variables spreading at high wavenumbers.} 
%\label{Fig:Sketch} 
%\end{figure}
\begin{figure}
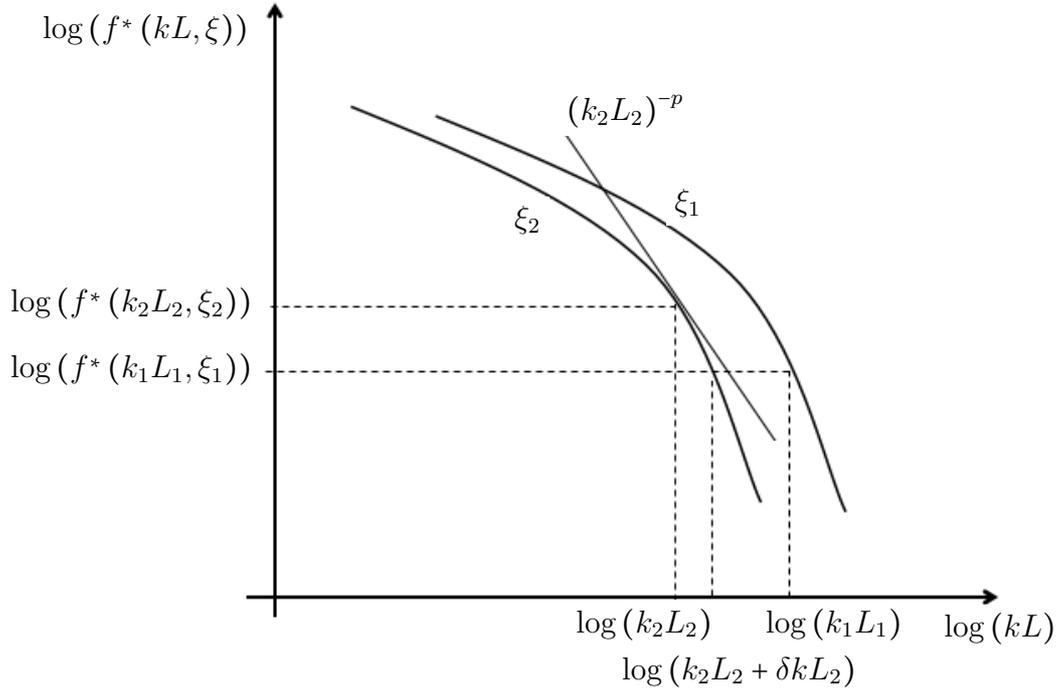

\centering
%trim option's parameter order: left bottom right top
   \begin{lpic}[l(-5mm),b(-3mm)]{SketchSpectralSpread(130mm)}
   \lbl[W]{205,13;$\log\left(k L\right)$}
   \lbl[W]{170,14;$\log\left(k_1 L_1\right)$}
   \lbl[W]{130,14;$\log\left(k_2 L_2\right)$}
   \lbl[W]{150,4;$\log\left(k_2 L_2+\delta kL_2\right)$}
   \lbl[W]{22,68;$\log\left( f^* \left( k_1 L_1,\xi_1\right)\right)$}
   \lbl[W]{22,82;$\log\left( f^* \left( k_2 L_2,\xi_2\right)\right)$}
   \lbl[W]{25,140;$\log\left( f^* \left( k L,\xi\right)\right)$}
   \lbl[W]{105.4,99.8;$\xi_2$}
   \lbl[W]{139.2,103.4;$\xi_1$}
   \lbl[W]{126,123;\hspace{2mm}$\left(k_2L_2 \right)^{-p}$\hspace{2mm}}
   \end{lpic}
\caption[Sketch of two spectra normalised with outer variables spreading at high wavenumbers]{Sketch of two spectra at two streamwise positions $x=\xi_1$ and $x=\xi_2$ normalised with outer variables spreading at high wavenumbers.} 
\label{Fig:Sketch} 
\end{figure}

Consider, for example, that the dissipation range of the longitudinal spectrum does scale with Kolmogorov variables, i.e. 
\begin{equation}
F_{11}^{(1)}(k,x)=\varepsilon^{2/3}\eta^{5/3}f(k\eta).
\label{eq:KolScaling}  
\end{equation}
(Note that, the same methodology can easily be extended to the energy containing range of the spectrum as well as to the case where the single-length scaling is assumed to be correct.)  

Take two streamwise distances $x=\xi_1$ and $x=\xi_2$ and write $\eta_1=\eta(\xi_1)$, $\eta_2=\eta(\xi_2)$, $\lambda_1=\lambda(\xi_1)$, $\lambda_2=\lambda(\xi_2)$, $L_1=L_{11}^{(1)}(\xi_1)$, $L_2=L_{11}^{(1)}(\xi_2)$, $u'_1=u'(\xi_1)$, $u'_2=u'(\xi_2)$, $\varepsilon_1=\varepsilon(\xi_1)$, $\varepsilon_2=\varepsilon(\xi_2)$ for the Kolmogorov scales, Taylor microscales, integral-length scales, r.m.s. turbulence velocities and dissipation rates at these two locations. It is arbitrarily chosen to take $\xi_2 > \xi_1$ so that $Re_{L_1} \equiv u'_1 L_1 / \nu > Re_{L_2} \equiv u'_2 L_2 / \nu$.

Choose two wavenumbers $k_1$ and $k_2$ in the dissipation range such that $k_1 \eta_1=k_2 \eta_2$ and $f(k_1 \eta_1)=f(k_2 \eta_2)$  by assumption. If one would normalise the same spectra in this range using $u'^{2}$ \& $L_{11}^{(1)}$, the dependence of the normalised spectra on $x$ would explicitly resurface, i.e. $F_{11}^{(1)}(k,x)=u'^{2}\, L_{11}^{(1)}\, f^{*}(kL_{11}^{(1)},x)$ (see figure \ref{Fig:Sketch}). 
Assume, also, that 
\begin{equation}
\varepsilon = C_{\varepsilon}^{1(1)} \frac{u'^3}{L_{11}^{(1)}}\,\Rightarrow\,\frac{L_{11}^{(1)}}{\lambda} = \frac{C_{\varepsilon}^{1(1)}}{15} Re_{\lambda}
\label{eq:Ceps}
\end{equation}
with $C_{\varepsilon}^{1(1)}$ independent of $x$.
It follows that $L_{11}^{(1)}/\eta=\left(C_{\varepsilon}^{1(1)}\right)^{1/4}Re_{L^{1(1)}}^{3/4}$ and it is possible to show from \eqref{eq:KolScaling} and \eqref{eq:Ceps} that
\begin{align}
f^{*}(k_1 L_1, \xi_1)=f^{*}(k_2 L_2, \xi_2)\left(\frac{\eta_1}{L_1}\frac{L_2}{\eta_2}\right)^{5/3}=f^{*}(k_2 L_2, \xi_2)\left(\frac{Re_{L_2}}{Re_{L_1}}\right)^{5/4}
\label{B1}
\end{align}
and
\begin{align}
k_1 L_1 = k_2 L_2 \left( \frac{\eta_2}{L_2}\frac{L_1}{\eta_1}\right)=k_2 L_2 \left(\frac{Re_{L_1}}{Re_{L_2}}\right)^{3/4},
\label{B2}
\end{align}
so that $f^{*}(k_1 L_1, \xi_1)\neq f^{*}(k_2 L_2, \xi_2)$ and $k_1 L_1 \neq k_2 L_2$.  

The spectral spread which characterises the degree of non-collapse by the form $F_{11}^{(1)}=u'^{2} L_{11}^{(1)} f^{*}\left(kL_{11}^{(1)}\right)$ is defined as 
\begin{equation}
\Psi = \log(k_1 L_1)-\log(k_2 L_2 + \delta kL_2),
\end{equation}
where $f^{*}(k_1 L_1, \xi_1)=f^{*}(k_2 L_2+ \delta kL_2, \xi_2)$, see figure \ref{Fig:Sketch}. There are two contributions to the spectral spread, one from the rescaling of the abscissae, $k_1 L_1 \neq k_2 L_2$, and another from the rescaling of the ordinates. From $Re_{L_1} > Re_{L_2} $ and equations \eqref{B1},  \eqref{B2} it follows that $k_1 L_1 > k_2 L_2$ and $f^{*}(k_1 L_1, \xi_1)< f^{*}(k_2 L_2, \xi_2)$ so that the two contributions to the spectral spread counteract each other and thus it misleadingly decreases the total spread. 
However, the second contribution depends on the functional form of $f^{*}\left(kL_{11}^{(1)}\right)$ and therefore it is not possible to quantify its spectral spread contribution without an analytical expression for $f^{*}\left(kL_{11}^{(1)},\xi\right)$. Nonetheless, as is shown below, it is possible to estimate a bound for this contribution, so that in the end one can estimate a upper and lower bound for the expected spectral spread $\Psi$ characterising the degree of non-collapse by the alternative scaling.

The contribution to the spread $\Psi$ from the abscissa's rescaling alone (which is the upper bound) is given by (using \eqref{B2})
\begin{align}
\Psi_{max}=\log(k_1 L_1)-\log(k_2 L_2)=\frac{3}{4} \log\left(\frac{Re_{L_1}}{Re_{L_2}}\right)=\frac{3}{2} \log\left(\frac{Re_{\lambda_1}}{Re_{\lambda_2}}\right).
\label{B5}
\end{align}
%(for the last equality, \eqref{Eq:LOverLambda} was used to relate the integral scale to the Taylor micro-scale with $C_{\varepsilon}=Const$ from Richardson-Kolmogorov phenomenology).

The contribution to the spread $\Psi$ from the ordinate's rescaling is measured as a fraction of the abscissa's rescaling
\begin{align}
\Phi \equiv \frac{\log(k_2 L_2+ \delta kL_2)-\log(k_2 L_2)}{\log(k_1 L_1)-\log(k_2 L_2)},
 \label{eq:Phi}
\end{align}
so that $\Phi=0$ for $\delta kL_2=0$ (ordinate rescaling has no effect) and $\Phi=1$ for $\delta kL_2=k_1 L_1-k_2 L_2$ (ordinate rescaling cancels the abscissas rescaling). It is possible to show using a first order Taylor expansion in logarithmic coordinates  that one can rewrite the function $\Phi$ to leading order as 
\begin{align}
\Phi = -\frac{5}{3}\left(\left.\frac{\partial\, \log\left(f^{*}\left(\log\left(kL_{11}^{(1)}\right), \xi\right)\right)}{\partial\, \log\left(kL_{11}^{(1)}\right)}\right|_{kL_{11}^{(1)}=k_2 L_2}\right)^{-1}.
\label{B4}
\end{align}

Since the spectra in the dissipation range roll-off faster than any power law one can always find a high enough wavenumber $k_t (p)L_{11}^{(1)}$ for which the tangent of the spectrum (in logarithmic coordinates) is steeper than $(kL)^{-p}$ given an exponent $p$ (see figure \ref{Fig:Sketch}). Consequently, for a given choice of $p$, one gets an upper bound for $\Phi$ for wavenumbers above $k_t (p) L_{11}^{(1)}$ which is $\Phi_{max}=5/(3p)$. Therefore one can estimate a lower bound for the spectral spread as $\Psi_{min} = \Psi_{max}(1-\Phi_{max})$ and thus 
\begin{align}
 \frac{3}{2} \log\left(\frac{Re_{\lambda_1}}{Re_{\lambda_2}}\right) \left(1- \frac{5}{3p}\right) < \Psi <  \frac{3}{2} \log\left(\frac{Re_{\lambda_1}}{Re_{\lambda_2}}\right).
 \label{eq:Psi}
\end{align}

Instead of assuming \eqref{eq:Ceps} with $C_{\varepsilon}^{1(1)}$ independent of $x$, one can consider the nonclassical dissipation scaling instead, $C_{\varepsilon}^{1(1)}\sim Re_{\lambda}^{-1}$. Repeating the same steps outlined above one arrives to
\begin{align}
 \frac{1}{2} \log\left(\frac{Re_{\lambda_1}}{Re_{\lambda_2}}\right) \left(1- \frac{5}{3p}\right) < \Psi <  \frac{1}{2} \log\left(\frac{Re_{\lambda_1}}{Re_{\lambda_2}}\right).
 \label{eq:PsiB}
\end{align}

One can also repeat the exact same analysis assuming the validity of $F_{11} (k,x)= u'^2 L_{11}^{(1)} f^{*}\left(kL_{11}^{(1)}\right)$, i.e. assuming that the spectra would effectively have a single dynamically relevant length-scale. 
In this case, the spread would surface if the normalisation with Kolmogorov variables was attempted and the spectral spread would also be given by \eqref{eq:Psi} and \eqref{eq:PsiB} for $C_{\varepsilon}^{1(1)}\sim \mathrm{constant}$ and $C_{\varepsilon}^{1(1)}\sim Re_{\lambda}^{-1}$, respectively.

\subsection{Spectral spread with classical and nonclassical \mbox{dissipation} behaviour} \label{sec:spreadproof}

In the previous section, it was shown that the spread of incorrectly normalised spectra (in logarithmic coordinates) at two downstream locations is proportional to the logarithm of the $Re_{\lambda}$ ratio straddled (see \eqref{eq:Psi} and \eqref{eq:PsiB}). 

\begin{figure}
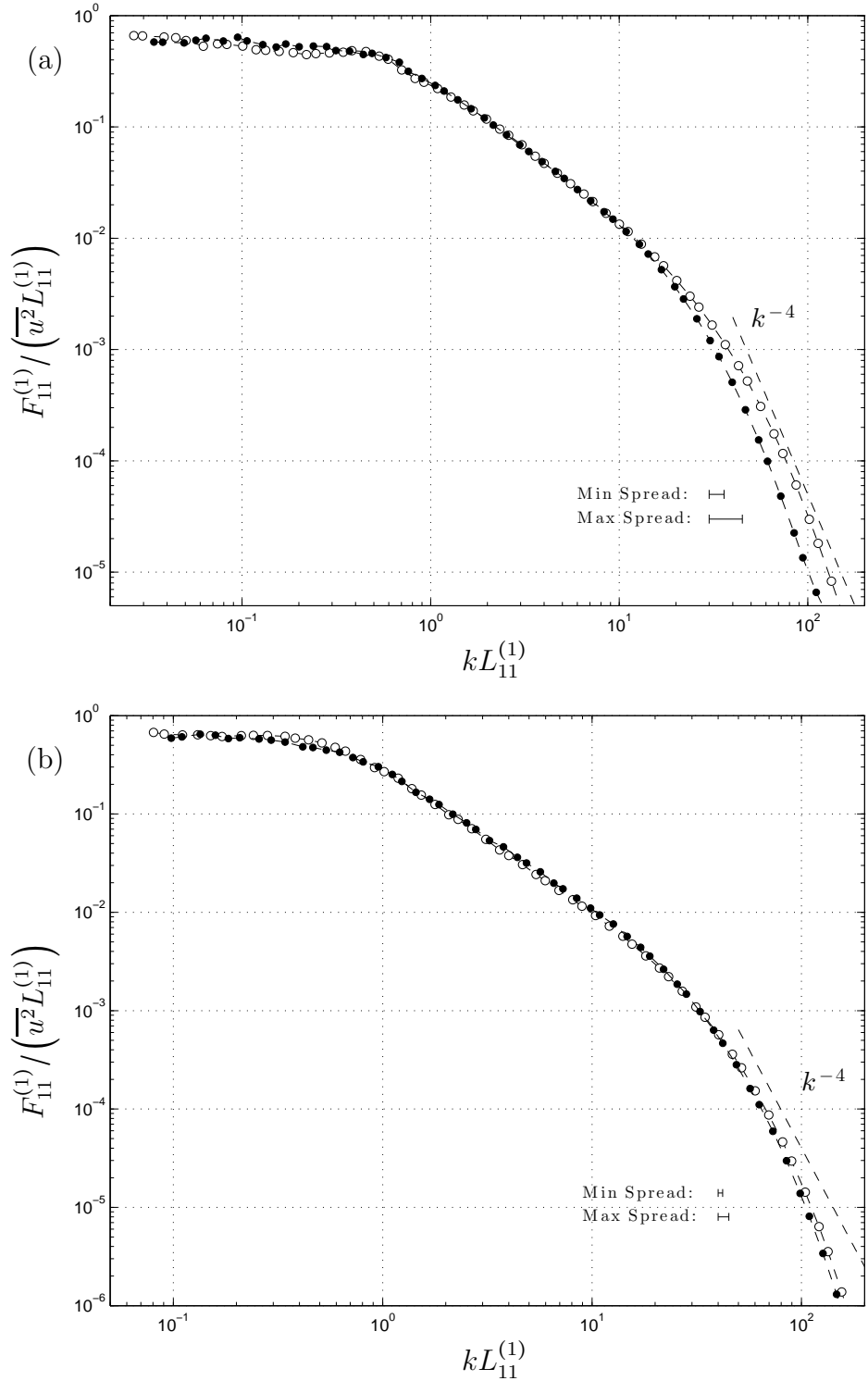

\centering
\begin{minipage}[c]{\linewidth}
   \centering
   \begin{lpic}[b(2mm)]{CompensatedSpectra-RG-ComparisonSFG(120mm)}
   \lbl{4,130;(a)}
   \lbl[W]{99,4;$k L_{11}^{(1)}$}
   \lbl[W]{2,74,90;$F_{11}^{(1)}/\left(\overline{u^2}L_{11}^{(1)}\right)$}
   \end{lpic}
\end{minipage}
\begin{minipage}[c]{\linewidth}
   \centering 
   \begin{lpic}{CompensatedSpectra-SFG-ComparisonRG(120mm)}
   \lbl[W]{99,4;$k L_{11}^{(1)}$}
   \lbl[W]{2,74,90;$F_{11}^{(1)}/\left(\overline{u^2}L_{11}^{(1)}\right)$}
   \lbl{4,130;(b)}
   \end{lpic}
\end{minipage}
\caption[$F_{11}^{(1)}$ normalised with $\overline{u^{2}}$ \& $L_{11}^{(1)}$ for RG60 and FSG18''x18''  data]{Longitudinal energy spectra normalised with $\overline{u^{2}}$ \& $L_{11}^{(1)}$ at two downstream locations along the centreline.
 (a) RG60-generated turbulence recorded for $U_{\infty}=20$ms$^{-1}$ at (\protect\raisebox{-0.5ex}{\SmallCircle}) $x/x_{\mathrm{peak}}\approx 9$ and  (\protect\raisebox{-0.5ex}{\FilledSmallCircle}) $x/x_{\mathrm{peak}}\approx 27$ corresponding to ${Re}^{\mathrm{iso}}_{\lambda_1} \approx 156$ and ${Re}^{\mathrm{iso}}_{\lambda_2} \approx 120$.
 (b) FSG18''x18''-generated turbulence recorded for $U_{\infty}=10$ms$^{-1}$ at (\protect\raisebox{-0.5ex}{\SmallCircle}) $x/x_{\mathrm{peak}}\approx 1.9$ and  (\protect\raisebox{-0.5ex}{\FilledSmallCircle}) $x/x_{\mathrm{peak}}\approx 3.0$ corresponding to ${Re}^{\mathrm{iso}}_{\lambda_1} \approx 231$ and  ${Re}^{\mathrm{iso}}_{\lambda_2} \approx 179$. 
Both plots have roughly the same Reynolds number ratio, ${Re}^{\mathrm{iso}}_{\lambda_1}/{Re}^{\mathrm{iso}}_{\lambda_2}\approx 1.3$.}
\label{Fig:ComparisonRG_SFG}
\end{figure}

The one-dimensional longitudinal spectra normalised with outer variables (i.e. $u'$ and $L_{11}^{(1)}$) taken at two locations within the classical dissipation region of RG60-generated turbulence are compared with the spectra taken at two locations within the nonclassical dissipation region of FSG18''x18''-generated turbulence (see \ref{Fig:ComparisonRG_SFG}). 
The ratio of $Re_{\lambda}$ straddled between the two downstream locations is approximately the same, ${Re}^{\mathrm{iso}}_{\lambda_1}/{Re}^{\mathrm{iso}}_{\lambda_2}\approx 1.3$.

Remarkably, it would seem that the spectra of the nonclassical dissipation  region of the FSG18''x18''-generated turbulence reasonably collapse with the outer variables, for both low and high wavenumbers, whereas for the classical dissipation region spectra spread at high wavenumbers and collapse at low, as expected.

%% Data from the JFM FSG 10m/s 5um wire versus RG60 20m/s 2.5um wire. Classical Taylor's hypothesis

%%%%%%
\begin{figure}[p]
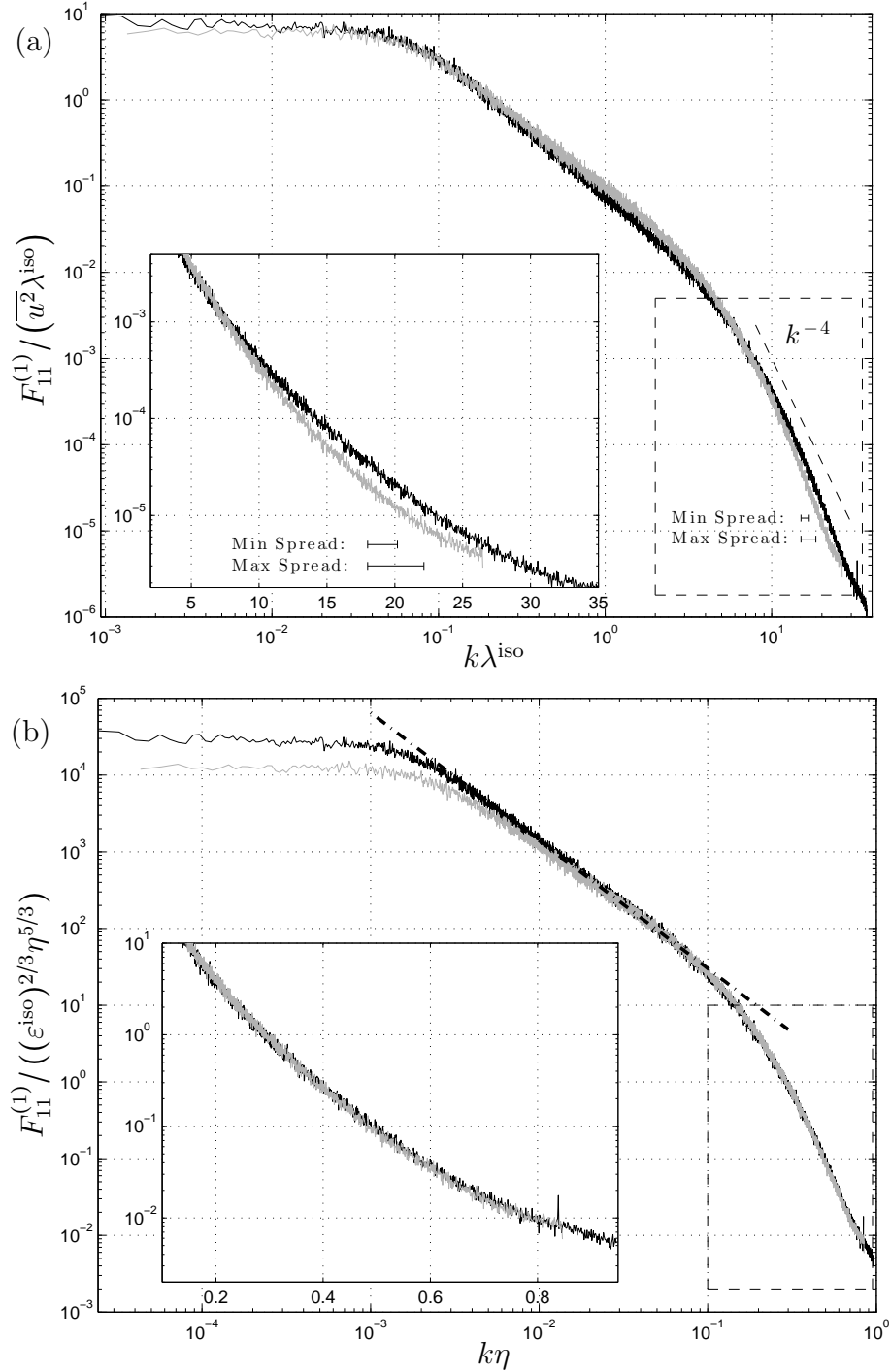

\centering
\begin{minipage}[c]{\linewidth}
   \centering
   \begin{lpic}[l(-2mm)]{FSG17_CollapseLambda(117mm)}
   \lbl{4,130;(a)}
   \lbl[W]{99,4;$k \lambda^{\mathrm{iso}}$}
   \lbl[W]{4,74,90;$F_{11}^{(1)}/\left(\overline{u^2} \lambda ^{\mathrm{iso}}\right)$}
   \end{lpic}
\end{minipage}
\begin{minipage}[c]{\linewidth}
   \centering 
   \begin{lpic}[b(-1mm)]{FSG17_CollapseEta(120mm)}
   \lbl{4,130;(b)}
   \lbl[W]{98,4;$k \eta$}
   \lbl[W]{4,74,90;$F_{11}^{(1)}/\left((\varepsilon^{\mathrm{iso}})^{2/3}\eta^{5/3}\right)$}
   \end{lpic}
\end{minipage}
\caption[$F_{11}^{(1)}$ normalised with  $\overline{u^{2}}$ \& $\lambda^{\mathrm{iso}}$ and $\varepsilon^{\mathrm{iso}}$ \& $\eta$ for  FSG18''x18''  data]{Longitudinal energy spectra normalised with (a) $\overline{u^{2}}$ \& $\lambda^{\mathrm{iso}}$ and (b) $\varepsilon^{\mathrm{iso}}$ \& $\eta$ at two streamwise locations within the nonclassical dissipation region in the lee of FSG18''x18'' recorded at $U_{\infty}=15$ms$^{-1}$. (black)  $x/x_{\mathrm{peak}}\approx 1.0$ and ${Re}^{\mathrm{iso}}_{\lambda_1} \approx 390$ and (light grey) $x/x_{\mathrm{peak}}\approx 2.5$ and ${Re}^{\mathrm{iso}}_{\lambda_2} \approx 253$.
The data within the dash-dotted box are re-plotted in the inset with linear abscissae and logarithmically spaced ordinates. 
The minimum and maximum high wavenumber spread indicated in figure (a) are computed using \eqref{eq:PsiB} setting $p=4$. 
The dash-dotted straight line in (b) represents a Kolmogorov-Obukhov spectrum, $F_{11}^{(1)}=\alpha\,\varepsilon^{2/3}k^{-5/3}$ with $\alpha=0.65$.}
\label{Fig:ComparisonLambdaEtaFSG}
\end{figure}

\begin{figure}[p]
\centering
\begin{minipage}[c]{\linewidth}
   \centering
   \begin{lpic}[l(-2mm)]{RG230_CollapseLambda(117mm)}
   \lbl{4,130;(a)}
   \lbl[W]{99,4;$k \lambda^{\mathrm{iso}}$}
   \lbl[W]{4,74,90;$F_{11}^{(1)}/\left(\overline{u^2} \lambda ^{\mathrm{iso}}\right)$}
   \end{lpic}
\end{minipage}
\begin{minipage}[c]{\linewidth}
   \centering 
   \begin{lpic}[b(-1mm)]{RG230_CollapseEta(120mm)}
   \lbl{4,130;(b)}
   \lbl[W]{98,4;$k \eta$}
   \lbl[W]{4,74,90;$F_{11}^{(1)}/\left((\varepsilon^{\mathrm{iso}})^{2/3}\eta^{5/3}\right)$}
   \end{lpic}
\end{minipage}
\caption[$F_{11}^{(1)}$ normalised with  $\overline{u^{2}}$ \& $\lambda^{\mathrm{iso}}$ and $\varepsilon^{\mathrm{iso}}$ \& $\eta$ for RG230 data]{Longitudinal energy spectra normalised with (a) $\overline{u^{2}}$ \& $\lambda^{\mathrm{iso}}$ and (b) $\varepsilon^{\mathrm{iso}}$ \& $\eta$ at two streamwise locations within the nonclassical dissipation region in the lee of RG230 recorded at $U_{\infty}=15$ms$^{-1}$. (black)  $x/x_{\mathrm{peak}}\approx 1.0$ and ${Re}^{\mathrm{iso}}_{\lambda_1} \approx 384$ and (light grey) $x/x_{\mathrm{peak}}\approx 1.9$ and ${Re}^{\mathrm{iso}}_{\lambda_2} \approx 300$.
The data within the dash-dotted box are re-plotted in the inset with linear abscissae and logarithmically spaced ordinates. 
The minimum and maximum high wavenumber spread indicated in figure (a) are computed using \eqref{eq:PsiB} setting $p=4$. 
The dash-dotted straight line in (b) represents a Kolmogorov-Obukhov spectrum, $F_{11}^{(1)}=\alpha\,\varepsilon^{2/3}k^{-5/3}$ with $\alpha=0.65$.}
\label{Fig:ComparisonLambdaEtaRG230}
\end{figure}

%However, let's assume, for the sake of the analyses, that the high frequency spectra scales with Kolmogorov variables, and let's estimate the spread due to the supposedly incorrect scaling of the spectra with outer variables. 
However, let us estimate the spread of the spectra normalised with outer variables in the case where the large wavenumbers  scale with Kolmogorov variables.
(Note that, should the spectra, conversely, scale with outer variables for all wavenumbers, this spread estimate would undoubtedly be larger than the collapse observed and the same estimate would quantify the spread of the normalisation with Kolmogorov variables.)
Applying \eqref{eq:Psi} and \eqref{eq:PsiB}, the estimates for these two cases are $9\%<\Psi<18\%$ and $3\%<\Psi<6\%$  for the  classical and nonclassical dissipation behaviour, respectively (see figures \ref{Fig:ComparisonRG_SFG}a and \ref{Fig:ComparisonRG_SFG}b where the estimated spreads are shown; the lower bound was estimated considering spectrum slopes steeper than $k^{-4}$).
Notice that the small spread observed in the nonclassical dissipation case is compatible with the estimated spread at large wavenumbers (see figure \ref{Fig:ComparisonRG_SFG}b), so it is conceivable that this collapse is only apparent.

It thus becomes clear that the nonclassical dissipation scaling, by itself, implies that the spread is three times smaller and that the reduced spread increases the difficulty to distinguish between large wavenumber scalings `\emph{\`a la}' Kolmogorov or George.
This is illustrated in figures \ref{Fig:ComparisonLambdaEtaFSG} and \ref{Fig:ComparisonLambdaEtaRG230} with spectra taken in the nonclassical dissipation region in the lee of RG230 and  FSG18''x18''.
The scaling with $\overline{u^2}$ and $\lambda$ (or $L_{11}^{(1)}$ since their ratio is constant in turbulence with a nonclassical dissipation behaviour)  misleadingly overlays the spectra for all wavenumbers (although not perfectly, see figures \ref{Fig:ComparisonLambdaEtaFSG}a and \ref{Fig:ComparisonLambdaEtaRG230}a), whereas for large wavenumbers the collapse with  Kolmogorov variables is nearly perfect (see figures \ref{Fig:ComparisonLambdaEtaFSG}b and \ref{Fig:ComparisonLambdaEtaRG230}b).
Changing the abscissae to linear coordinates emphasises the spread (insets of figures \ref{Fig:ComparisonLambdaEtaFSG} and \ref{Fig:ComparisonLambdaEtaFSG}) and, in these more stringent plots, it becomes evident that indeed the Kolmogorov variables collapse the spectra at  large wavenumber  whereas the single-length scale variables  $\overline{u^2}$ and $\lambda$ do not (compare figure \ref{Fig:ComparisonLambdaEtaFSG}a with figure \ref{Fig:ComparisonLambdaEtaFSG}b and  figure \ref{Fig:ComparisonLambdaEtaRG230}a with figure \ref{Fig:ComparisonLambdaEtaRG230}b). 

\section[``$-5/3$'' power-law spectra]{Nonclassical dissipation turbulence with ``$-5/3$'' power-law spectra}

\begin{figure}[t!]
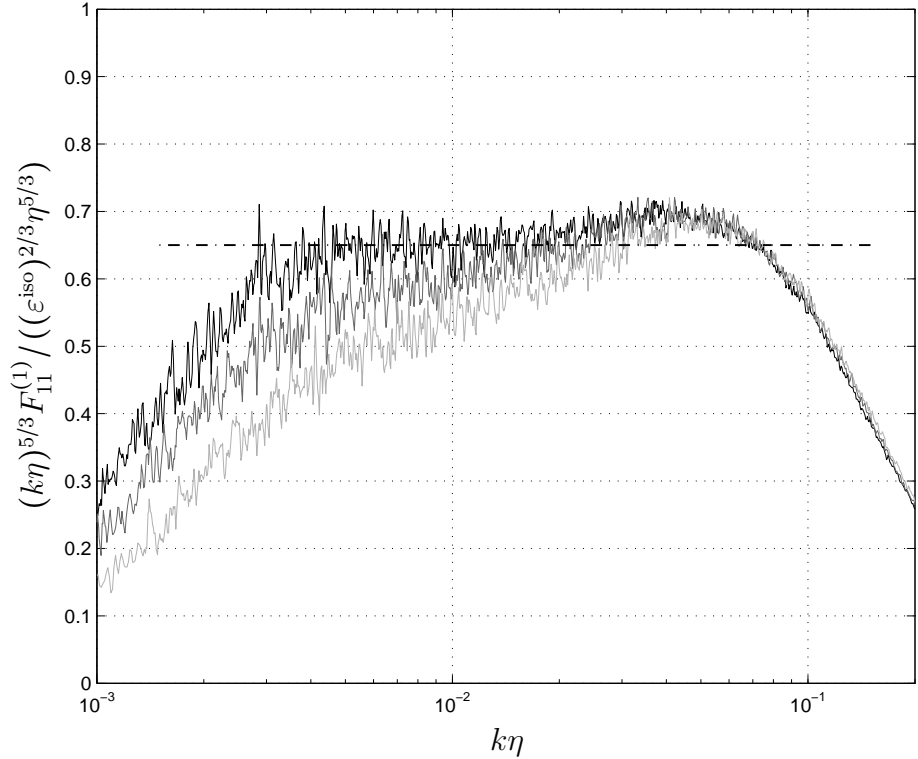

\centering
   \begin{lpic}[b(4mm)]{Thesis_PlotFSG18_FiveThirdsSpectra(120mm)}
   \lbl[W]{92,2;$k \eta$}
   \lbl[W]{4,78,90;$(k \eta)^{5/3}F_{11}^{(1)}/\left((\varepsilon^{\mathrm{iso}})^{2/3}\eta^{5/3}\right)$}
   \end{lpic}
\caption[Spectra of FSG18''x18''-generated turbulence  compensated by $k^{5/3}$]{Spectra of FSG18''x18''-generated turbulence  compensated by $k^{5/3}$ and normalised using $\varepsilon^{\mathrm{iso}}$ \& $\eta$ at three downstream locations along the centreline: (black) $x/x_{\mathrm{peak}} \approx 1.1$, $Re_{\lambda}^{\mathrm{iso}}\approx 418$, (dark grey) $x/x_{\mathrm{peak}} \approx 1.6$, $Re_{\lambda}^{\mathrm{iso}}\approx 350$ (light grey) $x/x_{\mathrm{peak}} \approx 2.5$, $Re_{\lambda}^{\mathrm{iso}}\approx 175$.}
\label{Fig:SpectraSlope}
\end{figure}

So far, it has been experimentally established that turbulence with both classical and nonclassical dissipation scalings present two pairs of dynamical scales characterising the spectrum at small and large wavenumbers, $\overline{u^2}$ \& $L_{11}^{(1)}$ and $\varepsilon$ \& $\nu$ respectively. 
For the highest $Re_{\lambda}^{\mathrm{iso}}$($\approx 400$) these spectra also exhibit a convincing power-law range with the  Kolmogorov-Obukhov exponent $-5/3$ (\citealt{K41a} and \citealt{O41}; see figures \ref{Fig:ComparisonLambdaEtaFSG}b, \ref{Fig:ComparisonLambdaEtaRG230}b and \ref{Fig:SpectraSlope}).  
Note that with the present experimental facility these $Re_{\lambda}^{\mathrm{iso}}$ can only be achieved with largest mesh-size grids at the highest $U_{\infty}$ and close to $x=x_{\mathrm{peak}}$. Therefore,  $-5/3$ power-law spectra are only obtained in the first part of the nonclassical dissipation region. As the turbulence (and $Re_{\lambda}^{\mathrm{iso}}$) decays the power-law region of the spectra become less defined and the exponents depart from the  $-5/3$ benchmark (\emph{cf.} figure \ref{Fig:SpectraSlope}). Note that the overshoot  between $0.02\lesssim k\eta \lesssim 0.08$ of the compensated spectra  presented in figure \ref{Fig:SpectraSlope} are usually denoted as pre-dissipative `bumps' \cite[]{Coantic99}.

Nevertheless, the celebrated prediction of \cite{K41a,K41c} and \cite{O41} of power-law spectra with a $-5/3$ exponent for high Reynolds number turbulence is typically associated with $C_{\varepsilon}\sim \mathrm{constant}$ (i.e. $C_{\varepsilon}$ both independent of local and global Reynolds numbers, e.g. $Re_{L^{1(1)}}$ and $Re_{M}$). 
In fact, \cite{Lumley92} suggests that the vast experimental support of the (one-dimensional longitudinal) Kolmogorov-Obukhov spectrum, $F_{11}^{(1)}(k)=\alpha\, \varepsilon^{2/3}k^{-5/3}$ is ``a more direct, and considerably less time consuming, demonstration'' of  $C_{\varepsilon}\sim \mathrm{constant}$. 
The present data, however, suggest that  $C_{\varepsilon}\sim \mathrm{constant}$ is not a necessary condition for the spectrum to exhibit a $-5/3$ power-law.\\

\begin{figure}[p]
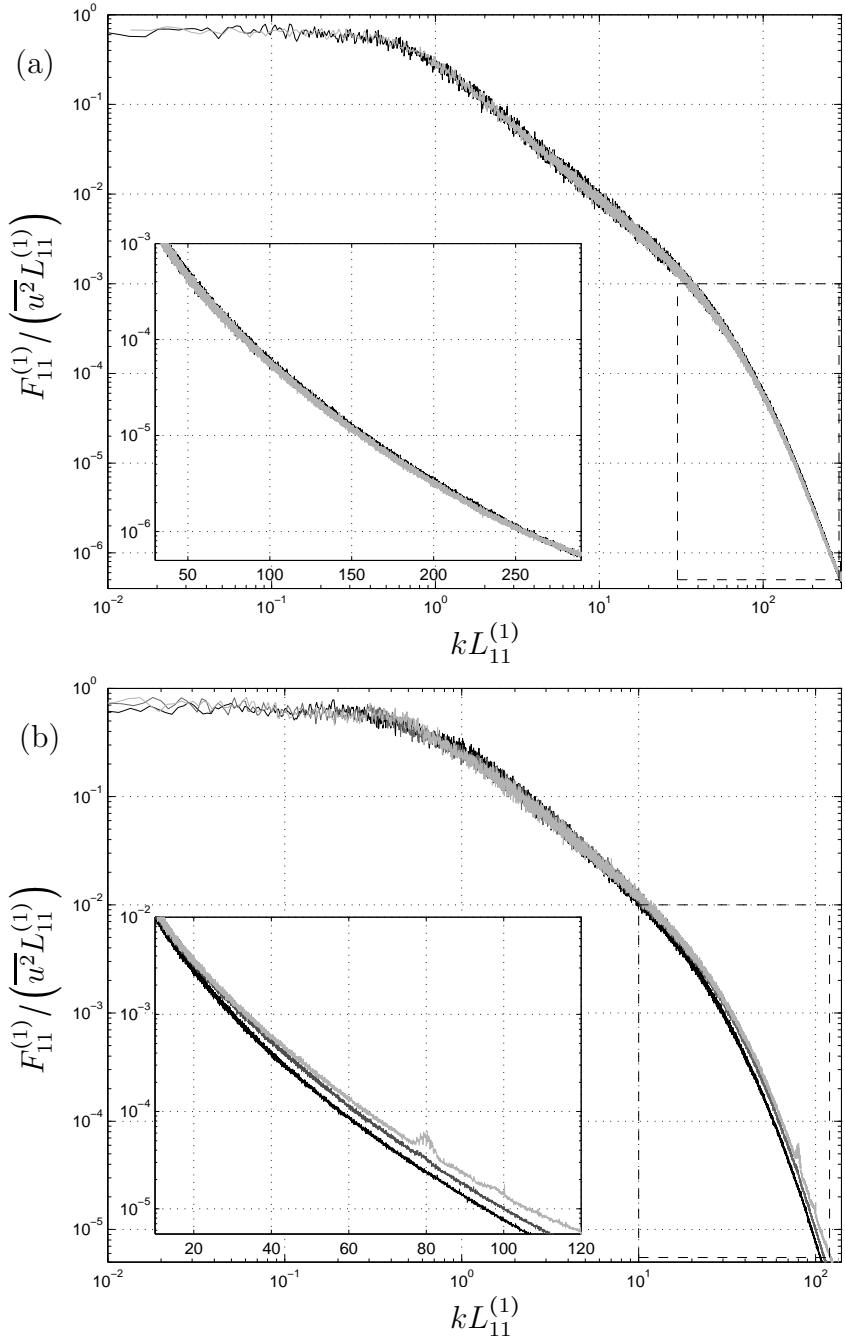

\centering
\vspace{-2mm}
\begin{minipage}[c]{\linewidth}
   \centering 
   \begin{lpic}[b(3mm)]{Thesis_ShowSpectraScaling(108mm)}
   \lbl{2,125;(a)}
   \lbl[W]{100,0;$k L_{11}^{(1)}$}
   \lbl[W]{2,74,90;$F_{11}^{(1)}/\left(\overline{u^2}L_{11}^{(1)}\right)$}
   \end{lpic}
\end{minipage}
\begin{minipage}[c]{\linewidth}
   \centering 
   \begin{lpic}[b(2mm)]{Thesis_ShowCepsMattersRG60(108mm)}
   \lbl{3,125;(b)}
   \lbl[W]{100,0;$k L_{11}^{(1)}$}
   \lbl[W]{2,74,90;$F_{11}^{(1)}/\left(\overline{u^2}L_{11}^{(1)}\right)$}
   \end{lpic}
\end{minipage}
\caption[Spectra at comparable $Re_{L^{1(1)}}$ and (a) comparable or (b) different $C_{\varepsilon}^{1(1)}$ ]{(a) Spectra of (black) FSG18''x18''- and (light grey) RG230-generated turbulence along the centreline normalised with $\overline{u^{2}}$ \& $L_{11}^{(1)}$ at comparable $Re_{L^{1(1)}}$ ($Re_{L^{1(1)}}=4091$ versus $Re_{L^{1(1)}}=4023$, respectively) and at comparable $C_{\varepsilon}^{1(1)}$ ($C_{\varepsilon}^{1(1)}\approx 0.50$ versus $C_{\varepsilon}^{1(1)}\approx 0.49$, respectively). FSG18''x18'' and RG230 data are acquired at $U_{\infty}=17.5\,\&\,20$ms$^{-1}$, respectively.
(b) Spectra of RG60-generated turbulence normalised with $\overline{u^{2}}$ \& $L_{11}^{(1)}$ at comparable $Re_{L^{1(1)}}$ but different $C_{\varepsilon}^{1(1)}$. This is achieved by acquiring data  at different $U_{\infty}$ and choosing the downstream position along the centreline where $Re_{L^{1(1)}}$ is similar (for $x/x_{\mathrm{peak}} \approx 1.8,\,4.8\,\&\,11.8$ recorded at $U_{\infty}=10,\,15\,\&\,20$ms$^{-1}$, $Re_{L^{1(1)}}=1302,\,1313\,\&\,1301$ and $C_{\varepsilon}^{1(1)}\approx 0.62,\,0.76\,\&\,0.86$, respectively). (black) $x/x_{\mathrm{peak}} \approx 1.8$, (dark grey) $x/x_{\mathrm{peak}} \approx 4.8$ and (light grey) $x/x_{\mathrm{peak}} \approx 11.8$. The data within the dash-dotted box are re-plotted in the inset with linear abscissae and logarithmically spaced ordinates. }
\label{Fig:SpectraVars}
\end{figure}

In the following it is attempted to formulate a simple phenomenological theory  to support the experimental observation of a $-5/3$ power-law spectrum in high Reynolds number turbulence exhibiting a nonclassical behaviour of the type $C_{\varepsilon}^{1(1)}\sim Re_M^{1/2}/Re_{\lambda}^{\mathrm{iso}}\sim Re_M/{Re}^{\mathrm{iso}}_{L^{1(1)}}$ (see ch. \ref{chp:4}). 
The starting point is the hypothesis that the longitudinal spectrum\footnote{This analysis is undertaken for $F_{11}^{(1)}$ for the sake of comparison with the data. However, the analysis is readily extendable for other components of the velocity spectrum as well as for the spherically averaged three-dimensional energy spectrum.}, $F_{11}^{(1)}$, only depends on five variables: (i) the wavenumber $k$, (ii) the mean square velocity $\overline{u^2}$, (iii) the mean square longitudinal velocity derivate  $\overline{\left(\partial u/\partial x\right)^2}$, (iv) the kinematic viscosity $\nu$ of the fluid and (v) the integral-length scale $L_{11}^{(1)}$. 
 This choice of variables is motivated by the facts that,
\begin{equation}
\overline{u^2} = \int\limits_0^{\infty}\! F_{11}^{(1)}\,dk; \hspace{5mm}
\overline{\left(\partial u/\partial x\right)^2} = \int\limits_0^{\infty}\! k^2\,F_{11}^{(1)}\,dk; \hspace{5mm} \mathrm{and }\hspace{5mm}
\lim_{k\rightarrow 0} F_{11}^{(1)} = \overline{u^2}\,L_{11}^{(1)}.
\end{equation}   
For convenience it is preferred to use the dissipation surrogate $\varepsilon^{\mathrm{iso}}=15\nu\overline{\left(\partial u/\partial x\right)^2}$ as a working variable and thus, 
\begin{equation}
F_{11}^{(1)} =F_{11}^{(1)}\left(k,\,\overline{u^2},\,L_{11}^{(1)},\,\varepsilon^{\mathrm{iso}},\,\nu\right).
\label{eq:F11vars}
\end{equation}   
From dimensional analysis it follows that 
\begin{equation}
F_{11}^{(1)} =(\varepsilon^{\mathrm{iso}})^{2/3}k^{-5/3}\,f\left(kL_{11}^{(1)},\,k\eta,\,C_{\varepsilon}^{1(1)}\right)=v_K^2 \,\eta\,\left(k\eta\right)^{-5/3}\,f\left(kL_{11}^{(1)},\,k\eta,\,C_{\varepsilon}^{1(1)}\right), 
 \label{eq:norm1}
\end{equation}   
choosing the $k$ and $\varepsilon^{\mathrm{iso}}$ as normalising variables ($\eta=(\nu^3/\varepsilon^{\mathrm{iso}})^{1/4}$ and $v_K=\left(\varepsilon^{\mathrm{iso}} \nu \right)^{1/4}$ are the usual Kolmogorov length and velocity scales), or
\begin{equation}
 F_{11}^{(1)} =\overline{u^2} L_{11}^{(1)}\,\, f'\left(kL_{11}^{(1)},\,{Re}^{\mathrm{iso}}_{L^{1(1)}},\,C_{\varepsilon}^{1(1)}\right),
 \label{eq:norm2}
 \end{equation}   
choosing $u'$ and $L_{11}^{(1)}$ instead ($f$ and $f'$ are two non-dimensional functions).

Note that relation \eqref{eq:F11vars} implies that the spectra do not directly depend on, for example, the mean velocity, the downstream location, the geometry of the grid and/or other initial/boundary conditions -- only indirectly via $\left(k,\,\overline{u^2},\,L_{11}^{(1)},\,\varepsilon^{\mathrm{iso}},\,\nu\right)$.
The present data can, nonetheless, be used to test \eqref{eq:norm2} and therefore the validity of the assumption expressed by \eqref{eq:F11vars}. 
In figure \ref{Fig:SpectraVars}a velocity spectra of turbulence generated by two geometrically different grids at similar $Re_{L^{1(1)}}$ and $C_{\varepsilon}^{1(1)}$ are presented. The fact that the spectra lay on top of each other remarkably well provides direct confirmation that the geometry of the grid does not have a direct role in shaping the spectra. Also, the fact that the data are recorded at different $U_{\infty}$ confirms the assumption that mean flow does not directly influence the spectra either. 
In figure \ref{Fig:SpectraVars}b a different test is performed by comparing velocity spectra in the lee of the same grid at similar  $Re_{L^{1(1)}}$ but at different $C_{\varepsilon}^{1(1)}$. The spread of the data at large wavenumber confirms that  there is a functional dependence on $C_{\varepsilon}^{1(1)}$ as anticipated by \eqref{eq:norm2}.
These observations lead to the presumption that  \eqref{eq:F11vars} may be a good approximation for the present class of turbulent flows.
Assumption  \eqref{eq:F11vars} is, nevertheless, stronger than stating that $F_{11}^{(1)}$ is explicitly independent of $U_{\infty}$, downstream location and grid geometry.\\
%All in all, the data presented in figures \ref{Fig:SpectraVars}a,b support  \eqref{eq:norm2}.

%One may now use  \eqref{eq:F11vars}, as a starting point for a phenomenological analysis, somewhat in the spirit of \cite{K41a}. 
%For high Reynolds numbers turbulent flows such that  $C_{\varepsilon}^{1(1)}$  becomes independent of $\nu$, even though it can take different values for different locations (see ch. \ref{chp:4}, where it is shown that the present high $Re_M$ data suggest that nonclassical dissipation scaling follows, to a first approximation at least, $C_{\varepsilon}^{1(1)}\sim Re_M/{Re}^{\mathrm{iso}}_{L^{1(1)}}$ consistent with the expectation that  $C_{\varepsilon}^{1(1)}$  becomes independent of $\nu$, even though it can take different values for the various downstream (and spanwise) locations).
%
%, the present turbulent flow as $Re_M\rightarrow\infty$ (or $\nu\rightarrow 0$) for a fixed downstream/spanwise location are considered. 
%
%, the wavenumbers much larger than those characteristic of the large-scales i.e. $kL_{11}^{(1)}\ll 1$ 
%
% together with the assumption that the high wavenumber spectrum . 

One may now repeat the matched-asymptotic expansion analysis of \cite{Lundgren2002} with the following revisions:
\begin{itemize}
\item The high $Re_M$ nonclassical dissipation scaling $C_{\varepsilon}^{1(1)}\sim Re_M/{Re}^{\mathrm{iso}}_{L^{1(1)}}$ implies that $C_{\varepsilon}^{1(1)}$ is, to a first approximation at least, independent of $\nu$ but varies with downstream (and spanwise) location (see ch. \ref{chp:4}). 
Consequently the matched-asymptotic expansion analysis needs to be repeated for each downstream/spanwise location. 
For a fixed location, $C_{\varepsilon}^{1(1)}$ takes a constant value and the analysis is done by matching the asymptotics as $\nu\rightarrow 0$\footnote{Note that this would not be necessarily the case if $C_{\varepsilon}^{1(1)}\rightarrow0$ or $C_{\varepsilon}^{1(1)}\rightarrow\infty$. Even though the present data suggest that $0.4<C_{\varepsilon}^{1(1)}<2$, the largest straddled Reynolds numbers are only moderately high.}. 

\item In the analysis of \cite{Lundgren2002}, $C_{\varepsilon}^{1(1)}$ ($P_1$ in his notation) is arbitrarily set to unity and does not explicitly enter his calculations. Here, the parameter $C_{\varepsilon}^{1(1)}$ is retained in the analysis and takes different numerical values for the various locations.

\item The analysis is done in wavenumber space based on the longitudinal spectrum $ F_{11}^{(1)} $ rather than in physical space with structure functions as in \cite{Lundgren2002}. In the present notation, $ F^{i} $ and $ F^{o} $ are the inner and outer expansions of  $ F_{11}^{(1)} $ for large and small wavenumbers, respectively. Recall that $\overline{u^2}$ \& $L_{11}^{(1)}$ and $v_{K}$ \& $\eta$ are, respectively, the outer and inner similarity variables.

\item The relations between the inner and outer scales \cite[eq. (18) in][]{Lundgren2002} are modified to take into account the dependence on $C_{\varepsilon}^{1(1)}$,
\[\eta = L_{11}^{(1)} \left(C_{\varepsilon}^{1(1)}\right)^{-1/4} \left({Re}^{\mathrm{iso}}_{L^{1(1)}}\right)^{-3/4}, \hspace{5mm} v^2_{K} = \overline{u^2} \left(C_{\varepsilon}^{1(1)}\right)^{1/2} \left({Re}^{\mathrm{iso}}_{L^{1(1)}}\right)^{-1/2}.\]

\item The one-term inner expansion ($f^{i,1}$) expressed in outer variables \cite[eq. (22) in][]{Lundgren2002} now reads,
\[ F^i = \overline{u^2} L_{11}^{(1)} \left(C_{\varepsilon}^{1(1)}\right)^{1/4} \left({Re}^{\mathrm{iso}}_{L^{1(1)}}\right)^{-5/4} f^{i,1} \left(kL_{11}^{(1)}\left(C_{\varepsilon}^{1(1)}\right)^{-1/4} \left({Re}^{\mathrm{iso}}_{L^{1(1)}}\right)^{-3/4}\right)  \]

\item The matching condition as $\nu\rightarrow 0$ (or ${Re}^{\mathrm{iso}}_{L^{1(1)}}\rightarrow \infty$ for a fixed spatial location) between the inner expansion evaluated for small values of its argument ($k\eta\ll 1$) and the outer expansion for large values of its argument ($kL_{11}^{(1)}\gg1$)  \cite[eq. (27) in][]{Lundgren2002} now reads (the inner and outer expansions for, respectively, small and large values of their arguments are denoted with an overtilde), 
\begin{equation*}
 \hspace{-5mm}\overline{u^2}  L_{11}^{(1)} \left(C_{\varepsilon}^{1(1)}\right)^{1/4} \left({Re}^{\mathrm{iso}}_{L^{1(1)}}\right)^{-5/4} \tilde f^{i,1} \left(kL_{11}^{(1)}\left(C_{\varepsilon}^{1(1)}\right)^{-1/4} \left({Re}^{\mathrm{iso}}_{L^{1(1)}}\right)^{-3/4}\right) = \overline{u^2}  L_{11}^{(1)} \tilde f^{o,1} \left(kL_{11}^{(1)}\right).
 \label{eq:matching}  
 \end{equation*}

\item From the matching condition it follows that  \cite[see][]{Lundgren2002} 
\begin{equation}
\begin{aligned}
\tilde f^{i,1} \left(k\eta\right) &= \alpha \left(k\eta\right)^{-5/3}\\
\tilde f^{o,1} \left(kL_{11}^{(1)}\right)&= \alpha \left(C_{\varepsilon}^{1(1)}\right)^{2/3} \left(kL_{11}^{(1)}\right)^{-5/3},
\end{aligned}
 \label{eq:UpdatedLundgrenResults}  
 \end{equation}
 where $\alpha$ is usually denoted as the Kolmogorov-Obukhov constant.
 \end{itemize}
Summarising, one may use matched-asymptotic analysis starting from the assumption expressed by \eqref{eq:F11vars}, together with the assumption that $C_{\varepsilon}^{1(1)}$ is independent of $\nu$ (even though it can take different numerical values at different locations), to establish that for the $\nu\rightarrow 0$ limit there is a range of wavenumbers 
(such that $k\eta\ll 1$ and $kL_{11}^{(1)}\gg1$) where the  turbulence spectrum follows,
\begin{equation}
F_{11}^{(1)}= \alpha (\varepsilon^{\mathrm{iso}})^{2/3} k^{-5/3} =  \alpha \left(C_{\varepsilon}^{1(1)}\right)^{2/3} \overline{u^2} L_{11}^{(1)}  \left(kL_{11}^{(1)}\right)^{-5/3}.
 \label{eq:KOLSpectra}  
 \end{equation}
It might be expected that one of the `constants' of the normalised spectrum, $\alpha$ or $A\equiv \alpha \left(C_{\varepsilon}^{1(1)}\right)^{2/3}$, is also invariant to the downstream/spanwise location (since all the position dependence of the spectrum is indirectly ascertained via the variables expressed in \eqref{eq:F11vars}) and there is no \emph{a priori} reason to prefer either. The present data are not recorded at sufficiently high Reynolds numbers for the spectra to exhibit $-5/3$ power-law throughout the decay  (\emph{cf.} figure \ref{Fig:SpectraSlope}) and therefore cannot be used to experimentally determine whether $\alpha$ or $A$  are invariant during decay of turbulence exhibiting a nonclassical dissipation behaviour.
Note that $C_{\varepsilon}^{1(1)} = (A/\alpha)^{3/2}$ is known result from \cite{K41b}, see his equation (28), which he actually used to conjecture that during decay $C_{\varepsilon}^{1(1)}\sim \mathrm{constant}$. In this thesis this is shown not to be always true, even for high Reynolds number ($Re_M$) turbulence.\\

The formal analysis presented above can be repeated in a more straightforward way which, in principle, retain all the essential physics. 
The starting point is, once more, the assumption expressed by \eqref{eq:norm1}.
Firstly, recall that one is considering very high Reynolds number turbulent flows ($Re^{\mathrm{iso}}_{L^{1(1)}}\gg1$) where $C_{\varepsilon}^{1(1)}$ is independent of $\nu$, even tough its numerical value can vary with downstream (and spanwise) location.
Note that  $Re^{\mathrm{iso}}_{L^{1(1)}}\gg1$ also implies that $L_{11}^{(1)}/\eta \gg 1$, since $L_{11}^{(1)}/\eta = \left(C_{\varepsilon}^{1(1)}\right)^{1/4}\left(Re^{\mathrm{iso}}_{L^{1(1)}}\right)^{3/4}$.
In the spirit of \cite{K41a} one expects the spectra,  for $kL_{11}^{(1)}\gg 1$ and $k\eta \ll 1$, to be independent of both $k\eta$ and $kL_{11}^{(1)}$, since these wavenumbers are associated with scales that are too small to be directly affected by the large-scale motions and too large to the directly affected by viscosity, i.e. \[F_{11}^{(1)}= (\varepsilon^{\mathrm{iso}})^{2/3} k^{-5/3} f(C_{\varepsilon}^{1(1)}).\]
Denoting $\alpha\equiv f(C_{\varepsilon}^{1(1)})$ immediately leads to \eqref{eq:KOLSpectra}.

\section{Summary}
In this chapter substantial evidence is presented that supports that turbulence with both classical and nonclassical behaviours of the dissipation rate have two sets of dynamically relevant length and velocity scales and, therefore, do not follow complete self-preserving solutions of the governing equations (e.g. the inhomogeneous von K\'{a}rm\'{a}n-Howarth equation, see ch. \ref{chp:6}). 
Nevertheless, it is shown that the nonequilibrium dissipation behaviour reduces the spread of improperly normalised spectra making it difficult to distinguish the collapse of the large wavenumber spectra normalised with $\eta$ \& $v_k$ versus $\overline{u^2}$ \& $\lambda$. 
This (mis)lead previous studies to relate the decay with nonclassical dissipation behaviour with self-preserving solutions of the von K\'{a}rm\'{a}n-Howarth or Lin equations \cite[see e.g.][]{MV2010,VV2011}.
It is also shown that for the highest $Re_{\lambda}$ data, which is acquired in the nonclassical dissipation region, the spectra exhibit a convincing $-5/3$ power-law. 
A matched-asymptotic expansion analysis of the type used by \cite{Lundgren2002} is used to demonstrate the compatibility between the $-5/3$ power-law and the observation that $C_{\varepsilon}^{1(1)}\sim Re_M/Re_{L^{1(1)}}$ for 
 the nonclassical dissipation region at large $Re_M$.

\clearemptydoublepage
\chapter{Scale-by-scale energy transfer budget}
\label{chp:6}

In this chapter the focus is directed to scale-by-scale budgets based on a dynamical equation similar to the  von K\'{a}rm\'{a}n-Howarth-Monin \cite[]{MY75}. The underlying physics of the nonclassical energy dissipation behaviour presented in the previous chapters are investigated.

\section{Scale-by-scale energy transfer budget equation} \label{sec:KHM}
A scale-by-scale energy transfer budget similar to the von K\'{a}rm\'{a}n-Howarth-Monin equation \cite[see (22.15) in][]{MY75}, but extended to inhomogeneous turbulent flows, can be derived directly from the Navier-Stokes \cite[see e.g.][ and references therein]{Deissler61,Casciola04,Danaila2012}.

The starting point is the incompressible Navier-Stokes decomposed into mean and fluctuating components at two  independent locations $\mathbf{x} \equiv \mathbf{X} - \mathbf{r}/2$ and $\mathbf{x'} \equiv \mathbf{X} + \mathbf{r}/2$ ($\mathbf{X}$ is the centroid of the two points and $r= |\mathbf{r}|$ their distance),
\begin{equation}
\left\lbrace
\begin{aligned}
\frac{\partial \,U_i + u_i}{\partial t} + U_k \frac{\partial u_i}{\partial x_k} + u_k \frac{\partial U_i}{\partial x_k} + U_k \frac{\partial U_i}{\partial x_k}  + u_k \frac{\partial u_i}{\partial x_k} = & -\frac{1}{\rho}\frac{\partial \,P + p}{\partial x_i} \,\,+ \nu \,\frac{\partial^2 \,U_i + u_i}{\partial x_k^2 }\\
\frac{\partial \,U'_i + u'_i}{\partial t} + U'_k \frac{\partial u'_i}{\partial x'_k} + u'_k \frac{\partial U'_i}{\partial x'_k} + U'_k \frac{\partial U'_i}{\partial x'_k} + u'_k \frac{\partial u'_i}{\partial x'_k} = & -\frac{1}{\rho}\frac{\partial \,P' + p'}{\partial x'_i} + \nu \frac{\partial^2\, U'_i + u'_i}{\partial x'^2_k},
\end{aligned}
\right.
\end{equation}
together with the continuity equations ($\partial U_k/ \partial x_k = \partial U'_k/ \partial x'_k =\partial u_k/ \partial x_k = \partial u'_k/ \partial x'_k = 0$). In the present notation  $U_i\equiv U_i(\mathbf{x})$, $u_i\equiv u_i(\mathbf{x})$, $P\equiv P(\mathbf{x})$, $U'_i\equiv U_i(\mathbf{x'})$, $u'_i\equiv u_i(\mathbf{x'})$ and $P'\equiv P(\mathbf{x'})$. 

The main steps in the derivation are to (i) subtract the two equations above and denote the velocity differences as $\delta u_i \equiv u_i - u'_i$, $\delta p \equiv p - p'$ and $\delta U_i \equiv U_i - U'_i$, (ii) multiply the resulting expression by $2\delta u_i$, (iii) ensemble average over an infinite number of realisations (denoted by overbars; in practice ergodicity is used on the basis of the time stationarity of the flow and time averages are performed) and (iv) change the coordinate system from ($\mathbf{x}$, $\mathbf{x'}$) to ($\mathbf{X}$, $\mathbf{r}$).
The resulting equation reads,
\begin{equation}
\begin{aligned}
\frac{\partial \, \overline{\delta q^2}}{\partial t} +  
\left( \frac{U_k + U'_k}{2} \right) \frac{\partial \, \overline{\delta q^2}}{\partial X_k} &+ 
\frac{\partial \, \overline{\delta u_k \delta q^2}}{\partial r_k} +
\frac{\partial \, \delta U_k \overline{\delta q^2}}{\partial r_k}  = \\
-2\overline{\delta u_i \delta u_k} \frac{\partial \, \delta U_i}{\partial r_k} &-
\overline{(u_k + u'_k)\delta u_i}\, \frac{\partial \, \delta U_i}{\partial X_k} -
\frac{\partial}{\partial X_k} \left( \overline{\frac{(u_k + u'_k) \delta q^2}{2}}  \right) -\\
\frac{2}{\rho} \frac{\partial \, \overline{\delta u_k \delta p} }{\partial X_k}  &+
\nu \left[ 2\frac{\partial^2 }{\partial r^2_k} +  \frac{1}{2}\frac{\partial^2 }{\partial X^2_k}  \right] \overline{\delta q^2}  - 
 2\nu \left[ \overline{ \left(  \frac{\partial u_i}{\partial x_k}  \right)^2} + \overline{ \left(  \frac{\partial u'_i}{\partial x'_k}  \right)^2}\right],
\end{aligned}
\label{eq:KHM}
\end{equation}
where $\overline{\delta q^2} \equiv \overline{(\delta u_i)^2}$. % and all the terms are functions of the centroid location, $X_j$, and the separation vector, $r_j$.
Equation \eqref{eq:KHM} is essentially an inhomogeneous von K\'{a}rm\'{a}n-Howarth-Monin equation with additional terms to account for the inhomogeneity of the turbulent flow field. 
Each of the terms can be interpreted as follows. 
\begin{enumerate}[(i)]
\item  $4\mathcal{A}_{t}^*(\mathbf{X},\mathbf{r})\equiv \partial \, \overline{\delta q^2}/\partial t$ results from the time dependence of $\overline{\delta q^2}(\mathbf{X},\,\mathbf{r})$.

\item  $4\mathcal{A}^*(\mathbf{X},\mathbf{r}) \equiv (U_k + U'_k)/2\,\, \partial \, \overline{\delta q^2}/\partial X_k$ represents an advection contribution to the change of  $\overline{\delta q^2}(\mathbf{X},\,\mathbf{r})$.

\item $4\Pi^*(\mathbf{X},\mathbf{r}) \equiv\partial \, \overline{\delta u_k \delta q^2}/\partial r_k$ represents a contribution which relates to nonlinear transfer of energy from a spherical shell centred at $\mathbf{X}$ with a radius $r$ at the orientation $\mathbf{r}/r$ to (a) concentric shells of larger radii (effectively to smaller radii since this term is typically negative) and (b) to other orientations within the same spherical shell. 
Notice that $\Pi^*$ is the divergence with respect to  $\mathbf{r}$ of the flux  $\overline{\delta u_k\delta q^2}$ and that owing to Gauss's theorem, $\iiint_{|\mathbf{r}|\leq r}\! \Pi^*\,dV = \oiint_{|\mathbf{r}|=r} \overline{\delta \mathbf{u} \delta q^2}\cdot\mathbf{r}/r \,dS$, i.e. the net contribution of $\Pi^*$ integrated over the sphere $|\mathbf{r}|\leq r$ is equal to the total radial flux over the spherical shell $|\mathbf{r}|=r$.
If the turbulence is homogeneous the radial flux is zero in the limit $r\rightarrow \infty$ and $4\Pi^*$ is indeed, unequivocally, a transfer term. 
Also note that (using a spherical coordinate system $(r,\,\theta,\,\phi)$ for $\mathbf{r}$) the integrals of the polar, $ \Pi^*_{\theta} $, and azimuthal, $\Pi^*_{\phi}$, components of the divergence $\Pi^*$ over the solid angle $\mathbf{r}/r$ are identically zero,  $\oiint_{|\mathbf{r}|=r} \Pi^*_{\theta} \,dS=\oiint_{|\mathbf{r}|=r} \Pi^*_{\phi} \,dS =0$, thus indicating a role of $\Pi^*$ in redistributing energy within a spherical shell.  

\item $4\Pi^*_U(\mathbf{X},\mathbf{r}) \equiv \partial \, \delta U_k\overline{\delta q^2}/\partial r_k$ represents  a contribution which relates to linear transfer of energy by mean velocity gradients from a spherical shell centred at $\mathbf{X}$ with a radius $r$ at the orientation $\mathbf{r}/r$ to concentric shells of larger radii. The motivation for this interpretation is analogous to that given for $\Pi^*$, where the turbulent flux is now $\delta U_k\overline{\delta q^2}$ \cite[see also][where the physical  interpretation of this term is given in wavenumber space]{Deissler61,Deissler81}.

\item $4\mathcal{P}^*(\mathbf{X},\mathbf{r}) \equiv-2\overline{\delta u_i \delta u_k} \, \partial \, \delta U_i / \partial r_k-\overline{(u_k + u'_k)\delta u_i}\, \partial \, \delta U_i/\partial X_k$ represents a contribution which relates to turbulent production. 
It is easiest to identify $\mathcal{P}^*$  as a production term by writing it in ($\mathbf{x}$, $\mathbf{x'}$)  coordinates, i.e.  $2\mathcal{P}^*=-\overline{u_i  u_k} \, \partial \,  U_i / \partial x_k - \overline{u'_i  u'_k} \, \partial \,  U'_i / \partial x'_k + \overline{u_i  u'_k} \, \partial \,  U_i / \partial x_k + \overline{u_i  u'_k} \, \partial \,  U'_i / \partial x'_k$, and recognising that the first two terms on the right-hand side  are the usual production terms of the single-point turbulent kinetic energy transport equation evaluated at $\mathbf{x}$ and $\mathbf{x'}$, respectively, cf. \eqref{eq:TKE}.

\item $4\mathcal{T}^*(\mathbf{X},\mathbf{r}) \equiv -\partial/\partial X_k \left(\overline{(u_k + u'_k) \delta q^2}/2+2/\rho\,\overline{ \delta u_k\delta p}\right)$ represents scale-by-scale turbulent transport from a spherical shell of radius $r$ centred at $\mathbf{X}$ at the orientation $\mathbf{r}/r$ to an adjacent  shell (centred at $\mathbf{X} + \delta \mathbf{X}$) with the same radius and at the same orientation. Notice that $\mathcal{T}^*$ is the divergence with respect to  $\mathbf{X}$ of the flux  $-\overline{(u_k + u'_k) \delta q^2}/2-2/\rho\,\overline{ \delta u_k\delta p}$ and thus, making use of Gauss's theorem, it follows that the net contribution of $\mathcal{T}^*$ integrated (with respect to $\mathbf{X}$ for each $\mathbf{r}$) over a volume $V$ is equal to the total flux over the bounding surface of $V$. This motivates the physical interpretation of this term as a scale-by-scale turbulent transport. 

\item $4\mathcal{D}^*_{\nu} (\mathbf{X},\mathbf{r}) \equiv 2\nu\,\partial^2 \overline{\delta q^2}/\partial r^2_k$ represents  viscous diffusion over a spherical shell of radius $r$ centred at $\mathbf{X}$  at the orientation $\mathbf{r}/r$ (note that $\lim_{r\rightarrow 0} \mathcal{D}^*_\nu (\mathbf{X},\mathbf{r}) = \varepsilon (\mathbf{X})$).

\item $4\mathcal{D}^*_{X,\nu} (\mathbf{X},\mathbf{r}) \equiv\nu/2\,\partial^2 \overline{\delta q^2}/\partial X^2_k$ represents scale-by-scale transport via viscous diffusion over a spherical shell of radius $r$ centred at $\mathbf{X}$  at the orientation $\mathbf{r}/r$. This can be seen as a transport term following the same reasoning  as that made for $\mathcal{T}^*$ by noticing that  $4\mathcal{D}^*_{X,\nu}$ can be written as a divergence of the viscous flux $\nu/2\,\partial \overline{\delta q^2}/\partial X_k$.

\item $4\varepsilon^*(\mathbf{X},\mathbf{r}) \equiv 2\nu \overline{ \left( \partial u_i/\partial x_k  \right)^2} + 2\nu\overline{ \left( \partial u'_i/\partial x'_k  \right)^2}$ represents the sum of twice the turbulent kinetic energy dissipation at the two locations, i.e. $2\varepsilon + 2\varepsilon' = 4\varepsilon^*$ with $\varepsilon^*\equiv (\varepsilon + \varepsilon')/2$. 

\end{enumerate}
For large $r$,  \eqref{eq:KHM} reduces to four times the average of two single-point turbulent kinetic energy transport equations \eqref{eq:TKE}, one evaluated at $\mathbf{x}$ and the other at $\mathbf{x'}$ \cite[see][]{Casciola04}.
Note that the dependence on the orientation $\mathbf{r}/r$ can be removed by averaging the terms over spherical shells of radius $r$, in the spirit of  \cite{NT99}.
The spherical shell averaged terms are denoted by removing the superscript asterisk.

%------------------------------------
\section{Experimental results}

\begin{figure}
\centering
\includegraphics[trim = 10mm 0mm 10mm 10mm, clip=true, width=160mm]{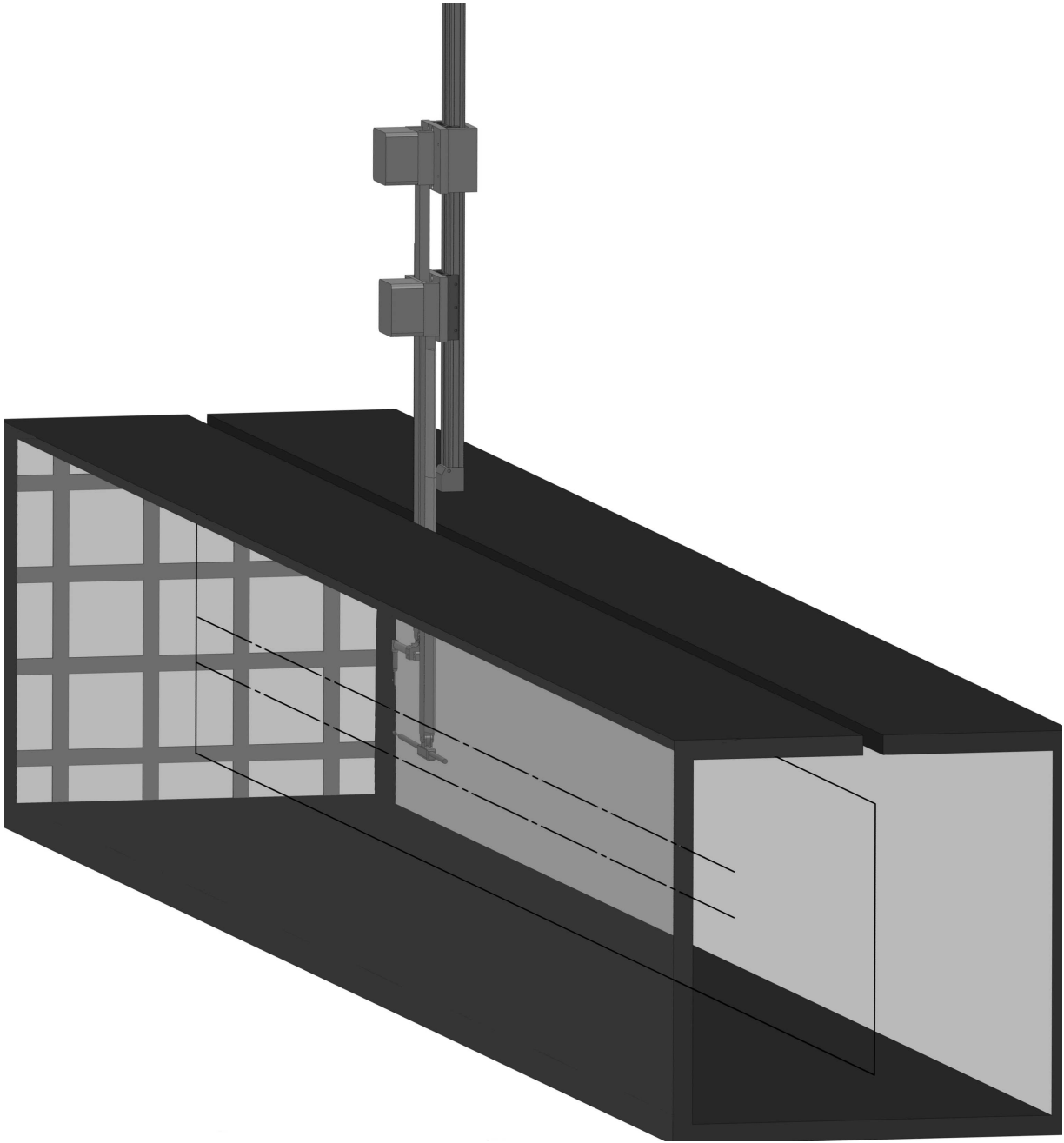}
\caption[Computer model of the  $2\times$XW measurement apparatus]{Computer model of the measurement apparatus for the present $2\times$XW experiments. For reference, a model of RG115 is added to the inlet of the test section. The upper dash-dotted line represents the centreline ($y=z=0$) and the lower  dash-dotted line represents the longitudinal line intercepting the lower bar of the grid ($y=-M/2$, $z=0$). The xy-plane at $z=0$ sketched in the figure represents the measurement plane. }
\label{fig:measloc}
\end{figure}

The experimental apparatus described in \S \ref{sec:WTB} and sketched in figure \ref{fig:measloc} is used to acquire data in the lee of the RG115 and RG60 grids in order to compute estimates of the terms in \eqref{eq:KHM} (except the pressure transport term).
The data are acquired with $\mathbf{X}$, the midpoint between the two X-probes,  along the centreline ($y=z=0$) at five downstream locations between  $x=1250$mm and $x=3050$mm
($X_1=1250,\,1700,\,2150,\,2600,\,3050\mathrm{mm}$ and $X_2=X_3=0$).
For two downstream locations of the centroid, $X_1=1250$mm and $X_1=2150$mm, additional datasets off-centreline at $X_2=-6$mm and $X_3=0$ are acquired so that derivatives of the statistics with respect to $X_2$ can be computed, particularly those needed to estimate $\partial/\partial X_2 \,\overline{(v + v') \delta q^2}$, see \eqref{eq:KHM}. 
The choice of  6mm as the distance to evaluate the $X_2$-derivative is based on the single-point data in the lee of RG115-turbulence used to estimate the lateral triple-correlation transport (i.e. $\partial/\partial y \,\overline{v  q^2}$ -- see \S \ref{sec:homo}). Based on those data it is found that  the spanwise derivative $\partial/\partial y \,\overline{v  q^2}$ is well approximated by $(\overline{v  q^2}(h_y)-\overline{v  q^2}(0))/h_y$ up to spacings of $h_y\approx 8$mm. Too small $h_y$ introduce unnecessary uncertainty to the estimates.

Recall that the X-probes are symmetrically traversed in the y-direction with respect to a fixed $\mathbf{X}$, thus enabling the measurement of the statistical correlations as a function of $r_2$, and that the dependence on $r_1$ is recovered using Taylor's hypothesis. 
On the other hand, the traverse mechanism does not allow displacements in the z-direction and the measurements are restricted to the  vertical xy-plane at $z=0$ and thus $X_3=0$ and $r_3=0$ (see figure \ref{fig:measloc} where the measurement plane is sketched). 

The downstream range of the measurements corresponds to $8-21x_{\mathrm{peak}}$ for RG60 and $1.5-3.7x_{\mathrm{peak}}$ for RG115, a stark difference in the streamwise range relative to $x_{\mathrm{peak}}$ owing to the geometrical differences between the grids (see ch. \ref{chp:2} and ch. \ref{chp:3}). 
In effect, the measurement range for the RG115 corresponds to a nonclassical energy dissipation region whereas for the RG60 it corresponds to a classical one (see ch. \ref{chp:4}), thus allowing their direct comparison. 
Recall that the $Re_{\lambda}$ range of the turbulence generated by the two grids, as well as the straddled Kolmogorov microscales $\eta$, are comparable at the measurement locations which is  beneficial since the same apparatus can be used for both experiments without penalising resolution ($77 \leq Re_{\lambda}\leq 91$ versus $88 \leq Re_{\lambda}\leq 114$ and $0.19\mathrm{mm} \leq \eta \leq 0.32\mathrm{mm}$ versus $0.16\mathrm{mm} \leq \eta \leq 0.28\mathrm{mm}$ for RG60- and  RG115-generated turbulence, respectively)\footnote{Even though the mesh sizes, $M$, of the two grids differ by a factor of about 2, the straddled $Re_{\lambda}$ is comparable because the blockage ratio, $\sigma$, of RG60 is considerably larger than $\sigma$ for RG115, see table \ref{table:grids}. On the other hand, the numerical value of  $\eta$ at $x_{\mathrm{peak}}$ is much smaller for the RG60 both due to the smaller $M$ and the higher $\sigma$. 
However, a heuristic explanation for $\eta$ to be comparable at the measurement locations ($1.2\mathrm{m} \lessapprox x\lessapprox 3$m) is that $\eta$ is a monotonically increasing function of the eddy turnover time, $\ell/u'$, during decay. Since $\ell/u'$ is proportional to $M$ and inversely proportional to $\sigma$ for a fixed $U_{\infty}$, the number of eddy turnovers is larger for RG60- than for RG115-generated turbulence at the same downstream locations, thus the additional growth of $\eta$ compensates for its smaller value at $x_{\mathrm{peak}}$.}. 

%--------------
\subsection[Estimation of the terms in the inhomogeneous  K\'{a}rm\'{a}n-Howarth-Monin equation]{Estimation of the terms in the inhomogeneous  K\'{a}rm\'{a}n-Howarth-Monin equation}\label{sec:KHMcomputed}

It is now described how the terms appearing in the inhomogeneous  K\'{a}rm\'{a}n-Howarth-Monin equation \eqref{eq:KHM}  are estimated from the present two-component, two-dimensional data using the statistical  characteristics of the flow and some additional assumptions. 

From the spatially-varying two-component turbulent signals, acquired simultaneously at the $23$ transverse separations, the second- and third-order structure functions ($\overline{(\delta u)^2}$, $\overline{(\delta v)^2}$, $\overline{(\delta u)^3}$, $\overline{(\delta v)^3}$, $\overline{\delta u(\delta v)^2}$, $\overline{\delta v(\delta u)^2}$) and the mixed structure functions ($\overline{(v + v')\delta u}$, $\overline{(u + u') (\delta u)^2}$, $\overline{(v + v') (\delta u)^2}$, $\overline{(u + u') (\delta v)^2}$, $\overline{(v + v') (\delta v)^2}$) are computed for all $(r_1,\,r_2)$. 
(Note that $r_2$ are just the $23$ transverse separations ($1.2\mathrm{mm}\leq \Delta y \leq 70$mm) and $r_1=n_i\,f_s/U_{\infty}$ where $n_i$ are the 23 integer multiples that yield $r_1\approx r_2$ and $f_s/U_{\infty}$ is the spatial sampling frequency by virtue of Taylor's hypothesis.)

\begin{figure}[t!]
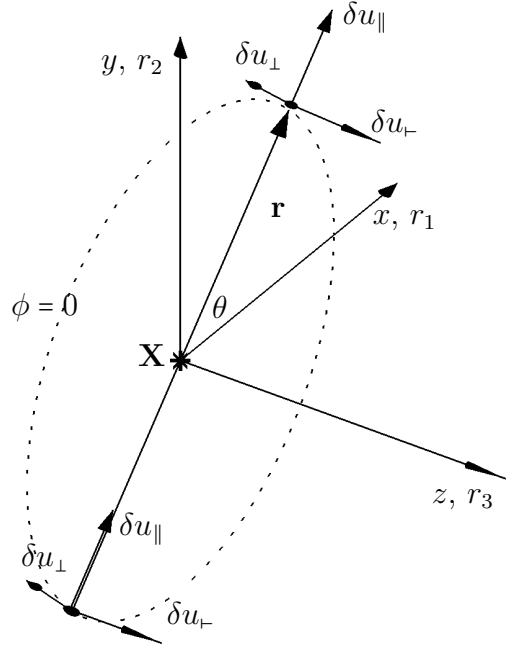

\centering
\begin{lpic}{Thesis_SketchAssumptions(65mm)}
   \lbl{39,44;$x$, $r_1$}
   \lbl{11,60;$y$, $r_2$}
   \lbl{45,15;$z$, $r_3$}
   \lbl{26,45;$\mathbf{r}$}
   \lbl{13,30;$\mathbf{X}$}
   \lbl{20,35;$\theta$}
   \lbl{2,35;$\phi=0$}
   \lbl{35,65;$\delta u_{\parallel}$}
   \lbl{24,61;$\delta u_{\perp}$}
   \lbl{38,54;$\delta u_{\vdash}$}
   \lbl{12,12;$\delta u_{\parallel}$}
   \lbl{2,9;$\delta u_{\perp}$}
   \lbl{17,3;  $\delta u_{\vdash}$}
\end{lpic}
\caption[Sketch of the three velocity-difference components, $\delta u_{\parallel}$, $\delta u_{\perp}$ and $\delta u_{\vdash}$]{Sketch of the three velocity-difference components ($\delta u_{\parallel}$, $\delta u_{\perp}$, $\delta u_{\vdash}$). $\delta u_{\parallel}$ and $\delta u_{\perp}$ are the velocity-difference components lying on the measurement plane ($r_3=0$) which are, respectively, parallel and perpendicular to the separation vector $\mathbf{r}$. $\delta u_{\vdash}$ is the velocity-difference component orthogonal to the other two ($\delta u_{\vdash}$). In the spherical coordinate system used here, $\theta$ is the angle between $\mathbf{r}$ and the $r_1$--axis and $\phi=0$ corresponds to the $r_1$--$r_2$ plane.}
\label{fig:sketch}
\end{figure}

The structure functions are then bi-linearly interpolated onto a spherical coordinate system $(r,\,\theta,\,\phi=0)$ such that $(r,\,0,\,0)$ is aligned with $r_1$ and $(r,\,\pi/2,\,0)$ with $r_2$ (see figure \ref{fig:sketch}). 
The grid points in the new coordinate system are located at the interceptions between the 23 circumferences of radius $r$ and $19$ equally spaced radial lines between the polar angles $\theta = [0\,\,\,\,\pi/2]$.
After the interpolation the data is smoothed with a weighted average between each data point at $(r,\,\theta)$ and its neighbours $(r \pm \Delta r,\,\theta \pm \Delta \theta)$ (the total weight of the neighbouring points amounts to $37.5\%$). 

The greatest limitation of the present measurements is lacking the data for the third velocity component, $w$.
\cite{Moisy2011} faced the same limitation in their PIV data which they negotiated by considering the two-component surrogates  of the structure functions, which may be sufficient to make qualitative inferences.
However, the aim here is to obtain quantitative estimates of the terms in \eqref{eq:KHM}.
This is achieved by assuming inter-component axisymmetry of the velocity-difference statistics relative to the $\mathbf{r}$-axis. 
I.e., it is assumed  that the statistics of the two velocity-difference components perpendicular to $\mathbf{r}=(r_1,\,r_2,\,0)$ ($\delta u_{\perp}$ and $\delta u_{\vdash}$, see figure \ref{fig:sketch}) are approximately equal.
For the second-order structure function this assumption leads to $\overline{(\delta q)^2}(\mathbf{r})=\overline{(\delta u_{\parallel})^2}+2\overline{(\delta u_{\perp})^2}$. 
Similarly, for the third-order structure functions,  $\overline{\delta u_i(\delta q)^2}(\mathbf{r}) = \overline{\delta u_i(\delta u_{\parallel})^2}+2\overline{\delta u_i(\delta u_{\perp})^2}$ and  $\overline{(u_i + u'_i)(\delta q)^2}(\mathbf{r}) = \overline{(u_i + u'_i)(\delta u_{\parallel})^2}+2\overline{(u_i + u'_i)(\delta u_{\perp})^2}$.  
Note that this assumption is weaker than complete isotropy as it allows for dependence on the orientation $\mathbf{r}/r$.
Nevertheless, there is no presently available data to substantiate this assumption  and therefore its validity requires further investigation. 
Even so, it has been verified against all the present data that the added component (i.e. the factor 2 in the equalities two sentences above) does not change the qualitative behaviour of the structure functions, only their magnitude.

% which the author found to marginally improve the accuracy of the numerical derivatives.
%This was checked by comparing the derivatives with respect to $r_1$ with all data points (i.e. $r_1$ equal to the convection velocity over the sampling frequency leading to equally spaced data points with higher resolution) with the derivatives taking the non-equally spaced and coarser points matching the vertical separations $r_2$. 

Using the processed data, each of the terms in \eqref{eq:KHM}, except the pressure transport, is estimated at the measurement plane  as follows. 
Note that the numerical derivatives, both first and second order are computed using a three-point, non-equally spaced central differences scheme \cite[]{SB09}. For equally spaced derivatives this algorithm returns the usual standard central differences scheme. 

\begin{itemize}
\item  $\mathcal{A}_{t}^* = 0$ since grid-generated turbulence is stationary in the Eulerian frame.

\item  $4\mathcal{A}^* \approx\! (U + U')/2\,\partial /\partial  X_1\, \overline{\delta q^2}$ since the mean flow is approximately parallel, $V \approx W \approx 0$ and consequently,  the advection in the y- and z-directions is negligible.
The streamwise derivatives, $\partial /\partial  X_1\,\overline{\delta q^2}$ are actually computed as  $\overline{\delta q^2}/X_1\,\,\partial /\partial \log X_1\, \left(\log\overline{\delta q^2}\right)$ using the various datasets at different $X_1$ ($X_1=1250$, $1700$, $2150$, $2600$, $3050$mm). 
Even though the various $X_1$ are coarsely spaced, it has been verified against the present data that the decay of $\overline{\delta q^2}$ can be reasonably approximated with a power-law  for all $r$ (even when the virtual origin coincides with the location of the grid) and therefore $\partial /\partial \log X_1\, \left(\log\overline{\delta q^2}\right)$ is expected to be a slowly varying function of $X_1$. 
Also, the longitudinal gradients of the mean velocity are small and therefore it is made use of $(U+U')/2 \approx \left(U(X_1,\,X_2 + r_2 /2,\,0)+U(X_1,\,X_2 - r_2 /2,\,0)\right)/2$ to calculate $(U+U')/2 = \left(U(X_1+r_1/2,\,X_2 + r_2 /2,\,0)+U(X_1-r_1/2,\,X_2 - r_2 /2,\,0^{})\right)/2$. 

\item $4\Pi^*  \approx 1/r^2\,\partial/\partial r \left(r^2 \overline{\delta u_{\parallel} \delta q^2} \right)+ 1/(r\sin\theta)\,\partial /\partial \theta\left( \overline{\delta u_{\perp} \delta q^2}\right)$, i.e. the divergence is computed in the spherical coordinate system and the azimuthal component is assumed to be negligible owing to the axisymmetry of the turbulence statistics with respect to the centreline (see \S \ref{sec:homo} and the discussion at the end of this subsection). Future work will be required to assess this assumption.

\item $\Pi^*_U\approx r_1\partial U /\partial x  \,\, \partial /\partial r_1\left(\overline{\delta q^2}\right)$ where the approximation that the flow is parallel has been used, $V \approx W \approx 0$. 
Note that $\delta U=U(X_1+r_1/2,\,r_2/2,\,0)-U(X_1-r_1/2,\,-r_2/2,\,0)$ and owing to the symmetry of the turbulence statistics $U(X_1,\,r_2/2)\approx U(X_1,\,-r_2/2)$ ($X_2=0$ since the centroid is located at the centreline). Therefore $\delta U$ is only non-zero for $r_1\neq0$. 
However, since the dependence of the turbulence statistics  on $r_1$ is recovered from Taylor's hypothesis, the gradients of mean quantities are zero and therefore one cannot measure $\delta U$ from a single dataset with a fixed $X_1$.
To negotiate this problem, a first order Taylor's expansion is used to write $\delta U\approx r_1\partial U /\partial x$ and the various data at different $X_1$ are now used to compute $\partial U /\partial x$ with a central differences scheme.
Even though  the various $X_1$ are coarsely spaced,  $U$ is a slowly varying function of $x$ (or $X_1$) as can be inferred from figures \ref{fig:Um}a and \ref{fig:UmvsX} by noticing that $U$, at any spanwise location $y$, varies less than about $5\%$ of $U_{\infty}$ throughout the streamwise extent of the measurements. 
It is shown in \S \ref{sec:neglect} that this term is negligibly small.

\item $4\mathcal{P}^*  \approx 2\overline{(\delta u)^2} \,\partial U/\partial x + 4\overline{(v + v')\delta u}\, \partial U/\partial y$ since $V \approx W \approx 0$ and $\partial U/\partial z = \partial U'/\partial z'\approx0$ due to the expected symmetry of the mean flow relative to the plane $z=0$. Also note that the symmetry of the mean flow relative to the centreline  (leading to $\partial U/\partial x \approx \partial U'/\partial x'$, $\partial U/\partial y \approx - \partial U'/\partial y'$) has been used to simplify $\partial\, \delta U/\partial r_k=1/2\left(\partial U/\partial x_k + \partial U'/\partial x'_k\right)$ as $\partial U/\partial x$ and $\partial\, \delta U/\partial X_k=\partial U/\partial x_k - \partial U'/\partial x'_k$ as $2\partial U/\partial y$.
 The transverse gradient $\partial U/\partial y$ is taken from a $12^{\mathrm{th}}$-order polynomial fit to the mean velocity data at each $X_1$ and the longitudinal gradient $\partial U/\partial x$ is computed as described in the previous item.

\item $4\mathcal{T}^* \approx -\partial/\partial X_1 \left(\overline{(u + u') \delta q^2}/2 \right)-\partial/\partial X_2 \left(\overline{(v + v') \delta q^2} \right) - 4\mathcal{T}_p^* $. 
The transverse derivative $\partial/\partial X_2(\overline{(v + v') \delta q^2}/2)$ ($\approx \partial/\partial X_3 \overline{(w + w')\delta q^2}/2$ owing to the symmetry of the turbulence statistics to $90^{\circ}$ rotations owing to grid's  geometry) is only computed where the additional off-centreline measurements  are acquired. 
The transverse derivative is simply taken as the difference between centreline and off-centreline data divided by their distance. 
The derivative with respect to $X_1$ is computed using the various datasets with different $X_1$. However, this can only be considered  as a rough approximation since the various $X_1$ are coarsely spaced. Nevertheless, the longitudinal turbulent transport is typically a small fraction of the lateral transport as was checked against the present two-point data as well as against the single-point transport data presented in \S \ref{sec:homo}. 
The pressure transport, $\mathcal{T}_p^*$, data cannot be directly estimated with the present apparatus.
However, there is no \emph{a priori} reason to consider it negligible and therefore it is retained in \eqref{eq:KHM} as an unknown. 
Nevertheless, the contribution from $\mathcal{T}_p^*$ can be inferred indirectly from the deviations of the measured terms' balance via \eqref{eq:KHM}. 

\item $4\mathcal{D}^*_{\nu} \approx 2\nu/r^2\,\partial/\partial r \left(r^2\,\partial/\partial r \left(\overline{\delta q^2}\right)\right)$, i.e. only the radial component of the laplacian is computed. Note that the integral of the polar, $ \mathcal{D}^*_{\nu,\theta} $, and azimuthal, $\mathcal{D}^*_{\nu,\phi}$, components of the laplacian over a spherical shell are identically zero,  $\oiint_{|\mathbf{r}|=r} \mathcal{D}^*_{\nu,\theta} \,dS=\oiint_{|\mathbf{r}|=r} \mathcal{D}^*_{\nu,\phi} \,dS=0$ and therefore these terms represent  the viscous diffusion across the different orientations $\mathbf{r}/r$. 
As will be seen below, only spherical shell averages (effectively circumferential averages) of this term are discussed and therefore the polar and azimuthal components are not computed.

\item $4\mathcal{D}^*_{X,\nu} \approx \nu/2\,\partial^2/\partial X_1^2\left( \overline{\delta q^2}\right)+\nu/2\,\partial^2/\partial X_2^2\left( \overline{\delta q^2}\right)$, where the component in the $z$-direction is taken to be zero since the turbulence statistics are symmetric with respect to the $z=0$ plane. The streamwise $2^{\mathrm{nd}}$-order derivative is computed from the various datasets at different $X_1$ (similar to what is done to compute the first-order longitudinal derivative in $\mathcal{T}^*$) and the spanwise  $2^{\mathrm{nd}}$-order derivative is computed at the two downstream locations, $X_1=1250$mm and $X_1=2150$mm, where the additional off-centreline datasets at $X_2=-6$mm are acquired. Symmetry of the turbulence statistics in the measurement plane with respect to the centreline is used to estimate the derivative $\partial^2/\partial X_2^2\left( \overline{\delta q^2}\right)$ with a $2^{\mathrm{nd}}$-order central differences scheme as $2\left( \overline{\delta q^2}(X_1,\,h,\,0;\mathbf{r})-\overline{\delta q^2}(X_1,\,0,\,0;\mathbf{r})\right)/h^2$ ($h=-6$mm).
This term  is shown to be negligibly small in \S \ref{sec:neglect}.

\item $4\varepsilon^*\approx 4\varepsilon^{\mathrm{iso,3}}$, i.e. the centreline energy dissipation estimate $\varepsilon^{\mathrm{iso,3}}$ is used as a surrogate for the average of the actual dissipation at $\mathbf{x}$ and $\mathbf{x'}$ (see \S \ref{sec:Eps} where the different dissipation estimates are discussed). Note that with the present data it is only possible to estimate $\varepsilon^{\mathrm{iso,3}}$ along the centreline. Nevertheless, the spanwise profiles of the (less suitable) surrogate $\varepsilon^{\mathrm{iso,1}}$ indicate that the departures from the centreline value are within $10\%$, see figure \ref{fig:ReyStress}d. 

\end{itemize}
%
%The simplified equation reads,
%\begin{equation}
% \begin{aligned}
%\overbrace{U(X_1,X_2 \pm r_2 /2,0)\frac{\partial \, \overline{\delta q^2}}{\partial x}}^{4\mathcal{A}^*} +  
%\overbrace{2\overline{(\delta u)^2} \frac{\partial U}{\partial x} + 4\overline{(v + v')\delta u}\, \frac{\partial U}{\partial y}}^{4\mathcal{P}^*}  &+  
%\overbrace{\frac{\partial \, \overline{\delta u_k \delta q^2}}{\partial r_k}}^{4\Pi^*} + \\
%\underbrace{\frac{\partial}{\partial X_k} \left( \overline{\frac{(u_k + u'_k) \delta q^2}{2}}  \right)}_{4\mathcal{T}^*} + 
%\underbrace{\frac{2}{\rho}\overline{\delta u_i \frac{\partial \, \delta p }{\partial X_i}}}_{4\mathcal{T}_p^*} &=  
%\underbrace{2 \nu \frac{\partial^2 \overline{\delta q^2} }{\partial r^2_k}}_{4\mathcal{D}_{\nu}^*}  -   4\,\varepsilon^*.
%\end{aligned}
%\label{eq:KHMSimplified}
%\end{equation}
%

Of particular importance to the subsequent discussions are the circumferential averages of the terms in \eqref{eq:KHM} in order to remove the dependence on orientation ($\mathbf{r}/r$) of the turbulence statistics.  The circumferential averages are expected to be good approximations to the averages over spherical shells considering the statistical axisymmetry of the turbulence with respect to the centreline\footnote{Recall that for most of the present data $\mathbf{X}$ (and therefore $r_1$) lies along the centreline. However, for the two datasets acquired off-centreline at $X_2=y=-6$mm one may expect the validity of this assumption to be more doubtful.} (see also \S \ref{sec:homo} and \S\ref{sec:Lu}).
The circumferential averages\footnote{Note that only one quarter of the domain is used due to the reflection symmetry of the structure functions around the $r_1$-- and $r_2$--axis, the former due to stationarity and the latter by construction.} are obtained by integration with respect to the polar angle $\theta$ as $\int_{0}^{\pi/2}\,A^*(r,\,\theta,\,0)\,\mathrm{sin}(\theta)\, d\theta$, where the integrand $A^*$  is any one of the measured terms in \eqref{eq:KHM}.
The wind tunnel measurements of \cite{Nagata2012} for the decay region in the lee of FSGs and the numerical data of \cite{Sylvain2011} for both FSGs and a RG gives substantial support to this assumption and therefore the circumferential averages are  interpreted as spherical shell averages throughout this thesis.
Recall that the spherical shell averaged terms are denoted by removing the superscript asterisk.

%--------
\subsection{Second- and third-order structure functions} \label{sec:aniso}
Turning to the data, the anisotropy of the structure functions $\overline{\delta q^2}(r,\,\theta,\,\phi=0)$ and $\overline{\delta u_i\delta q^2}(r,\,\theta,\,\phi=0)$  is qualitatively investigated from their dependence on $\theta$. (For notational simplicity and due to the assumed axisymmetry, $\phi$ is not explicitly used as an argument henceforth.)
Note that in the present context anisotropy refers to the dependence of the terms in \eqref{eq:KHM}  on the orientation $\mathbf{r}/r$  \cite[see also][]{Moisy2011,Danaila2012} and not to the kinematic relation between the components of the structure functions parallel and perpendicular to $\mathbf{r}$ (e.g. $\overline{(\delta u_{\parallel})^2}$ versus $\overline{(\delta u_{\perp})^2}$ and $\overline{\delta u_{\parallel} (\delta u_{\parallel})^2}$ versus $\overline{\delta u_{\parallel}(\delta u_{\perp})^2}$), except when clearly indicated. 
The latter anisotropy considerations are complementary to the first but pertain to, for example, the distribution of kinetic energy between the three orthogonal components and the inter-component energy transfer via pressure fluctuations \cite[see e.g. \citealt{SThesis}, ][and references therein]{SJ98}.

\begin{figure}
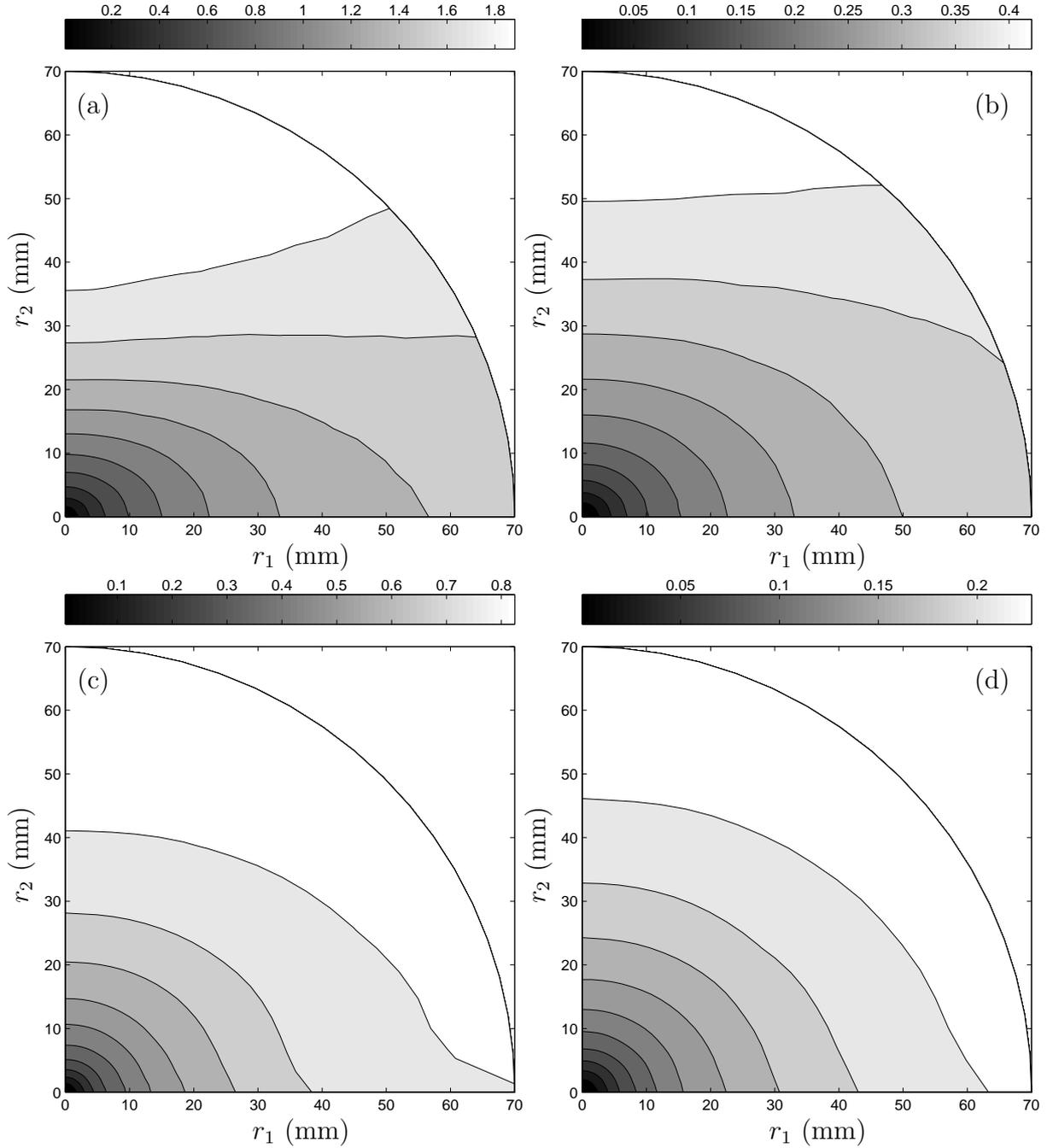

\centering
\begin{minipage}[c]{0.5\linewidth}
   \centering
   \begin{lpic}{RG115-dq2=1250(80mm)}
   \lbl{20,110;(a)}
   \lbl[W]{3,70,90;$r_2$ (mm)}
   \lbl[W]{70,3;$r_1$ (mm)}
  \end{lpic}
\end{minipage}%
\begin{minipage}[c]{0.5\linewidth}
   \centering 
   \begin{lpic}{RG115-dq2=3050(80mm)}
   \lbl{110,110;(b)}
   \lbl[W]{3,70,90;$r_2$ (mm)}
   \lbl[W]{70,3;$r_1$ (mm)}
  \end{lpic}
\end{minipage}
\begin{minipage}[c]{0.5\linewidth}
   \centering
   \begin{lpic}{RG60-dq2=1250(80mm)}
   \lbl{20,110;(c)}
   \lbl[W]{3,70,90;$r_2$ (mm)}
   \lbl[W]{70,3;$r_1$ (mm)}
   \end{lpic}
\end{minipage}%
\begin{minipage}[c]{0.5\linewidth}
   \centering 
   \begin{lpic}{RG60-dq2=3050(80mm)}
   \lbl{110,110;(d)}
   \lbl[W]{3,70,90;$r_2$ (mm)}
   \lbl[W]{70,3;$r_1$ (mm)}
   \end{lpic}
\end{minipage}
\caption[Second order structure functions, $\overline{\delta q^2}(r_1,r_2)$]{Contours of the second-order structure functions, $\overline{\delta q^2}(r_1,r_2)$ (m$^2$s$^{-2}$), at (a,c) $X_1=1250$mm and (b,d) $X_1=3050$mm for (top) RG115 and (bottom) RG60 data.  $X_1=1250$mm and $X_1=3050$mm correspond to $X_1/x_{\mathrm{peak}} = 1.5$ and $X_1/x_{\mathrm{peak}} = 3.7$ for the RG115 data and to $X_1/x_{\mathrm{peak}} = 8.5$ and $X_1/x_{\mathrm{peak}} = 20.7$ for the RG60 data.}
\label{fig:2ndStrut}
\end{figure}

The second-order structure functions $\overline{\delta q^2}(r,\,\theta)$ are presented in figures \ref{fig:2ndStrut}a-d for the furthermost upstream and downstream measurement locations and for turbulence generated by both RG115 and RG60.
Comparing the upstream data (figures \ref{fig:2ndStrut}a,c) with the downstream data (figures \ref{fig:2ndStrut}b,d) for both grids there seems to be a tendency for the contours to become increasingly circular as the turbulence decays, i.e. for the energy distribution to become increasingly isotropic.
Furthermore, comparing the RG115 with the RG60 data (figures \ref{fig:2ndStrut}a,b and \ref{fig:2ndStrut}c,d, respectively) it can be seen that the RG115 data, which are acquired closer to the grid in terms of $x_{\mathrm{peak}}$ multiples, is less isotropic.
Both these observations corroborate a tendency for the kinetic energy to become uniformly distributed over spherical shells for larger $x/x_{\mathrm{peak}}$. 
Nevertheless, for all cases the small scales seem to be more isotropic than the large ones, even though, at least at these moderate Reynolds numbers, the dissipation is still anisotropic (see  tables \ref{tab:RG60}  and \ref{tab:RG115}). 

\begin{figure}
\centering
\begin{lpic}{RG115-ModAndDiv_duidq2=1250(155mm)}
   \lbl{20,86;(a)}
   \lbl{168,86;(b)}
   \lbl[W]{3.5,54,90;$r_2$ (mm)}
   \lbl[W]{50,3.5;\hspace{10mm}$-r_1$ (mm)\hspace{10mm}}
   \lbl[W]{140,3.5;\hspace{10mm}$r_1$ (mm)\hspace{10mm}}
\end{lpic}
\begin{lpic}{RG115-ModAndDiv_duidq2=3050(155mm)}
   \lbl{20,86;(c)}
   \lbl{168,86;(d)}
   \lbl[W]{3.5,54,90;$r_2$ (mm)}
   \lbl[W]{50,3;\hspace{10mm}$-r_1$ (mm)\hspace{10mm}}
   \lbl[W]{140,3.5;\hspace{10mm}$r_1$ (mm)\hspace{10mm}}
\end{lpic}
\caption[Third-order structure functions for RG115 data]{(a,c) Third-order structure function vectors, $\overline{\delta u_i \delta q^2}$ and contours of their magnitude, $|\overline{\delta u_i \delta q^2}| \,(\times 10^{-3}\,\, \mathrm{m^3s^{-3}})$. (b,d) Contours of the radial component of the divergence of $\overline{\delta u_i \delta q^2}$, $\Pi^*_r \,(\mathrm{m^2s^{-3}})$. 
(top) $X_1=1250$mm and (bottom) $X_1=3050$mm. $X_1=1250$mm and $X_1=3050$mm correspond to $X_1/x_{\mathrm{peak}} = 1.5$ and $X_1/x_{\mathrm{peak}} = 3.7$.
Data are acquired in the lee of RG115.}
\label{fig:3rdStrutRG115}
\end{figure}

\begin{figure}
\centering
\begin{lpic}{RG60-ModAndDiv_duidq2=1250(155mm)}
   \lbl{20,86;(a)}
   \lbl{168,86;(b)}
   \lbl[W]{3.5,54,90;$r_2$ (mm)}
   \lbl[W]{50,3.5;\hspace{10mm}$-r_1$ (mm)\hspace{10mm}}
   \lbl[W]{140,3.5;\hspace{10mm}$r_1$ (mm)\hspace{10mm}}
\end{lpic}
\begin{lpic}{RG60-ModAndDiv_duidq2=3050(155mm)}
   \lbl{20,86;(c)}
   \lbl{168,86;(d)}
   \lbl[W]{3.5,54,90;$r_2$ (mm)}
   \lbl[W]{50,3;\hspace{10mm}$-r_1$ (mm)\hspace{10mm}}
   \lbl[W]{140,3.5;\hspace{10mm}$r_1$ (mm)\hspace{10mm}}
\end{lpic}
\caption[Third-order structure functions for RG60 data]{(a,c) Third-order structure function vectors, $\overline{\delta u_i \delta q^2}$ and contours of their magnitude, $|\overline{\delta u_i \delta q^2}| \,(\times 10^{-3}\,\, \mathrm{m^3s^{-3}})$. (b,d) Contours of the radial component of the divergence of $\overline{\delta u_i \delta q^2}$, $\Pi^{*}_r \,(\mathrm{m^2s^{-3}})$. 
(top) $X_1=1250$mm and (bottom) $X_1=3050$mm.
Data are acquired in the lee of RG60. $X_1=1250$mm and $X_1=3050$mm correspond to $X_1/x_{\mathrm{peak}} = 8.5$ and $X_1/x_{\mathrm{peak}} = 20.7$.}
\label{fig:3rdStrutRG60}
\end{figure}

Turning to the third-order structure function vectors $\overline{\delta u_i\delta q^2}(r,\,\theta)$, a similar tendency to isotropy is observed (figures \ref{fig:3rdStrutRG115}a,c and \ref{fig:3rdStrutRG60}a,c). The third-order structure function vectors, which for the RG115 data at $x = 1.5x_{\mathrm{peak}}$ are nearly aligned with the tangential direction (figure \ref{fig:3rdStrutRG115}a), progressively align with the radial direction and for the RG60 data at $x = 21x_{\mathrm{peak}}$ (figure \ref{fig:3rdStrutRG60}c) they are indeed nearly so. 
Note that the divergence of $\overline{\delta u_i\delta q^2}$ (i.e. $\Pi^*$) has a radial and a polar component (the azimuthal component is taken to be zero due to the assumed axisymmetry). 
As discussed in \S \ref{sec:KHM}, the radial component $\Pi_r^*$ relates to the interscale energy transfer, whereas the polar component $\Pi_{\theta}^*$ accounts for the redistribution of energy within a spherical shell.
The above mentioned tendency to isotropy as the flow decays is very likely linked to the redistribution of energy via $\Pi_{\theta}^*$.

%%%%%
\subsection{Estimates of  $\Pi^*_U$ and $\mathcal{D}^*_{X,\nu}$} \label{sec:neglect}
\begin{figure}
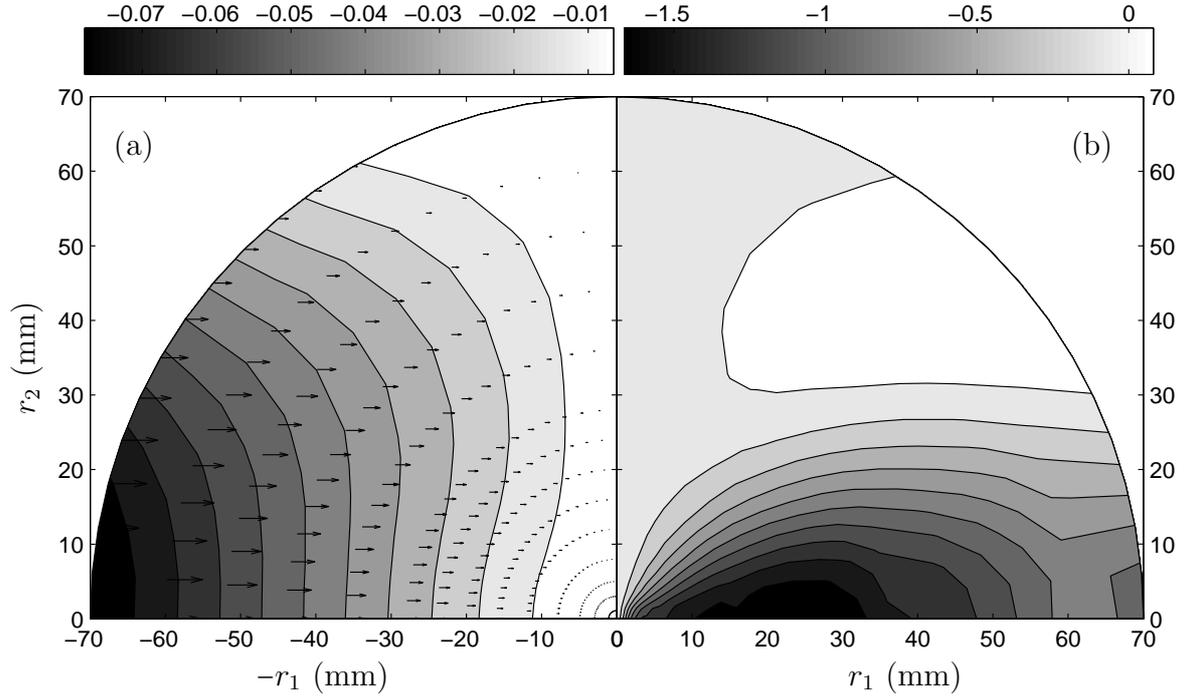

\centering
\begin{lpic}{Thesis_CountoursMeanVelocityTransf=1250(155mm)}
   \lbl{20,86;(a)}
   \lbl{168,86;(b)}
   \lbl[W]{3.5,54,90;$r_2$ (mm)}
   \lbl[W]{50,3.5;\hspace{10mm}$-r_1$ (mm)\hspace{10mm}}
   \lbl[W]{140,3.5;\hspace{10mm}$r_1$ (mm)\hspace{10mm}}
\end{lpic}
\caption[Energy transfer via mean velocity gradients for RG115 data]{Energy transfer via mean velocity gradients, $\Pi^*_{U}$; (a) vector field of the flux $\delta U \overline{\delta q^2}$ ($\mathrm{m^3s^{-3}}$) and contours of its magnitude, $|\delta U \overline{\delta q^2}| \, (\mathrm{m^3s^{-3}})$ and (b) contours of $\Pi^*_{U}/\varepsilon$ \,($\%$).
Data are acquired in the lee of RG115 at $X_1=1250$mm ($X_1/x_{\mathrm{peak}} = 1.5$).}
\label{fig:MeanUTransfer}
\end{figure}

\begin{figure}
\centering
\begin{lpic}{Thesis_NegligibleTerms(100mm)}
   \lbl[W]{-4,80,90;$-\Pi_{U}/\varepsilon$, $\mathcal{D}_{X,\nu}/\varepsilon$ ($\%$)}
   \lbl[W]{90,2;$r$ (mm)}
\end{lpic}
\caption[Negligible terms in \eqref{eq:KHM} averaged over spherical shells]{Negligible terms in \eqref{eq:KHM} averaged over spherical shells and normalised by the dissipation. (\protect\raisebox{-0.5ex}{\SmallCircle} $\!|\!\!$ \protect\raisebox{-0.5ex}{\SmallSquare})  $-\Pi_{U}/\varepsilon$ ($\%$) at $X_1=1250$mm and $X_1=2150$mm, respectively and (\protect\raisebox{-0.5ex}{\FilledSmallCircle} $\!|\!\!$ \protect\raisebox{-0.5ex}{\FilledSmallSquare}) $\mathcal{D}_{X,\nu}/\varepsilon$ ($\%$) at $X_1=1250$mm and $X_1=2150$mm, respectively. $X_1=1250$mm and $X_1=2150$mm correspond to $X_1/x_{\mathrm{peak}} = 1.5$ and $X_1/x_{\mathrm{peak}} = 2.6$, respectively.
Data are acquired in the lee of RG115.}
\label{fig:NegligibleTerms}
\end{figure}

In this section evidence is presented which suggests that the energy transfer via mean velocity gradients, $\Pi^*_{U}$, and the transport via viscous diffusion, $\mathcal{D}^*_{X,\nu}$, are negligible compared to the other terms in \eqref{eq:KHM}. 
These terms are computed from the acquired data as described in \S \ref{sec:KHMcomputed}.
In figure \ref{fig:MeanUTransfer}a it can be seen that the vectors representing the flux $\delta U \overline{\delta q^2}$ are horizontal, which stems from the assumption that the mean flow is parallel and therefore $\delta V  \overline{\delta q^2}\approx \delta W  \overline{\delta q^2} \approx 0 $. 
Taking the divergence of the flux one gets $\Pi_{U}^*$ for a given orientation $\mathbf{r}/r$. 
Notice that $\Pi_{U}^*$  is largest for small $\theta$ (i.e. closer to the $r_1$--axis) but is never larger than $\approx 1.5\%\, \varepsilon$, even at the furthermost upstream location $X_1/x_{\mathrm{peak}} = 1.5$ (\emph{cf.} figure \ref{fig:MeanUTransfer}b). 
By averaging this term over spherical shells it can be seen that  $\Pi_{U}$ represents less than $0.4\%$ of the dissipation and that further downstream, $X_1/x_{\mathrm{peak}} = 2.6$, it decreases to less than $0.05\%$ (see figure \ref{fig:NegligibleTerms}).
It is therefore confirmed that the inter-scale energy transfer due to the residual mean shear is negligible.
Turning now to the transport via viscous diffusion averaged over spherical shells, $\mathcal{D}_{X,\nu}$, it can be seen in figure \ref{fig:NegligibleTerms} that it represents less than $0.1\%$ for both downstream locations, $X_1/x_{\mathrm{peak}} = 1.5$ and  $X_1/x_{\mathrm{peak}} = 2.6$. This is not surprising given the moderately high Reynolds number of the turbulent flow ($Re_{\lambda}=\mathcal{O}(100)$).
These two observations support the neglect of these terms in the subsequent analyses.

\subsection{The role of turbulence production and transport} \label{sec:inhomo}
The effect of transport and production in the single-point kinetic energy balance was investigated in \S \ref{sec:homo} where it was found that, for the assessed region of the RG115-generated turbulence, both contributions are non-negligible by comparison with the energy dissipation. 
This region of the RG115-generated turbulence was also compared with an equivalent region of turbulence generated by FSGs and considerable differences were found in the downstream evolution (and transverse profiles) of  transport and production relative to the dissipation. 
Nevertheless, the two different turbulent flows were found to have a nonclassical dissipation behaviour and, consequently, the differences in the production and transport reinforced the conjecture that the nonclassical  behaviour is exhibited despite the inhomogeneity of the turbulent flow and not due to it. 
Indeed, based on the conceptual picture of turbulence (figure \ref{fig:Corrsin}), the turbulent transport and production are expected to  be large-scale phenomena that play no direct role in the scale-by-scale energy transfer mechanisms, even at these Reynolds numbers ($Re_{\lambda}=\mathcal{O}(100)$). 
Here, data are presented  which allow a precise quantification of the effect of production and transport on the scale-by-scale energy budget \eqref{eq:KHM}. 

One may average \eqref{eq:KHM} over spherical shells to eliminate the dependence of each term on the  orientation $\mathbf{r}/r$ yielding the average contribution of each scale to the balance. 
Retaining only the non-negligible terms, the spherical averaged scale-by-scale energy balance reads, 
\begin{equation}
\mathcal{A} +  \Pi -  \mathcal{P}  - \mathcal{T} - \mathcal{T}_p =   \mathcal{D}_{\nu}  -  \varepsilon,
\label{eq:KHMSimplified}
\end{equation}
where the $\mathcal{T}$ represents the measured component of turbulent transport and  $\mathcal{T}_p$ represents the unknown contribution from the pressure transport.

\begin{figure}
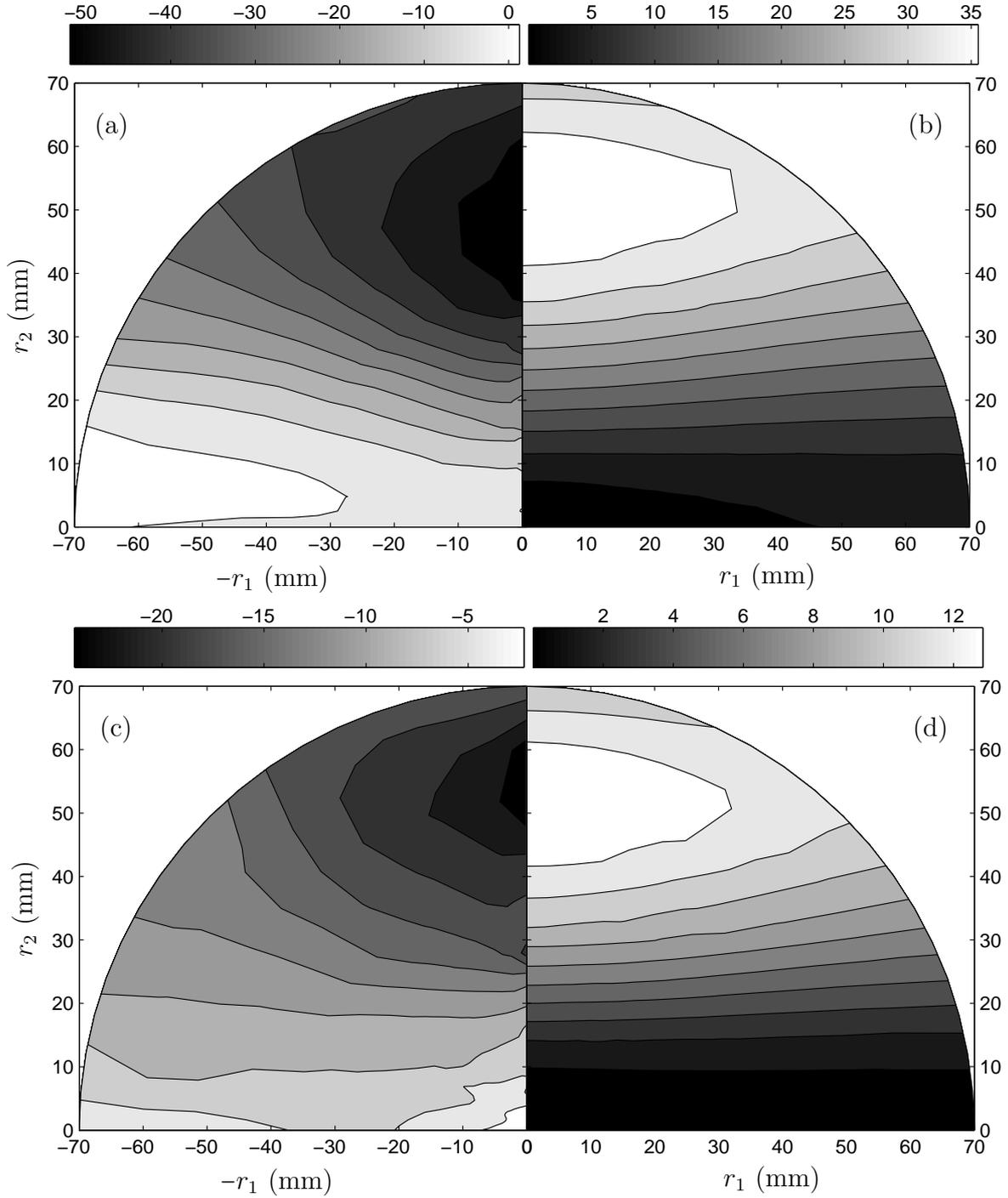

\centering
\begin{lpic}{TranspAndProd=1250(155mm)}
   \lbl{20,86;(a)}
   \lbl{168,86;(b)}
   \lbl[W]{3.5,54,90;\hspace{10mm}$r_2$ (mm)\hspace{10mm}}
   \lbl[W]{50,3;\hspace{10mm}$-r_1$ (mm)\hspace{10mm}}
   \lbl[W]{140,3.5;\hspace{10mm}$r_1$ (mm)\hspace{10mm}}
\end{lpic}
\begin{lpic}{TranspAndProd=2150(155mm)}
   \lbl{20,86;(c)}
   \lbl{168,86;(d)}
   \lbl[W]{3.5,54,90;\hspace{10mm}$r_2$ (mm)\hspace{10mm}}
   \lbl[W]{50,3;\hspace{10mm}$-r_1$ (mm)\hspace{10mm}}
   \lbl[W]{140,3.5;\hspace{10mm}$r_1$ (mm)\hspace{10mm}}
\end{lpic}
\caption[Normalised turbulence transport and production]{Normalised turbulence (a,c) transport, $\mathcal{T}^*/\varepsilon\,(\%)$ and (b,d) production, $\mathcal{P}^*/\varepsilon\,(\%)$ versus $(r_x,\,r_y)$ at $x/x_{\mathrm{peak}} = 1.5$ (top) and $x/x_{\mathrm{peak}} = 2.6$ (bottom) in the lee of RG115.}
\label{fig:inhomo}
\end{figure}

\begin{figure}
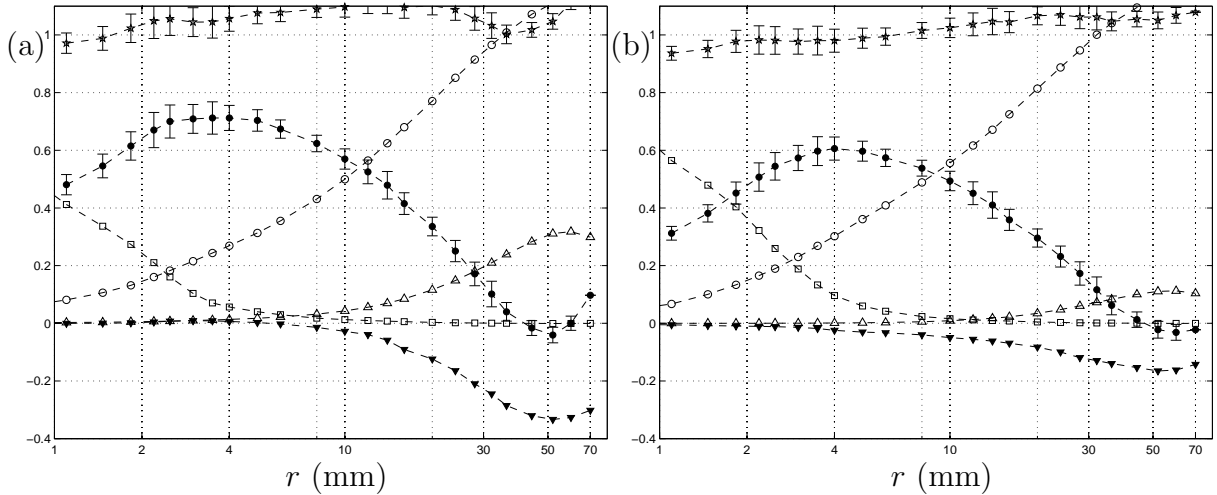

\centering
\begin{minipage}[c]{0.5\linewidth}
   \centering
   \begin{lpic}{RG115-KHMAzAve=1250(77mm)}
   \lbl{0,125;(a)}
   \lbl[W]{90,1;$r$ (mm)}
  \end{lpic}
\end{minipage}%
\begin{minipage}[c]{0.5\linewidth}
   \centering 
   \begin{lpic}{RG115-KHMAzAve=2150(77mm)}
   \lbl{0,125;(b)}
   \lbl[W]{90,1;$r$ (mm)}
   \end{lpic}
\end{minipage}
\caption[$\Pi/\varepsilon$, $\mathcal{A}/\varepsilon$, $\mathcal{D}_{\nu}/\varepsilon$,  $\mathcal{P}/\varepsilon$,  $\mathcal{T}/\varepsilon$ and $(\mathcal{D}_{\nu}  + \mathcal{P} + \mathcal{T} - \Pi - \mathcal{A})/\varepsilon$ for RG115 data]{Spherically averaged (\protect\raisebox{-0.5ex}{\FilledSmallCircle}) $-\Pi/\varepsilon$, (\protect\raisebox{-0.5ex}{\SmallCircle})  $-\mathcal{A}/\varepsilon$,  (\protect\raisebox{-0.5ex}{\SmallSquare}) $\mathcal{D}_{\nu}/\varepsilon$, ($\medtriangleup$) $\mathcal{P}/\varepsilon$, ($\filledmedtriangledown$) $\mathcal{T}/\varepsilon$ and (\small{\ding{73}}\normalsize) $(\mathcal{D}_{\nu}  + \mathcal{P} + \mathcal{T} - \Pi - \mathcal{A})/\varepsilon$ at (a) $x/x_{\mathrm{peak}} = 1.5$ and (a) $x/x_{\mathrm{peak}} = 2.6$ for RG115-generated turbulence. The size of the error bars in the energy transfer data are discussed in \S \ref{sec:DAQ}. Error bars of equal size are added to the datapoints representing $(\mathcal{D}_{\nu}  + \mathcal{P} + \mathcal{T} - \Pi - \mathcal{A})$, however, this underestimates the error margins as it does not take into account the uncertainty associated with the estimates of the other terms.}
\label{fig:KHMwInhomo}
\end{figure}

Turning to the data, the contour maps of the transport and production terms normalised by the dissipation, $\varepsilon$, 
indicate that most of the transport and production occur for $r \gtrapprox L_{11}^{(1)} \approx 30$mm and $\theta \approx \pi/2$ (figures \ref{fig:inhomo}a,c and \ref{fig:inhomo}b,d). 
At smaller values of $r$ both $\mathcal{T}$ and $\mathcal{P}$ are less than about $15\%$ of $\varepsilon$. 
Note that the production for large $r$ is much smaller for $\theta \approx 0$ than for $\theta \approx \pi/2$ because $\partial U/\partial y$ tends to zero at the centreline and the remaining production term, $2\overline{(\delta u)^2} \partial U/\partial x$, is small by comparison.  
Similarly, the transport for large $r$ and $\theta \approx 0$ is also smaller because the lateral transport overwhelms the longitudinal transport.

The spherical averaged contribution of these terms to the balance \eqref{eq:KHMSimplified} are plotted together with the spherical shell averaged advection, energy transfer and viscous diffusion\footnote{Recall that the dissipation estimates are compensated for the resolution of the sensor, see \ref{sec:resolution}.  The finite resolution of the sensor also biases $\mathcal{D}_{\nu}$ since  $\lim_{r\rightarrow 0}\mathcal{D}_{\nu}(r)=\varepsilon$. A rough compensation for this bias is applied by multiplying $\mathcal{D}_{\nu}$ with the ratio between the corrected and the measured $\varepsilon$} in figures \ref{fig:KHMwInhomo}a,b.
The radial distribution of the advection, energy transfer and viscous diffusion are similar to those found in the literature for data at comparable Reynolds numbers \cite[see e.g.][]{Antonia2006}. 
%However, to the author's best knowledge, the radial distribution of the scale-by-scale turbulent transport and production are presented here for the first time. 
From the data it is clear that these turbulent transport and production terms are significant for scales of the order of the integral-length scale but become negligible at scales smaller than $r \approx 10\mathrm{mm}\approx L_{11}^{(1)}/3$  and therefore cannot tamper with the scale-by-scale energy transfer around its maximum ($r \approx 4\mathrm{mm}\approx L_{11}^{(1)}/8$ for the present data).
This provides  quantitative evidence that the influence of the turbulence production and transport on the energy transfer mechanisms is negligible and consequently the nonclassical dissipation behaviour is very unlikely to be related with such effects.

Note that in figures \ref{fig:KHMwInhomo}a,b the balance of the measured terms is also presented. 
By virtue of \eqref{eq:KHMSimplified}, the scale-by-scale advection, energy transfer, production, transport and viscous diffusion should balance the dissipation plus the unknown contribution from scale-by-scale pressure transport, $\mathcal{T}_p$. 
Even though $\mathcal{T}_p$ is not accounted for, it can be seen that there is a reasonable balance between the measured terms, at least within the expected uncertainty of the data. 
Note that the error bars added to the balance ($\mathcal{D}_{\nu}  + \mathcal{P} + \mathcal{T} - \Pi - \mathcal{A}$, see figures \ref{fig:KHMwInhomo}a,b) underestimate the overall uncertainty of the data since they do not take into account  uncertainties associated with the measurements of the advection, transport and production terms and possible departures from the assumptions used to compute the terms in \eqref{eq:KHM}, see \S \ref{sec:KHMcomputed}. 
%Nevertheless, the present data still complies with \eqref{eq:KHMSimplified} within $\pm10\%$ to $\pm15\%$.

%%
\subsection{Advection, energy transfer and dissipation \mbox{scalings}}
It is now investigated how the stark differences in the scaling of the energy dissipation observed in the two measured regions (see ch. \ref{chp:4}) relate to the behaviour of the advection, energy transfer and viscous diffusion during decay (the remaining terms in \eqref{eq:KHMSimplified}  are negligible for small $r$, cf. figure \ref{fig:KHMwInhomo}).

\begin{figure}
\centering 
\begin{lpic}[b(0mm)]{Thesis_RG60KHMnormyEpsnormxNone(130mm)}
\lbl{5,130;(a)}
\lbl{70,82;$-\Pi/\varepsilon$}
\lbl{75,28;$\mathcal{D}_{\nu}/\varepsilon$}
\lbl{155,105;$-\mathcal{A}/\varepsilon$}
\lbl{45,130;$(\mathcal{D}_{\nu} - \Pi - \mathcal{A})/\varepsilon$}
\lbl[W]{4,80,90;\hspace{10mm}($\mathcal{A}$, $\Pi$, $\mathcal{D}_{\nu}$)/$\varepsilon$\hspace{10mm}}
\lbl[W]{90,4;\hspace{10mm}$r$ (mm)\hspace{10mm}}
\end{lpic}
\begin{lpic}[b(0mm)]{Thesis_RG60KHMnormyEpsnormxLamb(130mm)}
\lbl{5,130;(b)}
\lbl{74,82;$-\Pi/\varepsilon$}
\lbl{75,25;$\mathcal{D}_{\nu}/\varepsilon$}
\lbl{140,105;$-\mathcal{A}/\varepsilon$}
\lbl{45,130;$(\mathcal{D}_{\nu} - \Pi - \mathcal{A})/\varepsilon$}
\lbl[W]{4,80,90;\hspace{10mm}($\mathcal{A}$, $\Pi$, $\mathcal{D}_{\nu}$)/$\varepsilon$\hspace{10mm}}
\lbl[W]{90,4;\hspace{10mm}$r/\lambda$ \hspace{10mm}}
\end{lpic}
\caption[$\Pi/\varepsilon$, $\mathcal{A}/\varepsilon$ and $\mathcal{D}_{\nu}/\varepsilon$ throughout the decay of RG60-generated turbulence]{Normalised, spherical shell averaged scale-by-scale energy transfer ($-\Pi/\varepsilon$), advection ($-\mathcal{A}/\varepsilon$) and viscous diffusion ($\mathcal{D}_{\nu}/\varepsilon$) versus (a) $r$ and (b) $r/\lambda$, during the decay of turbulence generated by  RG60 at  (\protect\raisebox{-0.5ex}{\SmallCircle}) $x/x_{\mathrm{peak}} = 8.5$, (\protect\raisebox{-0.5ex}{\SmallSquare}) $x/x_{\mathrm{peak}} = 11.5$, ($\medtriangleright$) $x/x_{\mathrm{peak}} = 16.6$, (\protect\raisebox{-0.5ex}{\Diamondshape}) $x/x_{\mathrm{peak}} = 17.6$  and  (\ding{73}) $x/x_{\mathrm{peak}} = 21$.}% The size of the error bars in the energy transfer data are discussed in \S \ref{sec:DAQ}.}
\label{fig:DownEvoRG60}
\end{figure}

\begin{figure}
\centering
\begin{lpic}[b(0mm)]{Thesis_RG115KHMnormyEpsnormxNone(130mm)}
\lbl{5,130;(a)}
\lbl{80,92;$-\Pi/\varepsilon$}
\lbl{75,28;$\mathcal{D}_{\nu}/\varepsilon$}
\lbl{140,105;$-\mathcal{A}/\varepsilon$}
\lbl{70,132;$(\mathcal{D}_{\nu} - \Pi - \mathcal{A})/\varepsilon$}
\lbl[W]{4,80,90;\hspace{10mm}($\mathcal{A}$, $\Pi$, $\mathcal{D}_{\nu}$)/$\varepsilon$\hspace{10mm}}
\lbl[W]{90,4;\hspace{10mm}$r$ (mm)\hspace{10mm}}
\end{lpic}
\begin{lpic}[b(0mm)]{Thesis_RG115KHMnormyEpsnormxLamb(130mm)}
\lbl{5,130;(b)}
\lbl{80,95;$-\Pi/\varepsilon$}
\lbl{75,25;$\mathcal{D}_{\nu}/\varepsilon$}
\lbl{140,105;$-\mathcal{A}/\varepsilon$}
\lbl{70,132;$(\mathcal{D}_{\nu} - \Pi - \mathcal{A})/\varepsilon$}
\lbl[W]{4,80,90;\hspace{10mm}($\mathcal{A}$, $\Pi$, $\mathcal{D}_{\nu}$)/$\varepsilon$\hspace{10mm}}
\lbl[W]{90,4;\hspace{10mm}$r/\lambda$ \hspace{10mm}}
\end{lpic}
\caption[$\Pi/\varepsilon$, $\mathcal{A}/\varepsilon$ and $\mathcal{D}_{\nu}/\varepsilon$ throughout the decay of RG115-generated turbulence]{Normalised, spherical shell averaged scale-by-scale energy transfer ($-\Pi/\varepsilon$), advection ($-\mathcal{A}/\varepsilon$) and viscous diffusion ($\mathcal{D}_{\nu}/\varepsilon$) versus (a) $r$ and (b) $r/\lambda$, during the decay of turbulence generated by  RG115 at  (\protect\raisebox{-0.5ex}{\SmallCircle}) $x/x_{\mathrm{peak}} = 1.5$, (\protect\raisebox{-0.5ex}{\SmallSquare}) $x/x_{\mathrm{peak}} = 2.0$, ($\medtriangleright$) $x/x_{\mathrm{peak}} = 2.6$, (\protect\raisebox{-0.5ex}{\Diamondshape}) $x/x_{\mathrm{peak}} = 3.1$ and  (\ding{73}) $x/x_{\mathrm{peak}} = 3.7$.}% The size of the error bars in the energy transfer data are discussed in \S \ref{sec:DAQ}.}
\label{fig:DownEvoRG115}
\end{figure}

\begin{figure}[t]
\centering
\begin{lpic}[b(0mm)]{Thesis_RG115KHMnormyq3normxLamb(130mm)}
\lbl{70,85;$-\Pi\,L_{11}^{(1)}/(\overline{q^2})^{3/2}$}
\lbl{85,25;$\mathcal{D}_{\nu}\,L_{11}^{(1)}/(\overline{q^2})^{3/2}$}
\lbl{147,90;$-\mathcal{A}\,L_{11}^{(1)}/(\overline{q^2})^{3/2}$}
%\lbl{65,133;$(\mathcal{D}_{\nu} - \Pi - \mathcal{A})\,L_{11}^{(1)}/(\overline{q^2})^{3/2}$}
\lbl[W]{5,80,90;\hspace{15mm}($\mathcal{A}$, $\Pi$, $\mathcal{D}_{\nu}$)$\,L_{11}^{(1)}/(\overline{q^2})^{3/2}$\hspace{15mm}}
\lbl[W]{90,4;\hspace{10mm}$r/\lambda$ \hspace{10mm}}
\end{lpic}
\caption[$\Pi \,L_{11}^{(1)}/(\overline{q^2})^{3/2}$, $\mathcal{A} \,L_{11}^{(1)}/(\overline{q^2})^{3/2}$ and $\mathcal{D}_{\nu}\,L_{11}^{(1)}/(\overline{q^2})^{3/2}$ throughout the decay of RG115-generated turbulence]{Normalised, spherical shell averaged scale-by-scale energy transfer ($-\Pi \,L_{11}^{(1)}/(\overline{q^2})^{3/2}$), advection ($-\mathcal{A}\,L_{11}^{(1)}/(\overline{q^2})^{3/2}$) and viscous diffusion ($\mathcal{D}_{\nu} \,L_{11}^{(1)}/(\overline{q^2})^{3/2}$) versus $r/\lambda$, during the decay of turbulence generated by  RG115 at  (\protect\raisebox{-0.5ex}{\SmallCircle}) $x/x_{\mathrm{peak}} = 1.5$, (\protect\raisebox{-0.5ex}{\SmallSquare}) $x/x_{\mathrm{peak}} = 2.0$, ($\medtriangleright$) $x/x_{\mathrm{peak}} = 2.6$, (\protect\raisebox{-0.5ex}{\Diamondshape}) $x/x_{\mathrm{peak}} = 3.1$ and  (\ding{73}) $x/x_{\mathrm{peak}} = 3.7$.}
\label{fig:DownEvoRG115b}
\end{figure}

Starting with the RG60 data, the downstream decay/evolution of the scale-by-scale viscous diffusion, energy transport and advection normalised by the dissipation are shown in figure \ref{fig:DownEvoRG60}a. 
As the turbulence decays these terms seem to move to the right reflecting the increase in the turbulent scales. 
Normalising the abscissae by $\lambda$ seems to account for much of the spread (figure \ref{fig:DownEvoRG60}b).  
% although it may be expected that for the viscous term (and perhaps some of the remaining terms for $r\rightarrow 0$), normalising the abscissae by $\eta$ may improve the collapse and similarly, for large $r$, the better scaling variable may be the integral-length scale (see discussion in ch. \ref{chp:5}). 
The scaling of the abscissae is, however, secondary to the main discussion here which pertains to the relative magnitude of the advection, the energy transfer, the viscous diffusion and the dissipation. 
Of particular importance is the observation that the maximum absolute value of the energy transfer $\Pi |_{\mathrm{max}}$ is roughly a constant fraction of the dissipation throughout the downstream extent of the data corresponding to a range of local Reynolds numbers $Re_{\lambda}$ between 91 and 71 ($-\Pi |_{\mathrm{max}}\approx 0.55 \varepsilon$ with the peak located at $r\approx\lambda$, see figure \ref{fig:DownEvoRG60}b).

\begin{figure}
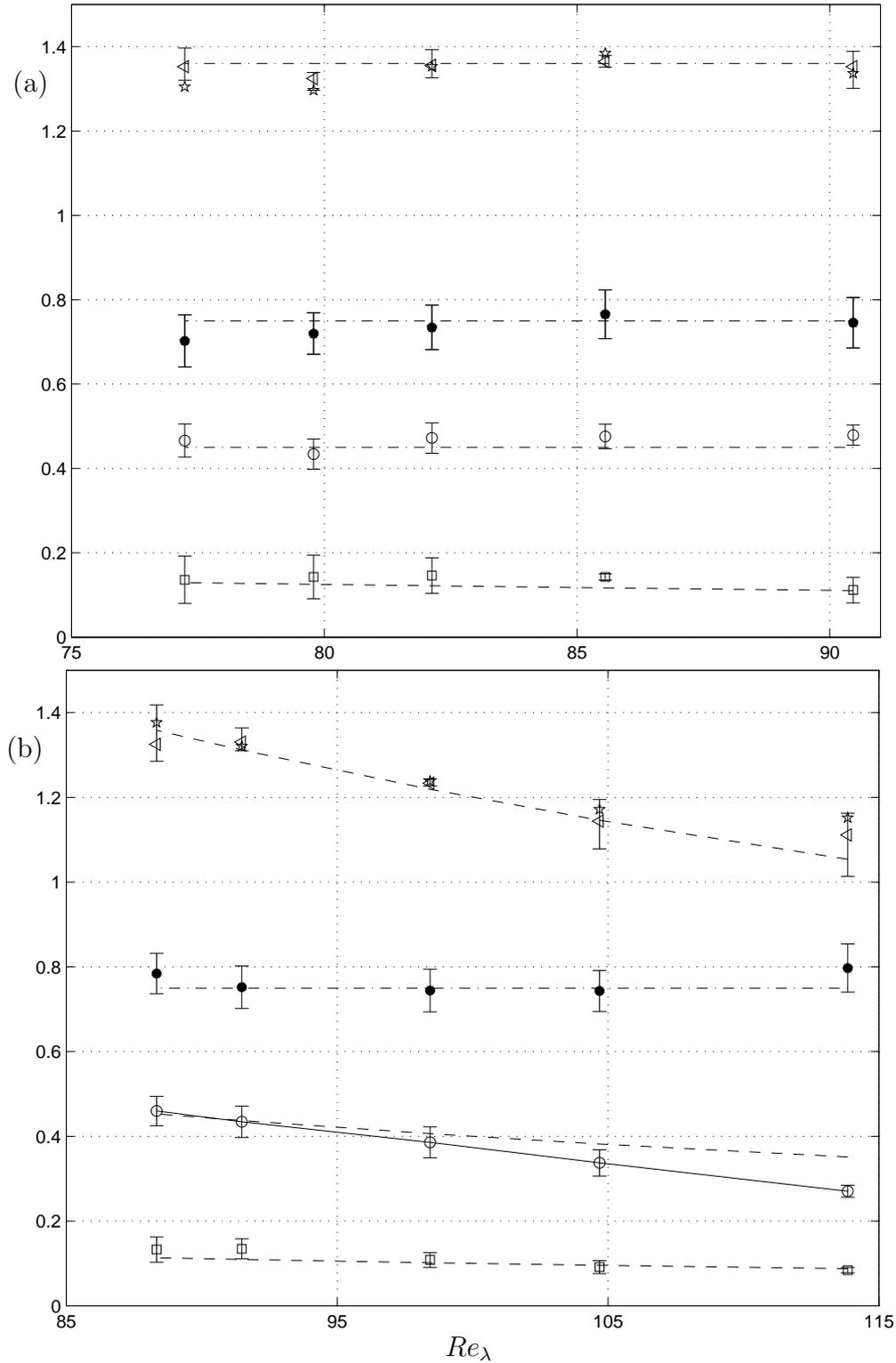

\centering
\begin{lpic}[b(-5mm)]{Thesis_RG60TermsNormTurnover(125mm)}
\lbl{-2,120;(a)}
\end{lpic}
\begin{lpic}[b(-1mm)]{Thesis_RG115TermsNormTurnover(128mm)}
\lbl{-2,120;(b)}
\lbl[W]{85,3;$Re_{\lambda}$}
\end{lpic}
\caption[$C_{\varepsilon}^{1(1)}$,   $C_{\Pi}^{1(1)}$,  $-\mathcal{A}|_{\mathrm{max}(\Pi)}L_{11}^{(1)}/(\overline{q^2}/3)^{3/2}$ and $\mathcal{D}_{\nu}|_{\mathrm{max}(\Pi)}L_{11}^{(1)}/(\overline{q^2}/3)^{3/2}$ versus $Re_{\lambda}$ during the decay of turbulence generated by RG60 and RG115]{Normalised energy dissipation, maximum scale-by-scale energy transfer and scale-by-scale advection and viscous diffusion at the maximum energy transfer versus $Re_{\lambda}$ during the decay of turbulence generated by (a) RG60 and (b) RG115;
 ($\medtriangleleft$) $C_{\varepsilon}^{1(1)}$,   (\protect\raisebox{-0.5ex}{\FilledSmallCircle}) $C_{\Pi}^{1(1)} \equiv  -\Pi |_{\mathrm{max}}L_{11}^{(1)}/(\overline{q^2}/3)^{3/2}$, (\protect\raisebox{-0.5ex}{\SmallCircle}) $-\mathcal{A}|_{\mathrm{max}(\Pi)}L_{11}^{(1)}/(\overline{q^2}/3)^{3/2}$,  (\protect\raisebox{-0.5ex}{\SmallSquare}) $\mathcal{D}_{\nu}|_{\mathrm{max}(\Pi)}L_{11}^{(1)}/(\overline{q^2}/3)^{3/2}$ and (\small{\ding{73}}\normalsize) $(\mathcal{D}_{\nu}|_{\mathrm{max}(\Pi)}\, - \,\Pi |_{\mathrm{max}}\,-\,\mathcal{A}|_{\mathrm{max}(\Pi)})L_{11}^{(1)}/(\overline{q^2}/3)^{3/2}$. Dash and dash-dot lines follow $\sim {Re}_{\lambda}^{-1}$ and $\sim {Re}_{\lambda}^{0}$, respectively.}
\label{fig:PeakValues}
\end{figure}

In fact, taking the numerical values of $\Pi |_{\mathrm{max}}$ and the numerical values of the advection at the separation $r$ where $\Pi(r) = \Pi |_{\mathrm{max}}$ and normalising the data with $(\overline{q^2}/3)^{3/2}/L_{11}^{(1)}$ it is clear that $-\mathcal{A}|_{\mathrm{max}(\Pi)}L_{11}^{(1)}/(\overline{q^2}/3)^{3/2}\sim C_{\Pi}^{1(1)}\left(\equiv - \Pi |_{\mathrm{max}}L_{11}^{(1)}/(\overline{q^2}/3)^{3/2}\right) \sim C_{\varepsilon}^{1(1)}\approx \mathrm{constant}$  (figure \ref{fig:PeakValues}a).
The viscous diffusion term, $\mathcal{D}_{\nu}|_{\mathrm{max}(\Pi)}$ is smaller than any of the other terms at this moderate $Re_{\lambda}$ ($<10\%$ of the dissipation) and it is difficult to discern whether $\mathcal{D}_{\nu}|_{\mathrm{max}(\Pi)}$ is constant or decreases with increasing $Re_{\lambda}$ as one might expect. 

Turning to the RG115 data presented in figure \ref{fig:DownEvoRG115}a two outstanding differences in the downstream evolution of these quantities can be registered: (i) the peak value of the energy transfer does not scale with the dissipation and (ii) the curves representing the advection term are moving from right to left, in the opposite direction than was the case for the RG60 data (figure \ref{fig:DownEvoRG60}a).
Normalising the abscissae with $\lambda$ takes into account most of the spread in the viscous diffusion term but now augments the spread of the advection term (see figure \ref{fig:DownEvoRG115}b and compare with  figure \ref{fig:DownEvoRG60}b).
(Note that for the RG115 data in this region, $L_{11}^{(1)}\sim \lambda$ as shown in ch. \ref{chp:4}, hence the normalisation of the abscissae with $L_{11}^{(1)}$ would yield an identical horizontal collapse as that presented in figure \ref{fig:DownEvoRG115}b). 
Concerning the scaling of the ordinates, it should be noted that, if instead of $\varepsilon$ one chooses to normalise the ordinates by $(\overline{q^2}/3)^{3/2}/L_{11}^{(1)}$ (figure \ref{fig:DownEvoRG115b}) the vertical spread of the energy transfer data is much reduced, but the spread of the advection is further augmented (as is the spread of the viscous diffusion term, since in the limit $r\rightarrow 0$ this term is equal to the dissipation).  

The procedure of normalising $\varepsilon$, $\Pi |_{\mathrm{max}}$, $\mathcal{A}|_{\mathrm{max}(\Pi)}$ and $\mathcal{D}_{\nu}|_{\mathrm{max}(\Pi)}$ with   $(\overline{q^2}/3)^{3/2}/L_{11}^{(1)}$ is repeated and the data are plotted in figure \ref{fig:PeakValues}b against $Re_{\lambda}$. 
Outstandingly, even though the dissipation follows $C_{\varepsilon}^{1(1)} = f(Re_M)/Re_{\lambda}$ in this region, as already reported in ch. \ref{chp:4}, it is clear that the behaviour of $C_{\Pi}^{1(1)}$ is strikingly different. 
In fact, $C_{\Pi}^{1(1)}$ is approximately constant and with the same numerical value ($C_{\Pi}^{1(1)}\approx 0.75$) than that reported for RG60 data in a region where $C_{\varepsilon}^{1(1)}$ is approximately constant.
On the other hand, the normalised advection term grows faster than $Re_{\lambda}^{-1}$ with decreasing $Re_{\lambda}$ and therefore adapts to cover most of the growing difference between the constant  $C_{\Pi}^{1(1)}$ and the increasing $C_{\varepsilon}^{1(1)}$ as the flow decays and $Re_{\lambda}$ decreases. 
The viscous diffusion term $\mathcal{D}_{\nu}|_{\mathrm{max}(\Pi)}$ is also small for the present data, similar to what is  found for the RG60 data.

\subsection{Discussion of the finite Reynolds number effects}
One of the main results attributed to Kolmogorov is the so-called `four-fifths law' which relates the energy transfer and the dissipation as $\partial \overline{(\delta u_{\parallel})^3}/\partial r = -4/5\,\varepsilon$ in the `inertial subrange' for very large Reynolds numbers \cite[]{K41c}.
This `law' is allegedly both general, in the sense that it holds regardless of the global homogeneity and isotropy of the turbulent flow, and exact in the sense that it is derived directly from the Navier-Stokes \cite[]{Frisch:book}. 
Note that the pre-factor of $4/5$ stems from the kinematic relation between the general form of the transfer term $\Pi^*$, see \eqref{eq:KHM}, and $\partial \overline{(\delta u_{\parallel})^3}/\partial r$ in the context of local isotropy, which in Kolmogorov's framework is achieved for sufficiently small $r$ in the very high Reynolds number limit regardless of the global homogeneity and isotropy of the turbulent flow.
The validity of the assumption of local isotropy is a research topic in itself and is secondary to the present discussion.
What does concern the present work is the validity of the general form of the `four-fifths law',
\begin{equation}
-\Pi(\mathbf{X},r) = \varepsilon(\mathbf{X}), 
\label{eq:PiEps}
\end{equation}
 i.e. derived without recurring to local isotropy\footnote{Note that local homogeneity is used to replace $\varepsilon^*(\mathbf{X},r)\equiv (\varepsilon(\mathbf{X}+\mathbf{r})+\varepsilon(\mathbf{X}-\mathbf{r}))/2$ with $\varepsilon(\mathbf{X})$. This is thought to be a good approximation for small $r$ since the numerical value of the dissipation changes slowly with the spatial location, see figure \ref{fig:ReyStress}d and note that in the case of RG115 the maximum value of the energy transfer occurs for $r\simeq 10$mm which corresponds to $y/M \simeq 0.05$. Note that \cite{NT99} negotiate the problem of local homogeneity by integrating over $\mathbf{X}$.} by averaging the quantities over spherical shells \cite[as was done by ][]{NT99}. 
 
Nevertheless, since measurements of $\overline{(\delta u_{\parallel})^3}$ are readily available with a SW whereas the measurements needed to estimate $\Pi$ are much more intricate, the overwhelming majority of the experimental measurements have been made to investigate the validity of Kolmogorov's four-fifths law and not the less stringent relation given by \eqref{eq:PiEps}. 
The result from these investigations is that at moderate to high Reynolds numbers, as those typically encountered in laboratory flows and engineering applications (say $\mathcal{O}(10^2) < Re_{\lambda} < \mathcal{O}(10^3)$), there are significant departures from the  four-fifths law, which are commonly labelled finite Reynolds number (FRN) effects, \cite[]{Qian99,Antonia2006,Bos2007,Cambon2012}. 
The data from experiments and simulations compiled by \cite{Antonia2006} (see also \citealt{Cambon2012}) illustrates that the approach to the four-fifths law with increasing $Re_{\lambda}$ is slow and becomes approximately valid only beyond $Re_{\lambda}\approx \mathcal{O}(10^3)$  for turbulent flows that are stationary and even higher, $Re_{\lambda}\approx \mathcal{O}(10^4)$ for spatial and/or time evolving turbulent flows. 
The dominant FRN effects are commonly attributed to the overlap between large and small scales which causes a non-negligible advection for $r\ll L_{11}^{(1)}$ and a non-negligible viscous diffusion for $r\gg \eta$; both contributions are thought to asymptotically vanish with increasing $Re_{\lambda}$ thus rendering \eqref{eq:PiEps} asymptotically exact. If the turbulence also becomes asymptotically isotropic for $r\ll L_{11}^{(1)}$ the four-fifths law is also recovered.

From data presented in figures \ref{fig:DownEvoRG60}, \ref{fig:DownEvoRG115} and \ref{fig:PeakValues} it can be seen that $\mathcal{D}_{\nu}|_{\mathrm{max}(\Pi)}$ is small compared to $\Pi |_{\mathrm{max}}$ and $\mathcal{A}|_{\mathrm{max}(\Pi)}$ and therefore the main FRN effect relates to small-scale advection.
In fact, if the turbulent flow was not decaying, the advection term would be identically zero and $\Pi |_{\mathrm{max}}$ would be within $10\%$ of the dissipation (since $-\mathcal{D}_{\nu}|_{\mathrm{max}(\Pi)}\lessapprox 10\%\,\varepsilon$).
This is not far from the DNS results for stationary turbulence included in \cite{Antonia2006}, see their figure 5.
For $100 \lesssim Re_{\lambda}^{\mathrm{iso}} \lesssim 150$ they show the peak value of the triple longitudinal structure function divided by $-4/5r$ to be within $10\%$ to $20\%$ of the dissipation.
Note that one should not necessarily expect the maximum value of $\overline{\delta u_{\parallel}\delta q^2}/r$ to be equal to $\Pi |_{\mathrm{max}}$, since the two are only equal if $\Pi$ reaches a plateau at very high $Re_{\lambda}$ \cite[see e.g.][]{K41c}.

It can also be advantageous to work with  $\Pi |_{\mathrm{max}}$ for the sake of comparison with data in wavenumber space\footnote{Nowadays the majority of numerical simulations of homogeneous turbulence are performed with spectral methods.} as it is straightforward to show that for homogenous turbulence $\Pi |_{\mathrm{max}}$ %for a homogeneous flow\footnote{For an inhomogeneous flow it may still be possible to cast the scale-by-scale energy transfer budget \eqref{eq:KHM} in spectral space as has been suggested by, e.g., \cite{Deissler61,Deissler81}.}
 is equal to its spectral space counterpart, $\Pi_K |_{\mathrm{max}}=\Pi |_{\mathrm{max}}$ ($\Pi_K \equiv \int_0^K\! T(k)\,dk$, where $T(k)$ is the spherical averaged non-linear spectral transfer term, see e.g. \citealt{Frisch:book}). 
This can be seen from (6.17) of \cite{Frisch:book}, noting that $\boldsymbol{\nabla_{\ell}}\cdot(\boldsymbol{\ell}/\ell^2 \,\,\Pi(\ell)\,)|_{\Pi(\ell) = \Pi |_{\mathrm{max}}} =  \Pi |_{\mathrm{max}} /\ell^2$ and that $\int_{\mathbb{R}^3} d\ell^3\, \mathrm{sin}(K\ell)/\ell^3 = 2\pi^2$ (using the book's notation and defining $\Pi(\ell) \equiv \boldsymbol{\nabla_\ell} \cdot \left<|\mathbf{\delta u}(\boldsymbol{\ell})|^2\mathbf{\delta u}(\boldsymbol{\ell})\right>/4$ and $\Pi |_{\mathrm{max}} \equiv \mathrm{max}\,(|\Pi(\ell)|)$).
It is less trivial to establish an equivalent relationship for $\overline{\delta u_{\parallel}\delta q^2}/r$ or $\overline{(\delta u_{\parallel})^3}/r$ and their spectral space counterparts \cite[see discussion in \S IV of][]{Cambon2012}. \\

It is very important to note that the growing relevance of the small-scale advection, as the nonclassical dissipation turbulence decays, is not to be confused with the FRN effects discussed by \cite{Qian99,Antonia2006,Cambon2012}. 
These particular FRN effects vary slowly with $Re_{\lambda}$ and for the Reynolds number range straddled in the present experiments the variations in the FRN effects can be considered negligible.
The RG60 and RG115 data presented in figures \ref{fig:PeakValues}a,b are an excellent case in point. 
Even though the range of Reynolds numbers for both datasets are large enough to clearly distinguish between $C_{\varepsilon}^{1(1)}\sim \mathrm{constant}$ and $C_{\varepsilon}^{1(1)}\sim 1/Re_{\lambda}$, they are nevertheless  small enough for $\mathcal{A}|_{\mathrm{max}(\Pi)}L_{11}^{(1)}/(\overline{q^2}/3)^{3/2} \approx \mathrm{constant}$ to be a good approximation  in the RG60 data. 
This would not have been possible if the effect of having a finite Reynolds number changed drastically with small variations in $Re_{\lambda}$.
Furthermore, the $C_{\varepsilon}^{1(1)}$ data presented in ch. \ref{chp:4} for higher Reynolds numbers ($Re_{\lambda}^{\mathrm{iso}}=\mathcal{O}(400)$ -- whereas for the present $2\times$XW data $Re_{\lambda}=\mathcal{O}(100)$)  suggest that the nonclassical dissipation behaviour persists with increasing Reynolds numbers.
In fact, $C_{\varepsilon}^{1(1)}$ is well fitted by $C_{\varepsilon}^{1(1)}\sim Re_M/Re_{L^{1(1)}}$ which follows from the assumption that $C_{\varepsilon}^{1(1)}$ is independent of $\nu$. 
FRN effects are such that their influence diminishes as Reynolds number increases and therefore not compatible with the functional form $C_{\varepsilon}^{1(1)}\sim Re_M/Re_{L^{1(1)}}$.

Nevertheless, one cannot entirely exclude the possibility that the functional form  $C_{\varepsilon}^{1(1)}\sim Re_M/Re_{L^{1(1)}}$, obtained with data at  Reynolds numbers up to $Re_{\lambda}^{\mathrm{iso}}=\mathcal{O}(400)$, may be no more than a rough approximation valid for the range of $Re_M$ values attainable until now. Consequently, the conclusions drawn here may not extrapolate to Reynolds numbers  orders of magnitude higher. This issue is further discussed in the concluding remarks of the thesis (ch. \ref{chp:7}).

\section{Nonequilibrium turbulence}

It has been shown that in the region of the decaying grid-generated turbulence where the dissipation follows a nonclassical behaviour of the type $C_{\varepsilon}^{1(1)}\sim f(Re_M)/Re_{\lambda}$ the behaviour of the peak energy transfer follows $\Pi |_{\mathrm{max}}\sim (\overline{q^2}/3)^{3/2}/L_{11}^{(1)}$ (i.e. $C_{\Pi}^{1(1)}\sim \mathrm{constant}$). 
This finding is both intriguing and enlightening. 
On the one hand, the fact that $C_{\Pi}^{1(1)}\sim \mathrm{constant}$ reinforces the classical arguments suggesting that the energy transfer should scale as $\Pi \sim (\overline{\delta q^2})^{3/2}/r$ \cite[see][ where it is shown that $C_{\Pi}\approx \mathrm{constant}$ even for very low $Re_{\lambda}$ decaying turbulence]{McComb2010}. In the same way, it gives support to the choice of $L_{11}^{(1)}$ as the characteristic length-scale (see discussion in \S \ref{sec:Lu}). 
On the other hand, the observation in ch. \ref{chp:4} that the nonclassical dissipation behaviour may persist at high Reynolds number in the form $C_{\varepsilon}^{1(1)}\sim {Re}^{1/2}_M/Re_{\lambda}$ (i.e. a function of the downstream location but independent of $\nu$) together with the observation that $C_{\Pi}^{1(1)}\sim \mathrm{constant}$ directly contradicts the four-fifths law (or generally $\Pi |_{\mathrm{max}}= -\varepsilon$).
Nevertheless, the present data do not allow the investigation of the behaviour at the very high Reynolds number limit, which is left for future research. 

What the data do allow the author to conclude is that at moderately high Reynolds numbers the dissipation and the peak energy transfer scale differently leading to an imbalance reflected in the small-scale advection. 
One may chose to denote this behaviour as nonequilibrium turbulence because the energy that is transferred to the small-scales is not in equilibrium with their dissipation. 
On the other hand, for the RG60 data beyond $x/x_{\mathrm{peak}}\approx 8$ it has been observed that  $\varepsilon \sim \Pi \sim \mathcal{A}|_{\mathrm{max}(\Pi)} \sim (\overline{q^2}/3)^{3/2}/L_{11}^{(1)}$ which suggests an equilibrium\footnote{At least to a first approximation, since the FRN effects are expected to vary, albeit slowly, with $Re_{\lambda}$.} between the energy that is transferred to the small scales, their advection and dissipation. 
Note that in a decaying flow at moderately high Reynolds numbers the flux of energy to small scales is always smaller than the dissipation, which reflects the change in the dissipation region of the velocity spectra as the flow decays (this is the physical meaning of the  wavenumber counterpart of the advection term).\\
 
%
%Considering, for example, \cite{Lumley92}, \cite{Frisch:book} and \cite{McComb2010} as representative of the literature reflecting the current understanding of the behaviour of the energy transfer and dissipation (and the modern articulation of Kolmogorov's ideas), one finds in the present results some pronounced departures. 
%For example, in \cite{Frisch:book} the starting point of the phenomenological analysis leading to the two-thirds power-law 2$^{\mathrm{nd}}$-order structure functions are actually the scaling arguments suggesting that $\Pi \sim (\overline{\delta q^2})^{3/2}/r$. 
%Then use is made of the four-fifths law to establish that $\varepsilon \sim \Pi \sim (\overline{\delta q^2})^{3/2}/r$ which is readily recognised if re-written as $\overline{\delta q^2} \sim \varepsilon^{2/3} r^{2/3}$. 
%
%The results of this chapter together with those presented in ch. \ref{chp:5} clearly show that even if $\varepsilon \nsim \Pi$ one may still have a Kolmogorov-Obukhov spectra of the type $F_{11}^{(1)}=C_K \varepsilon^{2/3} k^{-5/3}$ (which is the wavenumber counterpart of $\overline{\delta q^2} \sim \varepsilon^{2/3} r^{2/3}$) and that surprisingly $C_K$ takes values similar to those found in very many different turbulent flows \cite[]{Sreeni95b}. 
%This is strong evidence that the arguments presented by, for example, \cite{Lumley92} and \cite{Frisch:book} have to be revised, because even if in the very high Reynolds number limit they are self-consistent, for moderately high Reynolds numbers they are incorrect.
%

\section{Summary}
In this chapter a scale-by-scale energy transfer budget, based on an inhomogeneous  K\'{a}rm\'{a}n-Howarth-Monin equation, is experimentally assessed along the centreline of decaying turbulence with classical and nonclassical dissipation scalings.  
It is experimentally verified that, even though the turbulence with a nonclassical dissipation behaviour occurs in a streamwise region where there are non-negligible contributions from turbulent transport and production to the single-point kinetic energy balance (see ch. \ref{chp:3}), these terms are nevertheless negligible for small scales ($r\lessapprox L_{11}^{(1)}/3$) and do not affect the peak of the inter-scale energy transfer which occurs around $r\approx L_{11}^{(1)}/8$.
It is also confirmed that the inter-scale energy transfer due to the residual mean shear is negligible.
However, the most important result presented in this chapter is the fact that the inter-scale energy transfer always scales with $(\overline{q^2}/3)^{3/2}/L_{11}^{(1)}$, regardless whether the dissipation scales with $(\overline{q^2}/3)^{3/2}/L_{11}^{(1)}$ (i.e. $C_{\varepsilon}^{1(1)}\sim \mathrm{constant}$ -- the classical dissipation behaviour) or with  $U_{\infty}M\,\overline{q^2}/\left(L_{11}^{(1)}\right)^2$ (i.e. $C_{\varepsilon}^{1(1)}\sim Re_M/Re_{L^{1(1)}}$ -- the nonclassical dissipation behaviour). 
When the dissipation behaves in a nonclassical way, the fact that the inter-scale energy transfer scales differently leads to an imbalance between the two  and, as a result, the small-scale advection becomes an increasing proportion of the dissipation. 
Turbulence exhibiting this imbalance is denoted nonequilibrium turbulence.

\clearemptydoublepage
\chapter{Conclusion}
\label{chp:7}

In this thesis the nonclassical dissipation behaviour previously reported to occur in decaying turbulence in the lee of fractal square grids by \cite{SV2007} and \cite{MV2010}, i.e. $C_{\varepsilon}^{1(1)}\sim f(Re_M)/Re_{\lambda}$, is experimentally investigated in an attempt to advance the current understanding on this new and unexpected phenomenon. 
The main results of this investigation are summarised here, conveying their implications to the established concepts of turbulence phenomena and consequently on the current engineering models of turbulence.

% Describe the ceteris paribus experiments
The first step in this investigation is to confirm the previous results and to design a \emph{ceteris paribus} comparison with the classical benchmark of square-mesh regular grid-generated turbulence. 
Outstandingly, it is found that the nonclassical dissipation behaviour exhibited by the decaying FSG-generated turbulence along the centreline is also manifested in the lee of RGs for a comparable region along the centreline whose streamwise extent lies between the location of the turbulent kinetic energy peak, i.e. $x=x_{\mathrm{peak}}$, and its first few multiples. 
For one RG, where $x_{\mathrm{peak}}$ is small compared to the streamwise extent of the tunnel, the cross-over between the nonclassical and the classical (i.e. $C_{\varepsilon}^{1(1)}\sim \mathrm{constant}$) dissipation scalings is determined to be $x\approx 5x_{\mathrm{peak}}$. 
The finding that the nonclassical behaviour is not exceptional to the very special class of inflow conditions defined by FSGs renders this nonclassical behaviour of general scientific and engineering significance and therefore of much greater importance. 
On the other hand, it is surprising that this nonclassical behaviour has been overlooked in such a widely investigated turbulent flow such as square-mesh grid-generated turbulence.
It is speculated that this is due to the fact that $x_{\mathrm{peak}}$, and therefore the extent of this region, is small for the typical grids investigated in the literature (see \S \ref{sec:survey}), whereas owing to the chosen design of the FSGs the nonclassical region extends beyond the streamwise extent of typically sized laboratory wind tunnels.
Note that it would be misleading to denote this behaviour as merely `transient' since decaying turbulence is in itself a transient phenomenon and this region can be made as long as desired via the geometry of the grid.
Moreover, this region is permanently present in a particular region of space downstream of the grid and, in that sense,
clearly not transient.
%Homogeneity and influence from walls
From the comparison between turbulence generated by RGs and FSGs it is also observed how the grid geometry can influence the profiles of single-point statistics, including turbulence production and transport, in the region beyond $x_{\mathrm{peak}}$ where the wakes of the different bars have interacted and the turbulence is decaying. 
This reveals the potential of FSGs as passive flow controllers. 
Nevertheless, it is also shown how the confining wind tunnel walls can tamper with the turbulence statistics in this region of the flow. 
This  is likely attributable to the influence the walls have in the wake-interaction mechanism whenever the mesh-size is too large. 
It is therefore suggested that the mesh-size for both RGs and FSGs never exceeds 1/4 of the tunnel's width (such as is the case for the RG115 and RG60 grids studied here) in subsequent experimental investigations.

%Decay
In any case, the power-law fits to $\overline{u^2}$ along the centreline in the nonclassical dissipation region of both RG- and FSG-generated turbulence ($1\lessapprox x/x_{\mathrm{peak}}\lessapprox 4$) return much larger decay exponents, $2.4\lessapprox n \lessapprox 3.0$, than the usual exponents found in the literature, $1.0\lessapprox n \lessapprox 1.5$. 
The larger decay exponents of FSG-generated turbulence had previously been reported by \cite{MV2010}, which updated the exponential fit of \cite{HV2007,SV2007}.
Here different power-law fitting methods, fitting data obtained over a larger streamwise extent are used to provide a more accurate estimate of the exponents.
Note that, even though the homogenous kinetic energy balance between advection and dissipation is not satisfied for present flows in this region, for the FSG18''x18'' data the advection is approximately proportional to the dissipation along the centreline. 
Given the approximate proportionality between advection and dissipation it is argued that the power-law exponent of the decaying turbulence along the centreline can be compared with the values for homogeneous freely decaying turbulence.
It is also shown how the functional form of $C_{\varepsilon}^{1(1)}$ during the turbulence decay directly influence the decay exponent if a large-scale quantity is conserved. 
In particular, the \cite{Saffman67} and the \cite{Loitsyansky} invariants respectively lead to $n=3/2$ and $n=5/2$ whenever $C_{\varepsilon}^{1(1)}\sim f(Re_{M})/Re_{\lambda}$ instead of the classical prediction of $n=6/5$ and $n=10/7$ (taking $C_{\varepsilon}^{1(1)}\sim \mathrm{constant}$). 
The decay exponent measured for the FSG18''x18'' centreline data, $n\approx 2.5$, is consistent with the observation that $C_{\varepsilon}^{1(1)}\sim f(Re_{M})/Re_{\lambda}$  and the (hypothetical) conservation of the Loitsyansky invariant.
Nevertheless, there are no guarantees that such a quantity is indeed conserved in the present decaying turbulent flows and the consistency between the measured exponent and the prediction using the Loitsyansky invariant may be no more than a coincidence.

% small and large scale isotropy and behaviour behind the bar
The nonclassical dissipation region of RG115-generated turbulence  is further investigated using an experimental apparatus with two X-probes which allow the measurement of four components of the mean square velocity gradient tensor ($\overline{\left(\partial u/\partial x\right)^2}$, $\overline{\left(\partial u/\partial y\right)^2}$, $\overline{\left(\partial v/\partial x\right)^2}$, $\overline{\left(\partial v/\partial y\right)^2}$). 
%small scale
These data, acquired along the centreline, indicate that the small-scales do not follow the isotropic relations between the four measured components. 
Nevertheless, these ratios stay approximately constant during the assessed region of the decay, in-line with the findings of  \cite{gomesfernandesetal12} for a FSG similar to FSG18''x18'' and with the present RG60-generated turbulence data acquired for much larger $x/x_{\mathrm{peak}}$. 

% Large scale
Four velocity correlations functions are also measured within the nonclassical dissipation region of RG115-generated turbulence,  both along the centreline and  along the longitudinal line intercepting the lower bar of the grid ($y=- M/2$, $z=0$). 
Four integral-length scales are computed from the measured correlation functions, namely, $L_{11}^{(1)}$, $L_{22}^{(1)}$, $L_{11}^{(2)}$, $L_{22}^{(2)}$ -- see nomenclature. 
These data are used to compute the various ratios $L_{ii}^{(k)}/\lambda$ which directly relate to the behaviour of $C_{\varepsilon}^{i(k)}$, i.e. the dissipation normalised using the different integral-length scales $L_{ii}^{(k)}$ (\emph{cf.} \eqref{eq:CepsB}). 
It is shown that the behaviour of $C_{\varepsilon}^{1(1)}$ behind the bar follows the classical scaling, which is in stark contrast with the nonclassical behaviour along the centreline. 
(Additional FSG3'x3' data indicate that this dichotomy of behaviours along the centreline and behind the bar occurs for both RG- and FSG-generated turbulence.) 
Remarkably, if instead of using $L_{11}^{(1)}$ to normalise the dissipation one chooses $L_{22}^{(1)}$ or $L_{22}^{(2)}$, it is shown that  $C_{\varepsilon}^{2(1)}$ and $C_{\varepsilon}^{2(2)}$ indicate a nonclassical behaviour ($C_{\varepsilon}^{2(1)}\sim C_{\varepsilon}^{2(2)}\sim Re_{\lambda}^{-1}$) both along the centreline and behind the bar. 
Clearly the large eddies become less anisotropic as they decay along the longitudinal line intercepting the bar $(y=- M/2, z=0)$ because $L_{11}^{(1)}/L_{22}^{(1)}$ and $L_{11}^{(1)}/L_{22}^{(2)}$ decrease proportionally to $Re_{\lambda}$  as $Re_{\lambda}$ decreases towards the isotropic benchmark.  
The author is not aware of any other relation such as $L_{11}^{(1)}/L_{22}^{(1)} \sim Re_{\lambda}$ or $L_{11}^{(1)}/L_{22}^{(2)} \sim Re_{\lambda}$ in the literature to describe the large-scale anisotropy's dependence on $Re_{\lambda}$. 
It will be worth revisiting canonical free shear flows such as wakes and jets in future studies because, to the author's knowledge, only measurements of $C_{\varepsilon}^{1(1)}$ have been reported in such flows in support of $C_{\varepsilon}^{1(1)} \sim \mathrm{constant}$ for high enough Reynolds numbers. % \cite[e.g. see][]{Sreeni95,Pearson,Burattini2005}.
It will be interesting to know whether $C_{\varepsilon}^{2(1)}\sim C_{\varepsilon}^{2(2)} \sim Re_{\lambda}^{-1}$  also hold in such flows or whether these relations are only valid in grid-generated turbulence. 
Lastly, normalising the dissipation with the fourth estimated integral scale, i.e. $L_{11}^{(2)}$, it is observed that  $C_{\varepsilon}^{1(2)}$ grows faster than $Re_{\lambda}^{-1}$ with decreasing $Re_{\lambda}$.
No definitive explanation for this behaviour can be given at this point, but as discussed in \S \ref{sec:Lu} it may be related to periodic shedding from the bars which is contaminating the correlation functions, in particular $B_{11}^{(2)}$.

A finding of great importance in this thesis is that the peak of the nonlinear energy transfer to the small-scales  follows $C_{\Pi}^{1(1)}\sim \mathrm{constant}$ regardless whether the dissipation scales as $C_{\varepsilon}^{1(1)}\sim \mathrm{constant}$, i.e. the classical dissipation behaviour, or as   $C_{\varepsilon}^{1(1)}\sim f(Re_M)/Re_{L^{1(1)}}$, the nonclassical behaviour.
Whenever the dissipation behaves in a nonclassical way,  there is an increasing gap (as turbulence decays) between the energy transferred to the small scales and their dissipation  leading to a growth in small-scale advection to cover this increasing gap. 
This phenomenon is denoted here as nonequilibrium turbulence.
Nonequilibrium turbulence contrasts with turbulence in equilibrium where the dissipation, the energy transfer and the small scale advection are all in approximate proportion during decay. 
Note that, even though nonequilibrium turbulence occurs in a streamwise region where there are non-negligible contributions of turbulent transport and production (for the present flows at least), it is demonstrated that these do not tamper with the balance between advection, energy transfer and dissipation for the scales where the energy transfer is at a maximum.

% Ceps independent on viscosity
Another finding of general interest is that high Reynolds number nonequilibrium turbulence appears to be consistent with the expectation that $C_{\varepsilon}$ becomes independent of $\nu$, to a first approximation at least, by assuming the functional form $C_{\varepsilon} \sim Re_M/Re_L \sim {Re}_M^{1/2}/Re_{\lambda}$. This is accompanied by the onset of a $-5/3$ power-law range for the highest Reynolds number data. 
These findings suggest the possibility to extend the Kolmogorov-Obukhov phenomenology to include the nonequilibrium behaviour.  
In this thesis the matched-asymptotic expansion analysis used by \cite{Lundgren2002} is repeated and the analysis leads to the prediction of a power-law of the spectra (over a range such that $k\eta\ll 1$ and $kL_{11}^{(1)}\gg1$) following, $F_{11}^{(1)}= \alpha (\varepsilon^{\mathrm{iso}})^{2/3} k^{-5/3} =  A\, \overline{u^2} L_{11}^{(1)}  \left(kL_{11}^{(1)}\right)^{-5/3}$. 
This result is very similar to the Komogorov-Obukhov spectrum in the inertial subrange with an outstandingly important difference.    
The `constants' of the spectrum, $\alpha$ and $A$, are related by  $C_{\varepsilon}^{1(1)} = (A/\alpha)^{3/2}$
and consequently only one of the `constants' $\alpha$ or $A$ can be invariant during the decay of nonequilibrium turbulence (since the numerical value of $C_{\varepsilon}^{1(1)}$ is changing). 
In other words, the question is really whether the power-law range of the spectra follow a Komogorov-Obukhov spectrum, i.e. $F_{11}^{(1)}= \alpha (\varepsilon^{\mathrm{iso}})^{2/3} k^{-5/3}$ with $\alpha$ invariant during decay or conversely,  $F_{11}^{(1)}= A^* \Pi^{2/3} k^{-5/3}$ with $A*$ invariant instead ($A^*\equiv A/C_{\Pi}$).
The present data are not recorded at sufficiently high Reynolds numbers for the spectra to exhibit $-5/3$ power-law throughout the decay and therefore cannot be used to experimentally determine whether $\alpha$ or $A$  are invariant. This is left for subsequent research. 

Overall, these results bear important consequences for  turbulence research, in particular for the turbulence modelling community. 
Firstly, it has been shown that nonequilibrium turbulence is manifested in a relatively simple flow such as square-mesh grid-generated turbulence. 
It is likely that subsequent research will discover these phenomena in specific regions of other turbulent flows that have been so far neglected. 
Furthermore, one cannot expect the current RANS models (such as two-equation models, Reynolds stress closures, etc.), commonly used in engineering, which have inbuilt the assumption that $C_{\varepsilon}\sim\mathrm{constant}$, to be able to predict the transport of turbulent quantities in nonequilibrium flows. 
Even the more sophisticated LES models, commonly used in engineering and geophysics, have inbuilt the assumption that $C_{\varepsilon}\sim C_{\Pi}\sim\mathrm{constant}$ which was shown to breakdown in nonequilibrium turbulence, at least for the moderately high Reynolds numbers measured.
These modelling techniques may require a profound update if they are expected to cope with flows where a nonequilibrium region is present.

Lastly, the present work raises a fundamental question.
The present data at the highest Reynolds numbers, up to $Re_{\lambda}=\mathcal{O}(400)$, suggest that, in the nonequilibrium region, $C_{\varepsilon}\sim Re_M/Re_{L}$ whereas the $2\times$XW data (although acquired at lower Reynolds numbers, $Re_{\lambda}=\mathcal{O}(100)$) suggest that  $C_{\Pi}\sim\mathrm{constant}$. 
This is fundamentally incompatible with Kolmogorov's four-fifths law (or generally $\Pi |_{\mathrm{max}}=-\varepsilon$). 
Furthermore, in the nonequilibrium region the importance of the small-scale advection relative to the dissipation increases as the turbulence decays and $Re_{\lambda}$ decreases. It is thus reasonable to expect that when the dissipation transitions to the equilibrium behaviour (and owing to the turbulence decay the Reynolds number is smaller than that straddled in the nonequilibrium region), the small-scale advection will be, at least, as important. 
Therefore  it also reasonable to expect that the lack of validity of $\Pi |_{\mathrm{max}}=-\varepsilon$ in the nonequilibrium region extends to the equilibrium region.

It would be of upmost importance to investigate whether:
\begin{enumerate}[(i)]
\item  $C_{\varepsilon}\sim Re_M/Re_{L}$ and  $C_{\Pi}\sim\mathrm{constant}$ hold at Reynolds numbers much higher than those of the present data, implying the very important consequence that $\Pi |_{\mathrm{max}}\neq-\varepsilon$  for decaying flows even at overwhelmingly high Reynolds numbers;
\item $C_{\varepsilon} = C_{\Pi}\sim Re_M/Re_{L}$  implying that $\Pi |_{\mathrm{max}}=-\varepsilon$ is recovered for very high Reynolds numbers, but also implying that $\Pi$ would follow a nonclassical scaling;
\item The nonequilibrium phenomenon asymptotically vanishes for increasingly high $Re_M$, either because $C_{\varepsilon}\sim Re_M/Re_{L}$  is no more than a crude approximation or alternatively because the downstream extent of the nonequilibrium region asymptotically vanishes.
\end{enumerate}
These research questions, and those presented above, deserve further investigation both due to the fundamental implications in turbulence theory and the practical importance in turbulence modelling.

%% add paper as an appendix
%\clearemptydoublepage
%\chapter*{Appendix}
%\clearemptydoublepage
%\pagestyle{plain}
%\begin{figure}[t!]
%\centering
%\includegraphics[width=\textwidth]{chapters/Paper/PaperAppendix1}
%\end{figure}
%\pagestyle{plain}
%\begin{figure}[t!]
%\centering
%\includegraphics[width=\textwidth]{chapters/Paper/PaperAppendix2}
%\end{figure}
%\pagestyle{plain}
%\begin{figure}[t!]
%\centering
%\includegraphics[width=\textwidth]{chapters/Paper/PaperAppendix3}
%\end{figure}
%\pagestyle{plain}
%\begin{figure}[t!]
%\centering
%\includegraphics[width=\textwidth]{chapters/Paper/PaperAppendix4}
%\end{figure}
%\newpage
%${}$
%\pagestyle{plain}
%\begin{figure}[th!]
%\centering
%\includegraphics[width=\textwidth]{chapters/Paper/PaperAppendix5}
%\end{figure}

%% bibliography
\addcontentsline{toc}{chapter}{Bibliography}
\clearemptydoublepage
\bibliography{mybib}

\end{document}